%% file: WH_ME_PRD.tex
\documentclass[aps,prd,showpacs,superscriptaddress,twocolumn,nofootinbib]{revtex4} % UTOBS From STPRD

\usepackage{graphicx}% Include figure files
\usepackage{dcolumn}% Align table columns on decimal point
\usepackage{bm}% bold math
\usepackage{slashed}

\usepackage{subfigure}

\usepackage{natbib}
\def\ttbar {$t{\bar{t}}$}
\def\ppbar {$p{\bar{p}}$}
\def\bbbar {$b{\bar{b}}$}
\def\qqbar {$q{\bar{q}}$}

\def\Wbbbar {$Wb{\bar{b}}$}
\def\Wccbar {$Wc{\bar{c}}$}

\def\Wc {$Wc$}

\def\tev {TeV}
\def\gevcc {GeV/$c^2$}
\def\gevc {GeV/$c$}

\def\fb    {\ensuremath{\mathrm{fb}}}
\def\lumi {\mathcal{L}_\mathrm{int}}
\def\mhz     {\ensuremath{\mathrm{MHz}}}
\def\hz    {\ensuremath{\mathrm{Hz}}}

\def\ET {$E_{\rm T}$}
\def\pt {$p_{\rm T}$}

\def\EtMiss {\slashed{E}_{\rm T}}  

\def\EtMissVec {\vec{\slashed{E}}_{\rm T}}

\newcommand{\kikemet} {\hbox{E\kern-0.5em\lower-0.1ex\hbox{/}}_T}

\begin{document}

% Show the line numbers
%\pagewiselinenumbers

%\preprint{CDF/PUB/EXOTIC/CDFR/10617}
%\preprint{PRD Draft v 1.2: \today}

\title{\boldmath Search for Standard Model Higgs Boson Production in Association with a 
$W$ Boson Using a Matrix Element Technique at CDF in $p{\bar{p}}$
Collisions at $\sqrt{s}$ = 1.96 TeV}
% Force line breaks with \\

%We still use an include here because it would be UGLY otherwise
\input{September2011_Authors}

\date{\today}% It is always \today, today,
             % but any date may be explicitly specified

%-----------------------------
% ABSTRACT
%-----------------------------
%\input{./includes/prd_abstract}
\begin{abstract}
This paper presents a search for standard model Higgs boson production
in association with a $W$ boson using events recorded by the CDF
experiment in a dataset corresponding to an integrated luminosity of
5.6~fb$^{-1}$. The search is performed using a matrix element
technique in which the signal and background hypotheses are used to
create a powerful discriminator.  The discriminant output
distributions for signal and background are fit to the observed events
using a binned likelihood approach to search for the Higgs boson
signal. We find no evidence for a Higgs boson, and 95\% confidence
level (C.L.) upper limits are set on $\sigma(p\bar{p}\rightarrow
WH)\times {\cal{B}}(H\rightarrow b\bar{b})$.  The observed limits
range from $3.5$ to $37.6$ relative to the standard model expectation
for Higgs boson masses between $m_H$~=~100~\gevcc~and
$m_H$~=~150~\gevcc. The 95\% C.L. expected limit is estimated from the
median of an ensemble of simulated experiments and varies between
$2.9$ and $32.7$ relative to the production rate predicted by the
standard model over the Higgs boson mass range studied.
\end{abstract}



\maketitle

%\tableofcontents
%\newpage

%-----------------------------
% INTRODUCTION
%-----------------------------
%\input{./includes/prdIntrod}
%-------------------------------------------------------------------------------
\section{Introduction}
\label{sec:introd}
%\setlength{\footskip}{2cm}
%-------------------------------------------------------------------------------

In the standard model (SM), the Higgs mechanism~\cite{bib:Higgs_mech,
bib:Hagen, bib:Englert} is responsible for the spontaneous breaking of
the SU(2) x U(1) gauge symmetry which generates the masses of the
gauge bosons and more indirectly allows for the fermion masses.  This
theory predicts the existence of a scalar particle, the Higgs boson,
which remains the only SM particle that has not been observed by
experiment. Although the Higgs boson mass is not predicted by theory,
direct searches done at LEP and Tevatron collider experiments have set
limits that constrain the Higgs boson mass to be between 114.4 and
156~\gevcc~or above 175~\gevcc~ at 95\%
C.L.\cite{Barate:2003sz,bib:comb_TEV_result}. On the other hand,
precision electroweak measurements indirectly constrain its mass to be
less than 158~\gevcc~at 95\% C.L.~\cite{bib:LEPgroup}.

At the Tevatron \ppbar~collider, the Higgs boson is expected to be
produced mainly by gluon fusion, while the next most frequent
production channel is the associated production of Higgs and $W$
bosons, $WH$. For Higgs boson masses lower than 135 \gevcc, the Higgs
boson decay $H~\rightarrow$~\bbbar~has the largest branching
fraction~\cite{Djouadi:1997yw}. The production rate of \bbbar~pairs
from QCD processes is many orders of magnitude larger than Higgs boson
production, making the analysis of the process
$gg~\rightarrow~H~\rightarrow$~\bbbar~nonviable. Associated production
\qqbar~$\rightarrow~WH$ with the $W$ boson decaying leptonically gives
a cleaner signal because requiring a lepton helps to distinguish it
from the multijet QCD background~\cite{Han:1991ia}.

  Several searches for a low-mass Higgs boson at the CDF and D0
  experiments are combined in order to maximize
  sensitivity~\cite{bib:comb_TEV_result}. In that combination, the
  search in the $\ell\nu$\bbbar~final state has proven to be the most
  sensitive input and therefore carries the most weight in the
  combination.  So, optimizations in this analysis can have an
  important impact on the ultimate sensitivity of the Tevatron
  experiments to the Higgs boson.

  Recently, the experiments at the Large Hadron Collider (LHC) have
  obtained enough data to produce search results of similar
  sensitivity to the Tevatron experiments in the low mass
  region~\cite{CMS:combo}.  However, at the LHC the most sensitive low
  mass search is in the diphoton final state~\cite{ATLAS:hgg} and
  searches for $H~\rightarrow$~\bbbar~will take some time before they
  reach the sensitivity of the Tevatron combination in this
  channel~\cite{CMS:hbb}.  In that sense, the Tevatron and LHC are
  quite complementary in that both will provide important information
  in the search for a low-mass Higgs boson over the next few years.

In this Letter, we describe a search for the Higgs boson in the final
state where the $H$ is produced in association with a $W$ boson, the
Higgs boson decays to $b\bar{b}$, and the $W$ decays to an electron or
muon and its associated neutrino.  This final state has been
investigated before by both Tevatron experiments, CDF and
D0~\cite{bib:PRLWHCDF, bib:PRLWHD0}.  Here we present a new search in
a data sample corresponding to an integrated luminosity of 5.6
fb$^{-1}$ and using an optimized discriminant output distribution.

Finding evidence for Higgs boson production in association with a $W$
boson is extremely difficult since the expected production rate is
much lower than that of other processes with the same final state, for
example $W$~+~\bbbar~and top quark processes.  Some of the main
challenges of the analysis are the identification and the estimation
of these and other background processes and the development of
strategies to reduce their contribution while retaining high signal
efficiency.

The background processes contributing to the $WH$ final states are
$W$~+~\bbbar, $W$~+~$c\bar c$, \ttbar, single top, $Z$~+~jets,
dibosons ($WW$, $WZ$, and $ZZ$), $W$~+~jets events, where a jet not
originating from a $b$ quark has been misidentified as a heavy flavor
jet, and non-$W$ events where a jet is misidentified as a lepton.
These processes have characteristics which differ from those of $WH$
production that will be used to discriminate them from the signal.
The background rates are estimated from a combination of simulated and
observed events.  To distinguish signal from background events a
matrix element technique~\cite{bib:canelli,bib:brian_mohr} is applied,
in which event probability densities for the signal and background
hypotheses are calculated and used to create a powerful discriminator.
This method was used as part of the observation of single top
production~\cite{Aaltonen:2009jj} and many other analyses within the
CDF collaboration, such as the measurement of the $WW+WZ$ cross
section~\cite{bib:WWWZxsec}, the measurement of the top quark
mass~\cite{bib:top_mass}, the search for SM Higgs boson production in
the $WW$ decay channel ~\cite{Aaltonen:2010cm}, and the measurement of
the $WW$ production cross section~\cite{:2009us}.

This paper is organized as follows. Section~\ref{sec:cdf} briefly
describes the CDF II detector~\cite{bib:cdfII-tdr, bib:cdfII}, the
apparatus used to collect the observed events used in this analysis.
In Section~\ref{sec:dataSample}, the identification of the particles
and observables that make up the $WH$ final state is presented.
Section~\ref{sec:reconstr} describes the event selection. Identifying
\hyphenation{long-lived} $b$ hadrons in jets is essential, and the two
algorithms used to identify $b$ jets are presented in
Section~\ref{sec:btag}.  The signal and background signatures are
discussed in Section~\ref{sec:accept} and~\ref{sec:bkgEstim}
respectively, together with the method to estimate the total number of
events and also the background composition.  The matrix element method
is described in detail in Section~\ref{sec:meMethod}.  A discussion of
systematic uncertainties is included in Section~\ref{sec:systs}.
Finally, in Section~\ref{sec:results} and~\ref{sec:concl} the results
and conclusions of the analysis are presented.

%-----------------------------
% CDF DETECTOR
%-----------------------------
%\input{./includes/prdCDFdetector}
%-------------------------------------------------------------------------------
\section{The CDF II detector}
\label{sec:cdf}
\begin{figure}[h]
\begin{center}
\includegraphics[width=1.\columnwidth,clip=]
        {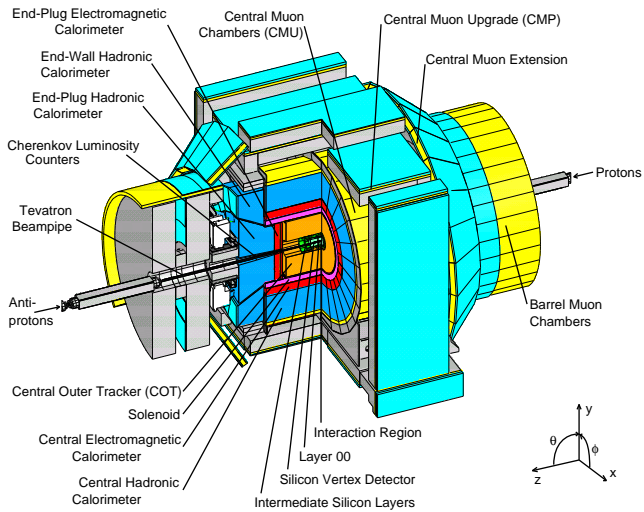}
\caption{A cutaway view of the CDF~II detector with quadrant cut to
expose the different subdetectors.  \label{fig:cdf_detector}}
\end{center}
\end{figure}

%\setlength{\footskip}{2cm}
%-------------------------------------------------------------------------------
The Collider Detector at Fermilab (CDF~II)~\cite{bib:cdfII-tdr,
bib:cdfII} is situated at one of the two collision points of the
Tevatron \ppbar~collider. It is a general purpose detector designed to
study the properties of these collisions.  The detector has both
azimuthal and forward-backward symmetry.  Since the CDF~II detector
has a barrel-like shape, we use a cylindrical coordinate system ($r$,
$\phi$, $z$). The origin is located at the center of the detector, $r$
is the radial distance from the beamline and the $z$-axis lies along
the nominal direction of the proton beam (toward east).  Spherical
coordinates ($\phi$, $\theta$) are also commonly used, where $\phi$ is
the azimuthal angle around the beam axis and $\theta$ is the polar
angle defined with respect to the proton beam direction.
Pseudorapidity $\eta$ is defined as $\eta \equiv
-\ln\left[\tan(\theta/2)\right]$.  The transverse energy and momentum
of a particle are defined as $E_{\rm T}=E\sin\theta$ and $p_{\rm
T}=p\sin\theta$, respectively.  A diagram of the CDF~II detector is
shown in Fig.~\ref{fig:cdf_detector}. A quadrant of the detector is
cut out to expose the different subdetectors.
 
The CDF~II detector consists of three primary subsystems: The
innermost part of the detector is the tracking system, which contains
silicon microstrip detectors and the Central Outer Tracker (COT), an
open cell drift chamber, inside a superconducting solenoid which
generates a $1.4~T$ magnetic field parallel to the beam axis. These
detector systems are designed to reconstruct the trajectories of
charged particles and precisely measure their momenta.  The silicon
detectors provide excellent impact parameter, azimuthal angle, and
$z$~resolution~\cite{Hill:2004qb,Sill:2000zz,Affolder:2000tj}. For
example, the typical intrinsic hit resolution of the silicon detector
is 11~$\mu$m. The transverse impact parameter (distance of closest
approach of a track to the beam line in the transverse plane)
resolution is $\sim$~40~$\mu$m, of which approximately 35~$\mu$m is
due to the transverse size of the Tevatron interaction region. The
entire system reconstructs tracks in three dimensions with the
precision needed to identify displaced vertices associated with $b$
and $c$ hadron decays. The COT~\cite{Affolder:2003ep} provides
excellent curvature and angular resolution, with coverage for
$|\eta|\leq$ 1. The COT has a transverse momentum resolution of
$\frac{\sigma_{P_T}}{P^2_{T}}$ = 0.0015 [GeV/$c$]$^{-1}$ which
improves to 0.0007~[GeV/$c$]$^{-1}$~\cite{bib:cdfII} including the
silicon detectors. The tracking efficiency of the COT is nearly 100\%
in the range $|\eta| < 1$, and the coverage is extended to $|\eta| <
1.8$ by including the silicon detectors.

Outside of the solenoid are the
calorimeters~\cite{Balka:1987ty,Bertolucci:1987zn,Albrow:2001jw},
which measure the energy of particles that shower when interacting
with matter.  The calorimeter is segmented into projective towers, and
each tower is divided into an inner electromagnetic and outer hadronic
sections.  This facilitates separation of electrons and photons from
hadrons by the energy deposition profiles as particles penetrate from
inner to outer sections.  The full array has an angular coverage of
$|\eta|~<$ 3.6. The central region, $|\eta|~<$ 1.1, is covered by the
central electromagnetic calorimeter and the central hadron
calorimeter. The central calorimeters have resolutions of $\sigma(E)/E
= 13.5\% / \sqrt{E\cdot\sin\theta}\oplus 2\%$~[GeV] and $\sigma(E)/E =
50\% / \sqrt{E}\oplus 3\%$~[GeV] for the electromagnetic and hadronic
calorimeters, respectively. The forward region, 1.1 $<~|\eta|~<$ 3.6,
is covered by the end-plug electromagnetic calorimeter and the
end-plug hadron calorimeter, with resolution of $\sigma(E)/E = 16\% /
\sqrt{E} \oplus 1\%$~[GeV] and $\sigma(E)/E = 80\% / \sqrt{E} \oplus
5\%$~[GeV] for the plug electromagnetic and hadronic calorimeters,
respectively.

Finally, outside of the calorimeters are the muon chambers, which
provide muon detection in the range $|\eta|~<$ 1.5.  The muon
detectors at CDF~\cite{bib:cdfII-tdr} make use of single wire drift
chambers as well as scintillator counters for fast timing.  For the
analyses presented in this article, muons are detected in four
separate subdetectors. Muons with $p_{\rm T}>1.4$~GeV/$c$ penetrating
the five absorption lengths of the calorimeter are detected in the
four layers of planar multi-wire drift chambers of the central muon
detector (CMU)~\cite{Ascoli:1987av}. Behind an additional 60~cm of
steel, a second set of four layers of drift chambers, the central muon
upgrade (CMP)~\cite{Dorigo:2000ip}, detects muons with $p_{\rm
T}>2.2$~GeV/$c$. The CMU and CMP cover the same part of the central
region $|\eta|<0.6$. The central muon extension
(CMX)~\cite{Dorigo:2000ip} extends the pseudorapidity coverage of the
muon system from 0.6 to 1.0 and thus completes the coverage over the
full fiducial region of the COT.  Muons in the $|\eta|$-range from 1.0
to 1.5 of the forward region are detected by the barrel muon chambers.

%-----------------------------
% DATA SAMPLE
%-----------------------------
%\input{./includes/prdDataSample}
%-------------------------------------------------------------------------------
\section{Data sample and event reconstruction}
\label{sec:dataSample}
%\setlength{\footskip}{2cm}
%-------------------------------------------------------------------------------

The data set used in this analysis comes from \ppbar~collisions at a
center-of-mass energy of $\sqrt{s}\,$~=~1.96~\tev~recorded by the CDF
II detector between March 2002 and February 2010. The CDF experiment
utilizes a three-level trigger
system~\cite{Thomson:2002xp,Downing:2006xb,GomezCeballos:2004jk} to
reduce the 1.7 $\mhz$ beam crossing rate to $\sim$200 $\hz$.  The
first two levels of the trigger system are custom hardware (the second
level also has a software component) and the third consists of a farm
of computers running a fast version of the offline event
reconstruction algorithms.

\begin{figure}[ht]
\begin{center}
\includegraphics[width=0.75\columnwidth]{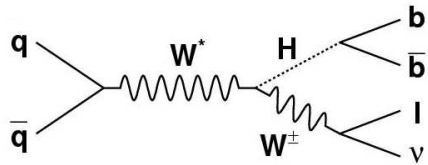}
\caption{Feynman diagram showing the final states of the
$WH$ process, with leptonic $W$ boson decays.  The final state
contains a charged lepton, a neutrino, and two $b$ quarks.}
\label{fig:WHdiagram}
\end{center}
\end{figure}

$WH$ events in the lepton~+~jets channel are characterized by the
presence of an electron or muon with high transverse energy, large
missing transverse energy resulting from the undetected neutrino, and
two high energy $b$ jets (see Fig.~\ref{fig:WHdiagram}).

The data sample used was collected by two trigger strategies, one
based on the selection of a high transverse momentum lepton (electron
or muon~\footnote{Note that leptonically decaying tau leptons make up
a small fraction of our signal acceptance since in this case the tau
can be identified as an isolated electron or muon.} and another one
based on missing transverse energy ($\EtMiss$, defined in
Section~\ref{sec:met})~+~jets.

The total integrated luminosity is 5.6~$\fb^{-1}$ for lepton-based
triggered events and 5.1~$\fb^{-1}$ for muon candidates collected by
the $\EtMiss$~+~jets trigger.  The different luminosities arise from
the different detector conditions necessary for each trigger.
Electrons reconstructed in the central and end-plug electromagnetic
calorimeters are referred to as CEM and PHX electrons,
respectively. Muons reconstructed in the central region by the CMU and
the CMP detectors are referred to as CMUP muons. Muons detected by the
CMX detector are referred to as CMX muons.  CEM, PHX, CMUP, and CMX
leptons are commonly known as tight leptons and the muons collected by
the $\EtMiss$~+~jets trigger are known as extended muon coverage (EMC)
muons.  In this section, we briefly discuss the lepton identification
requirements, the reconstruction of jets, and the calculation of
$\EtMiss$.

\subsection{Electron identification}
High-{\it p$_T$} electrons traversing the CDF II detector are expected
to leave a track in both the silicon detector and the
COT. Subsequently, the electrons will deposit most of their energy
into the central or plug electromagnetic calorimeters. The central
electron trigger begins by requiring a COT track with \pt~$>$~9~\gevc\
that extrapolates to an energy cluster of three central
electromagnetic calorimeter towers with \ET~$>$~18~GeV.  Several cuts
are then successively applied in order to improve the purity of the
electron selection.  The reconstructed track with
\pt~$>$~9~\gevc\ must match to an electromagnetic calorimeter cluster
with \ET~$>$~20~GeV. Furthermore, we require the ratio of hadronic
energy to electromagnetic energy $E_{\rm HAD}~/~E_{\rm EM}$ to be less
than 0.055~+~0.00045~$\times~E$/GeV and the ratio of the energy of the
cluster to the momentum of the track $E/pc$ to be smaller than 2.0 for
track momenta~$\le~50$~\gevc.

Electron candidates in the forward direction ($|\eta|~>$~1.1, PHX) are
defined by a cluster in the plug electromagnetic calorimeter with
$E_{\rm T}~>$~20~GeV and $E_{\rm HAD}~/~E_{\rm EM}~<$~0.05.  The
cluster position and the primary vertex position are combined to form
a trajectory on which the tracking algorithm utilizes hits in the
silicon tracker.

CEM candidates are rejected if an additional high-{\it p$_T$} track is
found which forms a common vertex with the track of the electron
candidate and has the opposite electric charge since these events are
likely to stem from the conversion of a photon.

Figure~\ref{fig:Triggertypes}(a) shows the $(\eta,\phi)$ distributions
of CEM and PHX electron candidates.

\subsection{Muon identification}
Muons are characterized by a track in the tracking system, energy
deposited in the calorimeter consistent with that of a minimum
ionizing particle, and in cases where they are fiducial to muon
chambers they will often leave a track, called a stub, in these
detectors.  The third-level muon trigger requires a COT track with
\pt~$>$~18~\gevc\ matched to a track segment in the muon chambers.

Muon identification requires an isolated \hyphenation{high-momentum}
COT track (\pt~$>$~20~GeV$/c$) that extrapolates to a track segment in
the muon chambers. Track segments must be detected either in the CMU
and the CMP simultaneously (CMUP muons), or in the CMX (CMX muons) for
triggered muons.  Several additional requirements are imposed in order
to minimize contamination from hadrons punching through the
calorimeter, decays in flight of charged hadrons, and cosmic rays.
The energy deposition in the electromagnetic and hadronic calorimeters
has to be small, as expected from a minimum-ionizing particle.  To
reject cosmic-ray muons and muons from in-flight decays of long-lived
particles such as $K_{S}^0$ and $\Lambda$, the impact parameter of the
track is required to be less than 0.2 cm if there are no silicon hits
on the muon candidate's track, and less than 0.02 cm if there are
silicon hits. The remaining cosmic rays are reduced to a negligible
level by taking advantage of their characteristic track timing and
topology.

\begin{figure*}
\begin{center}
\subfigure[]{
\includegraphics[width=0.75\columnwidth]{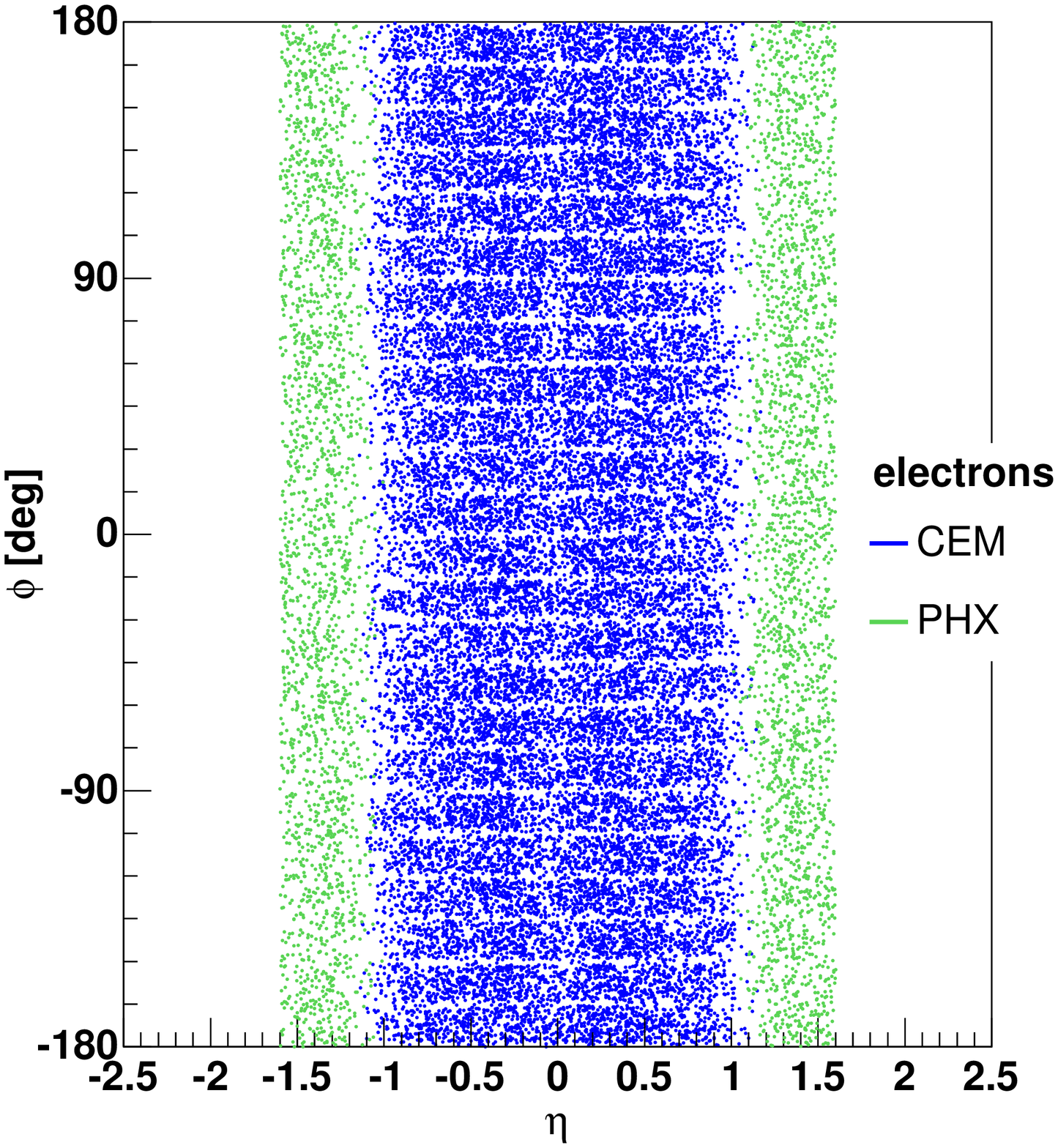}
\label{fig:eletypes}}
\subfigure[]{
\includegraphics[width=0.75\columnwidth]{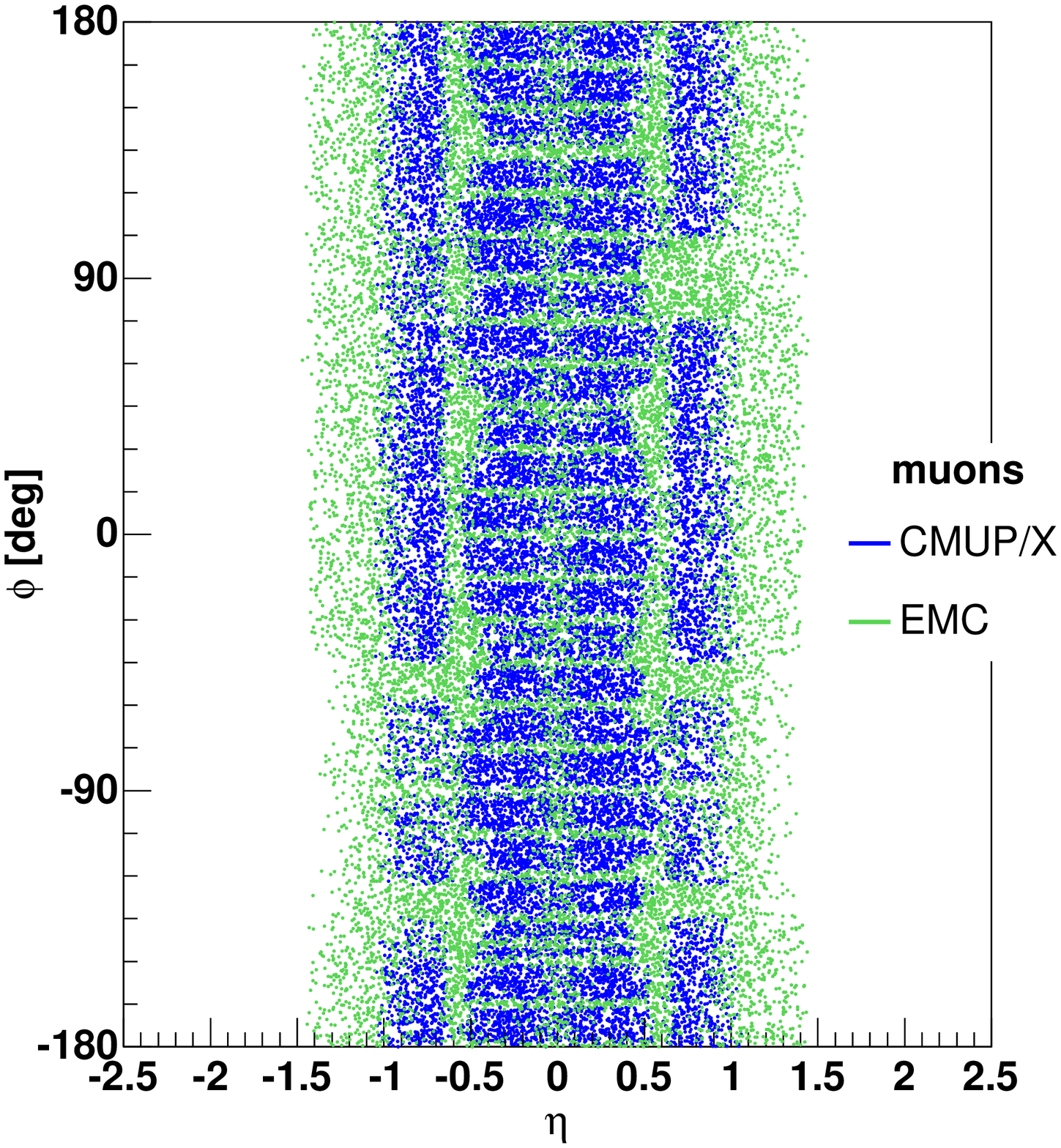}
\label{fig:mutypes}}
\end{center}
\caption{\label{fig:Triggertypes}Distributions in ($\phi-\eta$) space
of the electron (a) and muon (b) selection categories, showing the
coverage of the detector that each lepton type provides.  The trigger
based on $\EtMiss$ plus jets is used to fill in the gaps in the muon
trigger coverage.}
\end{figure*}

 In order to add acceptance for events containing muons which are not
triggered on directly, several additional muon types are taken from
the extended muon coverage (EMC) provided by triggers based on
$\EtMiss$ + jets requirements ($\EtMiss~>$~35~GeV and the presence of
at least two jets).  Events passing the $\EtMiss$~+~jets trigger are
also required to have two sufficiently-separated jets: $\Delta
R_{jj}~>~$1, where $\Delta R \equiv \sqrt{(\Delta \eta)^2 + (\Delta
\phi)^2}$.  Furthermore, one of the jets must be central, with
$|\eta|~<~$0.9, and jets are required to have transverse energies
above 25 GeV. These additional jet-based requirements remove the
dependence of the trigger efficiency to jet observables so that it can
be modeled by the $\EtMiss$ alone. The details of the EMC types and
selection are included in Ref.~\cite{bib:bruno_thesis}.
Figure~\ref{fig:Triggertypes}(b) shows the $(\eta,\phi)$ distribution
of all muon candidates.

\subsection{Lepton identification efficiencies}
The efficiency of lepton identification is measured using $Z
\rightarrow e^+ e^-$ and $Z \rightarrow \mu^+ \mu^-$ samples. A pure
sample of leptons can be obtained by selecting events where the
invariant mass of two high-\pt~track is near the mass of the $Z$ boson
and one track passed the trigger and tight lepton identification
selection. The other track can then be examined to see if it also
passed the identification cuts to study the efficiency.  The same
procedure can be applied to simulated Monte Carlo (MC) events and to
observed events in the detector and small differences in the
efficiencies are observed due to imperfect detector modeling.  To
correct for this difference, a correction factor is applied to the
efficiencies of Monte Carlo events based on the ratio of lepton
identification efficiencies calculated from observed events to the
efficiency found in Monte Carlo events. The correction factors for the
lepton identification are shown in Table~\ref{tab:scleptonid}.

\begin{table}
\begin{tabular}{cc}\\ \hline\hline
Lepton type & Correction factor \\ \hline CEM & 0.977 $\pm$ 0.001\\
PHX & 0.919 $\pm$ 0.002\\ CMUP & 0.894 $\pm$ 0.002\\ CMX & 0.952 $\pm$
0.002\\
%EMC   & 0.882 $\pm$ 0.003 - 1.070 $\pm$ 0.020~\cite{stprd}\\
EMC & 0.882 $\pm$ 0.003 - 1.070 $\pm$ 0.020\\
%CMU   & 0.891 $\pm$ 0.003\\ 
%CMP   & 0.882 $\pm$ 0.003\\ 
%BMU   & 1.070 $\pm$ 0.020\\ 
%CMIO  & 1.028 $\pm$ 0.030\\ 
%SCMIO & 0.952 $\pm$ 0.002\\ 
%CMXNT & 0.952 $\pm$ 0.002\\ 
\hline\hline
\end{tabular}
\caption{\label{tab:scleptonid} Correction factors applied to the
Monte Carlo events to correct the lepton identification
efficiencies. Since there are different sub-categories within the EMC
category, we quote the range of variation.}
\end{table}

\subsection{Jet reconstruction and corrections}
\label{sec:jets}
Jets consist of a shower of particles originating from the
hadronization of highly energetic quarks or gluons. Jets used in this
analysis are reconstructed using a cone algorithm~\cite{Bhatti:2005ai}
by summing the transverse calorimeter energy \ET\ in a cone of radius
$\Delta R \le 0.4$, for which the \ET\ of each tower is calculated
with respect to the primary vertex $z$ coordinate of the event.  The
calorimeter towers belonging to any electron candidate are not used by
the jet clustering algorithm.  The energy of each jet is
corrected~\cite{Bhatti:2005ai} for the $\eta$ dependence and the
nonlinearity of the calorimeter response.  The jet energies are also
adjusted by subtracting the average extra deposition of energy from
additional inelastic \ppbar\ collisions on the same beam crossing as
the triggered event.

\subsection{Missing transverse energy reconstruction}
\label{sec:met}
The presence of neutrinos in an event is inferred by an imbalance in
the transverse components of the energy measurements in the
calorimeter. The missing \ET\ vector ($\EtMissVec$) is defined by:
\begin{eqnarray}
\not\!\! \vec{E}_{\rm T} & = & - \sum_{i} E_{\rm T}^i \hat{n}_i,
\end{eqnarray}
where {\it i} is the index for calorimeter tower number with
$|\eta|<3.6$, and $\hat{n}_i$ is a unit vector perpendicular to the
beam axis and pointing at the $i^{\rm th}$ calorimeter
tower. $\EtMiss$ also refers to the magnitude $|\EtMissVec|$.  The
$\EtMiss$ calculation is based on uncorrected tower energies and is
then corrected based on the jet energy corrections of all of the jets
in the event. Also the $\EtMiss$ is corrected for the muons, since
they traverse the calorimeters without showering.  The transverse
momenta of all identified muons are added to the measured transverse
energy sum and the average ionization energy is removed from the
measured calorimeter energy deposits.

%-----------------------------
% EVENT SELECTION
%-----------------------------
%\input{./includes/prdEventSel}
%-------------------------------------------------------------------------------
\section{Event selection}
\label{sec:reconstr}
%\setlength{\footskip}{2cm}
%-------------------------------------------------------------------------------
The selection before identifying any jet as a $b$ jet is referred to
as pretag and only requires the presence of an electron or muon,
$\EtMiss~>$~20~GeV (25~GeV in the case of forward electrons) and two
or three jets with corrected \ET~$>$~20~GeV and $|\eta|~<$~2.0.  At
leading order one would expects to have only two high-\pt~jets in the
final state of $WH$ signal events. However, by allowing for the
presence of a third jet, signal acceptance is improved by about 25~\%
due to extra jets mostly produced by gluon radiation in the initial or
final state.

In order to reduce the $Z$~+~jets, top, and $WW$/$WZ$ background
rates, events with more than one lepton are removed.  If one of the
leptons is not identified correctly, $Z\rightarrow \ell^+\ell^-$
events still remain. To remove such events, the invariant mass of the
lepton and any track with opposite charge must not be in the $Z$ boson
mass window 76~$<~m_{l, \rm track}~<$~106~GeV/$c^2$.

The non-$W$ background consists of multijet events which do not
contain $W$ bosons; a description of these background events can be
found in Section~\ref{sec:nonw}.  This non-$W$ background is reduced
by applying additional selection requirements which are based on the
assumption that these events do not have large $\EtMiss$ from an
escaping neutrino, but rather the $\EtMiss$ that is observed comes
from lost or mismeasured jets. This requirement has been developed in
the framework of the single top observation and is described in detail
in~\cite{stprd}.

%-----------------------------
% B TAGGING STRATEGY
%-----------------------------
%\input{./includes/prdBTagging}
%-------------------------------------------------------------------------------
\section{$b$-jet tagging algorithms}
\label{sec:btag}
%\setlength{\footskip}{2cm}
%-------------------------------------------------------------------------------
The events selected by the above criteria are dominated by the
production of $W$ bosons in association with jets. In order to improve
the signal to background ratio for $WH$ events, at least one of the
jets in the event is required to be produced by a $b$ quark.
Identifying jets originating from $b$ quarks helps to reduce the
background from non-$W$ and $W$~+~light flavor ($W$~+~LF)
events. Therefore, the last step of the event selection is the
requirement of the presence of at least one $b$-tagged jet identified
using the {\sc SecVtx} algorithm~\cite{Acosta:2004hw}. In order to
increase the acceptance for events with two tagged $b$ jets, an
additional $b$-tagging algorithm that relies on high-impact-parameter
tracks within jets, {\sc Jet Probability}~\cite{jetprob}, is
used. These two tagging algorithms are based on the same principle:
the fact that $b$ quarks have a relatively long lifetime and high
mass. Therefore, $b$ hadrons formed during the hadronization of the
initial $b$ quark can travel a significant distance (on the order of a
few millimeters) before decaying to lighter hadrons.  Then, the
displacement of the $b$ hadron decay point can be detected either
directly by vertexing the tracks or indirectly by studying the impact
parameters of tracks.

\subsection{Secondary Vertex Tagger}
The {\sc SecVtx} algorithm looks inside the jet cone to construct
secondary vertices using tracks displaced from the primary vertices.
The tracks are distinguished by their large impact parameter
significance ($|d_0/\sigma_{d_0}|$), where $d_0$ and $\sigma_{d_0}$
are the impact parameter and its overall uncertainty.  The tracks are
fit to a common vertex using a two-pass approach. In the first pass,
applying loose track selection criteria ({\it p$_T$} $>$~0.5~GeV/$c$
and $|\frac{d_0}{\sigma_{d_0}}|~>$~2.5), the algorithm attempts to
reconstruct a secondary vertex which includes at least three tracks
(at least one of the tracks must have {\it p$_T$} $>$~1~GeV/$c$). If
no secondary vertex is found, the algorithm uses tighter track
selection requirements ({\it p$_T$} $>$~1~GeV/$c$ and
$|\frac{d_0}{\sigma_{d_0}}|~>$~3.0) and attempts to reconstruct a
two-track vertex in a second pass. If either pass is successful, the
transverse distance ($L_{xy}$) from the primary vertex of the event is
calculated along with the associated uncertainty $\sigma_{L_{xy}}$,
which includes the uncertainty on the primary vertex position.  Jets
are considered as tagged by requiring a displaced secondary vertex
within the jet.  Secondary vertices are accepted if the transverse
decay length significance ($L_{xy} / \sigma_{L_{xy}}$) is greater than
or equal to 7.5.

$L_{xy}$ is defined to be positive when the secondary vertex is
displaced in the same direction as the jet, and the jet is positively
tagged. A negative value of $L_{xy}$ indicates an incorrect $b$-tag
assignment due to mis-reconstructed tracks.  In this case the tag is
called negative. These negative tags are useful for estimating the
rate of incorrectly $b$-tagged jets as explained in
Section~\ref{sec:mistag}.

\subsection{ {\sc \bf {Jet Probability}} Tagger}
The {\sc Jet Probability} $b$-tagging algorithm is also used.  Unlike
{\sc SecVtx}, this algorithm does not explicitly require that the
tracks form a vertex.  Instead, it uses tracks associated with a jet
to determine the probability for these to come from the primary vertex
of the interaction~\cite{jetprob}.  The calculation of the probability
is based on the impact parameters of the tracks in the jet and their
uncertainties.  The impact parameter is assigned a positive or
negative sign depending on the position of the track's point of
closest approach to the primary vertex with respect to the jet
direction. It is positive (negative) if the angle $\phi$ between the
jet axis and the line connecting the primary vertex and the track's
point of closest approach to the primary vertex itself is smaller
(bigger) than $\pi/2$.  By construction, the probability for tracks
originating from the primary vertex is uniformly distributed from 0 to
1.  For a jet coming from heavy flavor hadronization, the distribution
peaks at 0, due to tracks from long lived particles that have a large
impact parameter with respect to the primary vertex.  To be considered
as tagged, the jets are required to have a value of the {\sc Jet
Probability} variable ($P_J$) less than 0.05 ($P_J<$ 5\%).

\subsection{Tagging efficiencies and mistag rates}
\label{sec:mistag}
The $b$-tagging efficiencies are needed to estimate the yields of
signal and background events, which are obtained from Monte Carlo
simulations.  The efficiency for identifying a heavy flavor jet is
different in simulated events and in observed events.  It is typically
overestimated by Monte Carlo models. To correct for this effect, a
scale factor is applied to the Monte Carlo tagging efficiency.

The method used to measure the tagging efficiency for heavy flavor
jets is described in detail in~\cite{Acosta:2004hw}.  To measure the
tagging efficiency in observed events, a calibration sample enriched
in heavy flavor is used. This sample is selected by requiring
electrons with $p_T$~$>$~8~GeV/c.  Along with the electron we require
the presence of two jets, the ``electron jet'' and the ``away
jet''. The electron jet is required to have $E_T$~$>$~15~GeV
(including the energy of the electron) and to be within 0.4 of the
electron in $\eta$-$\phi$ space (in other words the electron is within
the jet cone), and is presumed to contain the decay products of a
heavy flavor hadron. The away jet is required to have $E_T$~$>$~15~GeV
and $|\eta|$~$<$1.5, and it must be approximately back-to-back with
the electron jet ($\Delta \phi$~$>$2 rad).  To measure the tagging
efficiency of the heavy flavor electron jets we employ a double-tag
technique, requiring that the away jet be tagged by the corresponding
tagging algorithm. This enhances the heavy flavor fraction of the
electron jets and reduces the dependence on the heavy flavor fraction.
The tagging efficiency is also measured for simulated jets by using a
Monte Carlo sample similar to the calibration sample.  The tagging
efficiency ratio of observed events to Monte Carlo simulated events is
called the tagging scale factor ($SF$).  The tagging scale factors
used in this analysis are summarized in Table~\ref{tab:tageff1} for
$P_J<$ 5\%, and {\sc SecVtx}~\cite{bib:ChrisNeu}.  The uncertainties
shown are statistical and systematic.
\begin{table}[h]
\begin{center}
\caption{\label{tab:tageff1}
 Tagging scale factors and their uncertainties for $P_J<$ 5\%, and
 {\sc SecVtx}.}
\vspace{0.2cm}
\begin{tabular}{l@{\hspace{1.2cm}}c@{\hspace{1.2cm}}c}
\hline
\hline
          & $P_J<$ 5\% & {\sc SecVtx} \\
\hline
Scale factor & 0.806 $\pm$ 0.038 & 0.95 $\pm$ 0.04\\
\hline
\hline
\end{tabular}
\end{center}
\end{table}

The probability of misidentifying a light jet as a heavy-flavor jet
(``mistag") is closely related to the rate of negatively tagged jets.
The negative tag rate is measured in an inclusive-jet sample collected
by triggers with various jet $E_T$ thresholds. This tag rate is then
parametrized as a six-dimensional tag-rate matrix. The parametrization
of the mistag rate is done as a function of three jet variables:
transverse energy of the jet ($E_T$), the number of tracks in the jet
($N_{\rm {trk}}$), and the pseudorapidity of the jet ($\eta$) and
three event variables: the sum of the transverse energies of all jets
in the event ($\sum E_{T}^{\mathrm{jet}}$), the number of
reconstructed vertices in the event ($N_{\rm vtx}$), and the
$z$-position of the primary vertex ($z_{\rm vtx}$).  These
parametrized rates are used to obtain the probability that a given jet
will be negatively tagged.  It is assumed that the negative tags are
due to detector resolution effects only, while positive tags consist
of a mixture of heavy flavor tags, resolution-based mistags of
light-flavor jets, and mistags due to $K$'s, $\Lambda$'s and nuclear
interactions with the detector material.  The mistag rate is based on
the negative tag rate in the inclusive jet data, corrected for
estimations of the other contributions~\cite{bib:ChrisNeu}.
Typically, the mistag rate is of the order of a few percent.

\subsection{Splitting tagging categories}
As already mentioned above, the last step of the event selection is to
require the presence of at least one $b$-tagged jet using the {\sc
SecVtx} algorithm. In order to gain sensitivity, both $b$-tagging
algorithms are used to assign events to one of three non-overlapping
tagging categories, each with a different signal to background ratio.
The {\sc Jet Probability} tagger with the cut at 5\% is less
restrictive than {\sc SecVtx}.  This means that the selection
efficiency for real $b$ jets is higher, but it is accompanied by an
increase in the background contribution of light jets misidentified as
heavy flavor jets.  Some of the events that were not tagged by the
{\sc SecVtx} algorithm are recovered by {\sc Jet Probability}.  The
addition of these events translates into a 5\% improvement in the
final sensitivity of the analysis.  Events are selected in the
following order: events in which two or more jets are tagged by the
{\sc SecVtx} algorithm (SVSV events), events where only one jet is
tagged by {\sc SecVtx} and the other one is tagged by the {\sc Jet
Probability} algortihm (SVJP events), and events with only one jet
tagged by {\sc SecVtx} (in this case, none of the other jets is tagged
by any of the two algorithms, SVnoJP events).

%-----------------------------
% SIGNAL MODEL AND ACCEPTANCE
%-----------------------------
%\input{./includes/prdSignalAcc}
%-------------------------------------------------------------------------------
\section{Signal modeling and acceptance}
\label{sec:accept}
%\setlength{\footskip}{2cm}
%-------------------------------------------------------------------------------

Higgs boson events are modeled with the
{\sc{pythia}}~\cite{Sjostrand:2000wi} Monte Carlo generator using the
{\sc{cteq5l}}~\cite{Lai:1999wy} parton distribution functions
(PDFs). They are combined with a parametrized response of the CDF~II
detector~\cite{bib:geant3} and tuned to the Tevatron underlying event
data~\cite{bib:underlying}.

For this analysis, the Higgs boson mass region where the branching
ratio to $b\bar{b}$ is large is studied (Higgs boson masses between
100 and 150~GeV/$c^2$). Eleven signal MC samples are generated in this
range, 100~$<~m_H~<$~150~GeV/$c^2$ in 5~GeV/$c^2$ increments.

The number of expected $WH \rightarrow \ell\nu_{\ell}$\bbbar~events is
given by:
\begin{equation}
\label{eq:Nsig}
N = \sigma_{p \bar{p} \rightarrow WH} \cdot {\cal{B}}(H \rightarrow
b\bar{b}) \cdot \varepsilon_\mathrm{evt} \cdot \lumi
\end{equation}
where $\sigma_{p \bar{p} \rightarrow WH}$ is the theoretically
predicted cross section of the $WH$ process, $\cal{B}$ is the
branching ratio of a Higgs boson decaying to \bbbar,
$\varepsilon_\mathrm{evt}$ is the event detection efficiency, and
$\mathcal{L}_\mathrm{int}$ is the integrated luminosity.

The SM predicted cross sections for $WH$ production and the branching
ratios of a Higgs bosons decaying to \bbbar~for the different Higgs
boson masses are calculated to next-to-leading order
(NLO)~\cite{bib:higgs_br} and are quoted in
Table~\ref{tab:wr_xsectBR}.

\begin{table}[h]
  %\begin{center} 
\caption{\label{tab:wr_xsectBR} SM branching ratios
  ($H~ \rightarrow ~b\bar{b}$) and $WH$ production cross sections for
  all Higgs boson masses used in this analysis.}  \vspace{.2cm}
  \begin{tabular}{lcc} 
\hline \hline Higgs mass (\gevcc) &
  $\cal{B}$($H\rightarrow b\bar{b}$) & $\sigma$ (pb) \\ \hline 100 &
  0.812 & 0.286 \\ 105 & 0.796 & 0.253 \\ 110 & 0.770 & 0.219 \\ 115 &
  0.732 & 0.186 \\ 120 & 0.679 & 0.153 \\ 125 & 0.610 & 0.136 \\ 130 &
  0.527 & 0.120 \\ 135 & 0.436 & 0.103 \\ 140 & 0.344 & 0.086 \\ 145 &
  0.256 & 0.078 \\ 150 & 0.176 & 0.070 \\ \hline \hline 
\end{tabular}
  %\end{center} %\vspace{-0.5cm}
\end{table}

The event detection efficiency, $\varepsilon_\mathrm{evt}$, can be
broken down into several factors:
\begin{equation}
\label{eq:evt}
  \varepsilon_\mathrm{evt}= \varepsilon_\mathrm{z_0} \cdot
  \varepsilon_\mathrm{trigger} \cdot \varepsilon_\mathrm{lepton~Id}
  \cdot \varepsilon_\mathrm{tag} \cdot \varepsilon_\mathrm{acc} \cdot
  {\cal{B}}(W \rightarrow \ell\nu_{\ell})
\end{equation}
where each term corresponds, respectively, to the $z$ vertex cut
($|z|<60$ cm fiduciality), triggers, lepton identification, $b$
tagging, acceptance requirements, and the branching ratio of the $W$
boson decaying to a lepton and a neutrino.  The event detection
efficiency is estimated by performing the event selection on the
samples of simulated events. Control samples in the data are used to
calibrate the efficiencies of the trigger, the lepton identification,
and the $b$ tagging. These calibrations are then applied to the Monte
Carlo samples we use.

The predicted signal yields for the selected two- and three-jet events
for each tagging category are estimated by Eq.~\ref{eq:Nsig} at each
Higgs boson mass point. Tables~\ref{tab:EventYield_WH_2jets} (for
two-jet events) and~\ref{tab:EventYield_WH_3jets} (for three-jet
events) show the number of expected $WH$ events for each Higgs boson
mass for an integrated luminosity of 5.6~\fb.

\begin{table}[h]
  \caption{\label{tab:EventYield_WH_2jets} Summary of predicted number
of signal events based on 5.6~fb$^{-1}$ of integrated luminosity with
systematic and statistical uncertainties for each Higgs boson mass in
2-jet events passing all event selection requirements.}  \vspace{.2cm}
\begin{tabular}{lccc} \hline \hline Higgs mass & SVSV &
~~~~~~SVJP~~~~~~ & SVnoJP \\ (GeV/$c^2$) & & & \\ \hline 100 &
5.92$\pm$0.69 & 4.12$\pm$0.52 & 15.66$\pm$1.23 \\ 105 & 5.50$\pm$0.64
& 3.76$\pm$0.47 & 14.11$\pm$1.11 \\ 110 & 4.80$\pm$0.56 &
3.33$\pm$0.42 & 12.34$\pm$0.97 \\ 115 & 4.06$\pm$0.48 & 2.80$\pm$0.35
& 10.27$\pm$0.81 \\ 120 & 3.24$\pm$0.38 & 2.24$\pm$0.28 &
8.08$\pm$0.64 \\ 125 & 2.65$\pm$0.31 & 1.86$\pm$0.23 & 6.59$\pm$0.52
\\ 130 & 2.07$\pm$0.24 & 1.44$\pm$0.18 & 5.12$\pm$0.40 \\ 135 &
1.49$\pm$0.17 & 1.07$\pm$0.13 & 3.70$\pm$0.29 \\ 140 & 1.01$\pm$0.12 &
0.71$\pm$0.09 & 2.46$\pm$0.19 \\ 145 & 0.70$\pm$0.08 & 0.50$\pm$0.06 &
1.69$\pm$0.13 \\ 150 & 0.44$\pm$0.05 & 0.31$\pm$0.04 & 1.06$\pm$0.08
\\ \hline \hline \end{tabular}
\end{table}

\begin{table}[h]
  \caption{\label{tab:EventYield_WH_3jets} Summary of predicted number
  of signal events based on 5.6~fb$^{-1}$ of integrated luminosity
  with systematic and statistical uncertainties for each Higgs boson
  mass in 3-jet events passing all event selection requirements.}
  \vspace{.2cm}

    \begin{tabular}{lccc} \hline \hline Higgs mass & SVSV &
      ~~~~~~SVJP~~~~~~ & SVnoJP \\ (GeV/$c^2$) & & & \\ \hline 100 &
      1.43$\pm$0.17 & 1.10$\pm$0.15 & 3.36$\pm$0.27 \\ 105 &
      1.41$\pm$0.17 & 1.06$\pm$0.15 & 3.22$\pm$0.26 \\ 110 &
      1.29$\pm$0.15 & 0.98$\pm$0.13 & 3.00$\pm$0.24 \\ 115 &
      1.16$\pm$0.14 & 0.85$\pm$0.12 & 2.57$\pm$0.21 \\ 120 &
      0.95$\pm$0.11 & 0.71$\pm$0.10 & 2.11$\pm$0.17 \\ 125 &
      0.81$\pm$0.09 & 0.60$\pm$0.08 & 1.80$\pm$0.15 \\ 130 &
      0.68$\pm$0.08 & 0.49$\pm$0.07 & 1.44$\pm$0.12 \\ 135 &
      0.50$\pm$0.06 & 0.37$\pm$0.05 & 1.09$\pm$0.09 \\ 140 &
      0.35$\pm$0.04 & 0.26$\pm$0.04 & 0.76$\pm$0.06 \\ 145 &
      0.25$\pm$0.03 & 0.18$\pm$0.03 & 0.54$\pm$0.04 \\ 150 &
      0.16$\pm$0.02 & 0.12$\pm$0.02 & 0.35$\pm$0.03 \\ \hline \hline
      \end{tabular}
\end{table}

%-----------------------------
% Background Estimation
%-----------------------------
%\input{./includes/prdBkgEstimation}
%-------------------------------------------------------------------------------
\section{Background modeling and estimation}
\label{sec:bkgEstim}
%\setlength{\footskip}{2cm}
%-------------------------------------------------------------------------------
Other production processes can mimic the $WH \rightarrow
\ell\nu_{\ell}$\bbbar~final state.  The main contribution comes from
heavy-flavor production in association with a leptonic $W$ boson
(\Wbbbar, \Wccbar, \Wc). $W$~+~LF production also gives a significant
contribution due to mistagged jets. Smaller contributions come from
electroweak and top quark processes, \ttbar, single top, diboson
production ($WW$, $WZ$, $ZZ$), or $Z$~+~jets, and non-$W$ multijet
production with misidentified leptons.

In order to estimate the different background rates, a combination of
Monte Carlo samples and observed events are used.  The observed
lepton~+~jets events consist of electroweak, top (single top and
$t\bar{t}$), non-$W$ production, and $W$~+~jets processes.  Some
background processes are estimated based on Monte Carlo simulations
scaled to theoretical predictions of the cross section (such as
$t\bar{t}$); some are purely data-based (non-$W$); and some require a
combination of Monte Carlo and observed events ($W$~+~jets).  The
first step in the background estimate is to calculate the processes
that can be reliably simulated using Monte Carlo techniques.
Estimating the non-$W$ fraction is the next step. Finally, the
observed events that are not non-$W$, electroweak, or top quark
processes are considered to be all $W$~+~jets events where $b$-tag
rate estimates from the Monte Carlo are used to estimate the
contribution to the $b$-tagged signal region.  Details on each step of
this process are given in the sections below.

\subsection{Monte-Carlo based background processes} 
\begin{figure}[h]
\begin{center}
\includegraphics[width=0.95\columnwidth]{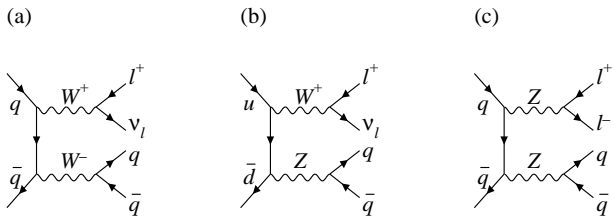}
\caption{Feynman diagrams for diboson production ($WW$, $WZ$, $ZZ$), which provides a
small background contribution to $WH$ production.}
\label{fig:diboson}
\end{center}
\end{figure}

\begin{figure}[t]
\begin{center}
\includegraphics[width=0.95\columnwidth]{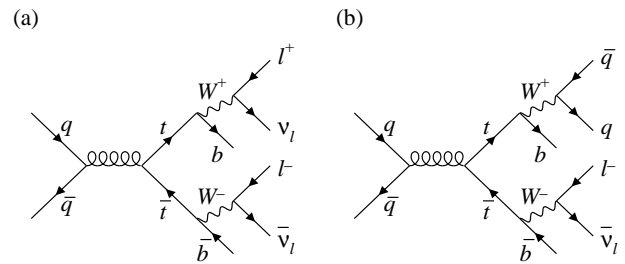}
\caption{Feynman diagrams of the $t\bar{t}$ background process to $WH$
production.  To pass the event selection, these events must have one
charged lepton (a) or two hadronic jets (b) that go undetected.}
\label{fig:ttbar}
\end{center}
\end{figure}

\begin{figure}[h]
\begin{center}
\includegraphics[width=0.95\columnwidth]{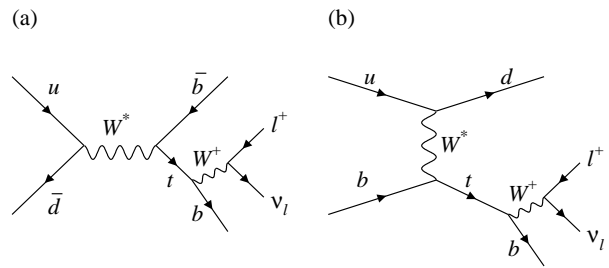}
\caption{Feynman diagrams showing the final states of the
$s$-channel (a) and $t$-channel (b) processes, with leptonic $W$ boson
decays.  Both final states contain a charged lepton, a neutrino, and
two jets, at least one of which originates from a $b$ quark.}
\label{fig:singletop}
\end{center}
\end{figure}

Diboson events ($WW$, $WZ$ and $ZZ$) can contribute to the tagged
lepton~+~jets sample when one boson decays leptonically and the other
decays into quarks (Fig.~\ref{fig:diboson}).  In addition, top pair
production in which one lepton (from Fig.~\ref{fig:ttbar} (a)) or two
jets (from Fig.~\ref{fig:ttbar} (b)) were not reconstructed also
constitutes an important background process. The diboson and
$t\bar{t}$ simulated events are generated using the
{\sc{pythia}}~\cite{Sjostrand:2000wi} Monte Carlo generator. There is
a contribution from single top quarks produced in association with a
$b$ quark, $s$-channel (Fig.~\ref{fig:singletop}(a)) and $t$-channel
(Fig.~\ref{fig:singletop}(b)) single top production. These events are
generated using the {\sc{madevent}}~\cite{Maltoni:2002qb} MC, and the
parton showering is done with {\sc{pythia}}.  Finally, the $Z$~+~jets
process in which one lepton from $Z$ boson decay is missed
(Fig.~\ref{fig:misc}(a)) can also contribute. $Z$~+~jets production is
simulated using a combination of {\sc{alpgen}}~\cite{Mangano:2002ea}
matrix element generation and {\sc{pythia}} parton showering.

The numbers of events from these processes are predicted based on
theoretical and measured cross sections, the measured integrated
luminosity, and the acceptances and tagging efficiencies derived from
Monte Carlo simulations in the same way as the $WH$ process described
in Section VI. The diboson cross sections are taken from the NLO
calculations with MCFM~\cite{PhysRevD.60.113006}.  For the $Z$~+~jets
background, the $Z$~+~jets cross section times the branching ratio of
$Z$ to charged leptons is normalized to the value measured by
CDF~\cite{Aaltonen:2007cp}.  Predictions based on NLO calculations are
also used for the $t\bar t$ and single top background
processes~\cite{bib:top_cross,bib:st_cross}.  Top cross section
predictions assume a top mass of 175 GeV/$c^2$.

The total diboson ($WW$, $WZ$, $ZZ$), $Z$~+~jets, \ttbar, and single
top quark predictions for each tagging category are shown in
Tables~\ref{tab:EventYield_2jets} (two-jet events)
and~\ref{tab:EventYield_3jets} (three-jet events). 

\subsection{Non-{\it W} multijet events} 
\label{sec:nonw}
\begin{figure}[h]
\begin{center}
\includegraphics[width=0.95\columnwidth]{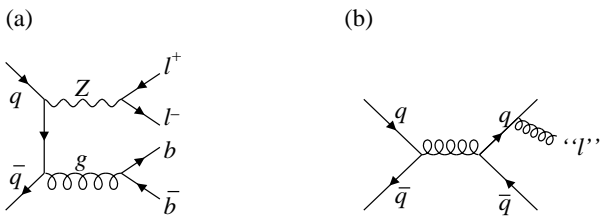}
\caption{Representative Feynman diagrams for (a) $Z$~+~jets 
production, where one lepton is missed, and (b) non-$W$ events, in
which a jet has to be misidentified as a lepton and $\EtMiss$\ must be
mismeasured to pass the event selection.  Because the cross section of
non-$W$ events is large, they still form a significant background
process.}
\label{fig:misc}
\end{center}
\end{figure}

The non-$W$ background process consists of events for which the
lepton~+~$\EtMiss$ signature is not due to the decay of a $W$ boson
but instead have a fake isolated lepton and mismeasured $\EtMiss$\
(Fig.~\ref{fig:misc}(b)).  The main contribution to this source of
background comes from QCD multijet production where a jet provides the
signature of a lepton and the missing transverse energy is due to a
mismeasurement of the jet energies. Semileptonic decays of $b$ hadrons
and misidentified photon conversions also contribute.  Due to their
instrumental nature, these processes can not be simulated reliably.
Therefore, samples of observed events are used to estimate the rates
of these processes and model their kinematic distributions.

Three different samples of observed events are used to model the
non-$W$ multijet contribution.  One sample is based on events that
fired the central electron trigger but failed at least two of the five
identification cuts of the electron selection requirements that do not
depend on the kinematic properties of the event, such as the fraction
of energy in the hadronic calorimeter.  This sample is used to
estimate the non-$W$ contribution from CEM, CMUP and CMX events.  A
second sample is formed from events that pass a generic jet trigger
with transverse energy \ET~$>$~20~GeV to model PHX events.  These jets
are additionally required to have a fraction of energy deposited in
the electromagnetic calorimeter between 80$\%$ and 95$\%$, and fewer
than four tracks, to mimic electrons.  A third sample, used to model
the non-$W$ background in EMC events, contains events that are
required to pass the $\EtMiss$~+~jets trigger (see
Section~\ref{sec:dataSample}) and contain a muon that passes all
identification requirements but failed the isolation requirement. In
this case, the isolation is defined as the ratio of the transverse
energy surrounding the muon to the transverse energy of the muon.  The
pseudorapidity distributions of the objects chosen to model the
falsely identified lepton must be consistent with that of the sample
it is modeling.  The first sample works well for central leptons, but
can't cover the PHX or EMC. Highly electromagnetic jets work well for
the PHX, while only non-isolated EMC muons give the correct
distribution for EMC non-$W$ events.

To estimate the non-$W$ fraction in both the pretag and tagged sample,
the $\EtMiss$ spectrum is fit to a sum of the predicted background
shapes, as described in detail elsewhere~\cite{stprd}. The fit has one
fixed component and two templates whose normalizations can float. The
fixed component is coming from the Monte Carlo based processes. The
two floating templates are a Monte Carlo $W$~+~jets template and a
non-$W$ template. The non-$W$ template is different depending on the
lepton category, as explained above.  The pretag non-$W$ fraction is
used to estimate the heavy flavor and light flavor fractions.

The total non-$W$ contribution for each tagging category is shown in
Tables~\ref{tab:EventYield_2jets} and~\ref{tab:EventYield_3jets}.

\subsection{{\it W} + heavy flavor contributions} 
\label{sec:wplushf}

\begin{figure}[h]
\begin{center}
\includegraphics[width=0.95\columnwidth]{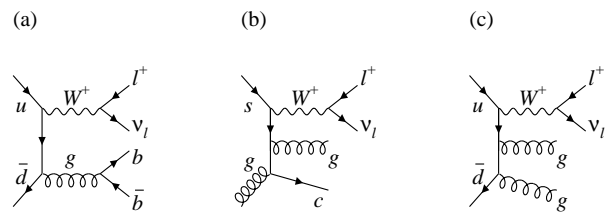}
\caption{Some representative diagrams of $W$~+~jets production. $Wc{\bar{c}}$ 
is the same process as $Wb{\bar{b}}$, but with charm quarks replacing
the $b$ quarks. }
\label{fig:Wjets}
\end{center}
\end{figure}

$W$~+~heavy flavor production is the main source of background in the
tagged lepton~+~jets sample.  $W$~+~jets production is simulated using
a combination of {\sc{alpgen}} matrix element generation and
{\sc{pythia}} parton showering (same as for $Z$~+~jets
events). Diagrams for some of the sample processes included in
{\sc{alpgen}} are shown in Fig.~\ref{fig:Wjets}.

The contribution of this background is estimated using the heavy
flavor fractions in $W$~+~jets production and the tagging efficiencies
for these processes. These quantities are derived from Monte Carlo
simulations as explained in~\cite{stprd}. The contribution of
$W$~+~heavy flavor events to our signal region is calculated by:

\begin{equation}
\label{wplushf}
N_{W+\mathrm{HF}}^{\mathrm{tag}}=(
N_{\mathrm{data}}^{\mathrm{\mathrm{pretag}}} \cdot (1 -
f_{non-W}^{\mathrm{pretag}}) - N_{\mathrm{MC}}^{\mathrm{pretag}})
\cdot f_{hf} \cdot k \cdot \varepsilon_{\mathrm{tag}},
\end{equation}
where $N_{\mathrm{data}}^{\mathrm{pretag}}$ is the number of observed
events in the pretag sample, $f_{non-W}^{\mathrm{pretag}}$ is the
fraction of non-$W$ events in the pretag sample, as determined from
the fits described in Section~\ref{sec:nonw}, and
$N_{\mathrm{MC}}^{\mathrm{pretag}}$ is the expected number of pretag
events in Monte Carlo based samples.  The fraction of $W$-boson events
with jets matched to heavy flavor quarks, $f_{hf}$, is calculated from
Monte Carlo simulation.  This fraction is multiplied by a scale
factor, $k$ = 1.4$\pm$0.4, to account for differences between the
heavy flavor fractions observed in data and the Monte Carlo
prediction.  The $k$-factor is primarily calculated in the one-jet
control sample and applied to all jet
multiplicities. $\varepsilon_{\mathrm{tag}}$ is the tagging selection
efficiency.  See Ref.~\cite{stprd} for more detail.
 
\subsection{Rates of events with mistagged jets} 
\label{sec:mistagrates}
The other $W$~+~jets contribution which can mimic the $\ell\nu_{\ell}
b\bar{b}$ final state is $W$~+~LF. In this case, jets from light
partons tagged as heavy flavor jets can contribute to the tagged
sample.  We count the events in the pretag sample and apply a mistag
matrix to calculate the fraction of $W$~+~light flavor events that
will be mistagged ($N_{\mathrm{mistag}} / N_{\mathrm{pretag}}$).  The
mistag rate parametrization is described in Section~\ref{sec:mistag}.
Then, in order to only use mistagged events from W+LF processes, we
subtract the fraction of pretag events which are due to non-$W$,
electroweak, top quark and $W$~+~heavy flavor processes from the
pretag sample.  The predicted number of background events from
$W$~+~LF processes is then calculated as:

\begin{equation}
\small
\label{wpluslf}
N_{W+\mathrm{LF}}^{\mathrm{tag}} = (
N_{\mathrm{data}}^{\mathrm{\mathrm{pretag}}} \cdot (1 -
f_{\mathrm{non-W}}^{\mathrm{pretag}}) -
N_{\mathrm{MC}}^{\mathrm{pretag}} -
N_{W+\mathrm{HF}}^{\mathrm{pretag}} ) \cdot \frac{ N_{\mathrm{mistag}}
}{ N_{\mathrm{pretag}}}.
\end{equation}

The total $Wb{\bar{b}}$, \Wccbar/$Wc$, and $W+LF$ contributions for
each tagging category are shown in Tables~\ref{tab:EventYield_2jets}
and~\ref{tab:EventYield_3jets}.

\subsection{Summary of background estimation}
\label{sec:bgsummary}

\begin{figure*}
\begin{center}
\subfigure[]{
\includegraphics[width=0.65\columnwidth]{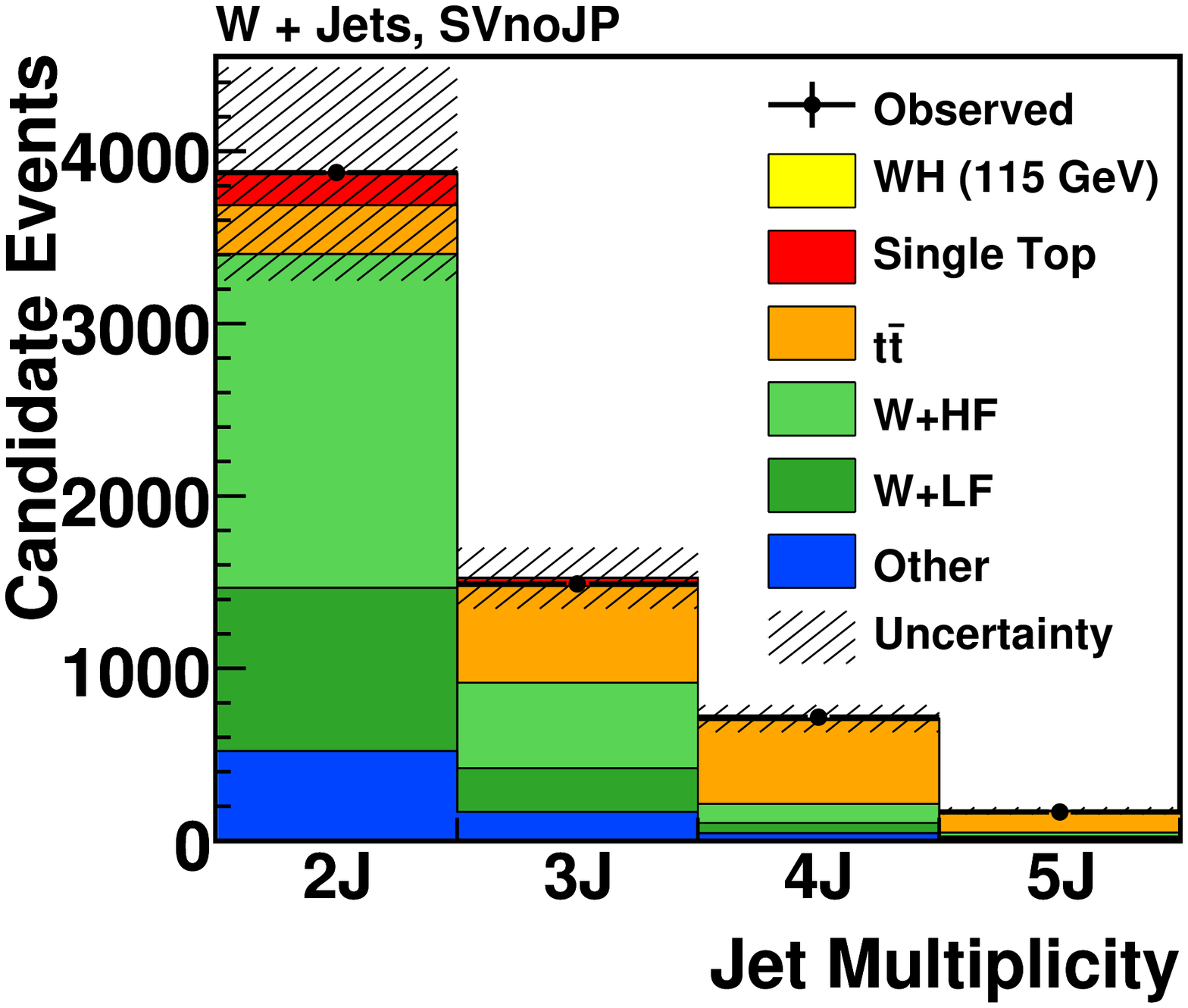}
\label{fig:Njets1}}
\subfigure[]{
\includegraphics[width=0.65\columnwidth]{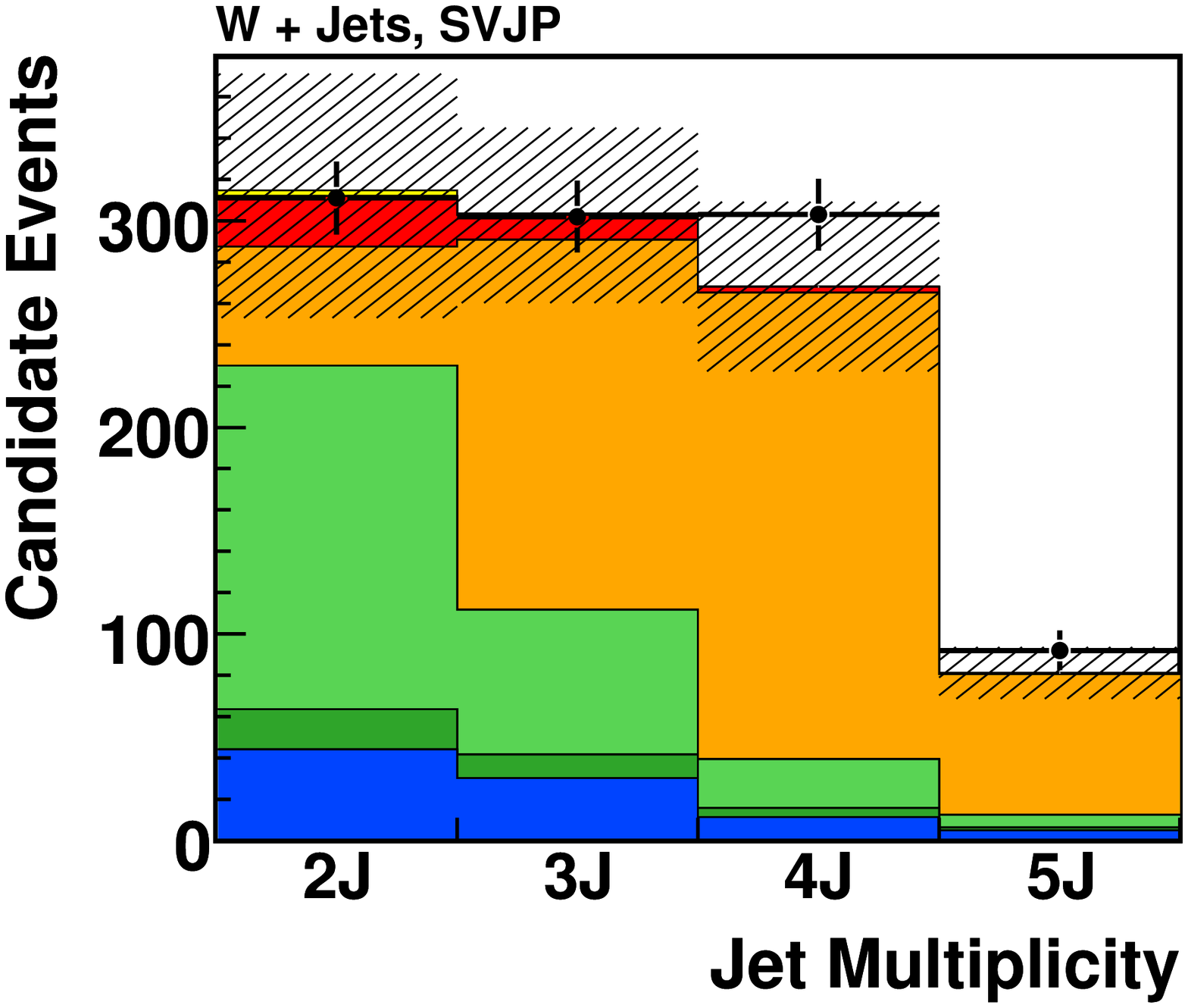}
\label{fig:Njets2}}
\subfigure[]{
\includegraphics[width=0.65\columnwidth]{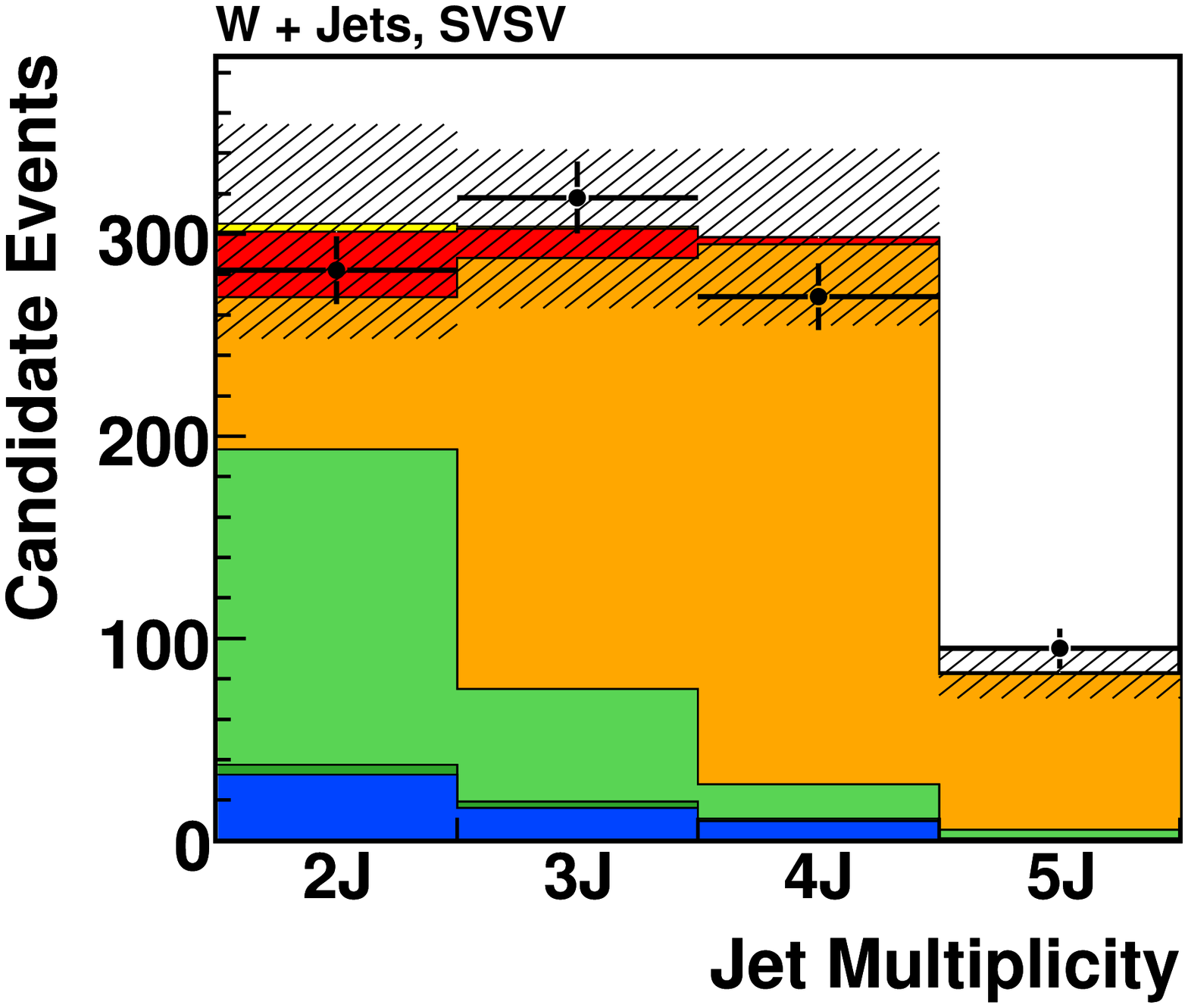}
\label{fig:Njets3}}
\end{center}
\caption{\label{fig:Njets}The predicted and observed number of events
for lepton~+~jets events.  The observed events are indicated with
points, and the shaded histograms show the signal and background
predictions which are stacked to form the total prediction. Other is
the sum of the $WW$, $WZ$, $ZZ$, and $Z$+jets contributions and $W$+HF
is the sum of the \Wbbbar, \Wccbar, and $Wc$ contributions. From left
to right: SVnoJP, SVJP, and SVSV events. }
\end{figure*}

The contributions of individual background sources have been described
in this section.  The summary of the background and signal
(m$_H$~=~115~GeV/$c^2$) estimates and the number of observed events
are shown for the three different tagging categories in
Tables~\ref{tab:EventYield_2jets} and~\ref{tab:EventYield_3jets}. The
numbers of expected and observed events are also shown in
Fig.~\ref{fig:Njets} as function of jet multiplicity. In these tables
and plots, all lepton types are combined. In general, the numbers of
expected and observed events are in good agreement within the
uncertainties on the background predictions.

\renewcommand{\arraystretch}{1.1}
\begin{table}[h]
  \caption{\label{tab:EventYield_2jets} Summary of predicted numbers
    of signal (m$_H$~=~115~GeV/$c^2$) and background $W$~+~2 jets
    events passing all the event selection requirements with
    systematic and statistical uncertainties. The total numbers of
    observed events passing the event selection are also shown.}
    \vspace{.2cm}

    \begin{tabular}{lccc} \hline \hline Process & SVSV & SVJP & SVnoJP
      \\ \hline $WW$ & 0.9$\pm$0.2 & 3.3$\pm$1.3 & 106$\pm$13 \\ $WZ$
      & 8.3$\pm$1.2 & 6.2$\pm$1.0 & 35.1$\pm$3.9 \\ $ZZ$ &
      0.30$\pm$0.05 & 0.3$\pm$0.1 & 1.4$\pm$0.2 \\ $t\bar t$
      (lepton+jets) & 47.0$\pm$7.8 & 37.6$\pm$6.8 & 205$\pm$29 \\
      $t\bar t$ (dilepton) & 28.2$\pm$4.6 & 20.0$\pm$3.4 & 77$\pm$11
      \\ Single top (t-channel) & 6.3$\pm$1.1 & 6.3$\pm$1.3 &
      116$\pm$17 \\ Single top (s-channel) & 26.2$\pm$4.3 &
      18.4$\pm$3.1 & 66.0$\pm$9.1 \\ $Z$+jets & 4.2$\pm$0.7 &
      5.1$\pm$1.3 & 80$\pm$12 \\ \Wbbbar & 142$\pm$46 & 121$\pm$39 &
      978$\pm$295 \\ \Wccbar/$Wc$ & 13.8$\pm$4.7 & 46$\pm$17 &
      959$\pm$296 \\ $W+LF$ & 4.7$\pm$1.5 & 19$\pm$11 & 946$\pm$138 \\
      Non-$W$ & 19.0$\pm$7.6 & 29$\pm$12 & 298$\pm$119 \\ \hline Total
      prediction & 301$\pm$53 & 312$\pm$59 & 3869$\pm$619 \\ \hline
      $WH$ (115 GeV/$c^2$) & 4.06$\pm$0.48 & 2.80$\pm$0.35 &
      10.27$\pm$0.81 \\ \hline Observed & 282 & 311 & 3878 \\ \hline
      \hline \end{tabular}
\end{table}

\begin{table}[h]
\caption{\label{tab:EventYield_3jets} Summary of
  predicted numbers of signal (m$_H$~=~115~GeV/$c^2$) and background
  $W$~+~3 jets events passing all the event selection requirements
  with systematic and statistical uncertainties. The total numbers of
  observed events passing the event selection are also shown.}
  \vspace{.2cm}

     \begin{tabular}{lccc} 
\hline \hline Process & SVSV
   & SVJP & SVnoJP \\ \hline $WW$ & 1.0$\pm$0.2 & 2.6$\pm$0.9 &
   32.8$\pm$4.0 \\ $WZ$ & 2.3$\pm$0.3 & 1.9$\pm$0.4 & 9.4$\pm$1.1 \\
   $ZZ$ & 0.19$\pm$0.03 & 0.15$\pm$0.03 & 0.6$\pm$0.1 \\ $t\bar t$
   (lepton+jets) & 188$\pm$31 & 161$\pm$29 & 504$\pm$70 \\ $t\bar t$
   (dilepton) & 25.4$\pm$4.1 & 18.2$\pm$3.1 & 57.6$\pm$8.0 \\ Single
   top (t-channel) & 5.6$\pm$0.9 & 5.0$\pm$0.9 & 26.1$\pm$3.7 \\
   Single top (s-channel) & 8.9$\pm$1.5 & 6.8$\pm$1.2 & 19.5$\pm$2.7
   \\ $Z$+jets & 3.0$\pm$0.5 & 4.0$\pm$1.1 & 29.7$\pm$4.4 \\ \Wbbbar &
   49$\pm$16 & 47$\pm$16 & 258$\pm$78 \\ \Wccbar/$Wc$ & 7.1$\pm$2.5 &
   22.9$\pm$8.6 & 237$\pm$73 \\ $W+LF$ & 3.2$\pm$1.1 & 11.3$\pm$5.9 &
   255$\pm$38 \\ Non-$W$ & 9.6$\pm$3.9 & 21.5$\pm$8.6 & 93$\pm$37 \\
   \hline Total prediction & 303$\pm$39 & 303$\pm$42 & 1522$\pm$177 \\
   \hline $WH$ (115 GeV/$c^2$) & 1.16$\pm$0.14 & 0.85$\pm$0.12 &
   2.57$\pm$0.21 \\ \hline Observed & 318 & 302 & 1491 \\ \hline
   \hline \end{tabular}
\end{table}

\subsection{Validation of the background model}
\label{sec:bgvalidation}
\begin{figure*}
\begin{center}
\subfigure[]{
\includegraphics[width=0.65\columnwidth]{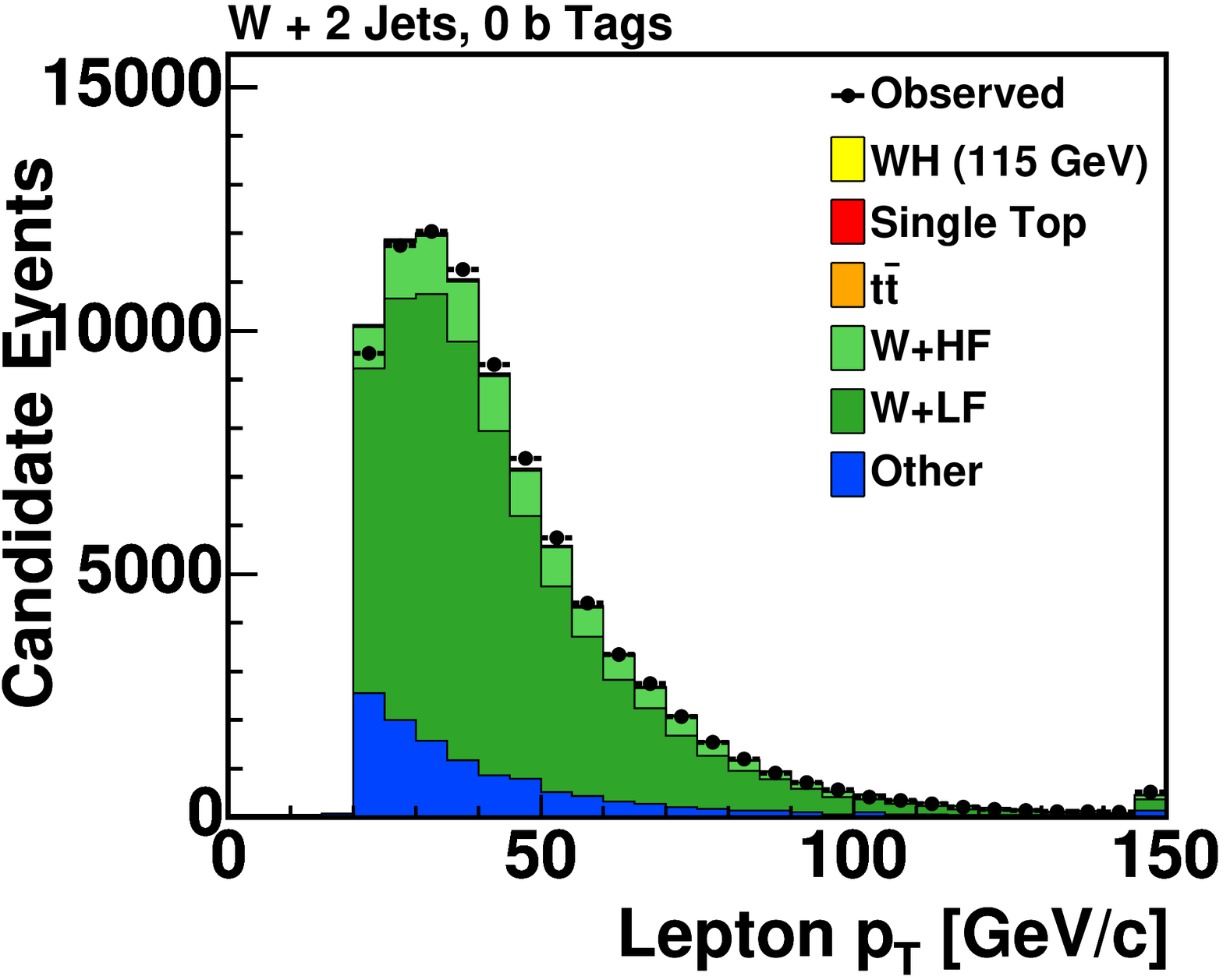}
\label{fig:LepPt2jet}}
\subfigure[]{
\includegraphics[width=0.65\columnwidth]{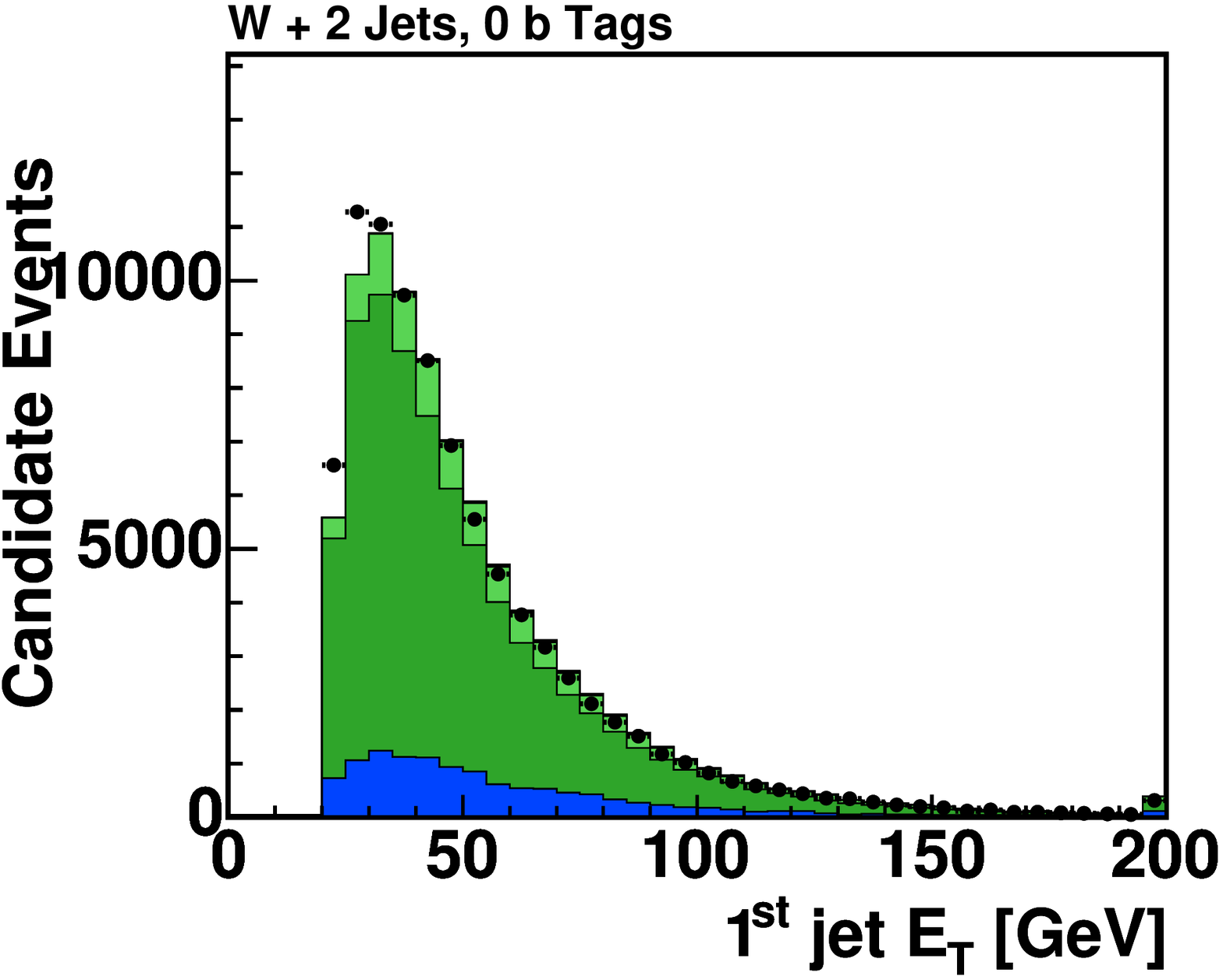}
\label{fig:LepEta2jet}}
\subfigure[]{
\includegraphics[width=0.65\columnwidth]{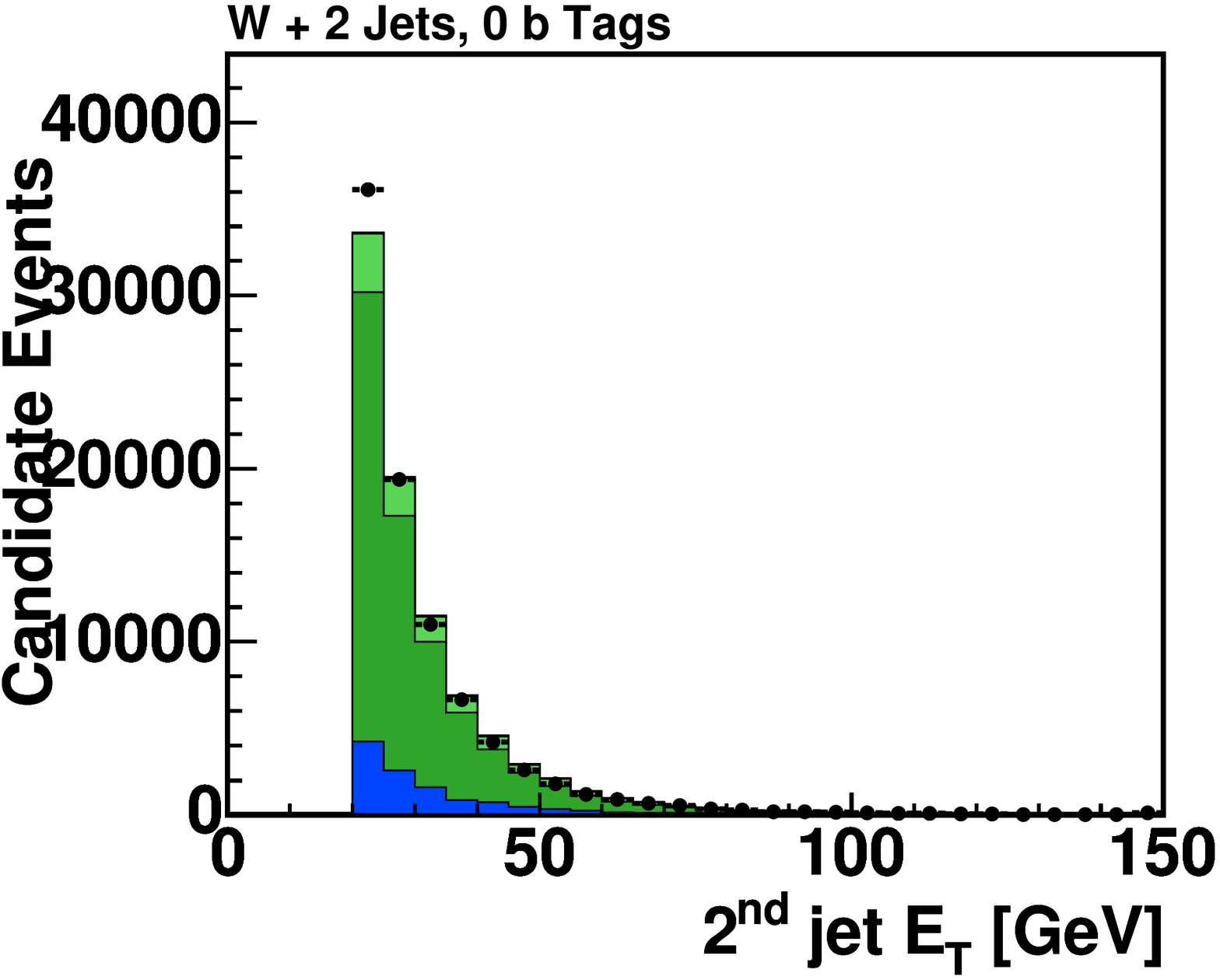}
\label{fig:J1Et2jet}}
\subfigure[]{
\includegraphics[width=0.65\columnwidth]{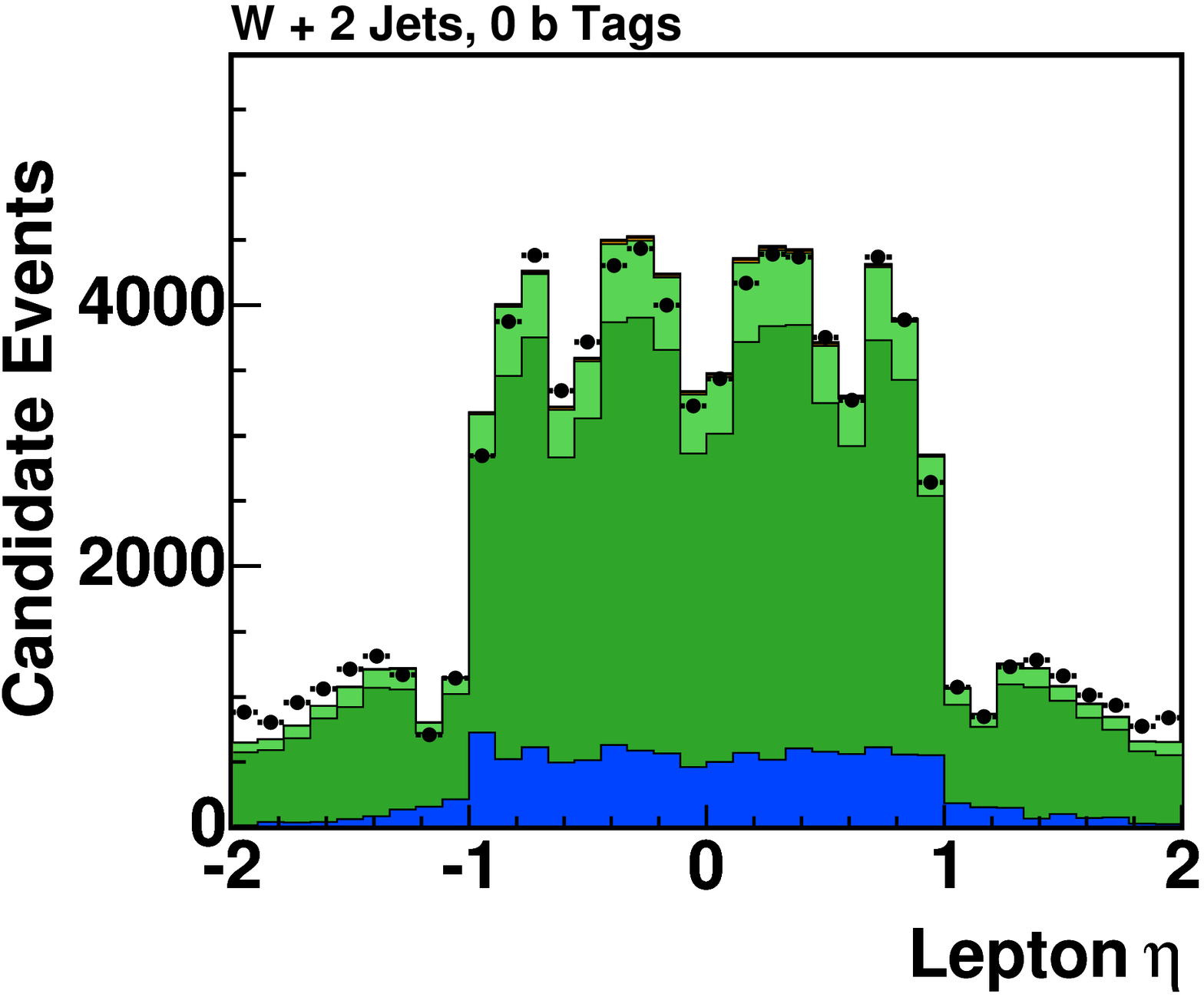}
\label{fig:J1Eta2jet}}
\subfigure[]{
\includegraphics[width=0.65\columnwidth]{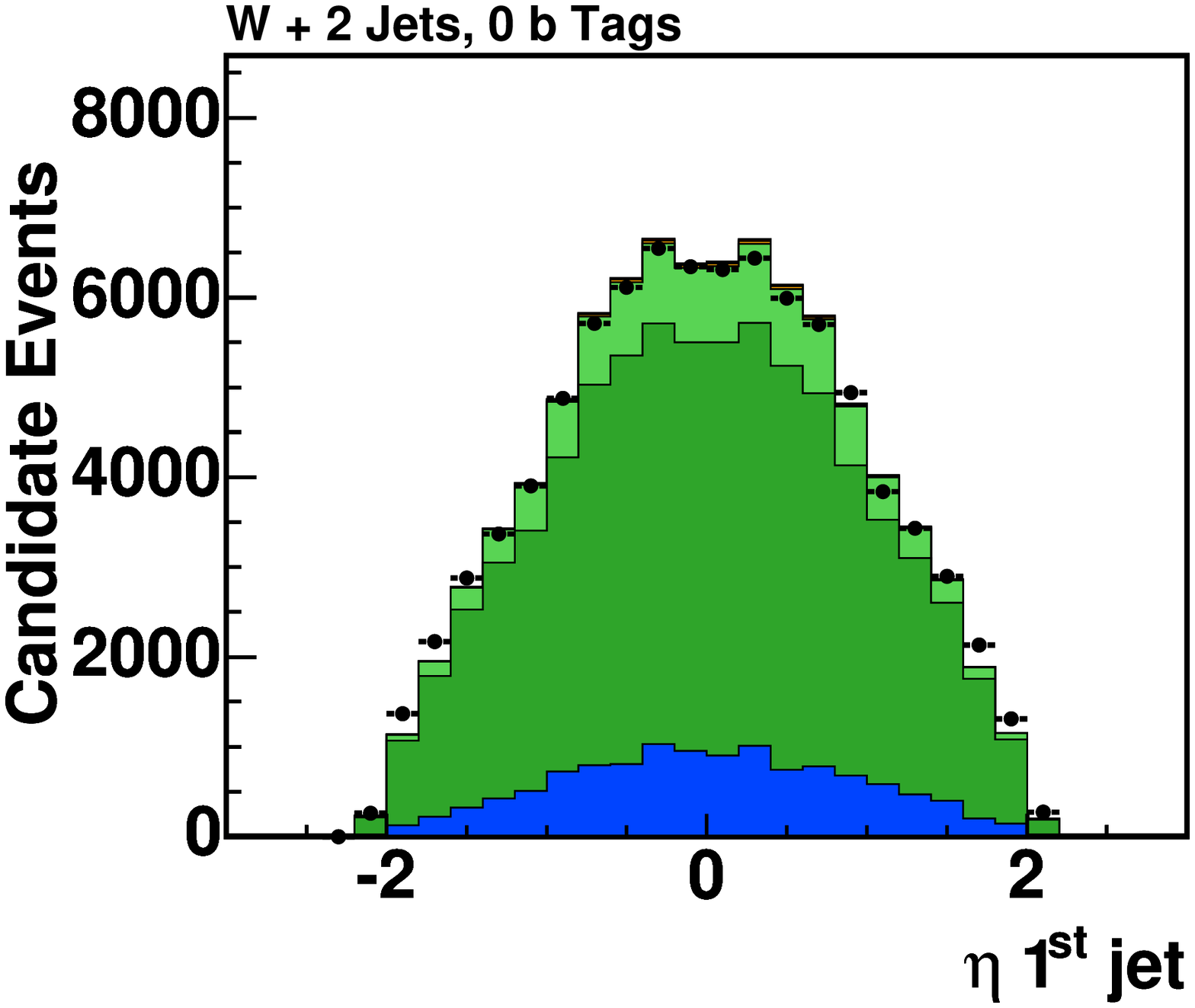}
\label{fig:J2ET2jet}}
\subfigure[]{
\includegraphics[width=0.65\columnwidth]{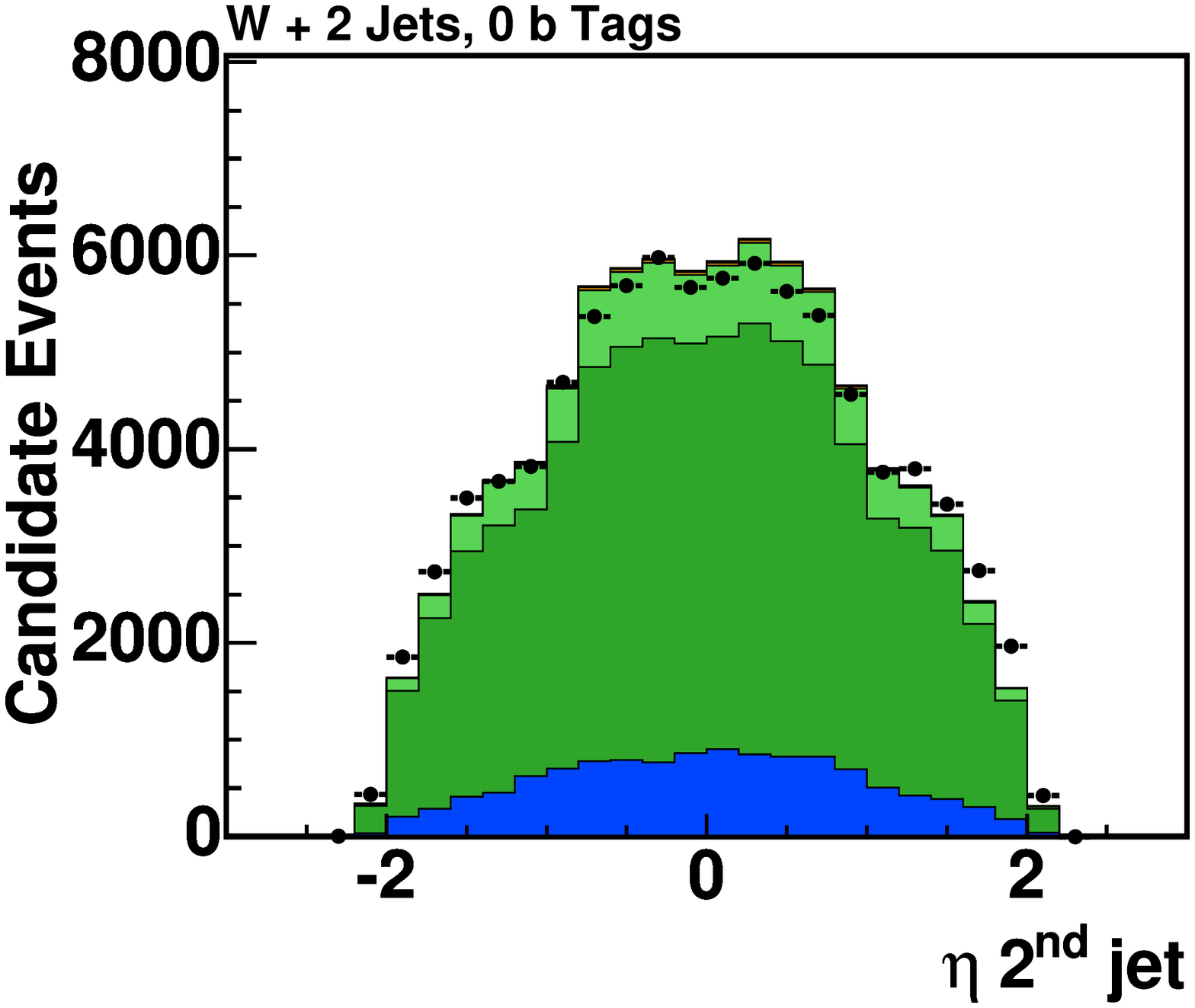}
\label{fig:J2Eta2jet}}

\subfigure[]{
\includegraphics[width=0.65\columnwidth]{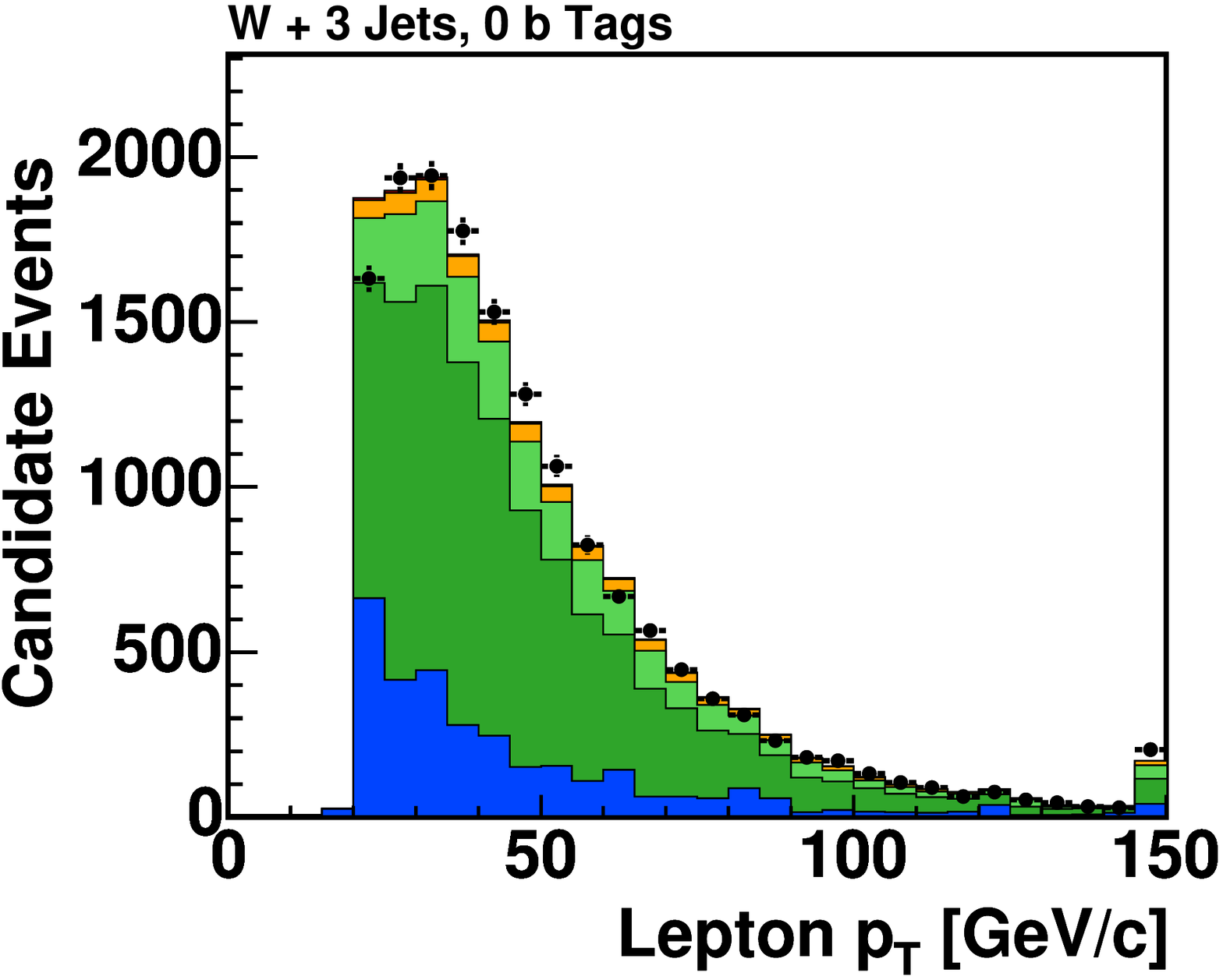}
\label{fig:LepPt3jet}}
\subfigure[]{
\includegraphics[width=0.65\columnwidth]{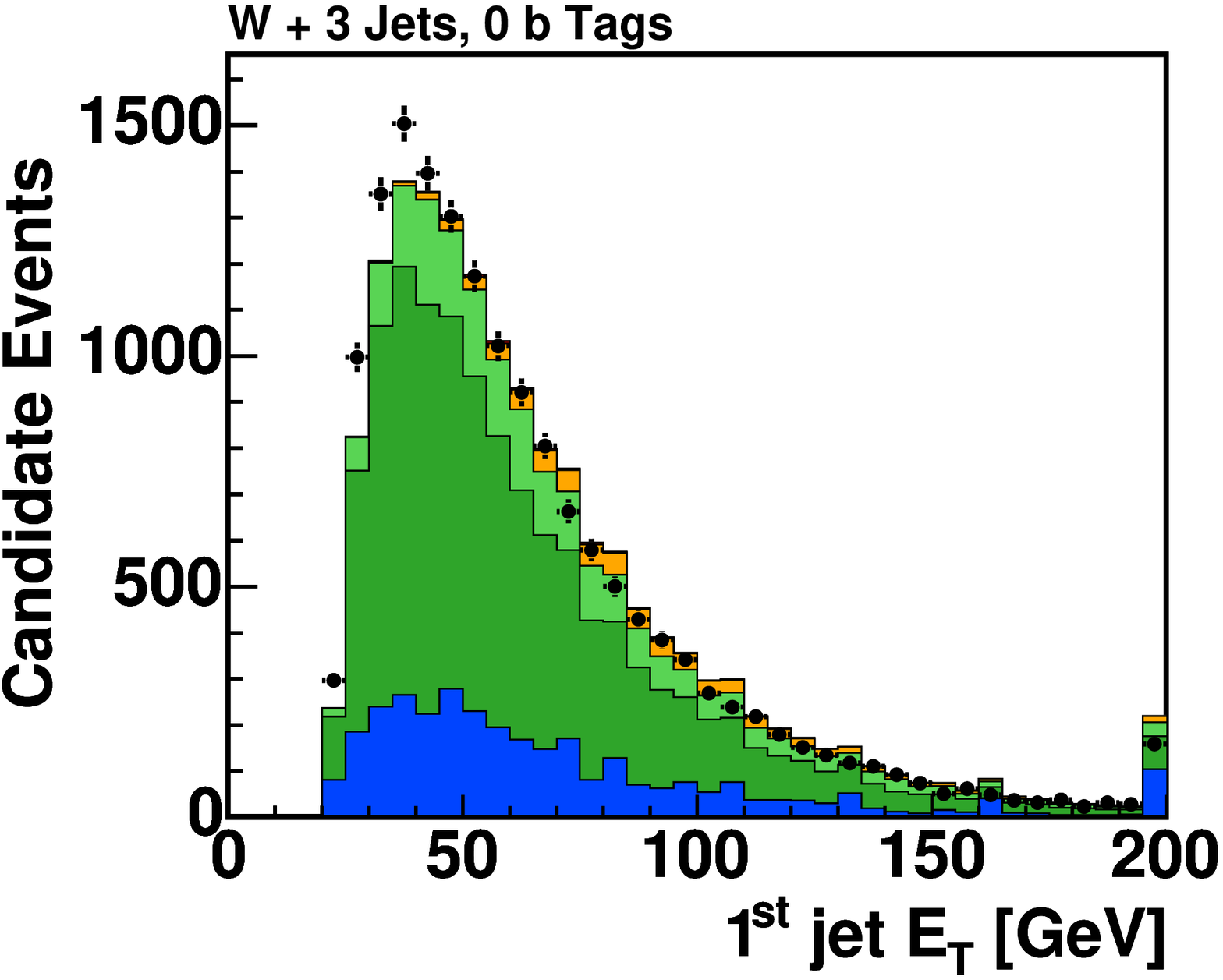}
\label{fig:LepEta3jet}}
\subfigure[]{
\includegraphics[width=0.65\columnwidth]{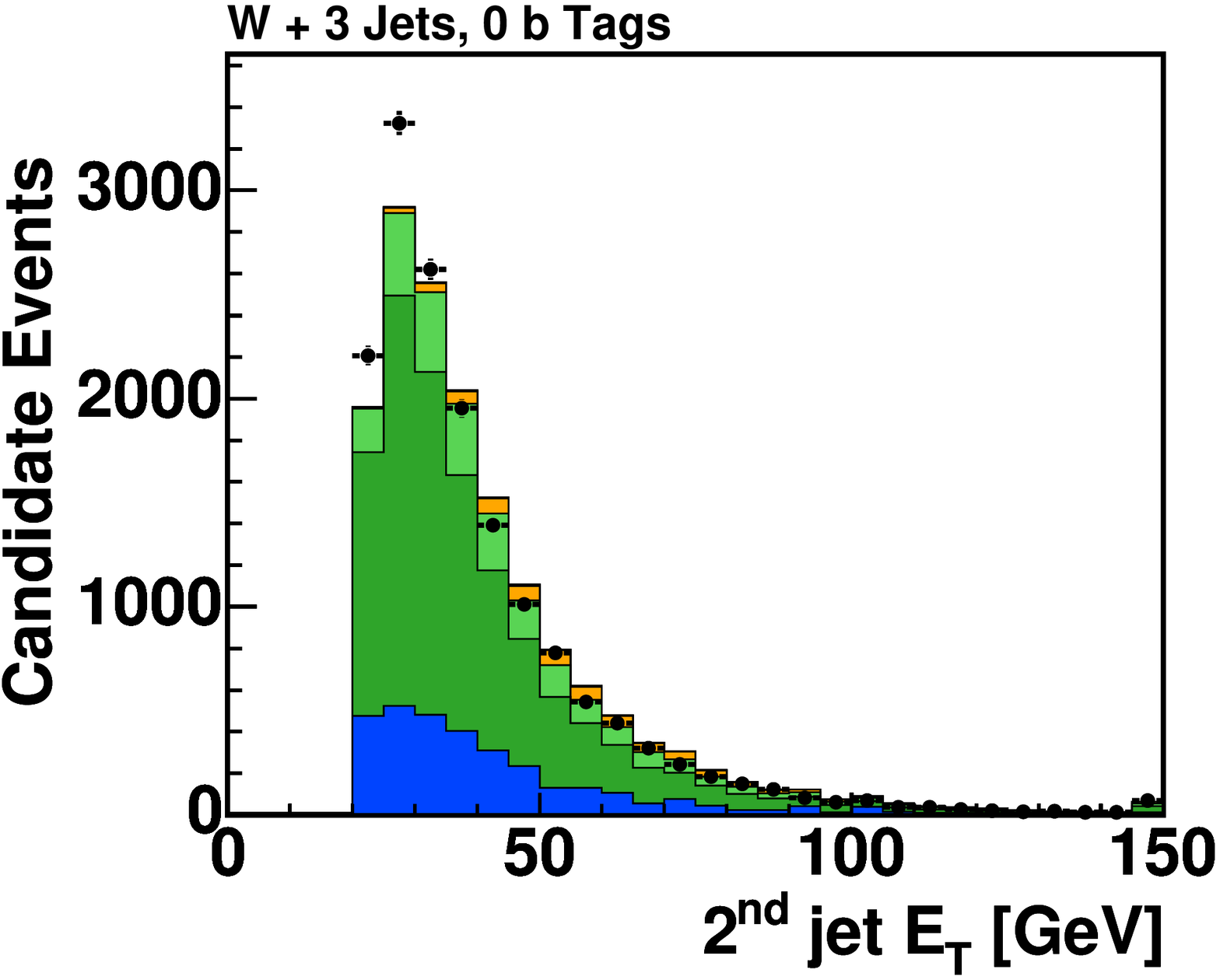}
\label{fig:J1Et3jet}}
\subfigure[]{
\includegraphics[width=0.65\columnwidth]{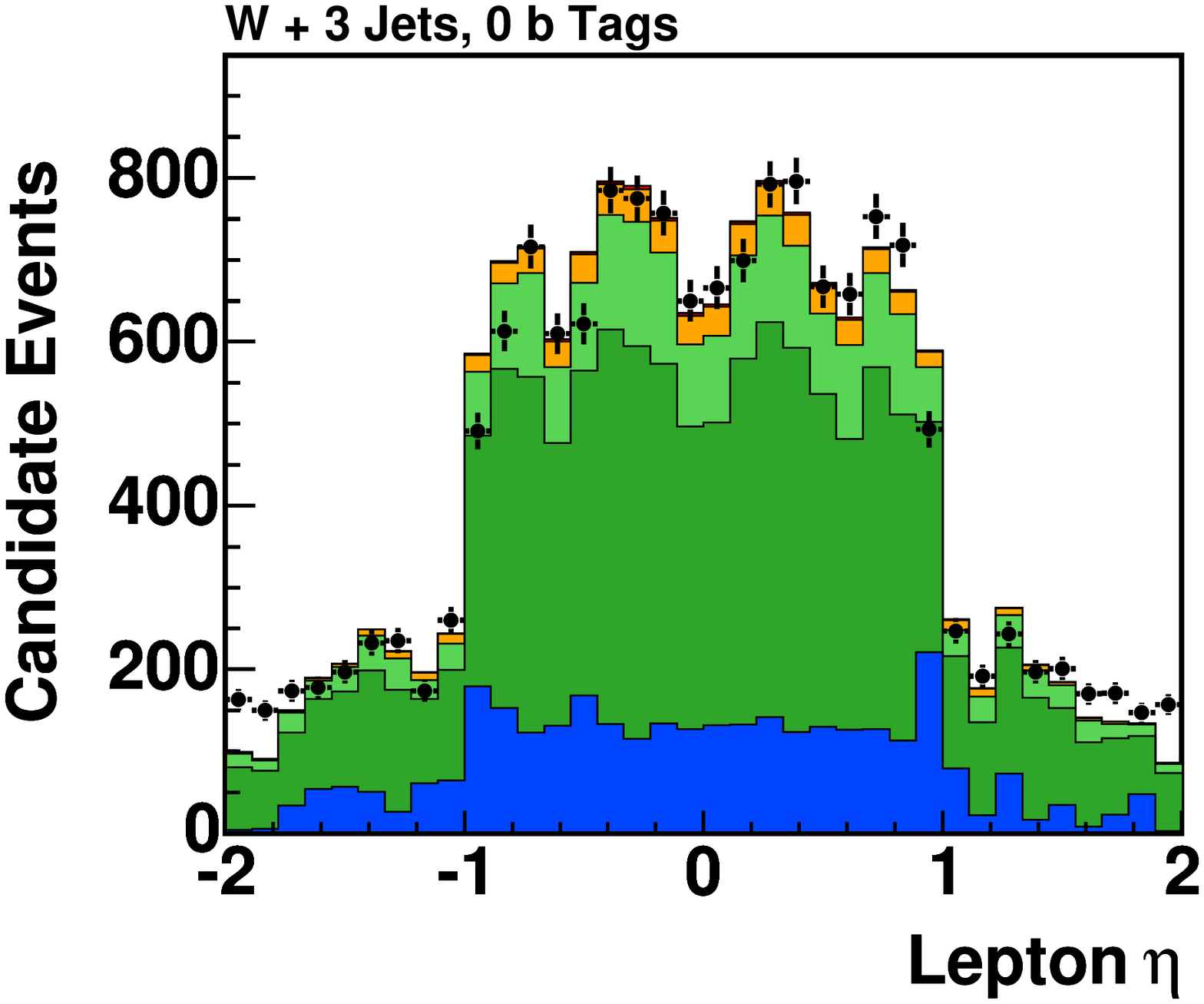}
\label{fig:J1Eta2jet}}
\subfigure[]{
\includegraphics[width=0.65\columnwidth]{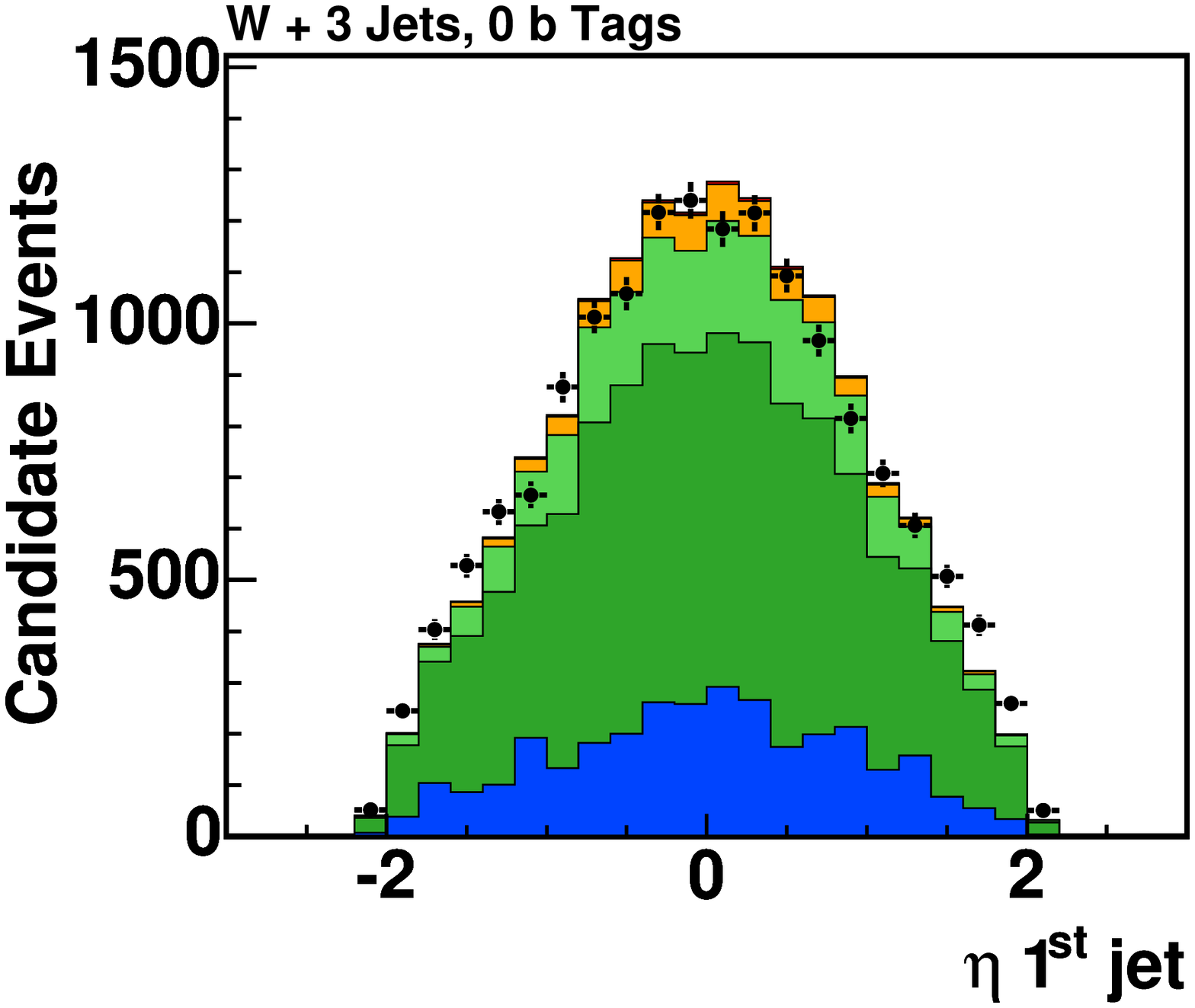}
\label{fig:J2Eta3jet}}
\subfigure[]{
\includegraphics[width=0.65\columnwidth]{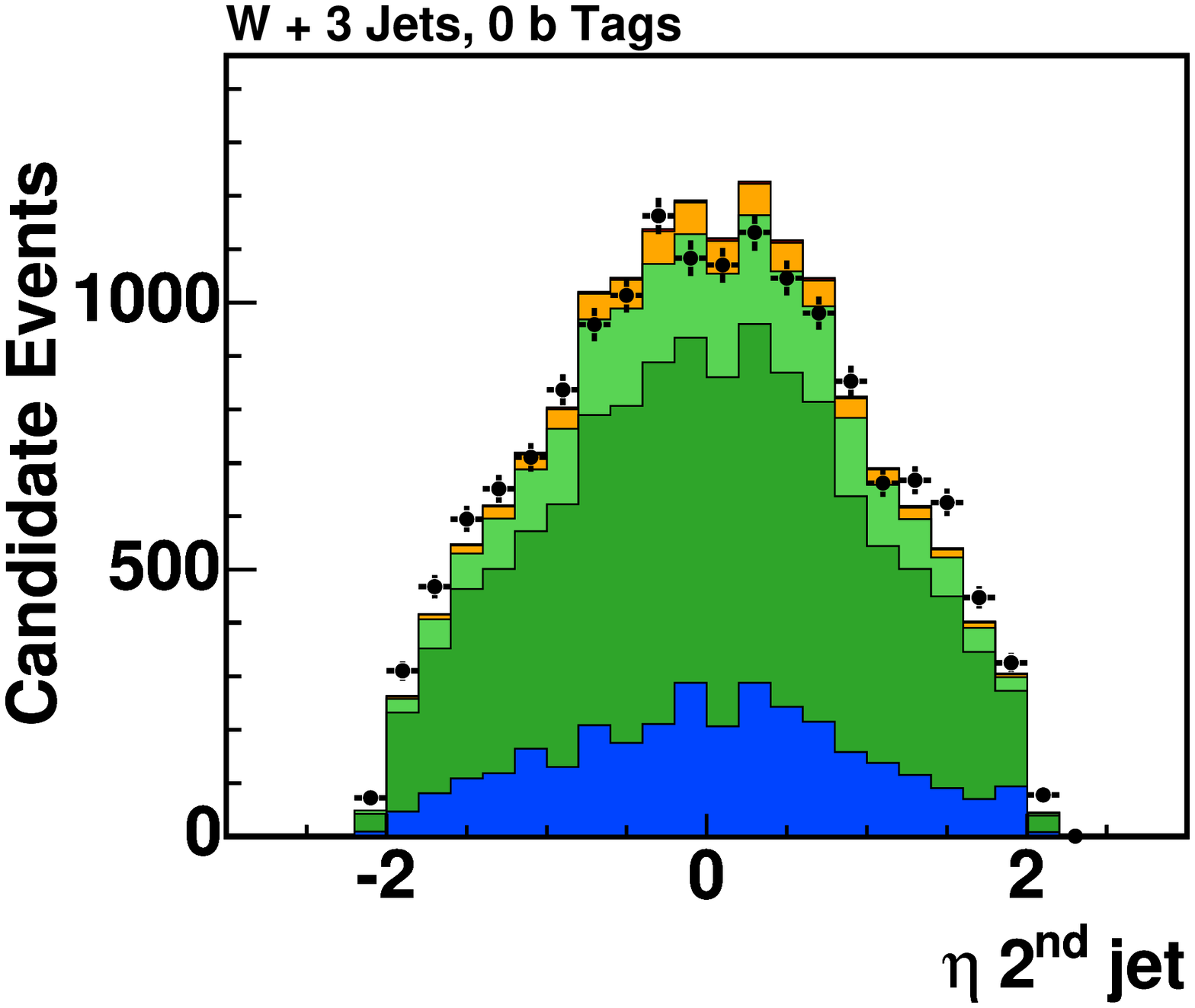}
\label{fig:J2ET3jet}}
\end{center}
\vspace{-0.85cm}\caption{\label{fig:Input_validation_3jet_0t}Validation
plots comparing observed events and Monte Carlo distributions for
basic kinematic quantities for events with two (a-f) and three (g-l)
jets and no $b$~tags.  The observed events are indicated with
points. }
\end{figure*}

\begin{figure*}
\begin{center}
\subfigure[]{
\includegraphics[width=0.65\columnwidth]{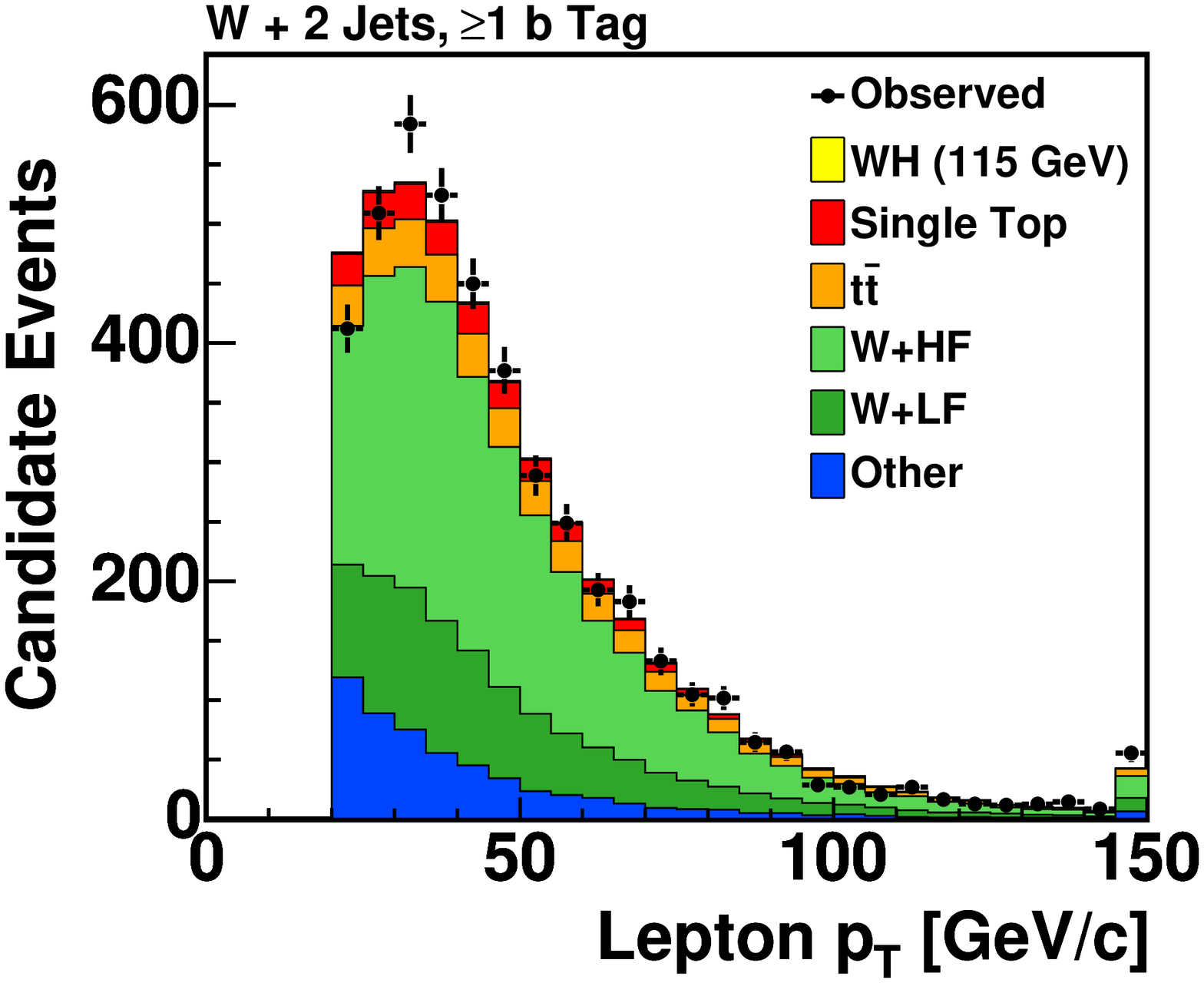}
\label{fig:LepPt2jet_gr1tag}}
\subfigure[]{
\includegraphics[width=0.65\columnwidth]{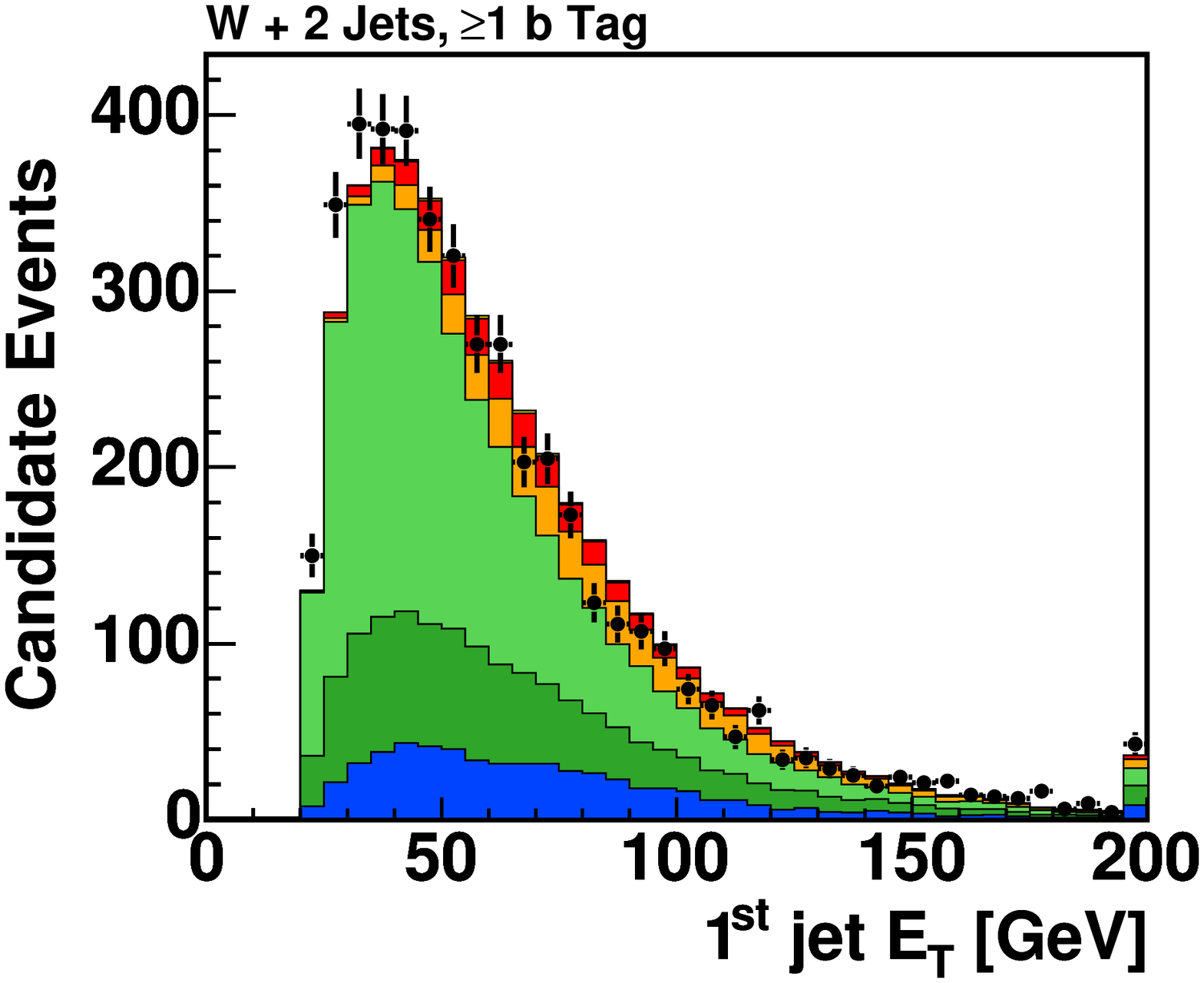}
\label{fig:LepEta2jet_gr1tag}}
\subfigure[]{
\includegraphics[width=0.65\columnwidth]{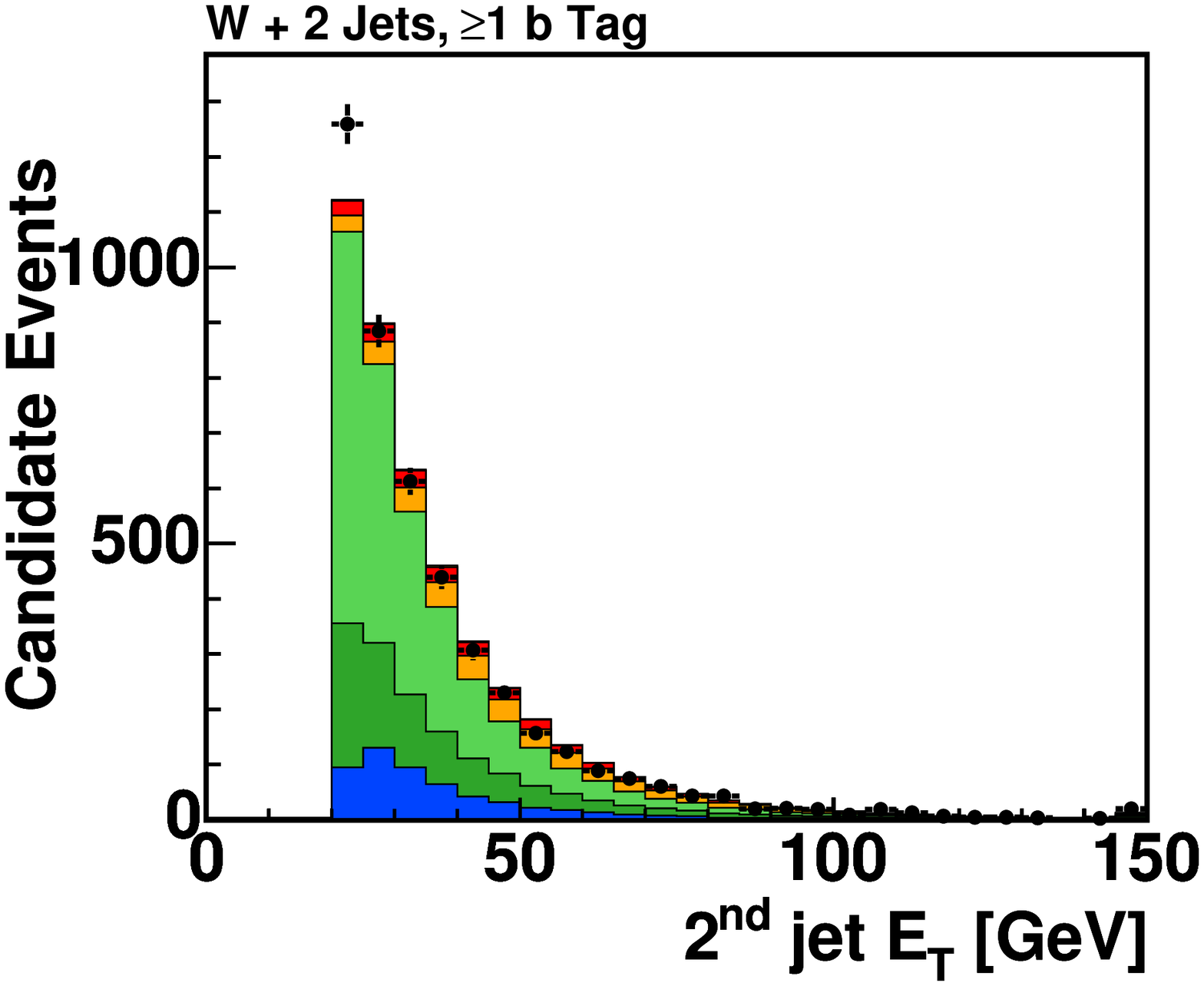}
\label{fig:J1Et2jet_gr1tag}}
\subfigure[]{
\includegraphics[width=0.65\columnwidth]{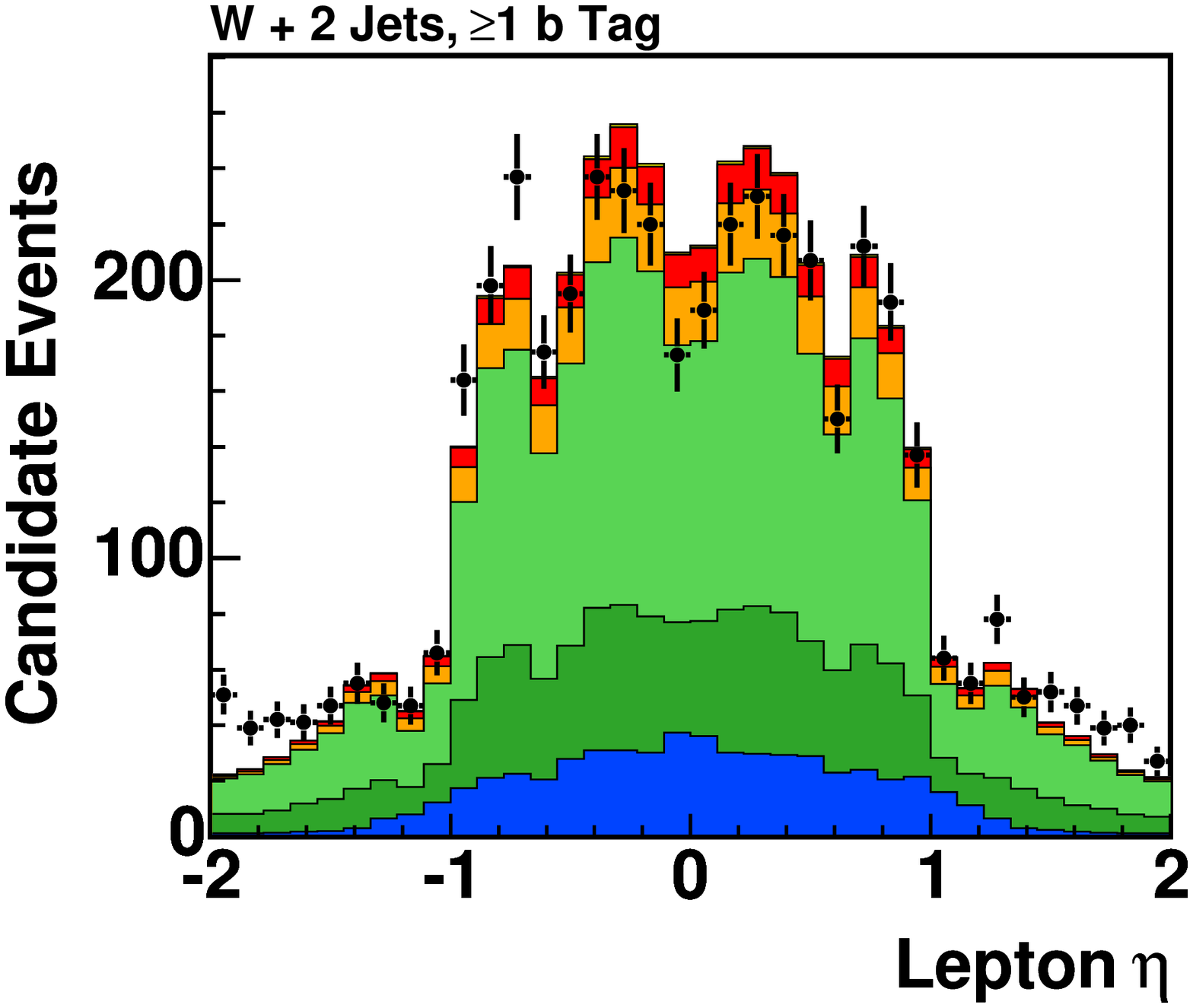}
\label{fig:J1Eta2jet_gr1tag}}
\subfigure[]{
\includegraphics[width=0.65\columnwidth]{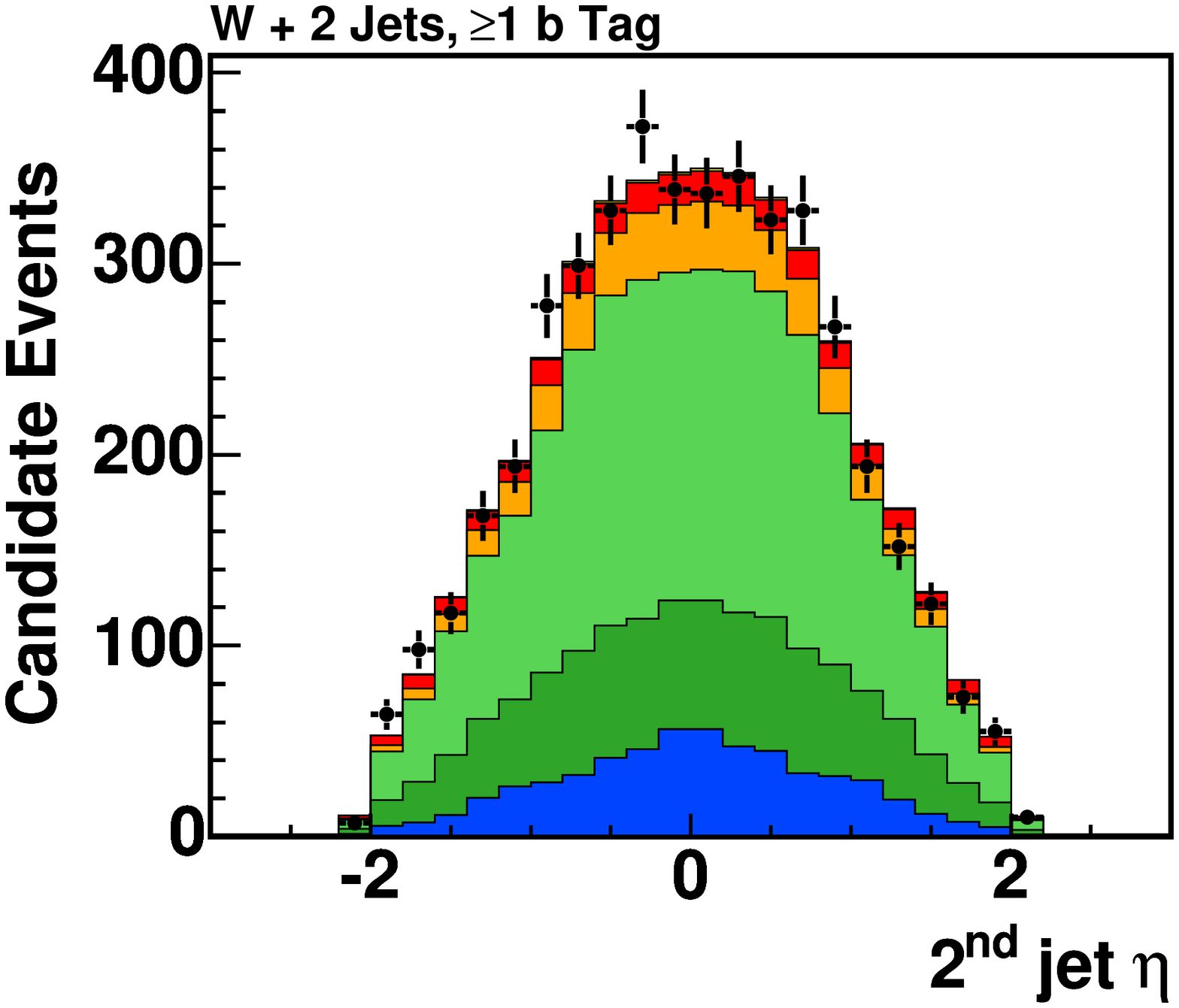}
\label{fig:J2ET2jet_gr1tag}}
\subfigure[]{
\includegraphics[width=0.65\columnwidth]{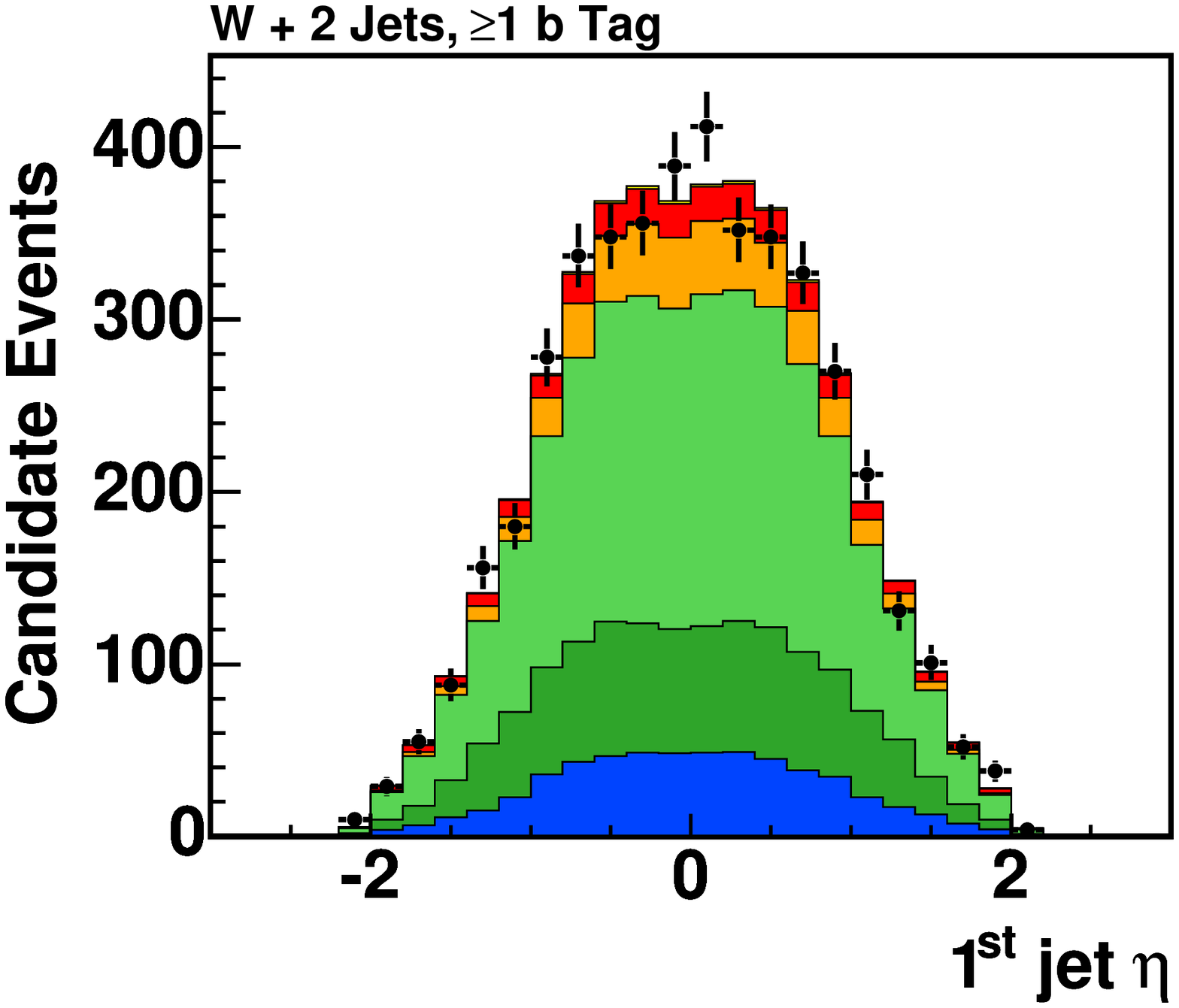}
\label{fig:J2Eta2jet_gr1tag}}

\subfigure[]{
\includegraphics[width=0.65\columnwidth]{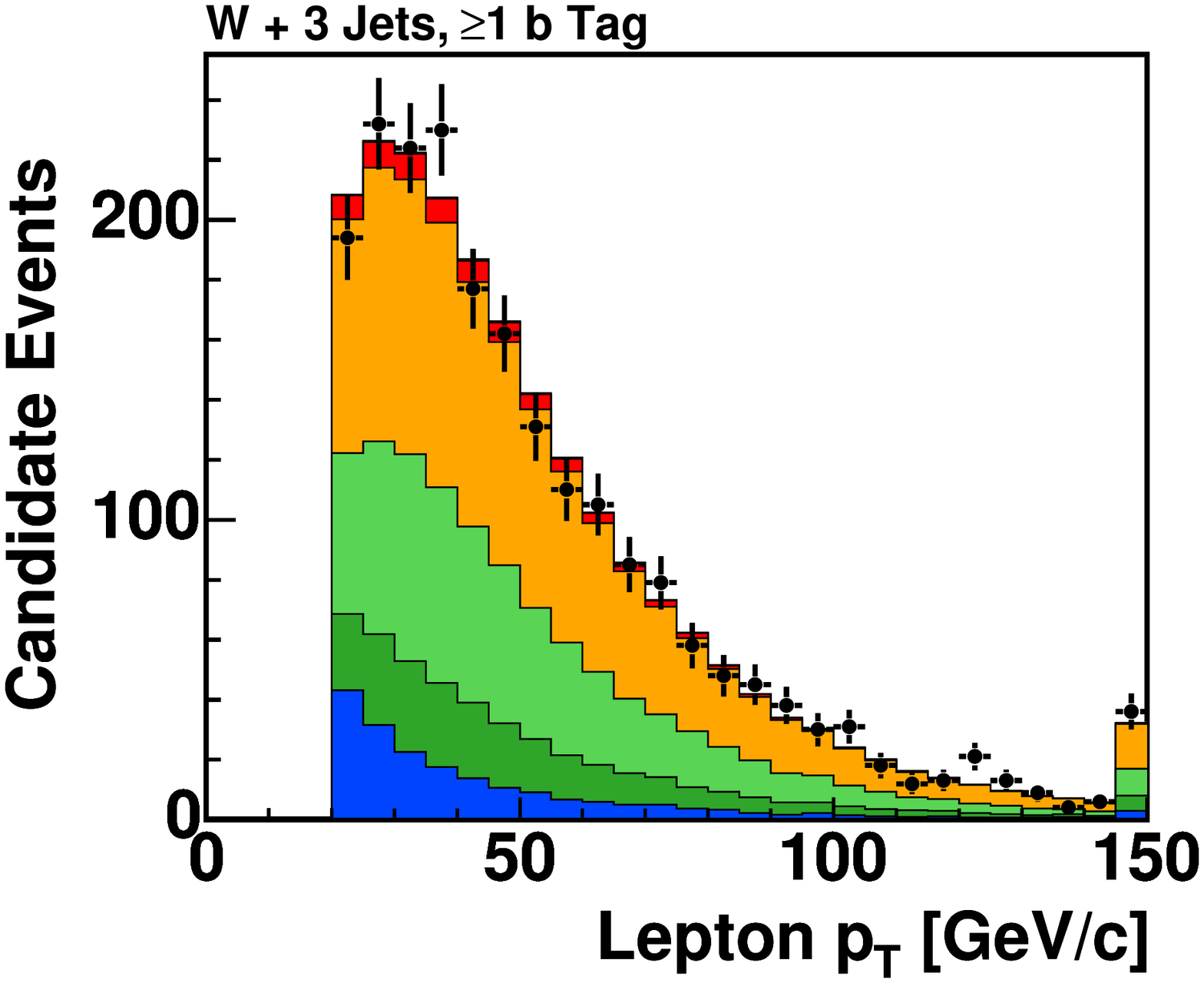}
\label{fig:LepPt3jet_gr1tag}}
\subfigure[]{
\includegraphics[width=0.65\columnwidth]{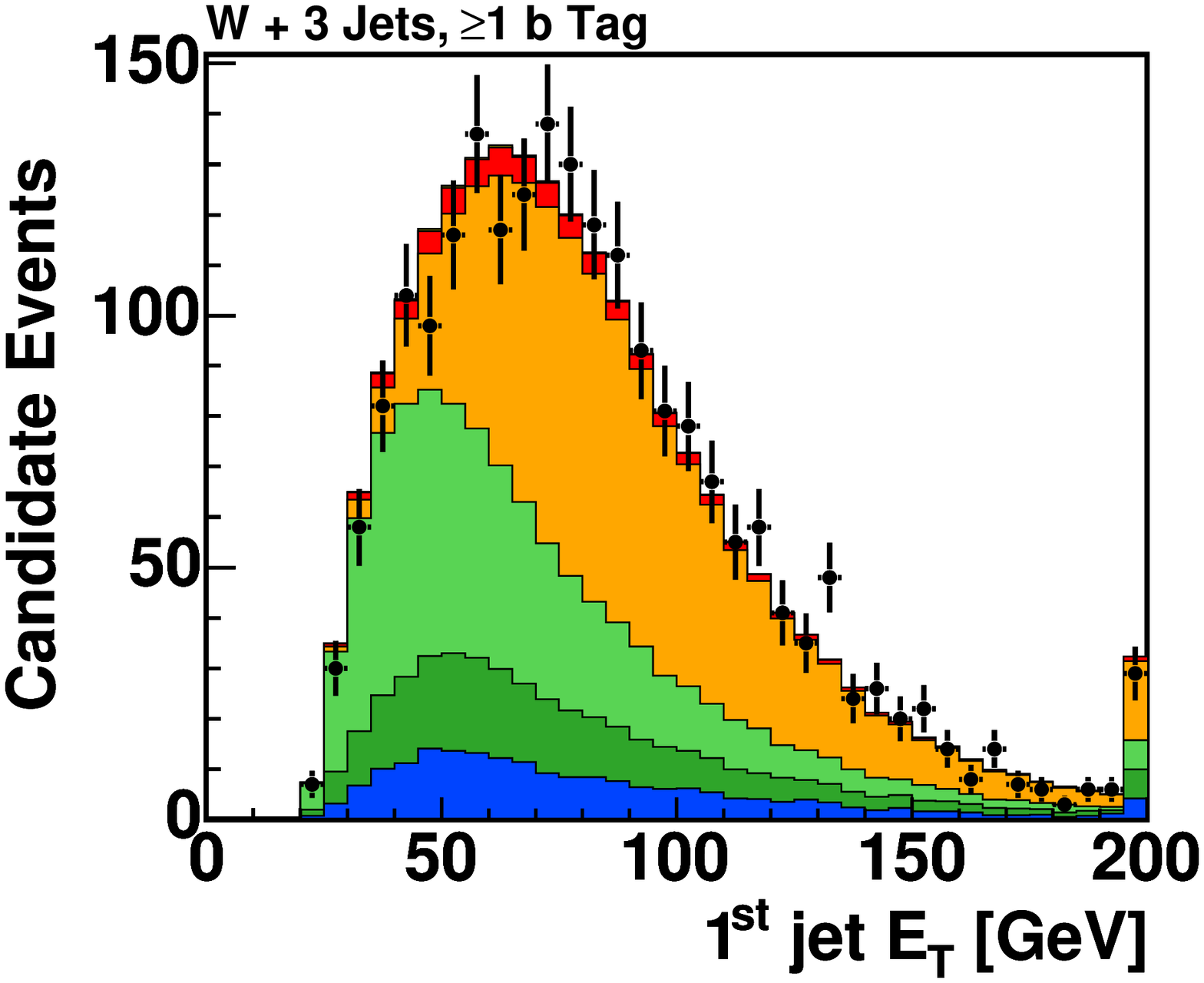}
\label{fig:LepEta3jet_gr1tag}}
\subfigure[]{
\includegraphics[width=0.65\columnwidth]{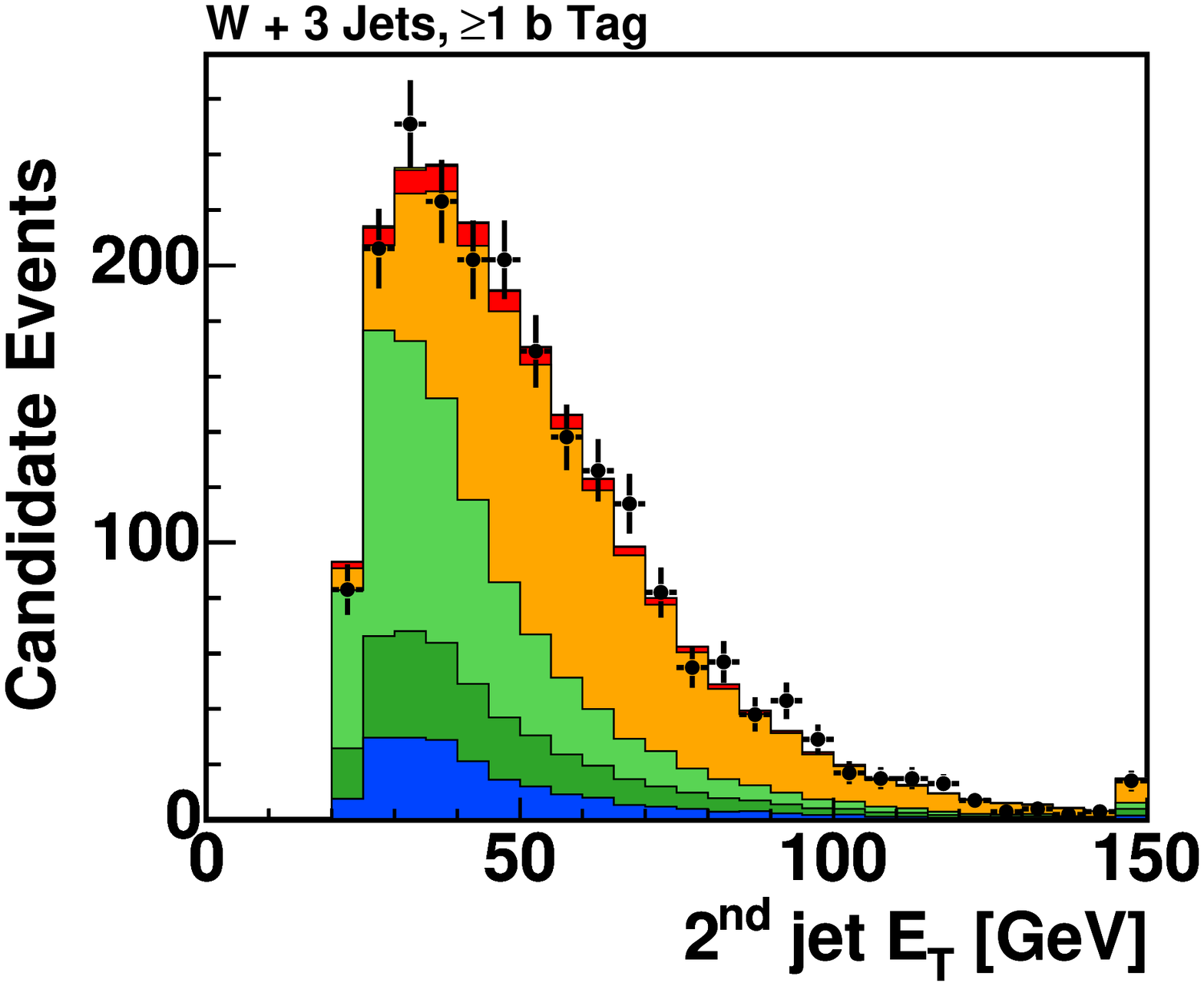}
\label{fig:J1Et3jet_gr1tag}}
\subfigure[]{
\includegraphics[width=0.65\columnwidth]{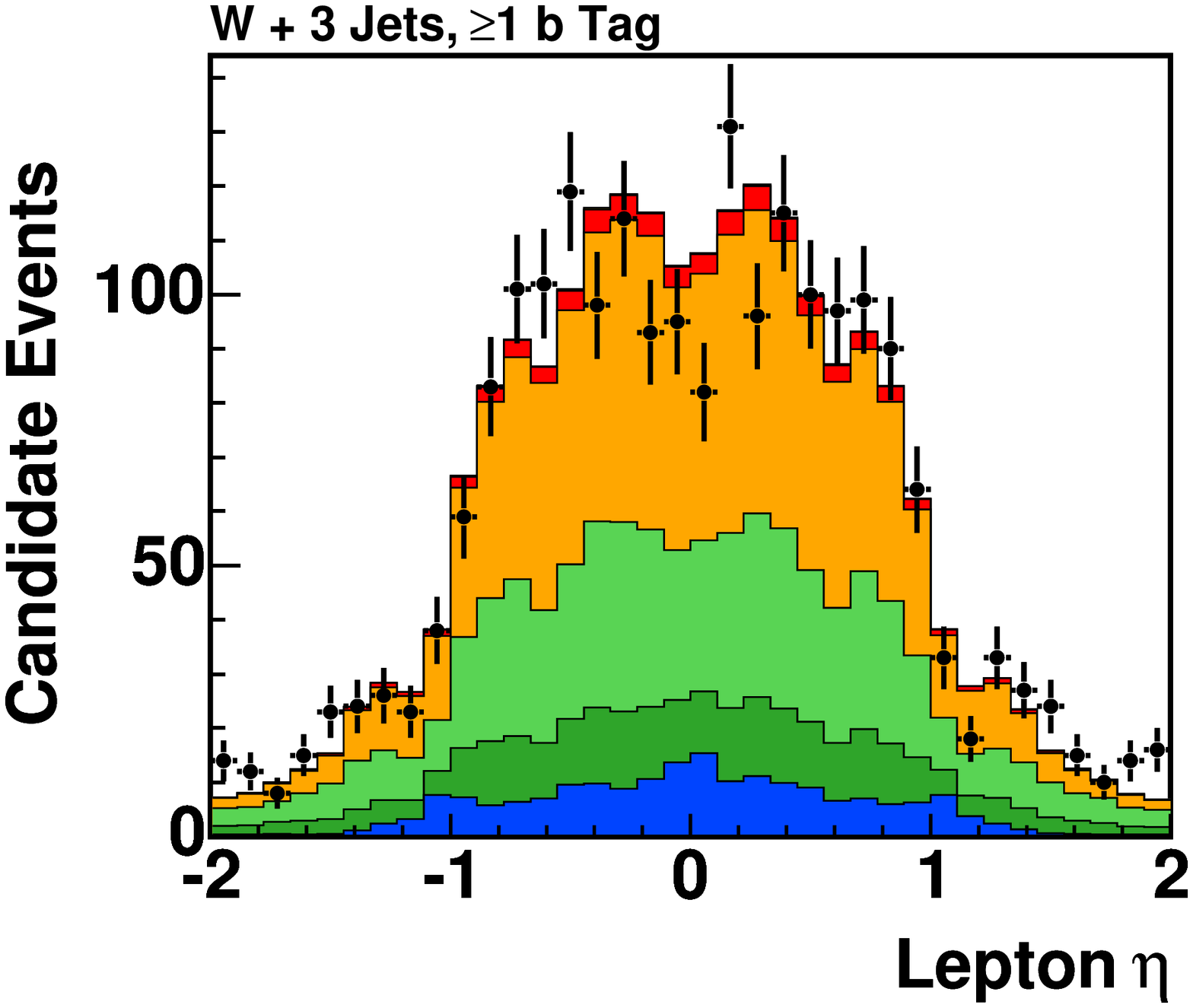}
\label{fig:J1Eta3jet_gr1tag}}
\subfigure[]{
\includegraphics[width=0.65\columnwidth]{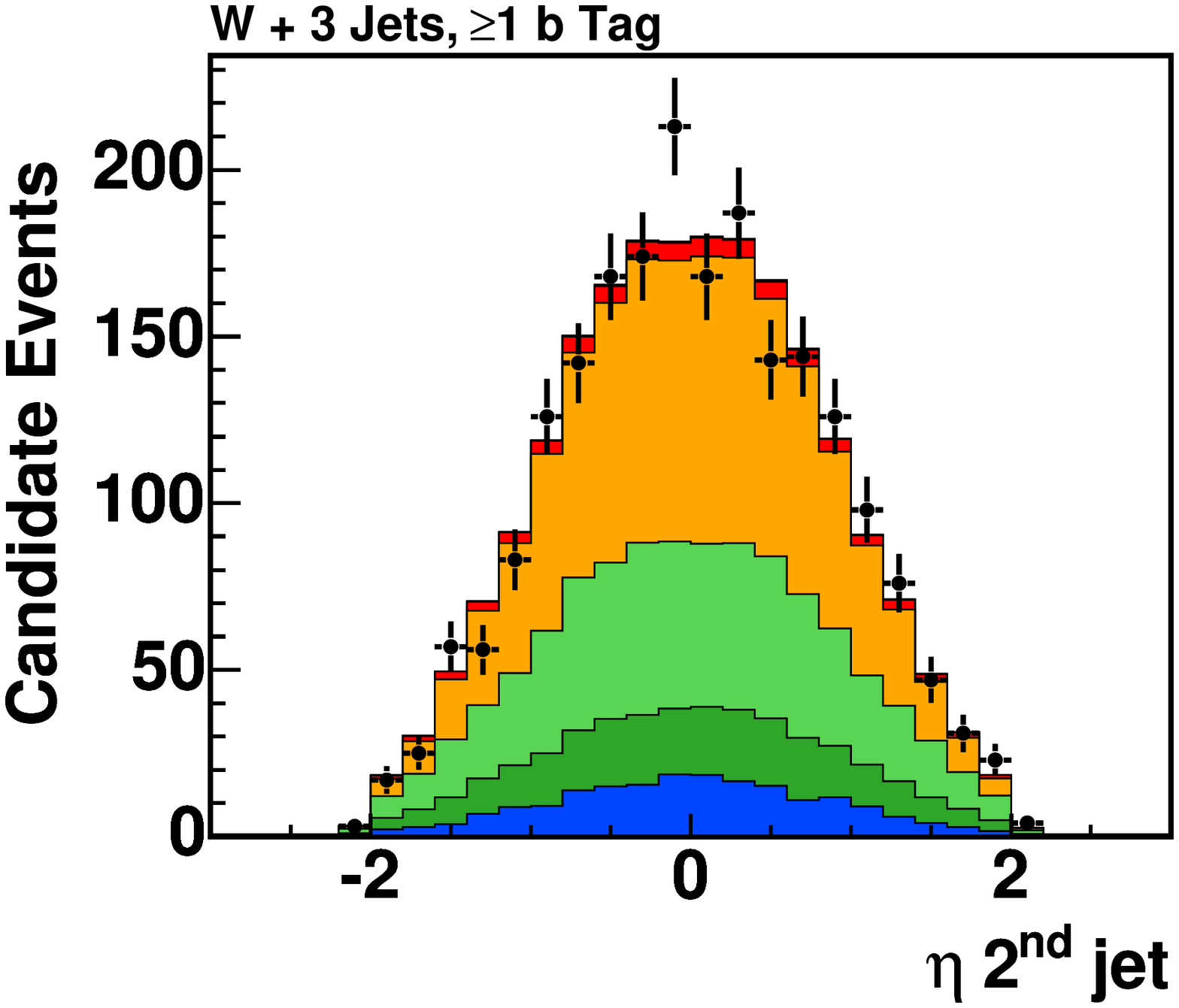}
\label{fig:J2ET3jet_gr1tag}}
\subfigure[]{
\includegraphics[width=0.65\columnwidth]{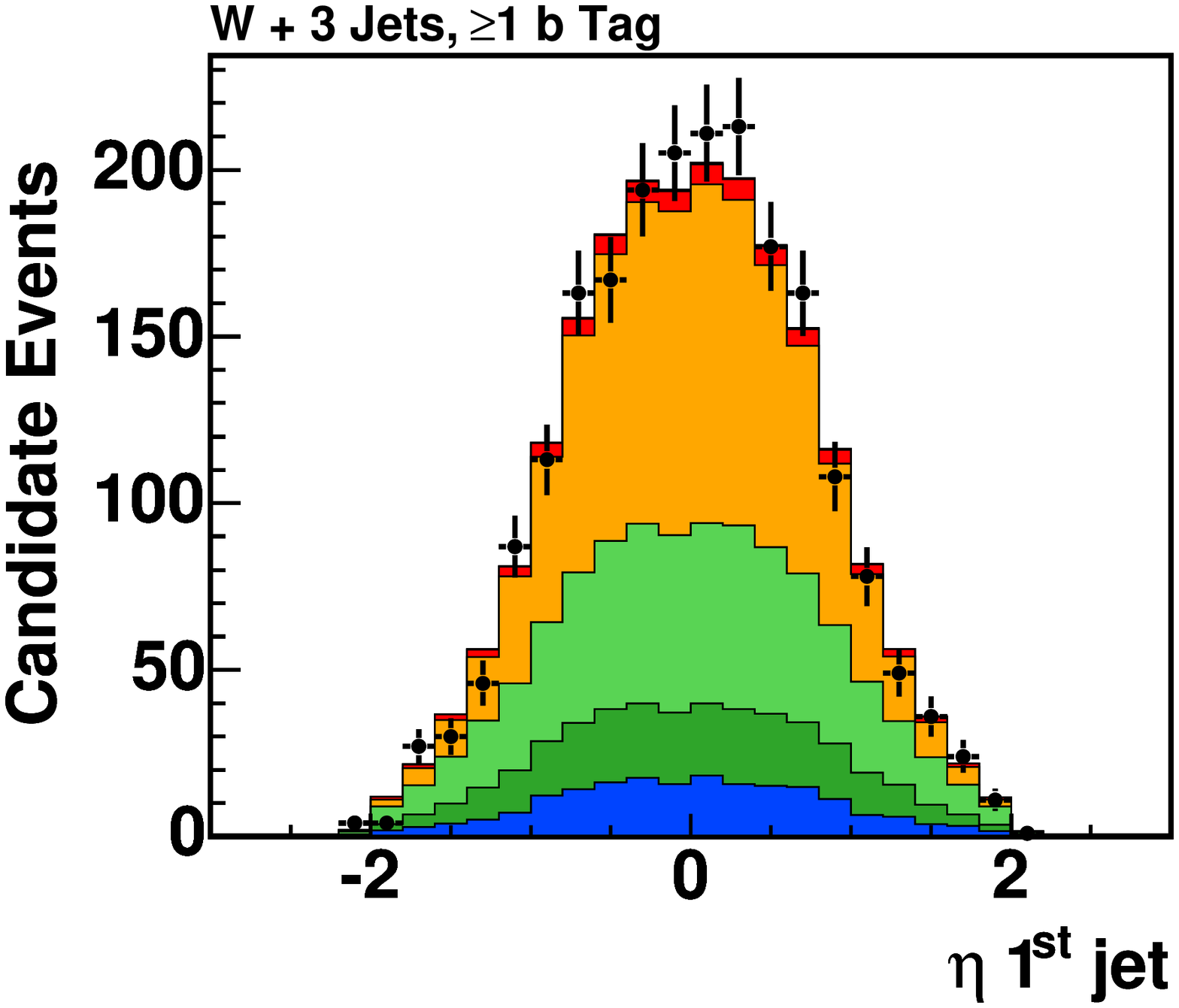}
\label{fig:J2Eta3jet_gr1tag}}
\end{center}
\vspace{-0.85cm}\caption{\label{fig:Input_validation_3jet}Validation
plots comparing observed events and Monte Carlo distributions for
basic kinematic quantities for events with two (a-f) and three (g-l)
jets and at least one $b$~tag.  The observed events are indicated with
points.}
\end{figure*}

\begin{figure*}
\begin{center}
\subfigure[]{
\includegraphics[width=0.7\columnwidth]{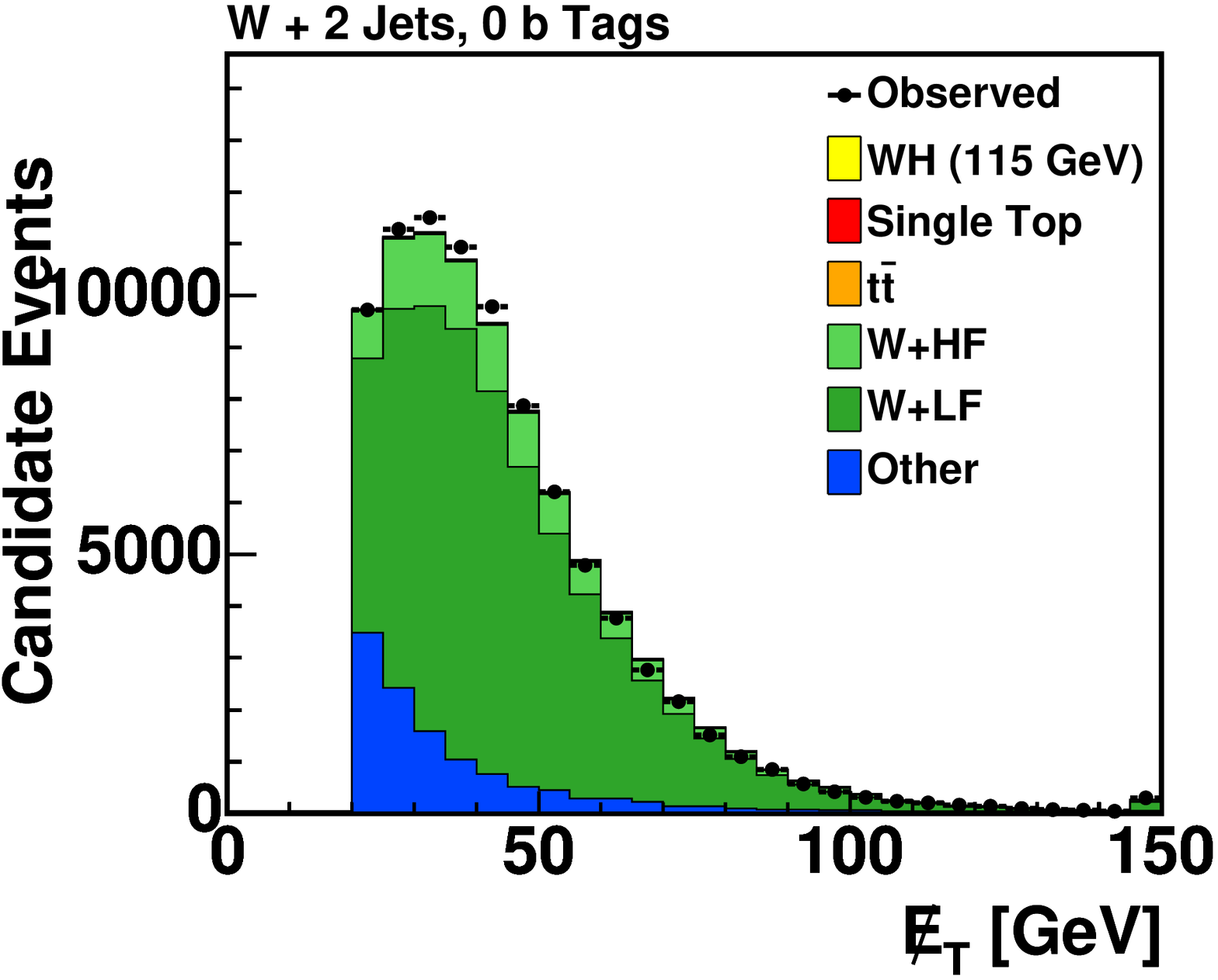}
\label{fig:MET2jet}}
\subfigure[]{
\includegraphics[width=0.7\columnwidth]{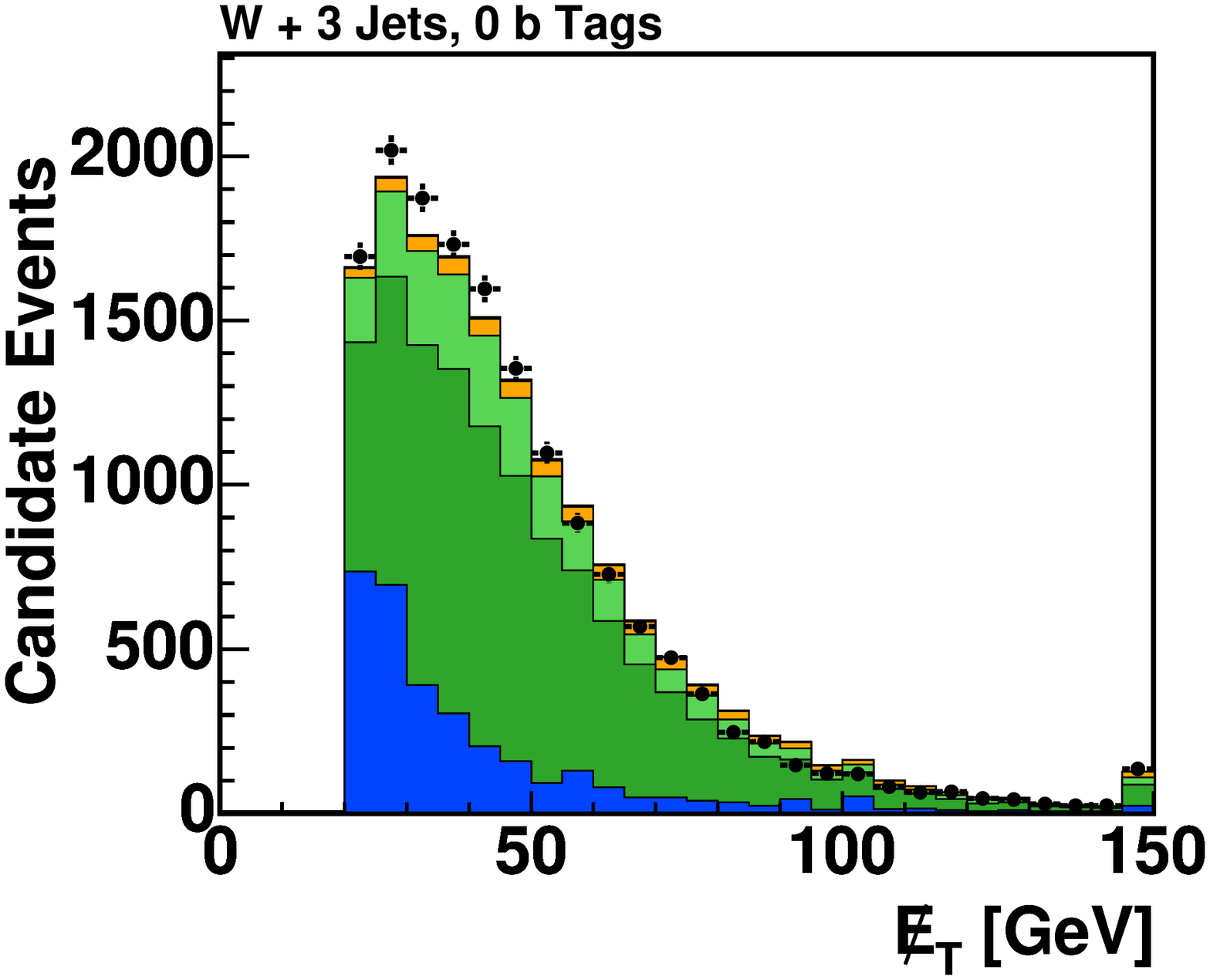}
\label{fig:MET3jet}}
\subfigure[]{
\includegraphics[width=0.7\columnwidth]{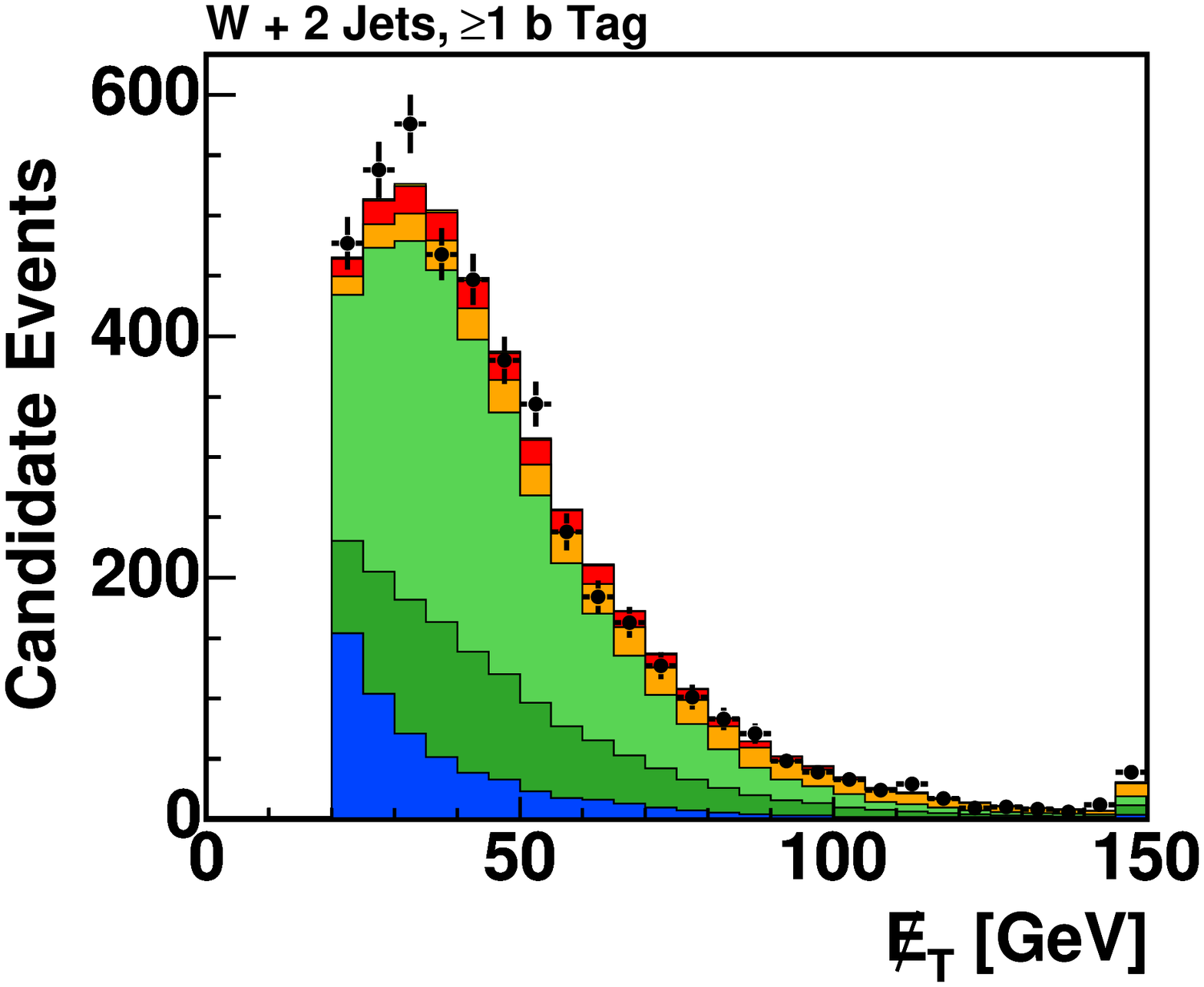}
\label{fig:MET2jet}}
\subfigure[]{
\includegraphics[width=0.7\columnwidth]{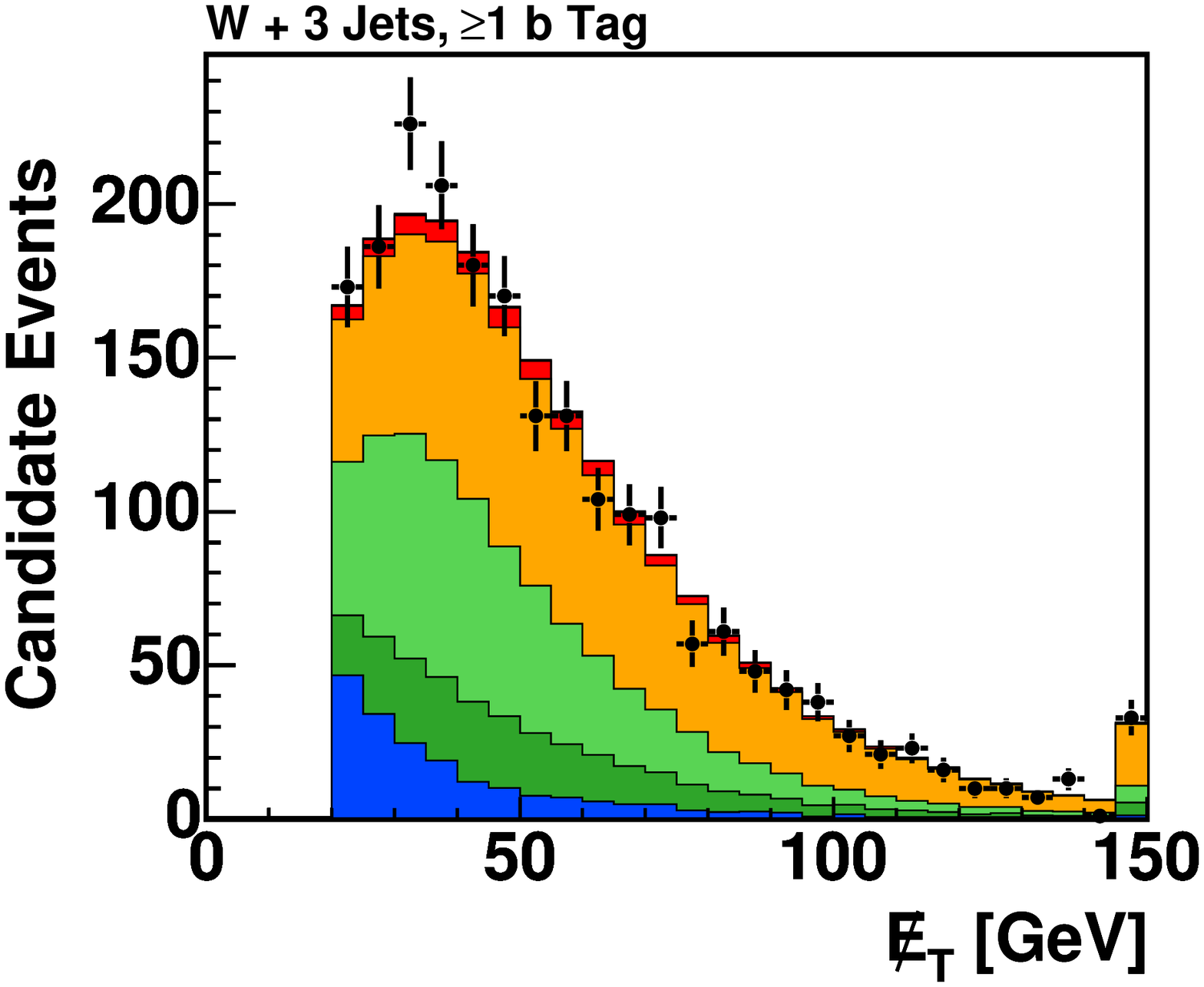}
\label{fig:MET3jet}}
\end{center}
\caption{\label{fig:Input_validation_met} Validation plots comparing observed events
and Monte Carlo distributions for missing transverse energy for events
with two (a and c) and three jets (b and d), with no $b$~tags (top)
and with at least one $b$~tag (bottom).  The observed events are
indicated with points. }
\end{figure*}

Since the analysis described here relies on Monte Carlo simulation,
the result depends on the proper modeling of the signal and the
background processes.  For that reason, the prediction of the
background model is compared with the observed events for hundreds of
distributions in the signal region and in different control regions.
Figs.~\ref{fig:Input_validation_3jet_0t},~\ref{fig:Input_validation_3jet},
and~\ref{fig:Input_validation_met} show examples of validation plots
for two and three jet bins, in a control region with no $b$-tagged
jets (to check the $W$~+~LF shapes) and in the signal region with at
least one tagged jet.  In general, the agreement is good.  The lepton
and jet transverse energy distributions are the least well modeled.
To check the effect of this mismodeling we derive weights from the
lepton and jet transverse energies in the control region, and we have
applied them to the discriminant variable in the signal region. We
check the effect of each variable one at a time by calculating the
expected limits in each case and found that the effect on the result
was not significant.  The validation of the modeling of other
observable quantities is shown later in this paper.

%-----------------------------
% MATRIX ELEMENT METHOD
%-----------------------------
%\input{./includes/prdMEmethod}
%-------------------------------------------------------------------------------
\section{Matrix element method}
\label{sec:meMethod}
\begin{figure}[h]
  \includegraphics[width=0.85\columnwidth]{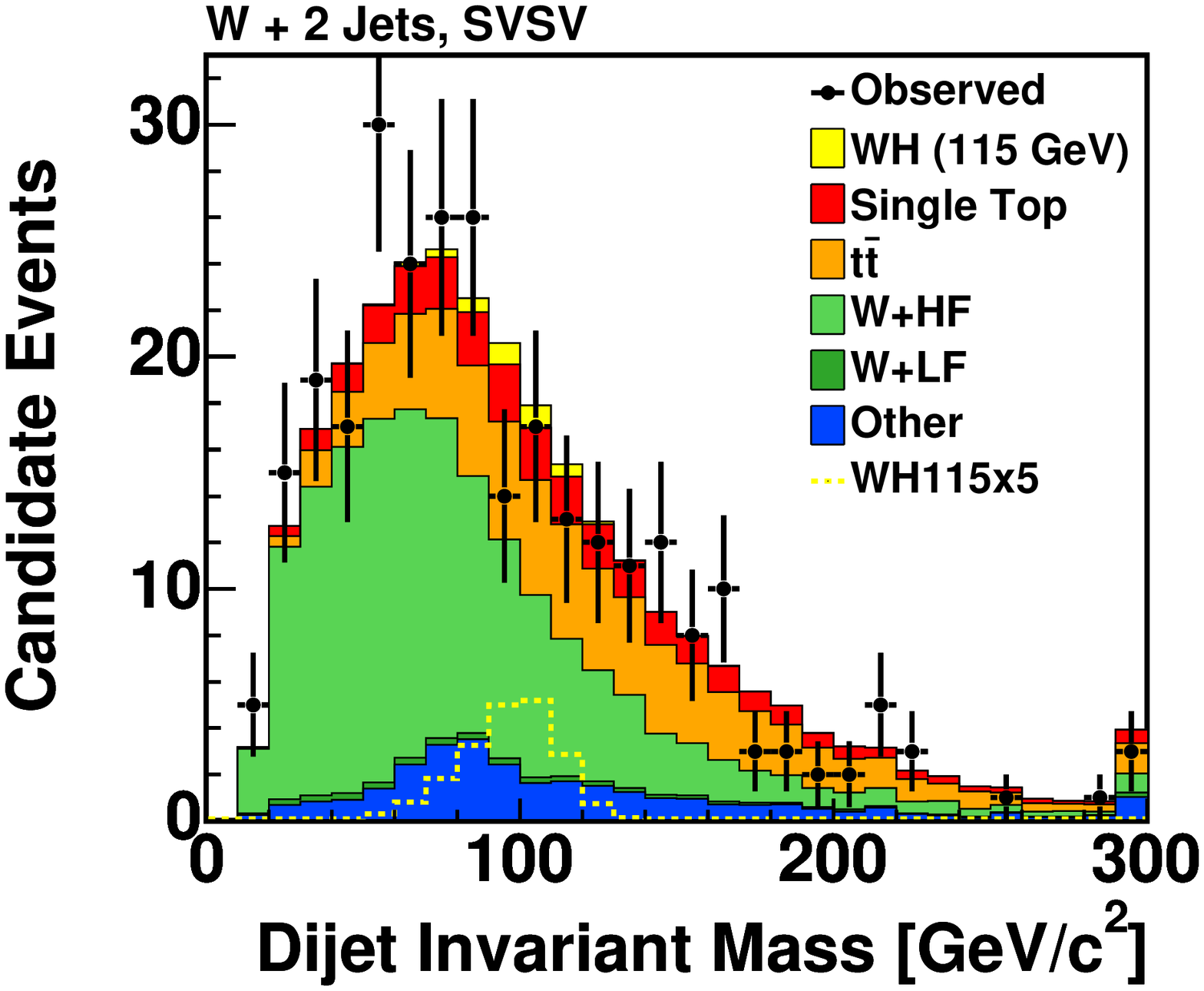}
  \caption{\label{fig:SVSV_mjj}Invariant mass distribution of the two
  leading jets for 2-jet SVSV events. The Higgs boson signal
  contribution (M$_H$ = 115 GeV/c$^2$) is multiplied by a factor 5 to
  make it visible. }
\end{figure}

%\setlength{\footskip}{2cm}
%-------------------------------------------------------------------------------
The number of expected signal events after the initial selection is
much smaller than the uncertainty in the background prediction. For
example, for a Higgs boson mass of 115 GeV/c$^2$ the
signal-to-background ratio is at best only about $1/70$ even in the
most signal rich $b$-tagging categories. Thus, a method based only on
counting the total number of events is unsuitable. The invariant mass
distribution of the two leading jets in the event is the most powerful
variable for discriminating signal from background, but it is limited
by the jet energy resolution.  Figure~\ref{fig:SVSV_mjj} shows the
invariant mass distribution of the two leading jets for two-jet SVSV
events.  Further discrimination between signal and background is
needed.

A matrix element (ME) method~\cite{bib:canelli,bib:brian_mohr} is used
in this search to discriminate signal from background events.  This
multivariate method relies on the evaluation of event probability
densities (commonly called event probabilities) for signal and
background processes based on calculations of the relevant standard
model differential cross sections. The ratio of signal and background
event probabilities is then used as a discriminant variable called the
event probability discriminant, EPD.  The goal is to maximize
sensitivity through the use of all kinematic information contained in
each event analyzed.  The discriminant distributions are optimized
separately for each Higgs boson mass hypothesis in order to extract
the maximum sensitivity.  Using the EPD as the discriminant variable
leads to an increase in sensitivity of $\sim$20\% with respect to only
using the invariant mass distribution of the two leading jets in the
event.

\subsection{Event probability}
If we could measure the four-vectors of the initial and final state
particles precisely, the event probability would be:
\begin{equation}
P_{evt}\sim \frac{d\sigma}{\sigma},
\end{equation}
where the differential cross-section is given by~\cite{bib:pdgstat}:
\begin{equation}
\label{eqn:dsigma}
d\sigma=\frac{(2\pi)^4|{\cal M}|^2}{4\sqrt{(q_1\cdot q_2)^2 -
m_{q_1}^2 m_{q_2}^2}}~d\Phi_n(q_1+q_2;p_1,..,p_n)
\end{equation}

\noindent where ${\cal M}$ is the Lorentz-invariant matrix element;
$q_1$, $q_2$ and $m_{q_1}$, $m_{q_2}$ are the four momenta and masses
of the incident particles; $p_1-p_n$ are the four momenta of the final
particles, and $d\Phi_n$ is the $n$-body phase space given
by~\cite{bib:pdgstat}:

\begin{equation}
\label{eqn:nphi}
d\Phi_n=\delta^4(q_1+q_2-\sum_{i=1}^n p_i)\prod_{i=1}^n
\frac{d^3p_i}{(2\pi)^32E_i}.
\end{equation}

However, several effects have to be considered: (1) the partons in the
initial state cannot be measured, (2) neutrinos in the final state are
not measured directly, and (3) the energy resolution of the detector
can not be ignored. To address the first point, the differential cross
section is weighted by parton distribution functions. To address the
second and third points, we integrate over all particle momenta which
we do not measure (the $p_z$ of the neutrino), or do not measure well,
due to resolution effects (the jet energies). The integration gives a
weighted sum over all possible parton-level variables $y$ leading to
the observed set of variables $x$ measured with the CDF detector. The
mapping between the particle variables $y$ and the measured variables
$x$ is established with the transfer function $W(y,x)$, which encodes
the detector resolution and is described in detail in
Section~\ref{sec:transferfunction}. Thus, the event probability now
takes the form:

\begin{equation}
\label{eqn:evtprob}
P(x)=\frac{1}{\sigma}\int d\sigma(y)dq_1dq_2f(y_1)f(y_2)W(y,x),
\end{equation}

\noindent where $d\sigma(y)$ is the differential cross section in 
terms of the particle variables; $f(y_i)$ are the parton distribution
functions, with $y_i$ being the fraction of the proton momentum
carried by the parton ($y_i=E_{q_i}/E_{beam}$); and $W(y,x)$ is the
transfer function. Substituting Eqs.~\ref{eqn:dsigma} and
\ref{eqn:nphi} into Eq.~\ref{eqn:evtprob}, and considering a final
state with four particles ($n$=4), transforms the event probability
to:

\begin{equation}
\label{eqn:evtprob2}
P(x)=\frac{1}{\sigma}\int {2\pi^4}|{\cal
M}|^2\frac{f(y_1)}{|E_{q_1}|}\frac{f(y_2)}{|E_{q_2}|}W(y,x)d\Phi_4
dE_{q_1}dE_{q_2},
\end{equation}

\noindent where the masses and transverse momenta of the initial partons are
neglected (i.e., $\sqrt{(q_1\cdot q_2)^2 - m_{q_1}^2 m_{q_2}^2}\simeq
2E_{q_1}E_{q_2}$).

The squared matrix element $|{\cal M}|^2$ for the event probability is
calculated at leading order by using the {\sc helas} (Helicity
Amplitude Subroutines for Feynman Diagram Evaluations)
package~\cite{Murayama:1992}. The subroutine calls for a given process
are automatically generated by {\sc madgraph}~\cite{Maltoni:2002qb}.
For events with two jets, event probability densities for the $WH$
signal (for 11 Higgs boson masses), as well as for the s-channel and
t-channel single top, $t\bar{t}$, $Wb\bar b$, $Wc\bar{c}$, $Wc$,
mistags ($Wgj$, and $Wgg$) and diboson ($WW$, $WZ$) background
processes are calculated.  The $WH$ channel is mainly produced in
two-jet events, but it can happen that an initial or final state
radiation jet is identified as the third jet of the event. Including
three-jet events increases signal acceptance and gains sensitivity to
the Higgs boson signal.  In the case of events with three jets in the
final state, event probability densities for the $WH$ signal, as well
as for the $s$-channel and $t$-channel single top, $t\bar{t}$, $Wb\bar
b$, and $Wc\bar c$ processes are calculated.  The $WH$ Feynman
diagrams include only those with initial and final state radiation,
and exclude those in which a $ggH$ coupling is present as these
contribute less than 1\% to the total cross section, but increase the
computation time by more than 20\%.

The integration performed in the matrix element calculation of this
analysis is identical to the one for the search for single top
production~\cite{stprd}.  The matrix elements correspond to
fixed-order tree-level calculations and thus are not perfect
representations of the probabilities for each process.  This
limitation of the matrix element calculations for the discriminant
affects the sensitivity of the analysis but not its correctness, as
the same matrix elements are calculated for both observed and Monte
Carlo events, which uses parton showers to approximate higher-order
effects on kinematic distributions.  The different combinations of
matching jets to quarks are also considered~\cite{bib:peter_thesis}.

\begin{figure*}
\begin{center}
\subfigure[]{
\includegraphics[width=0.65\columnwidth]{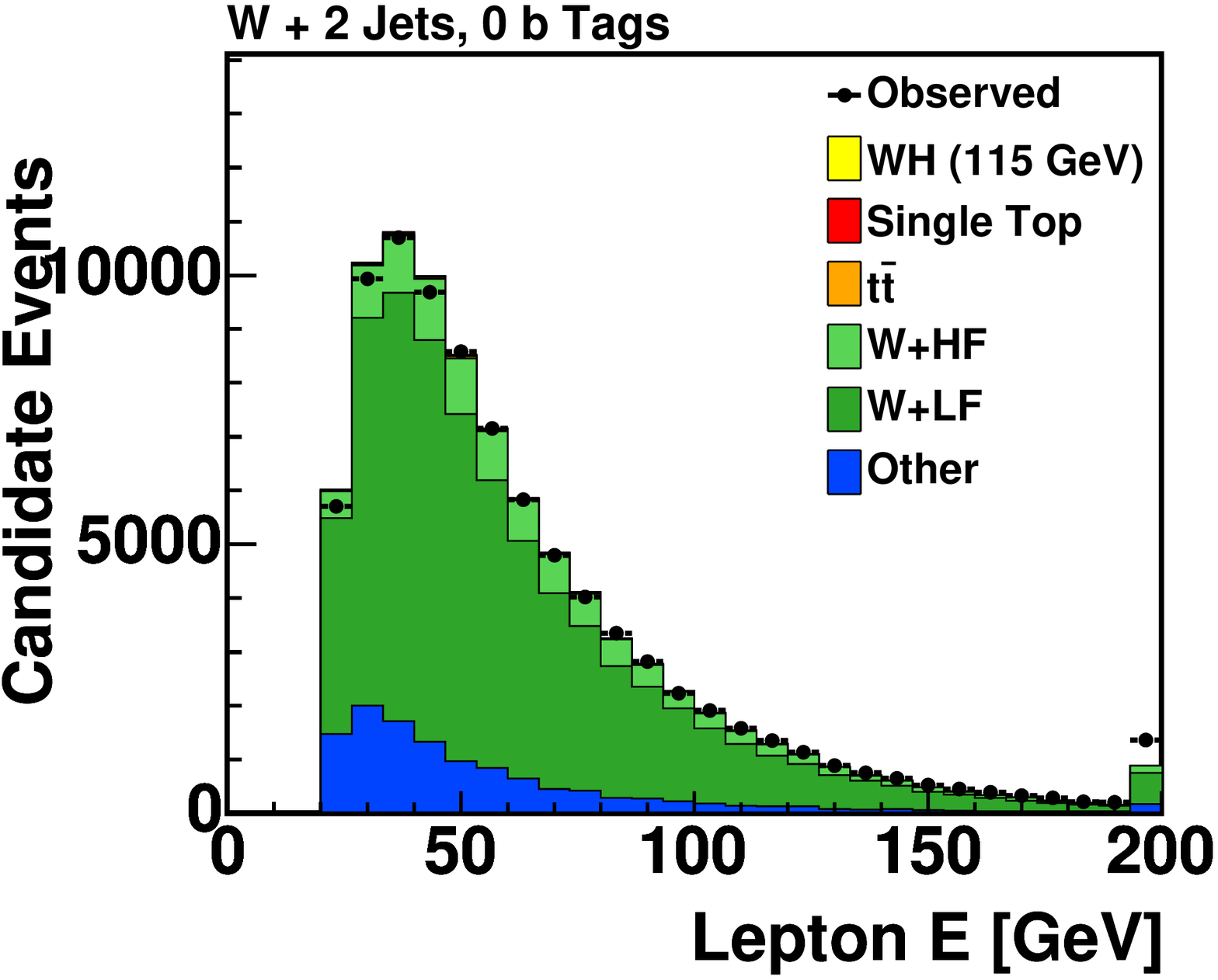}
\label{fig:e}}
\subfigure[]{
\includegraphics[width=0.65\columnwidth]{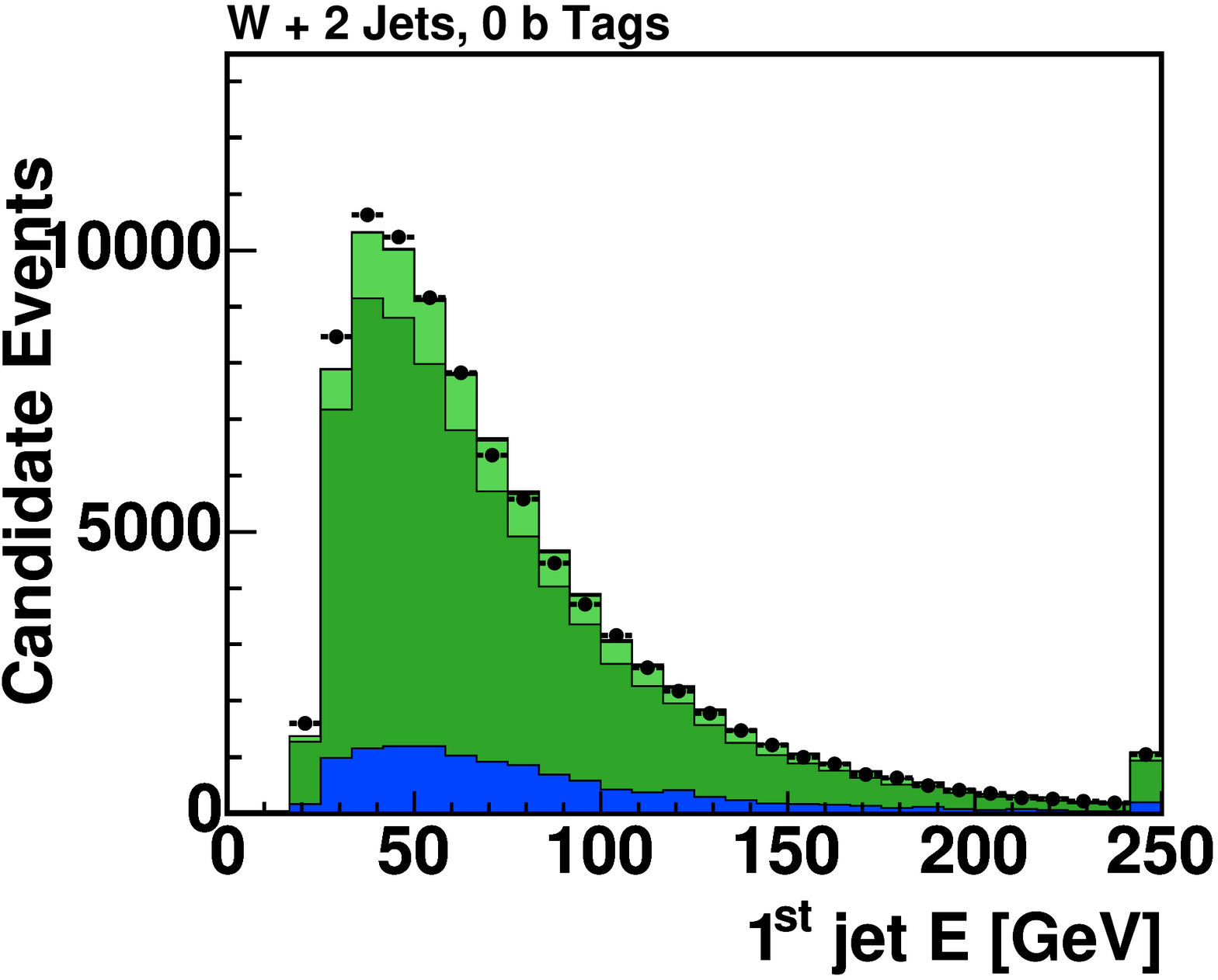}
\label{fig:e}}
\subfigure[]{
\includegraphics[width=0.65\columnwidth]{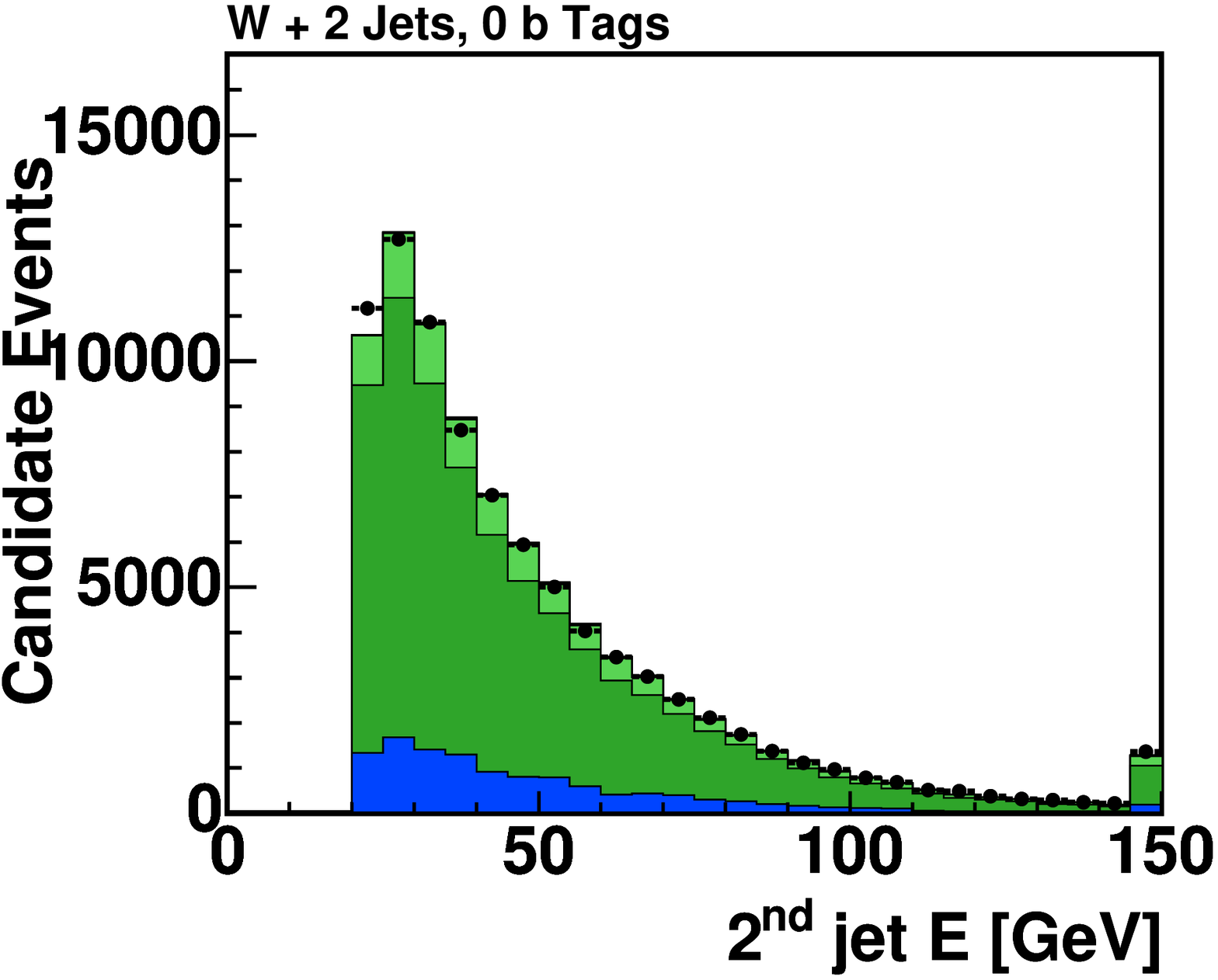}
\label{fig:e}}
\subfigure[]{
\includegraphics[width=0.65\columnwidth]{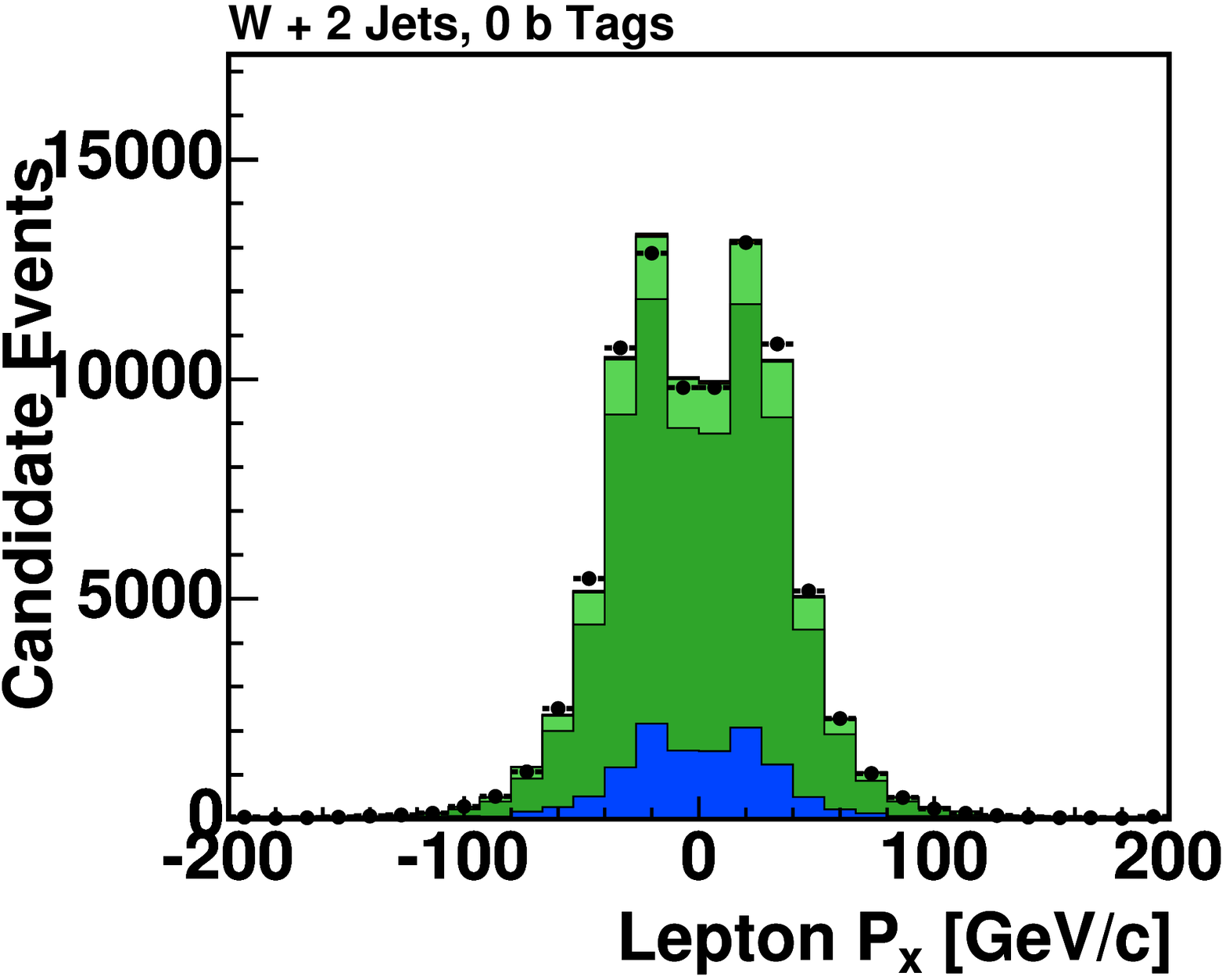}
\label{fig:e}}
\subfigure[]{
\includegraphics[width=0.65\columnwidth]{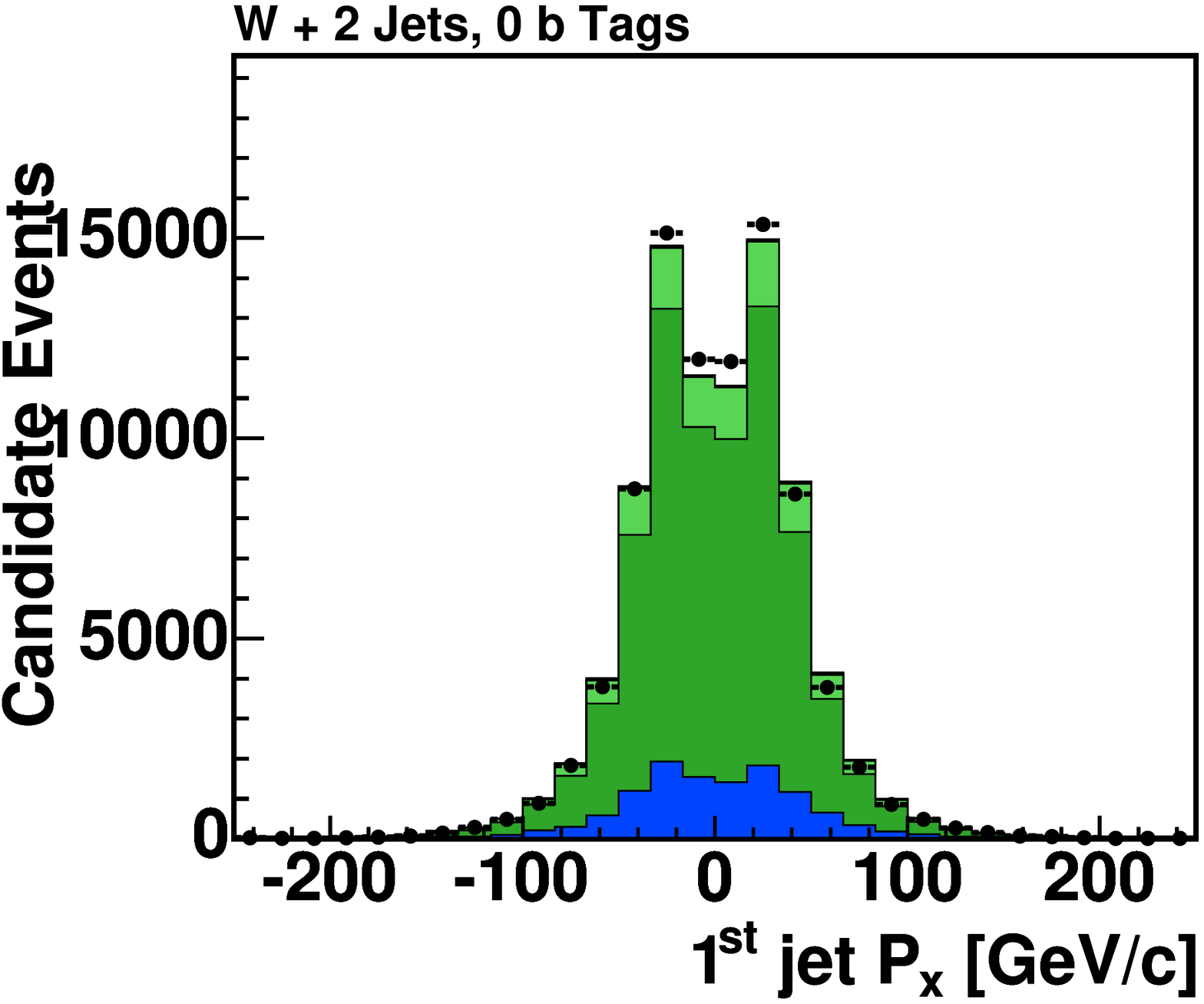}
\label{fig:e}}
\subfigure[]{
\includegraphics[width=0.65\columnwidth]{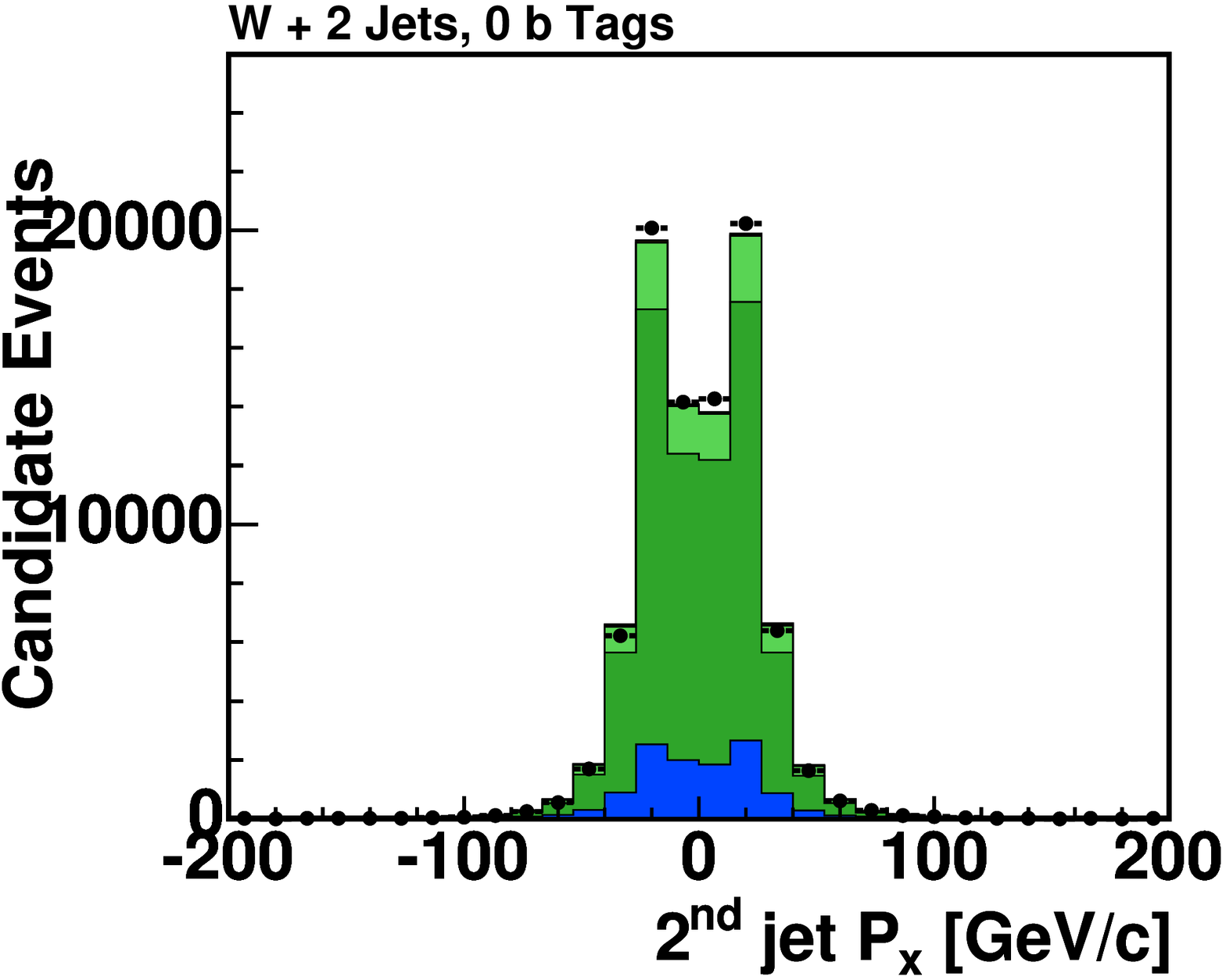}
\label{fig:e}}
\subfigure[]{
\includegraphics[width=0.65\columnwidth]{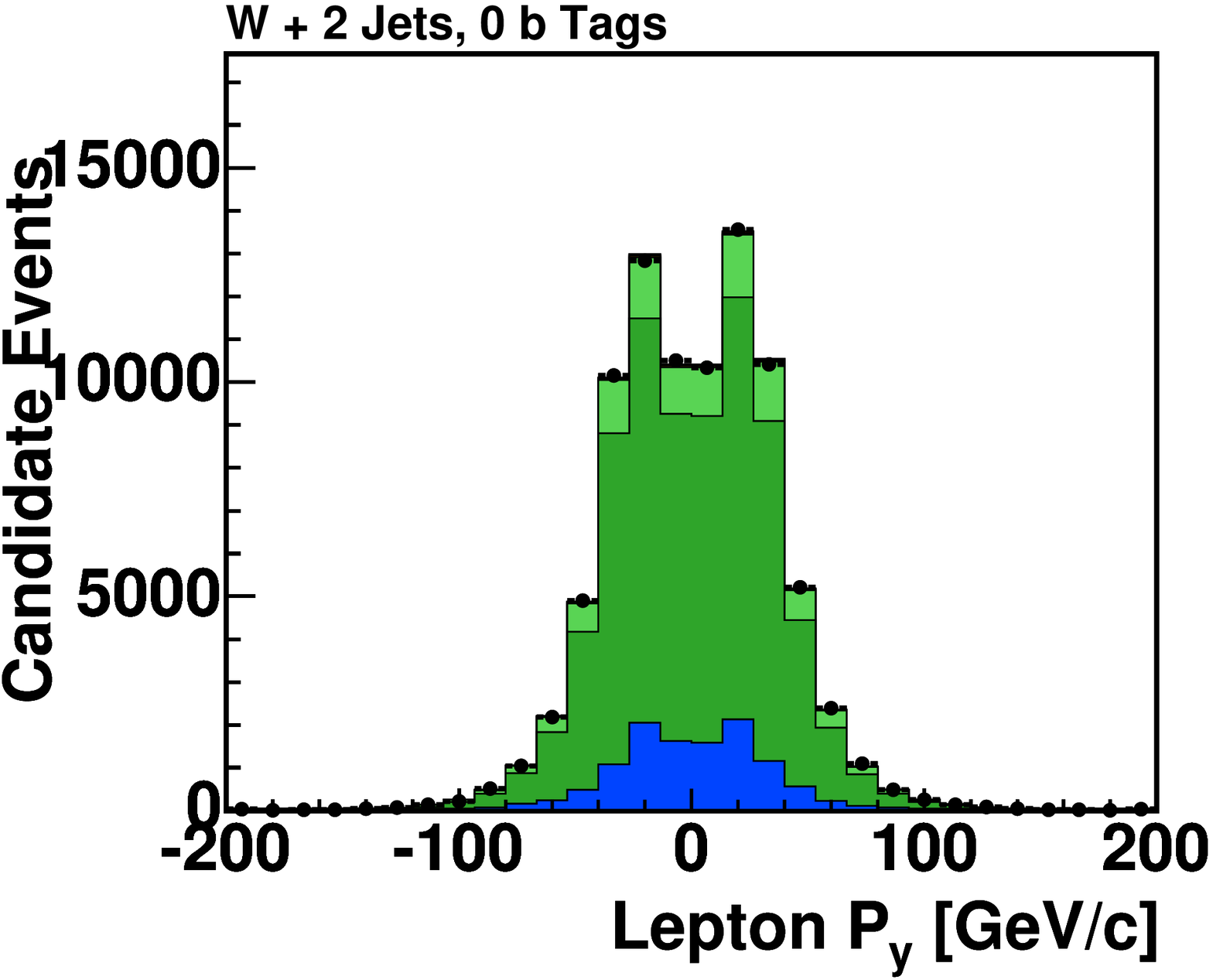}
\label{fig:e}}
\subfigure[]{
\includegraphics[width=0.65\columnwidth]{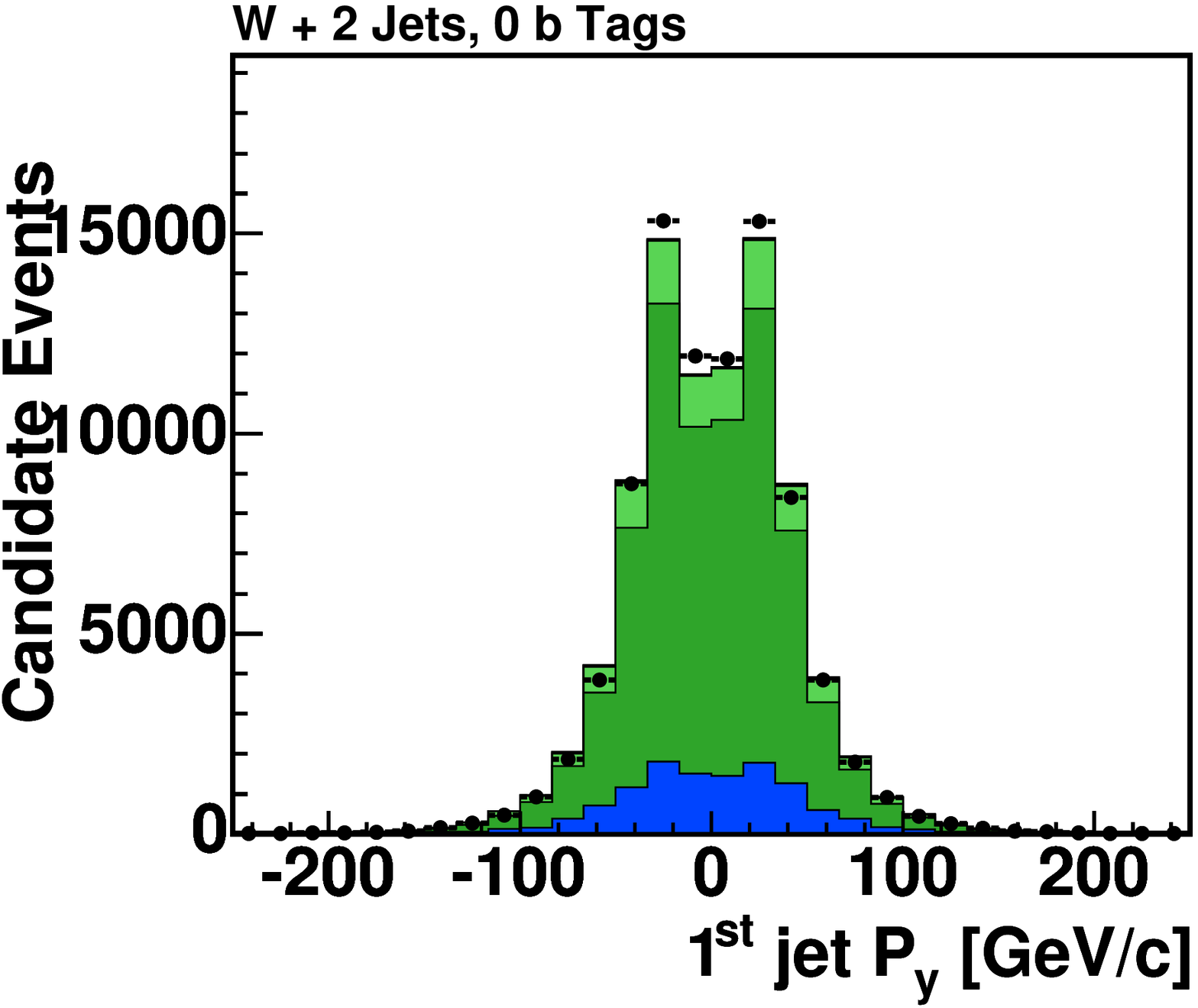}
\label{fig:e}}
\subfigure[]{
\includegraphics[width=0.65\columnwidth]{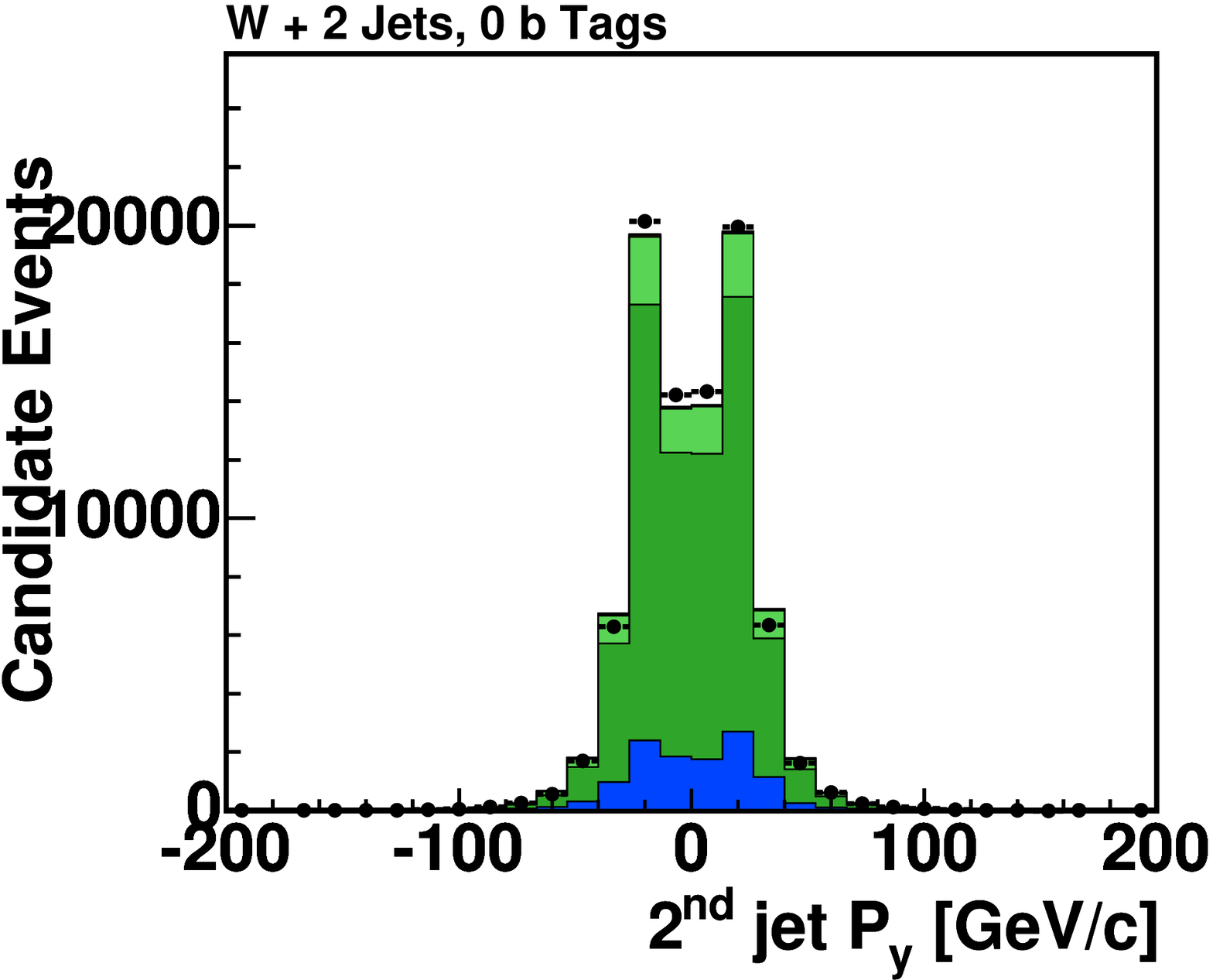}
\label{fig:e}}
\subfigure[]{
\includegraphics[width=0.65\columnwidth]{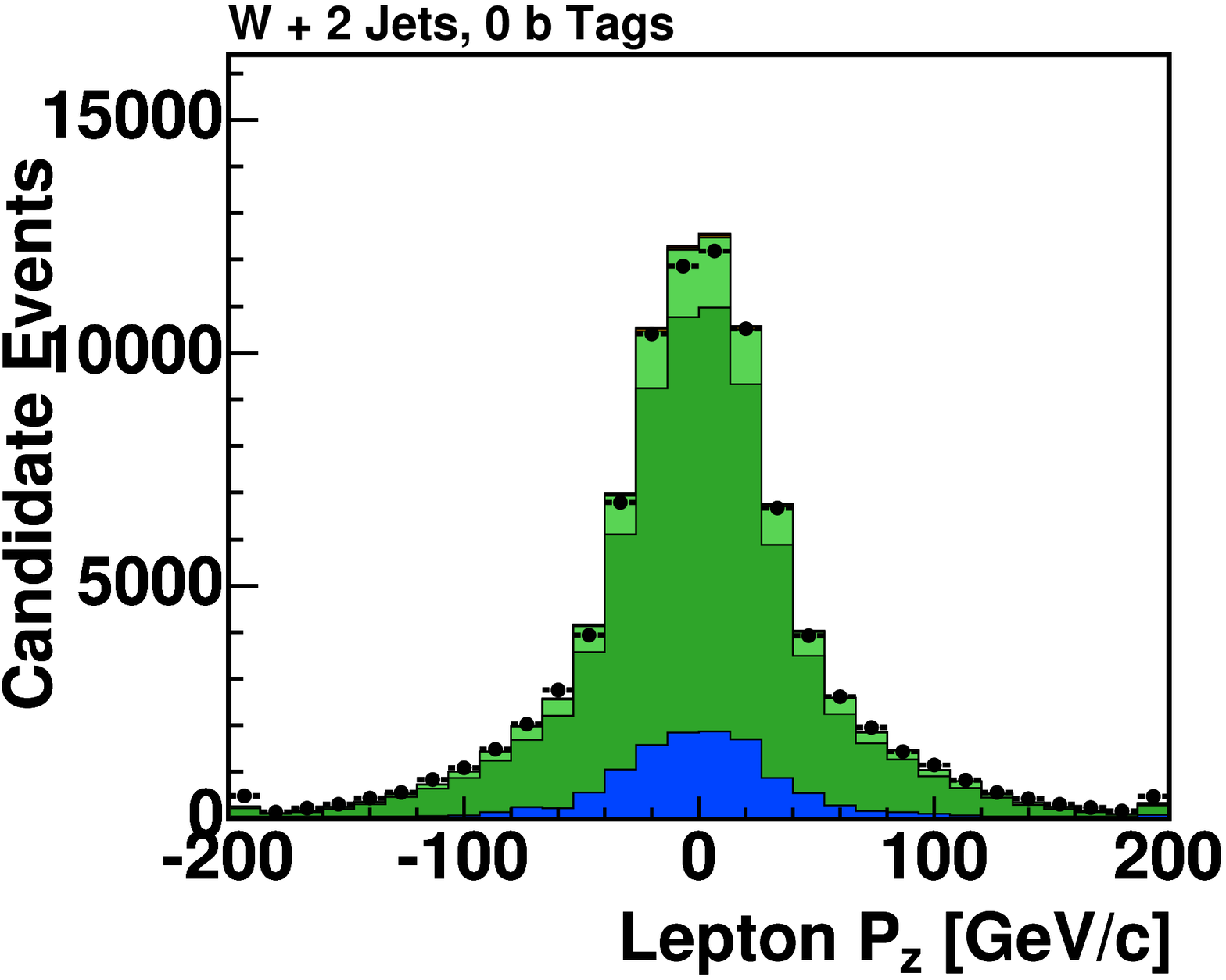}
\label{fig:e}}
\subfigure[]{
\includegraphics[width=0.65\columnwidth]{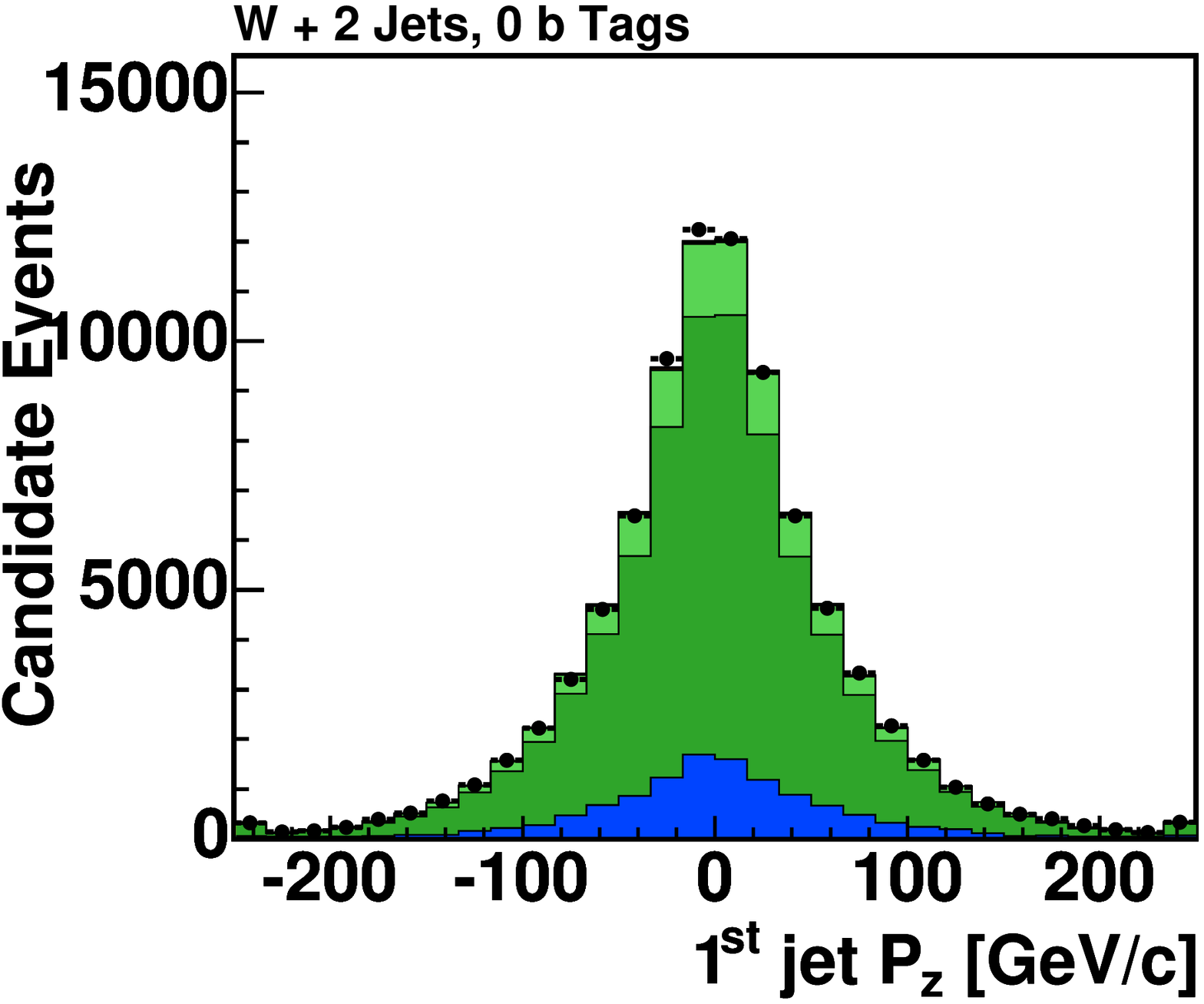}
\label{fig:e}}
\subfigure[]{
\includegraphics[width=0.65\columnwidth]{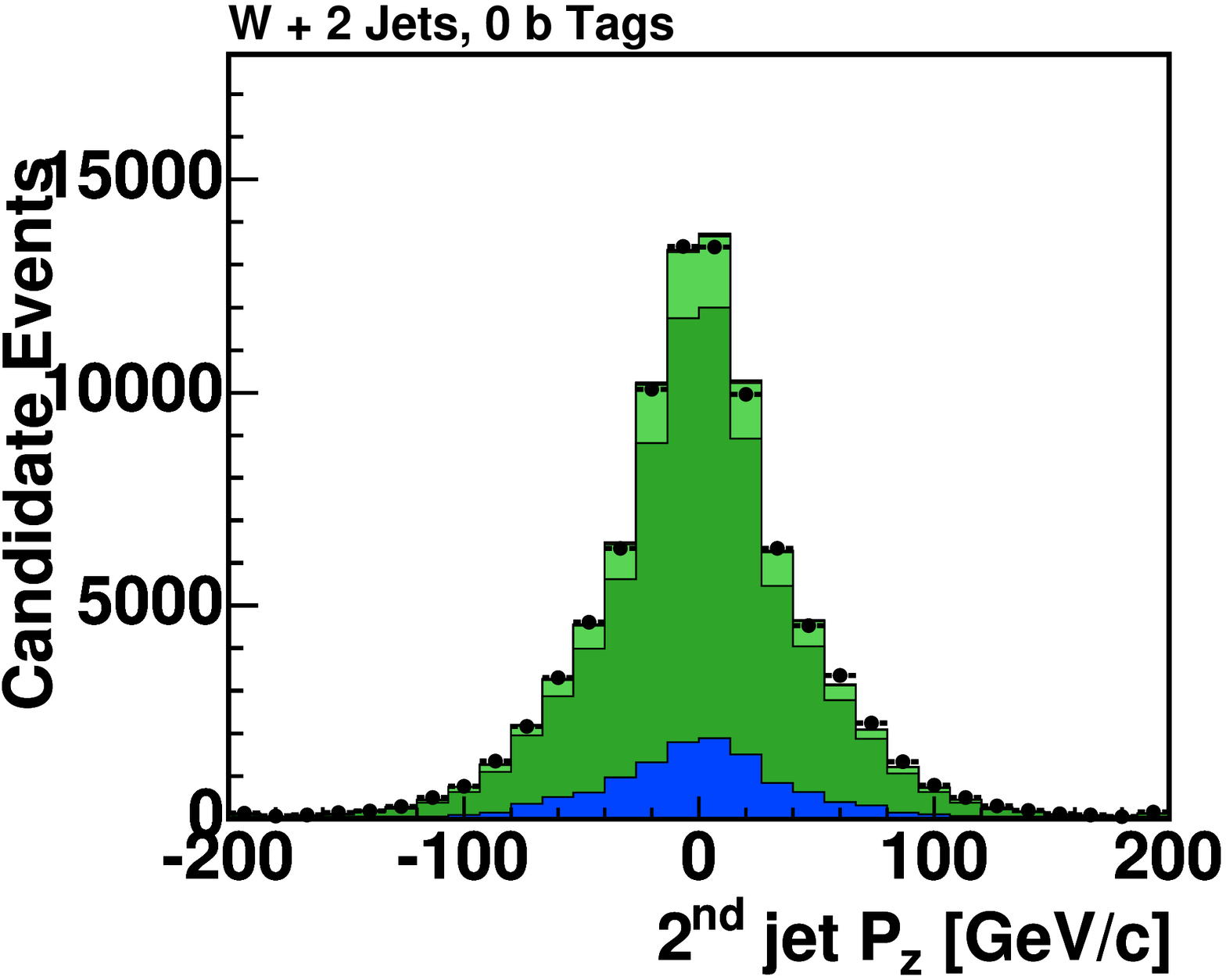}
\label{fig:e}}
\end{center}
\vspace{-0.85cm}\caption{\label{fig:4vect_2jet_0t}Validation plots comparing observed and MC simulated events for the 
  four-vector ($E$, $P_x$, $P_y$, $P_z$) of the lepton and the jets in
  2-jet untagged events.}
\end{figure*}

\begin{figure*}
\begin{center}
\subfigure[]{
\includegraphics[width=0.65\columnwidth]{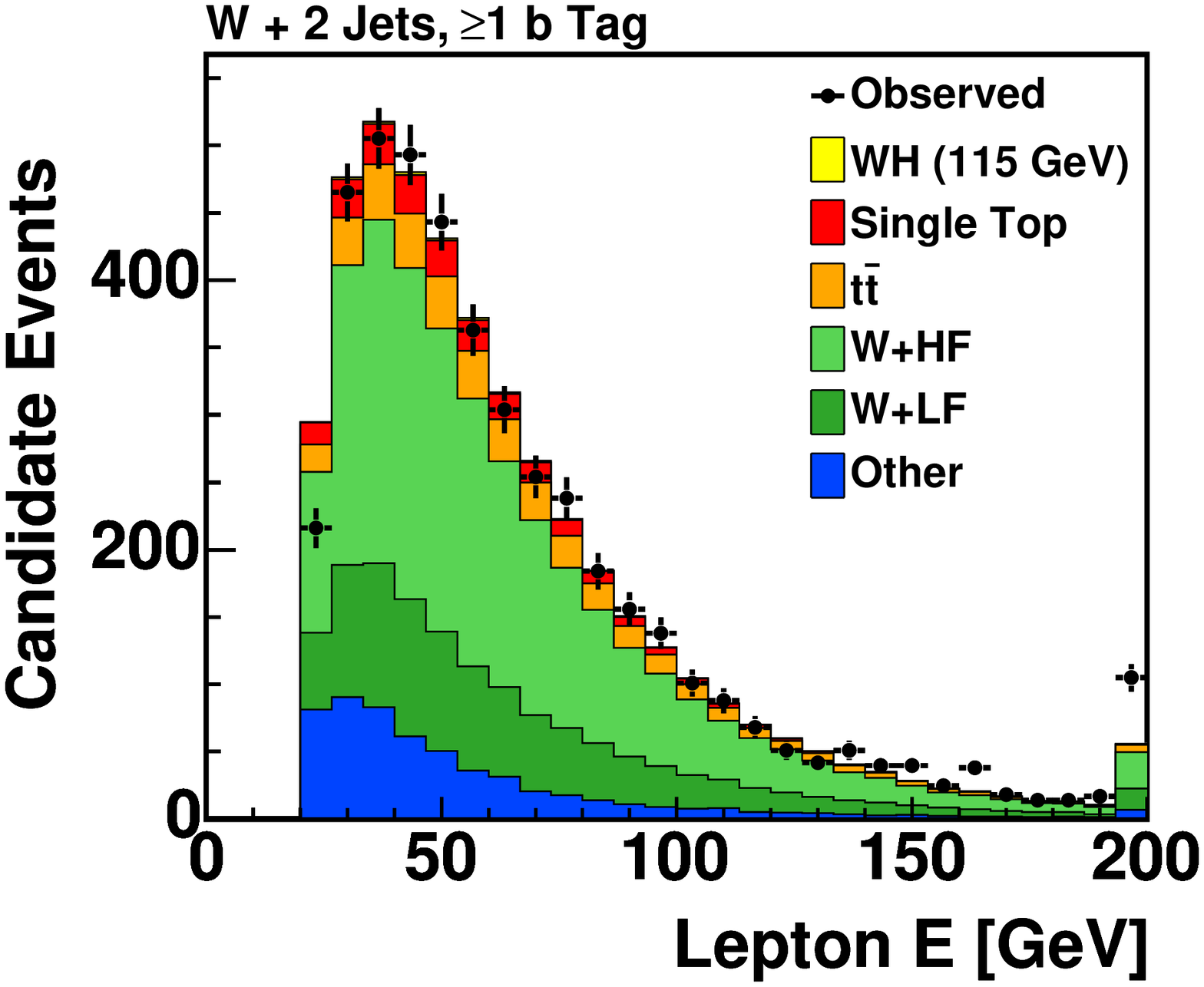}
\label{fig:e}}
\subfigure[]{
\includegraphics[width=0.65\columnwidth]{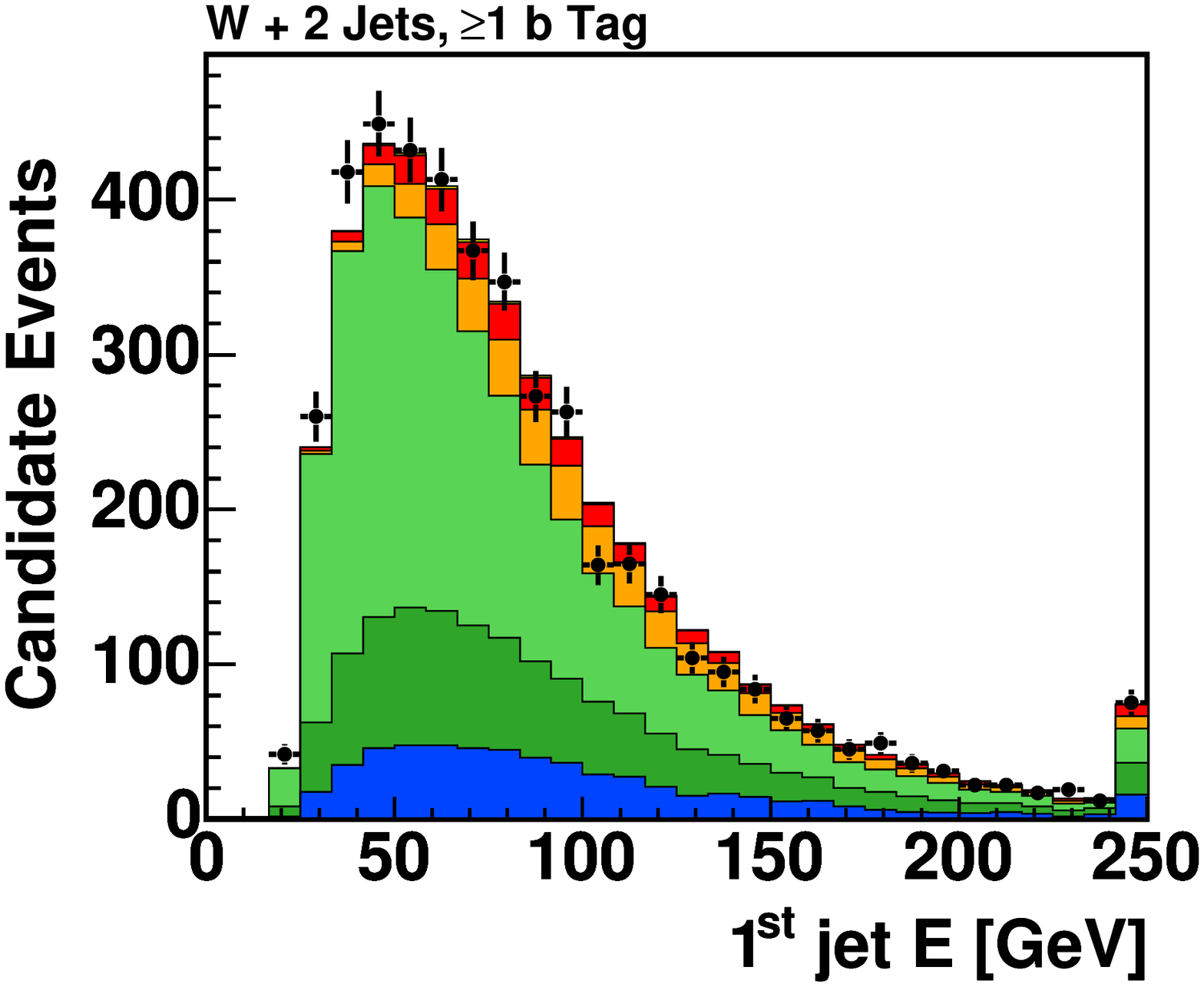}
\label{fig:e}}
\subfigure[]{
\includegraphics[width=0.65\columnwidth]{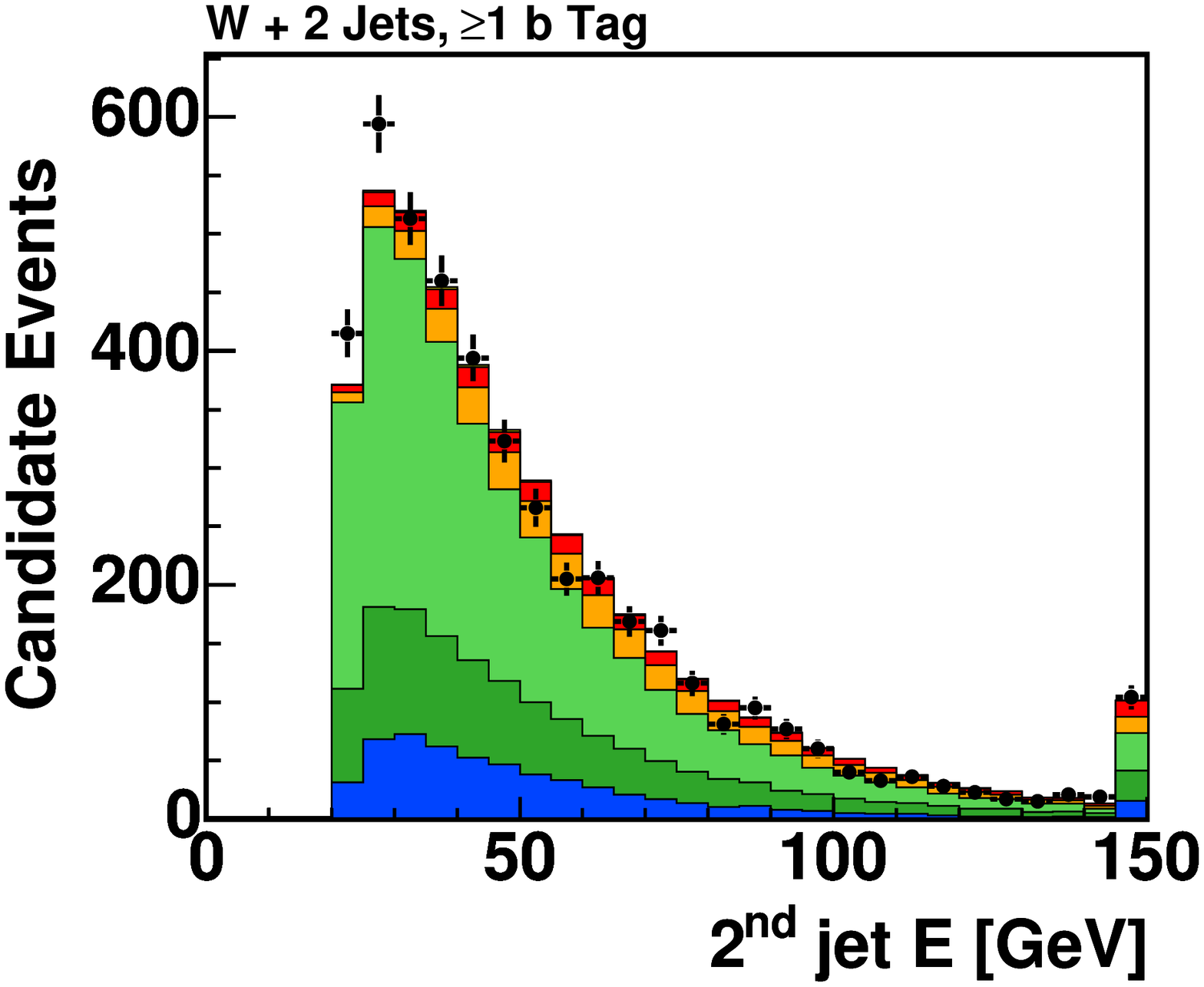}
\label{fig:e}}
\subfigure[]{
\includegraphics[width=0.65\columnwidth]{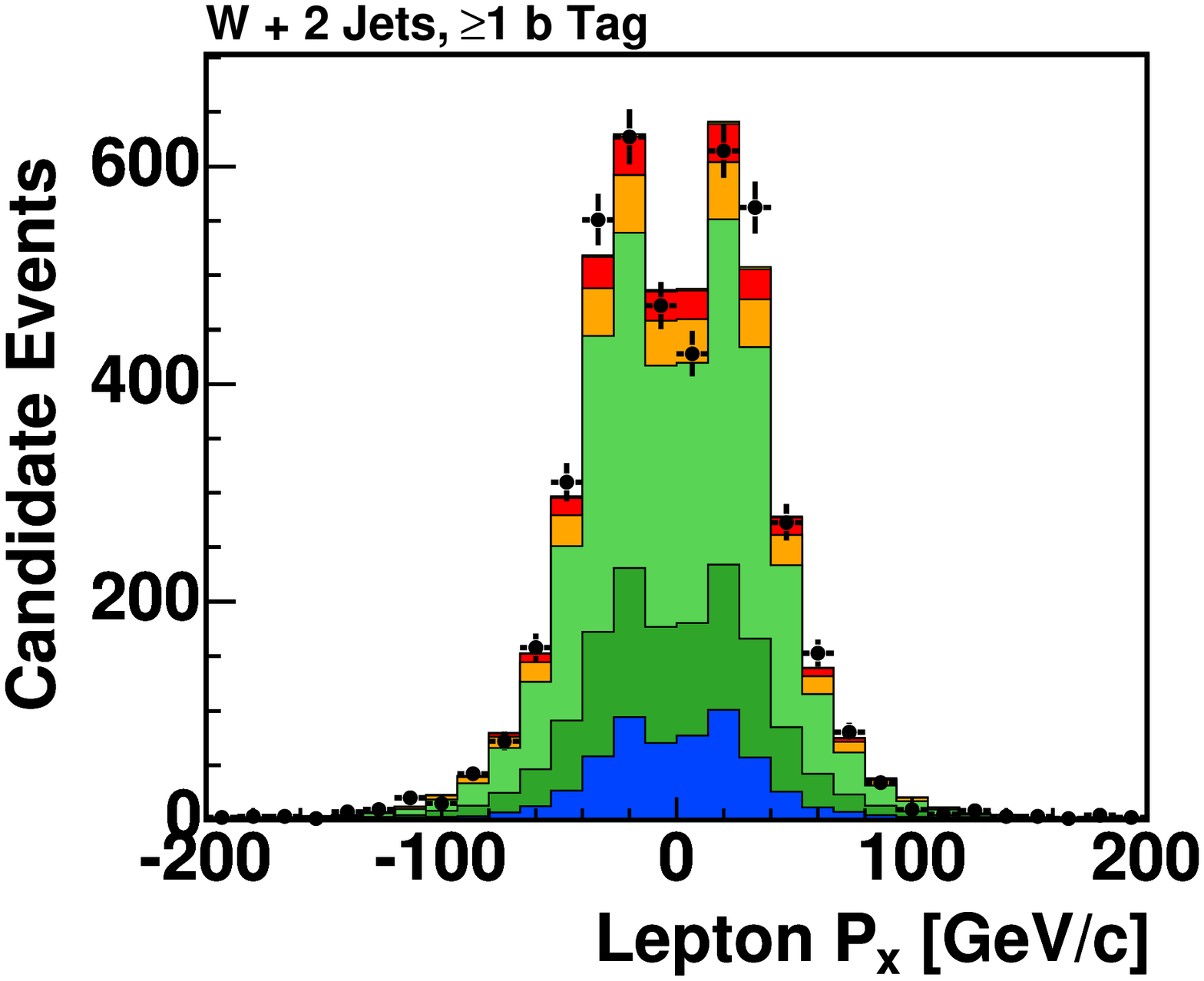}
\label{fig:e}}
\subfigure[]{
\includegraphics[width=0.65\columnwidth]{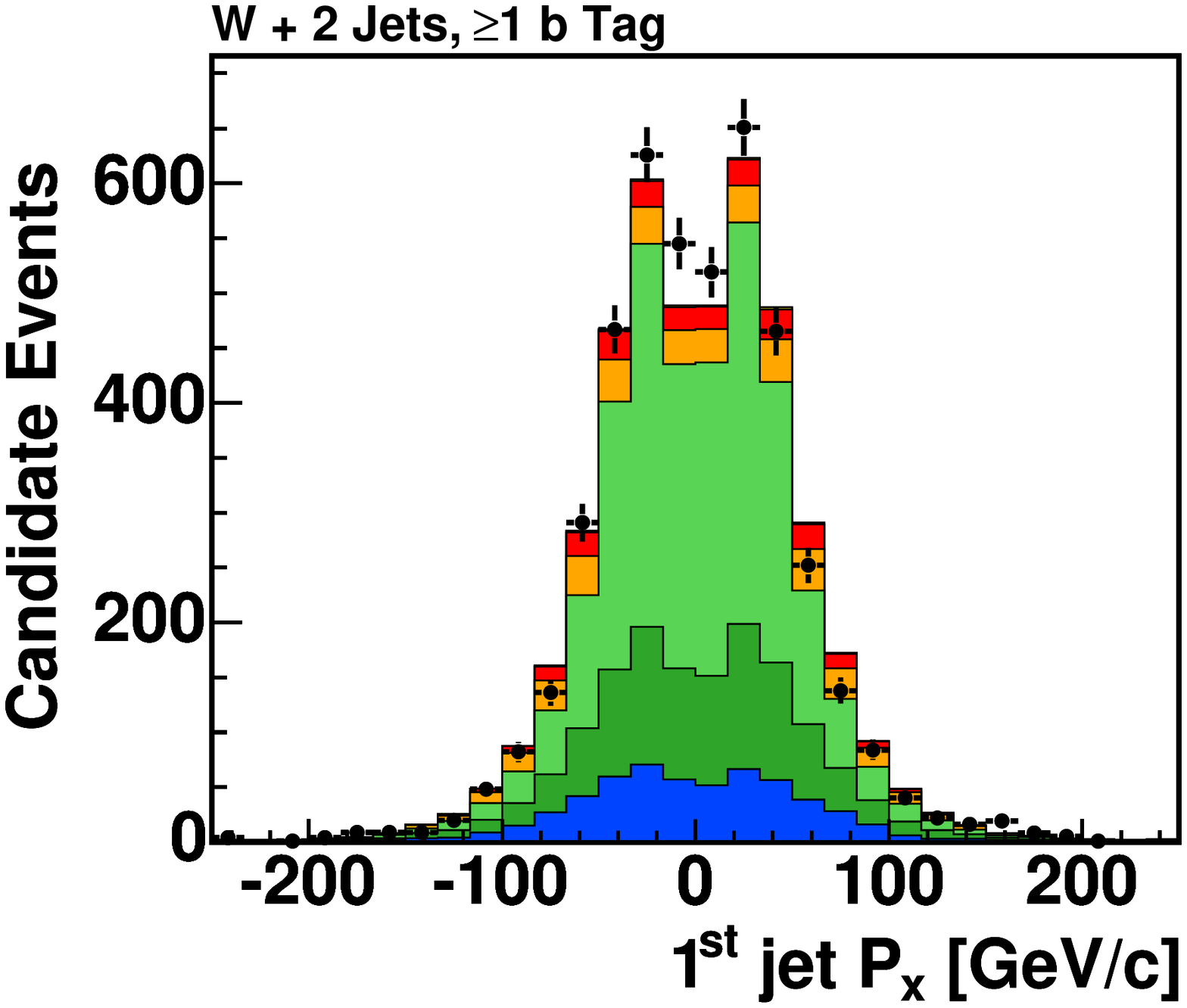}
\label{fig:e}}
\subfigure[]{
\includegraphics[width=0.65\columnwidth]{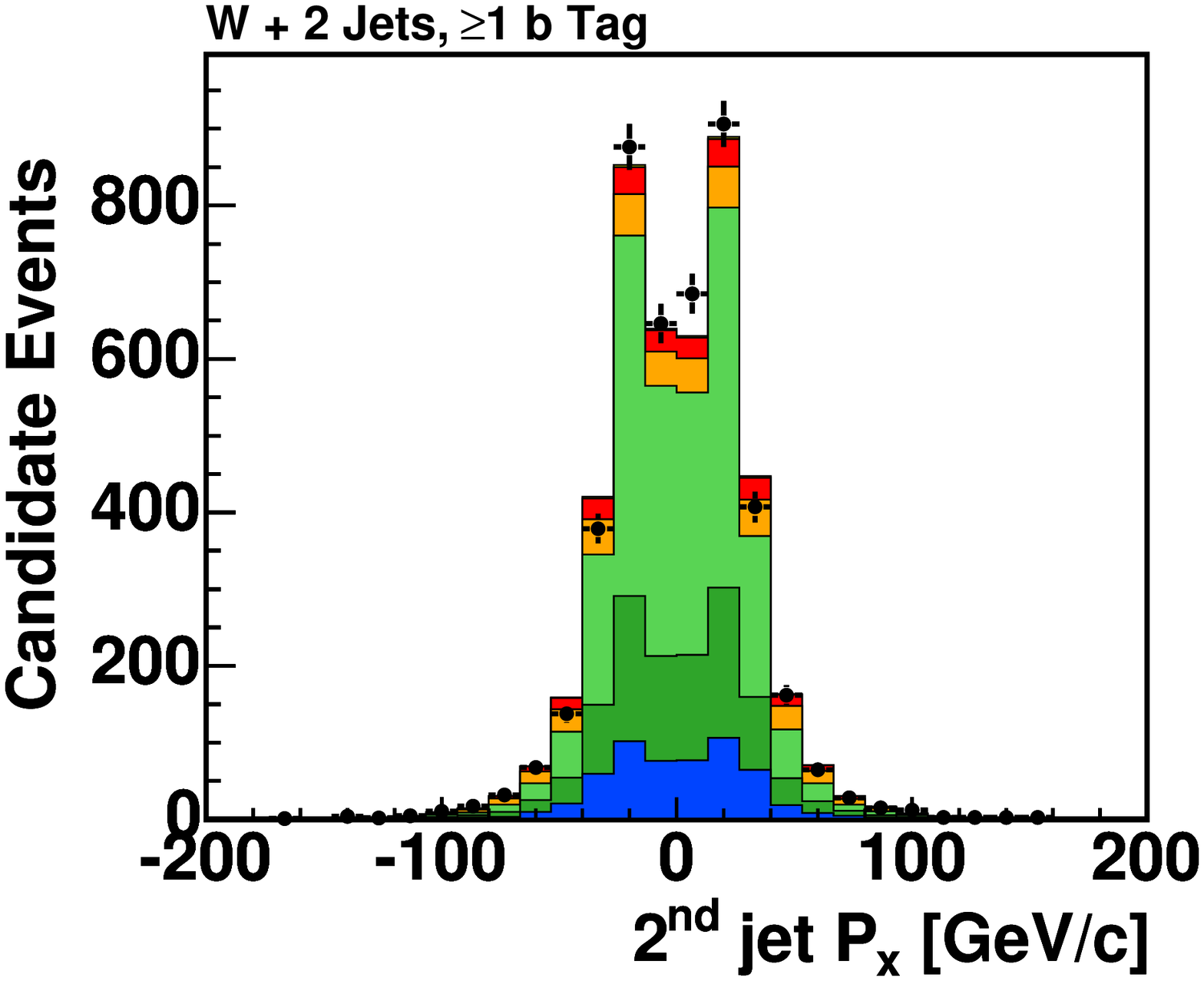}
\label{fig:e}}
\subfigure[]{
\includegraphics[width=0.65\columnwidth]{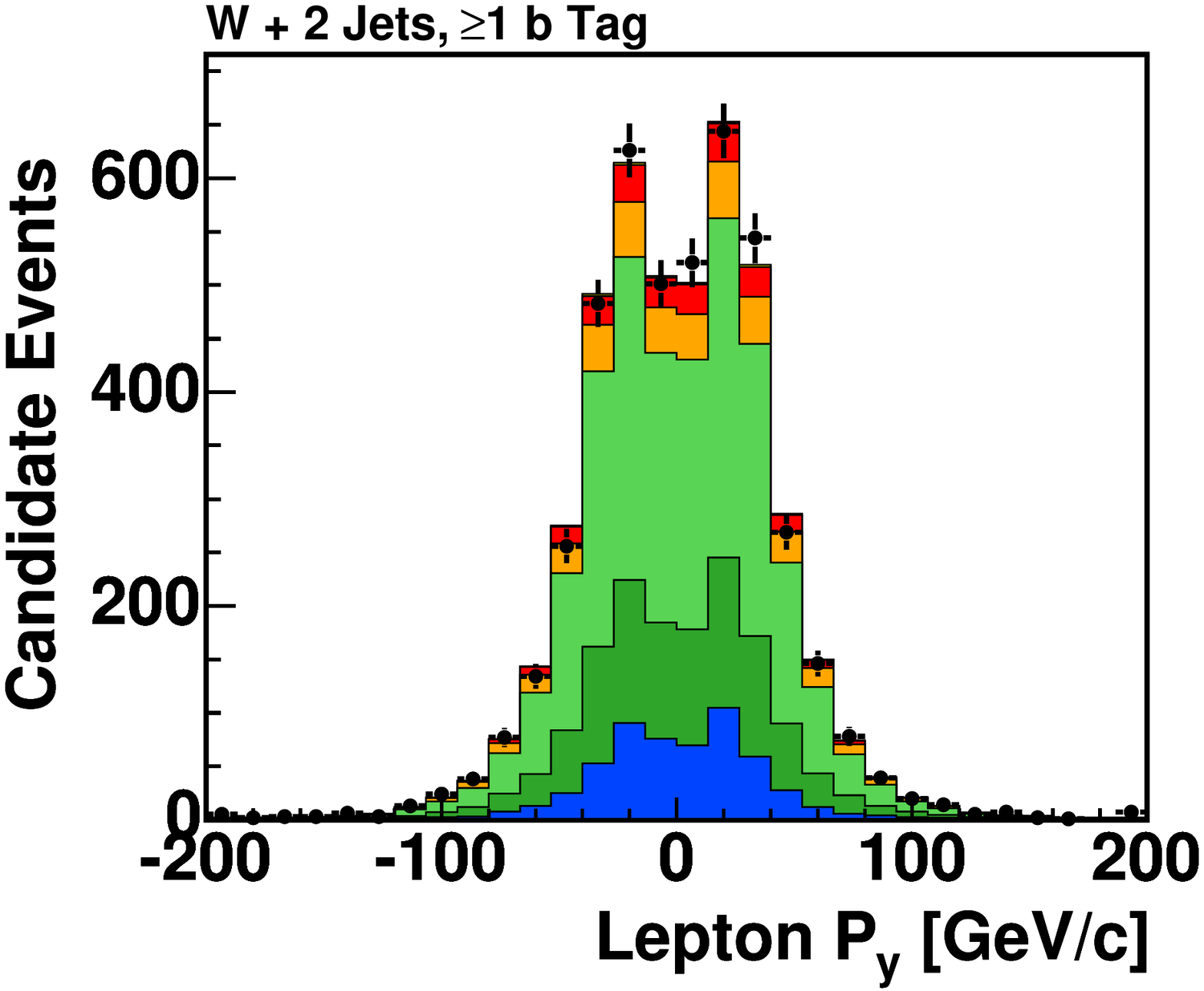}
\label{fig:e}}
\subfigure[]{
\includegraphics[width=0.65\columnwidth]{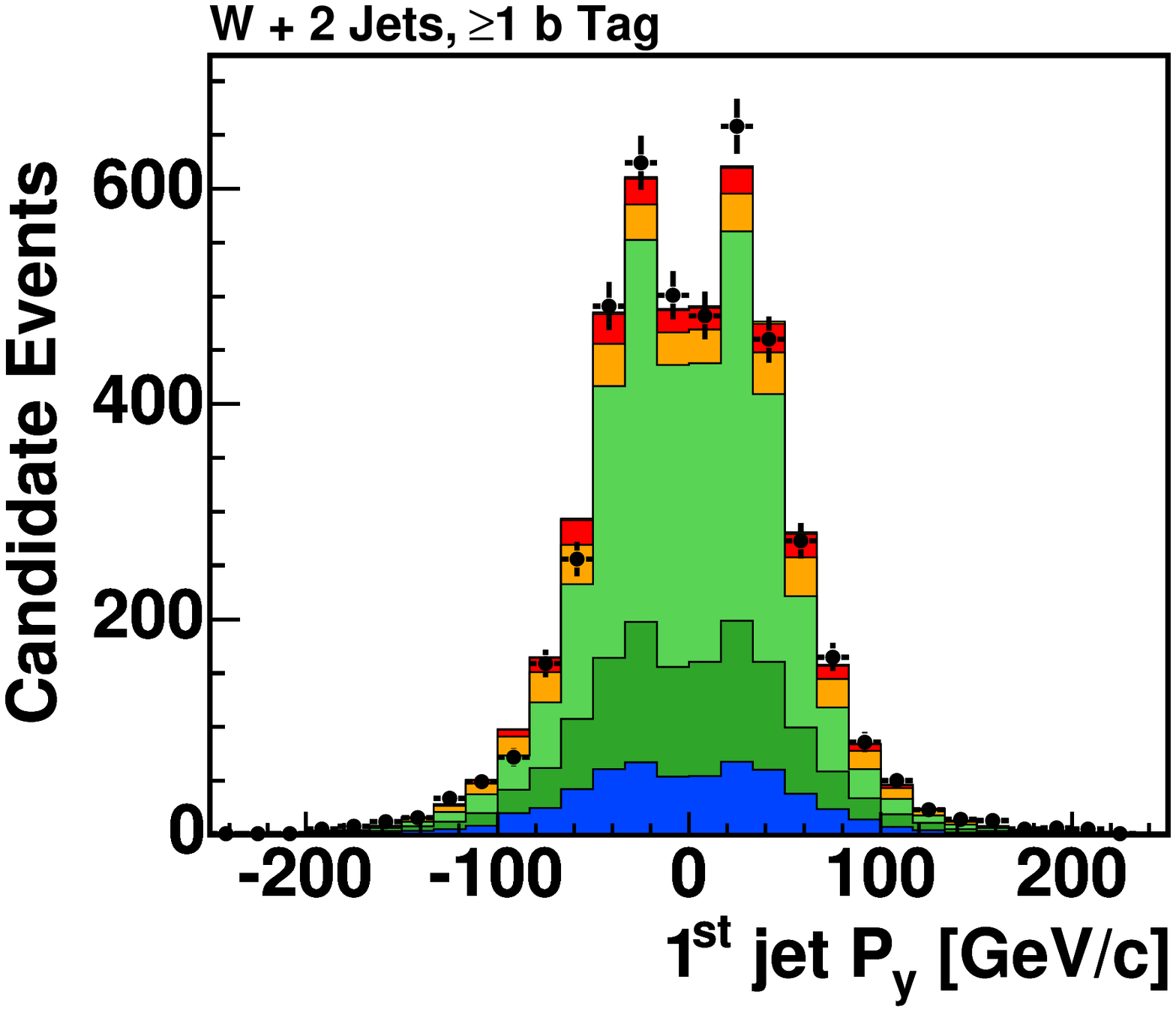}
\label{fig:e}}
\subfigure[]{
\includegraphics[width=0.65\columnwidth]{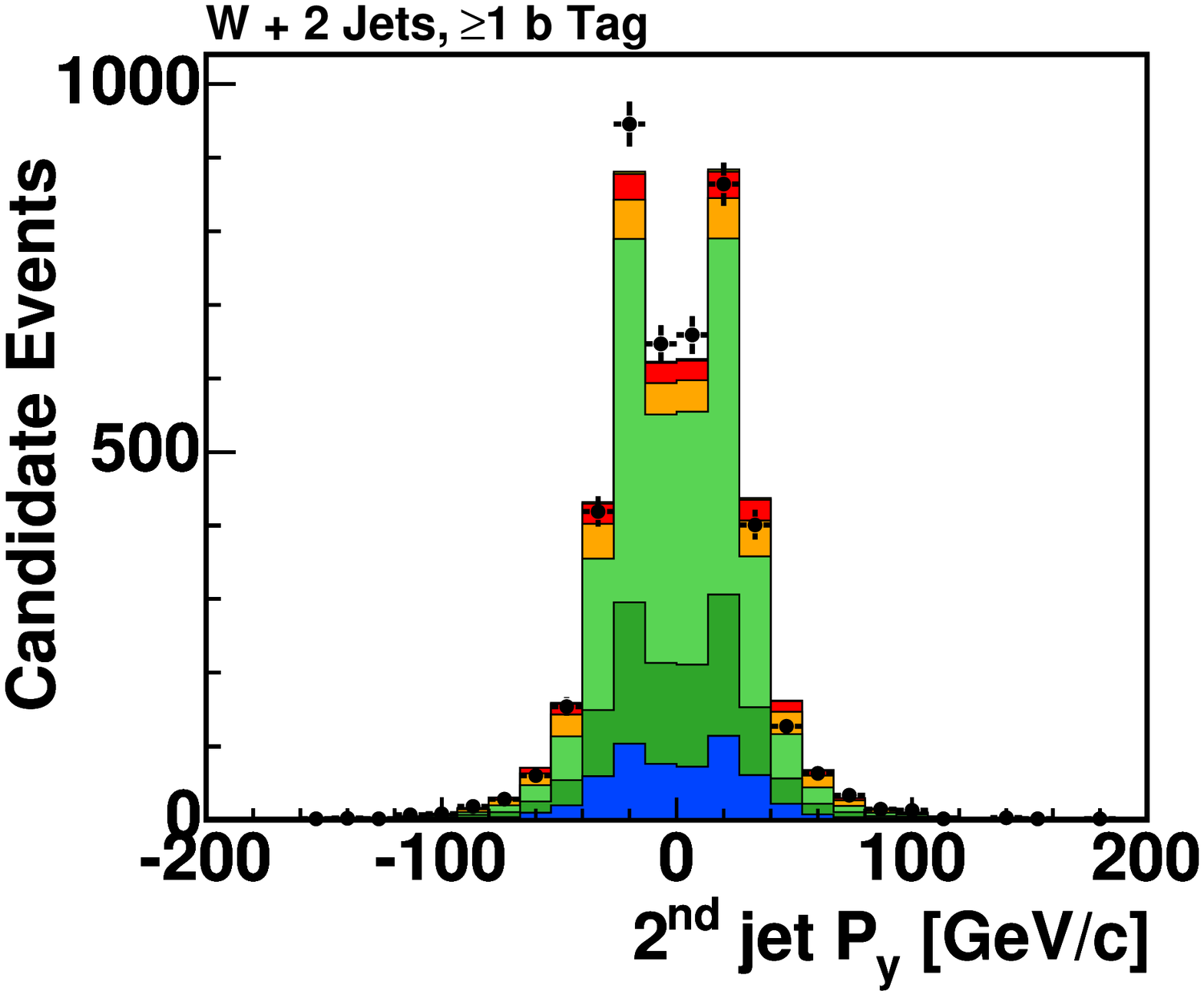}
\label{fig:e}}
\subfigure[]{
\includegraphics[width=0.65\columnwidth]{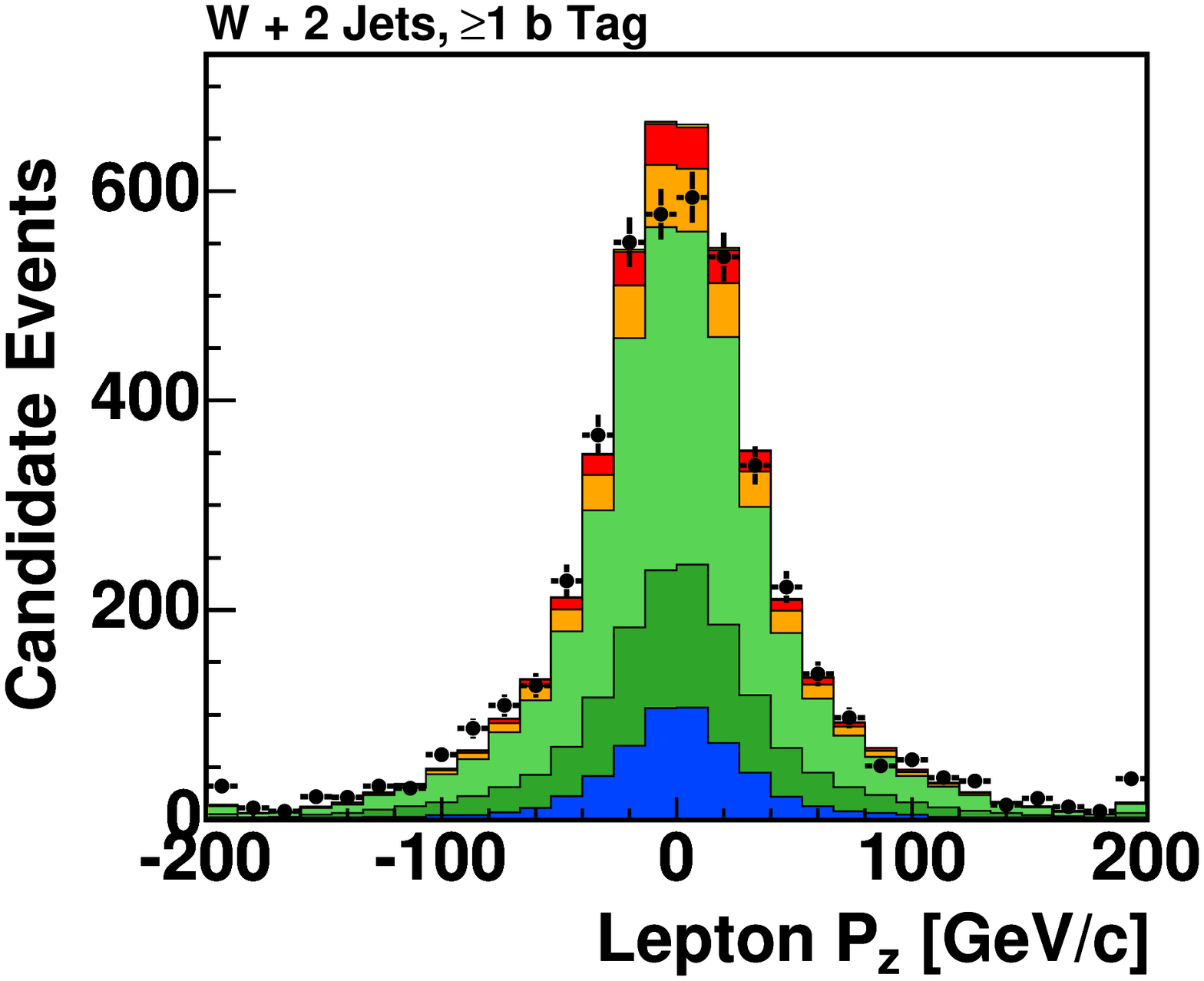}
\label{fig:e}}
\subfigure[]{
\includegraphics[width=0.65\columnwidth]{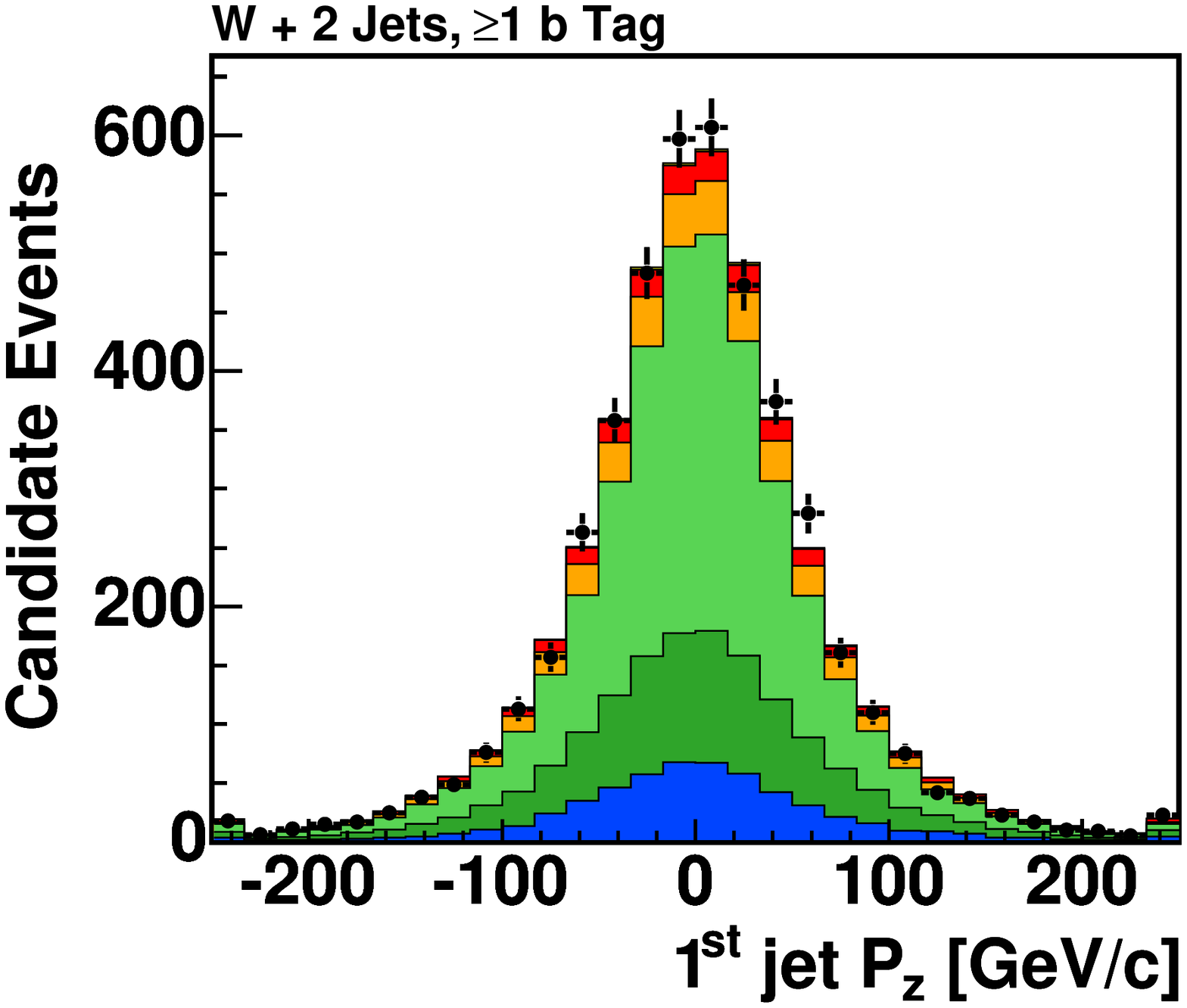}
\label{fig:e}}
\subfigure[]{
\includegraphics[width=0.65\columnwidth]{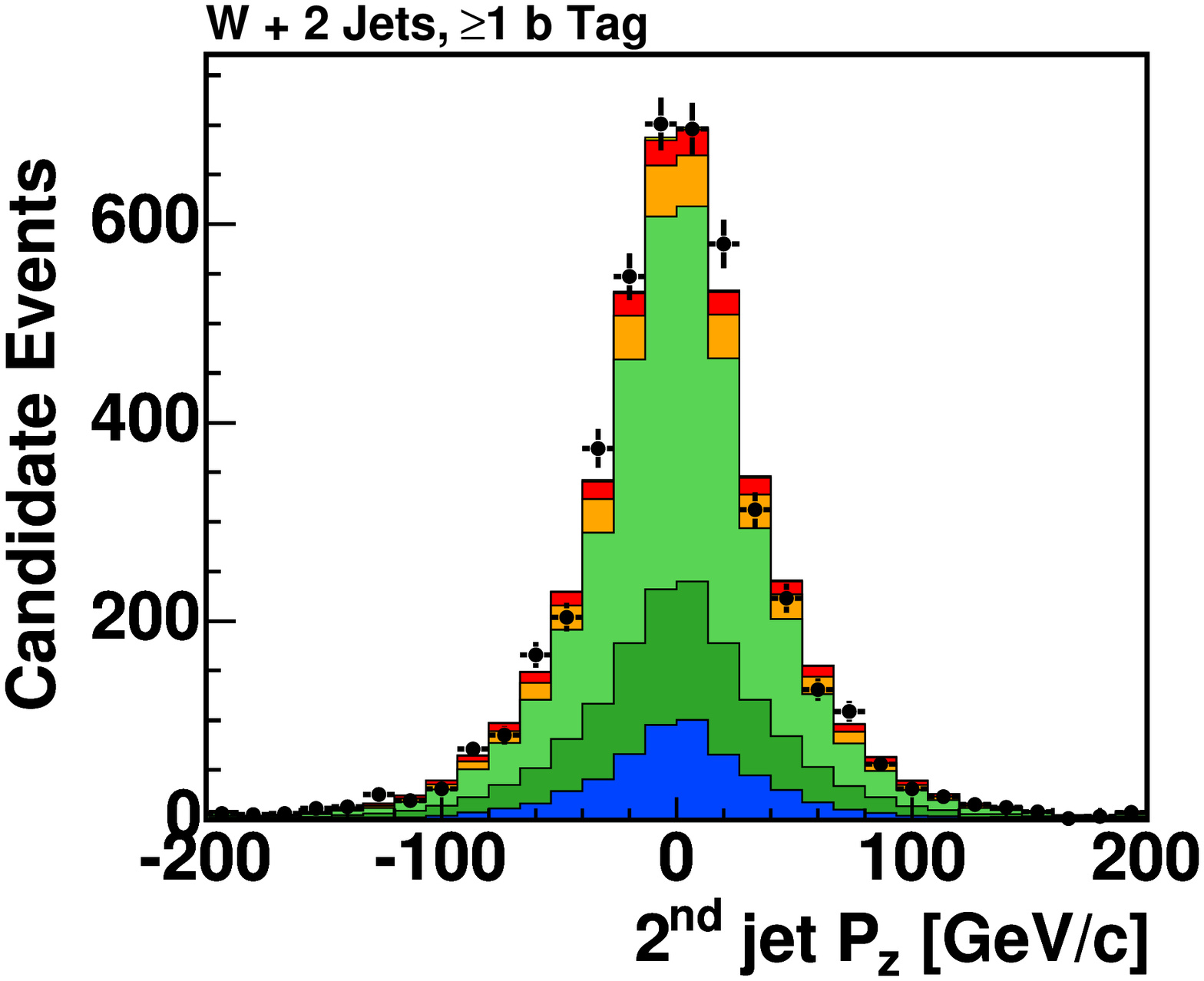}
\label{fig:e}}
\end{center}
\vspace{-0.85cm}\caption{\label{fig:4vect_2jet_1t}Validation plots comparing observed and MC simulated events for the 
  four-vector ($E$, $P_x$, $P_y$, $P_z$) of the lepton and the jets in
  events with 2-jets and at least one $b$-tagged jet.}
\end{figure*}

A data-MC comparison of the measured four vectors can be found in
Figs.~\ref{fig:4vect_2jet_0t} and \ref{fig:4vect_2jet_1t}.  This
comparison is done in the control (0 tag) and signal ($\geq$ 1 tag)
regions. In general, good agreement between observed data and MC
expectation is found.

\subsection{Transfer functions}
\label{sec:transferfunction}
The transfer function $W(y,x)$ gives the probability of measuring the
set of observable variables $x$ given specific values of the parton
variables $y$.  In the case of well-measured quantities, $W(y,x)$ is
taken as a $\delta$-function (i.e., the measured momenta are used in
the differential cross section calculation). When the detector
resolution cannot be ignored, $W(y,x)$ is a parametrized resolution
function based on fully simulated Monte Carlo events.  For unmeasured
quantities, such as the three components of the momentum of the
neutrino, the transfer function is constant. The choice of transfer
function affects the sensitivity of the analysis but not its
correctness, since the same transfer function is applied to both
observed and Monte Carlo events.

Lepton energies are measured well by the CDF detector and
$\delta$-functions are assumed for their transfer functions. The
angular resolution of the calorimeter and muon chambers is also
sufficient and $\delta$-functions are also assumed for the transfer
function of the lepton and jet directions. The resolution of jet
energies, however, is broad and it is described by a jet transfer
function $W_{\rm jet}(E_{\rm parton},E_{\rm jet})$.  Using these
assumptions, $W(y,x)$ takes the following form for the four final
state particles considered in the $WH$ search (lepton, neutrino and
two jets):
\begin{equation}
\label{eqn:wxy}
W(y,x)=\delta^3(\vec{p}_l^{~y}-\vec{p}_l^{~x})\prod_{i=1}^2\delta^2(\Omega_i^y-\Omega_i^x)\prod_{k=1}^2W_{j}(E_{p_k},E_{j_k})
\end{equation}
where $\vec{p}_l^{~y}$ and $\vec{p}_l^{~x}$ are the produced and
measured lepton momenta, $\Omega_i^y$ and $\Omega_i^x$ are the
produced quark and measured jet angles (cos$\Theta$, $\phi$), and
$E_{p_k}$ and $E_{j_k}$ are the produced quark and measured jet
energies.

\begin{figure}[h]
  \centering
  \includegraphics[width=0.9\columnwidth]{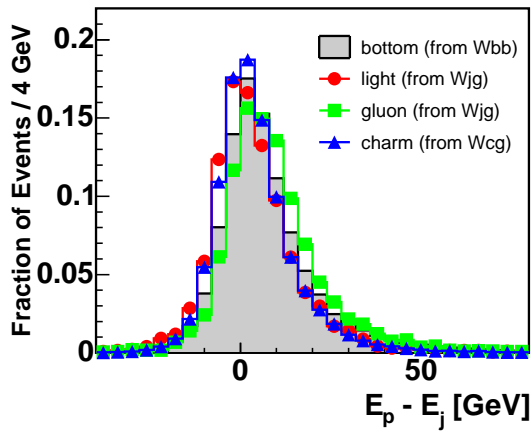}
  \caption[Distribution of
  $\delta_E~=~(E_{\mathrm{parton}}-E_{\mathrm{jet}})$]
  {\label{fig:wxy_all} Normalized
  $\delta_E=(E_{\mathrm{parton}}-E_{\mathrm{jet}})$ distributions for
  jets matched to partons in $WH$ with a Higgs boson mass of 115
  GeV/$c^2$ ($b$-jets), $W_{jg}$ (light-jets and gluons), and $W_{cg}$
  ($c$-jets) Monte Carlo events (passed through full detector
  simulation).}
\end{figure}

The jet energy transfer functions map parton energies to measured jet
energies after correction for instrumental detector
effects~\cite{Bhatti:2005ai}. This mapping includes effects of
radiation, hadronization, measurement resolution, and energy outside
the jet cone not included in the reconstruction algorithm. The jet
transfer functions are obtained by parametrizing the jet response in
fully simulated Monte Carlo events. The distributions of the
difference between the parton and jet energies,
$\delta_E=(E_{\mathrm{parton}}-E_{\mathrm{jet}})$, are parametrized as
a sum of two Gaussian functions:

\begin{widetext}
\begin{equation}
\label{eqn:wxyjet}
W_{\mathrm{jet}}(E_{\mathrm{parton}},E_{\mathrm{jet}})=\frac{1}{\sqrt{2\pi}(p_2+p_3p_5)}\left(\exp{\frac{-(\delta_E-p_1)^2}{2p_2^2}}+p_3~\exp{\frac{-(\delta_E-p_4)^2}{2p_5^2}}\right),
\end{equation}
\end{widetext}
one to account for the sharp peak and the other one to account for the
asymmetric tail, because the $\delta_E$ distributions (shown in
Fig.~\ref{fig:wxy_all} for different flavor jets) are asymmetric and
features a significant tail at positive $\delta_E$.

Different transfer functions are created depending on the physics
process and the flavor of the jet due to the different kinematics as
shown in Fig.~\ref{fig:wxy_all}. To take into account the different
kinematics of the physics processes used in this analysis ($WH$
[100-150] GeV/$c^2$, $Wb\bar b$, $t\bar t$, $s$-channel and
$t$-channel single top, $Wc\bar c$, $Wcg$, $Wjg$, $Wgg$, $WW$, and
$WZ$) and the different flavor of jet ($b$, $c$, light and gluons), 23
different transfer functions are created as explained below.

One of the novelties of this analysis is that, in order to better
reproduce the parton energy ($E_{\mathrm{parton}}$), a neural network
output ($O_{\mathrm{NN}}$) is used instead of the measured jet energy
($E_{\mathrm{jet}}$).  This output distribution is not a neural
network output event classifier distribution, but rather a functional
approximation to the parton energy. So $W_{\rm
jet}(E_{\mathrm{parton}}, E_{\mathrm{jet}})$ is substituted for
$W_{\rm jet}(E_{\mathrm{parton}}, O_{\mathrm{NN}})$, and it is
commonly referred as a neural network transfer function (or NN TF).
The $O_{\mathrm{NN}}$ used in the analysis is the result of training
neural networks (NNs) using the Stuttgart neural network simulator
(SNNS)~\cite{bib:snns}.  For each physics process considered, a
different NN is constructed for each type of jet in that process as
shown in Table~\ref{tab:nn_types}.  By using the jets from the
specific process to train the NN it is assured that the NN is
optimized for the kinematics of the jets associated with that process.

\begin{table}[h]
\caption{\label{tab:nn_types} Types of jets used to train the different NNs for each process.}
\begin{center}
\begin{tabular}{lcccc}
\hline
\hline
Process              & ~$b$ jets~ & ~$c$ jets~ &  ~light jets~  & ~~gluons~~ \\
\hline
$WH$ (11 $m_H$ values)& {X}   &            &           &     \\
$Wb\bar b$          & {X}   &            &           &     \\
$Wc\bar c$          &            & {X}   &           &     \\
$t\bar t$           & {X}   &            &           &     \\
$s$-channel           & {X}   &            &	          &     \\
$t$-channel           & {X}   &            &	{X}  &     \\
$Wcg$                &            & {X}   &           & {X}    \\
$Wjg$                 &            &            & {X}  & {X}    \\
$Wgg$                 &            &            &           & {X}    \\
$WW-WZ$               &            &            & {X}  &    \\
\hline \hline
\end{tabular}
\end{center}
\end{table}

The training of the NNs is based on MC simulated events. The MC
events used for the trainings are the remaining events after applying
the analysis event selection (see Section~\ref{sec:reconstr}) and the
jets are required to be aligned within a cone of $\Delta R<$ 0.4 with
the closest flavored parton ($b$ or $c$ depending on the physics
process) coming from the hard scattering process.

All the NN trainings have the same architecture and input variables.
Seven input variables related to the jet kinematics have been used:
the total corrected energy of the jet ($E$), the raw (measured)
transverse momentum of the jet ({\it p$_T$}), the azimuthal angle of
the jet ($\phi$), the pseudorapidity of the jet ($\eta$), the raw
(measured) energy of the jet, the total corrected energy of the jet in
a cone of radius $R\leq$0.7 (E cone 0.7), and the sum over the tracks
in the jet of the ratio of the transverse momentum of the track and
the sine of the $\theta$ of the track ($\sum_p$).

\begin{figure}[h]
\includegraphics[width=0.49\columnwidth]{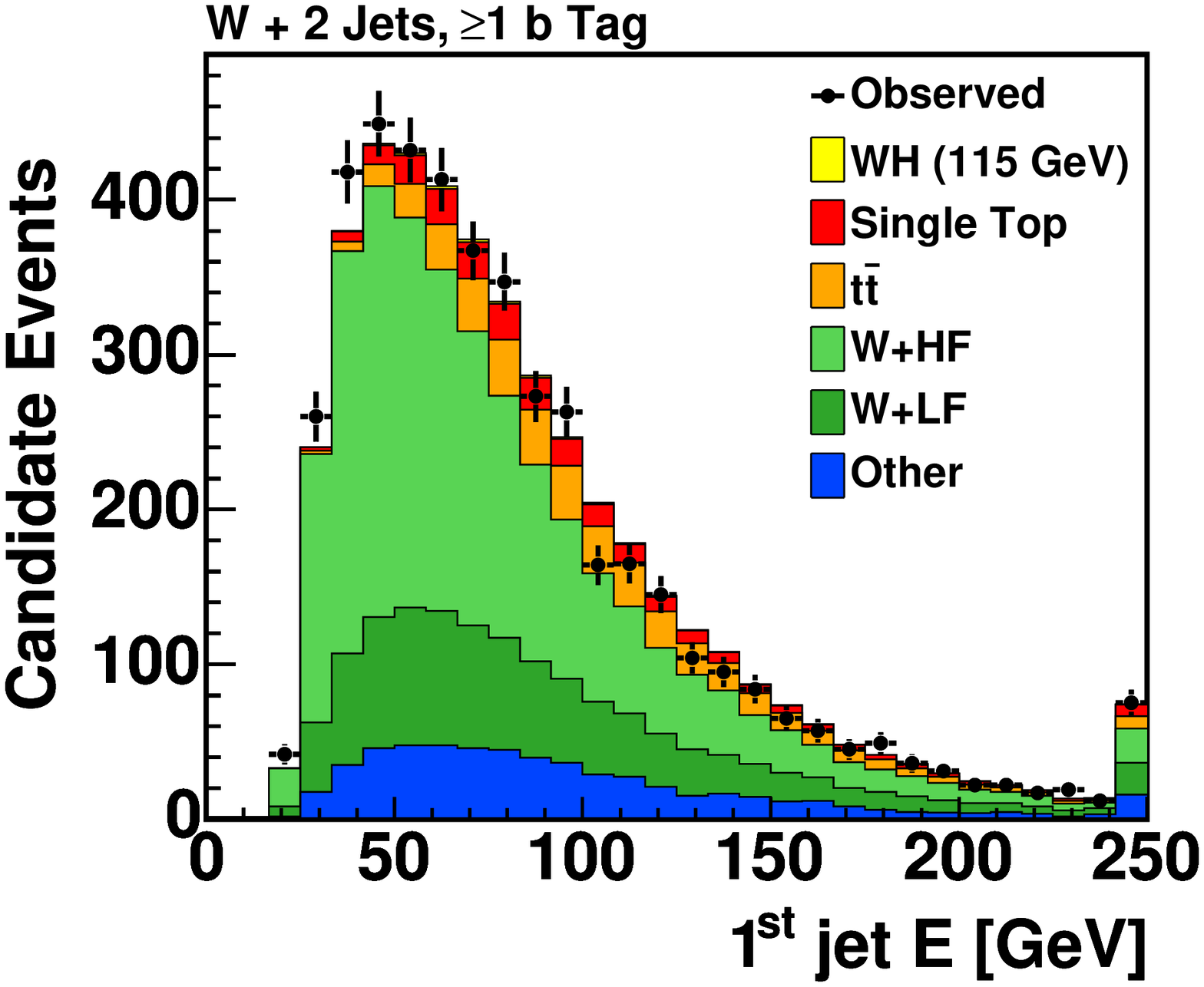}
\includegraphics[width=0.49\columnwidth]{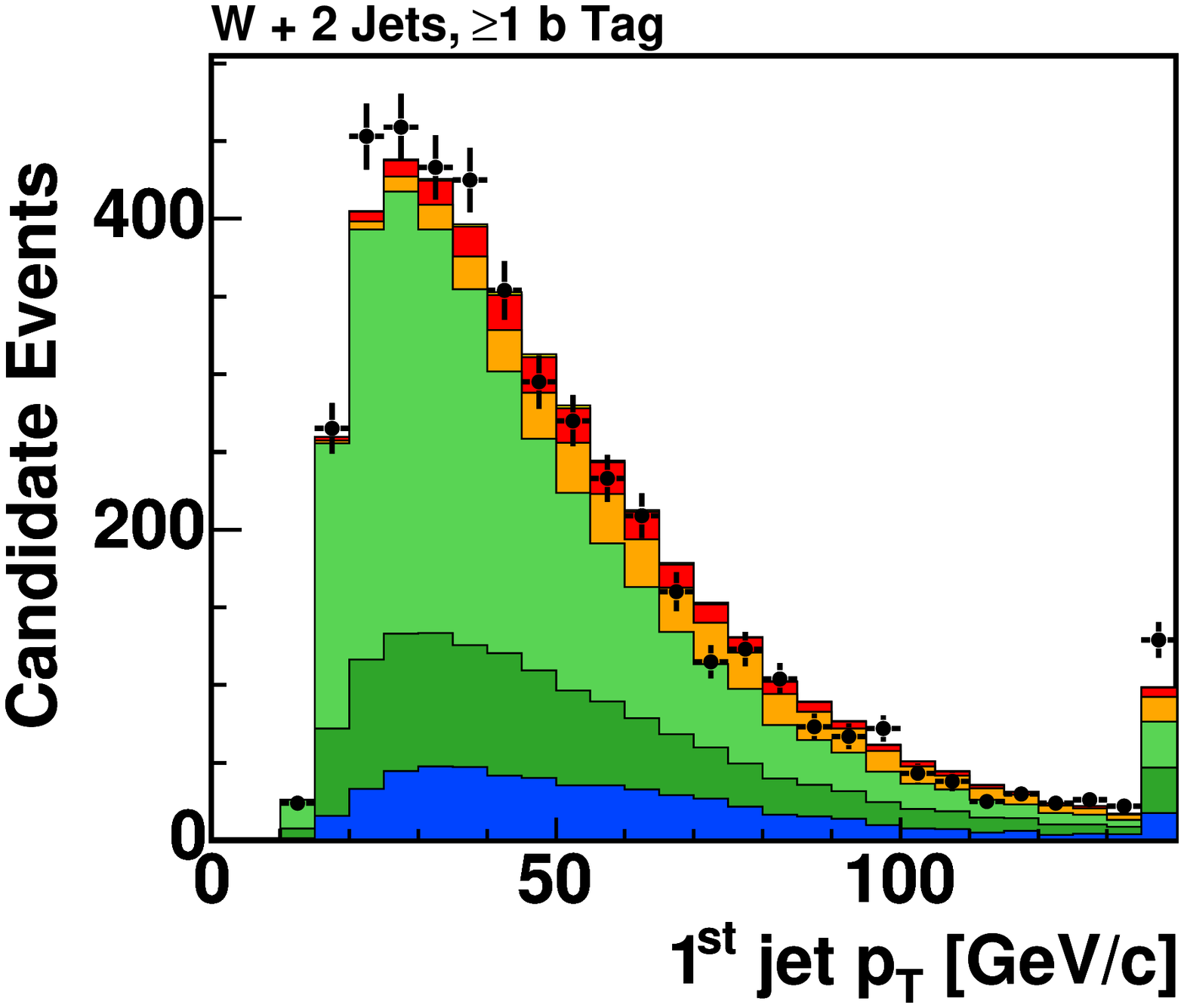}
\includegraphics[width=0.49\columnwidth]{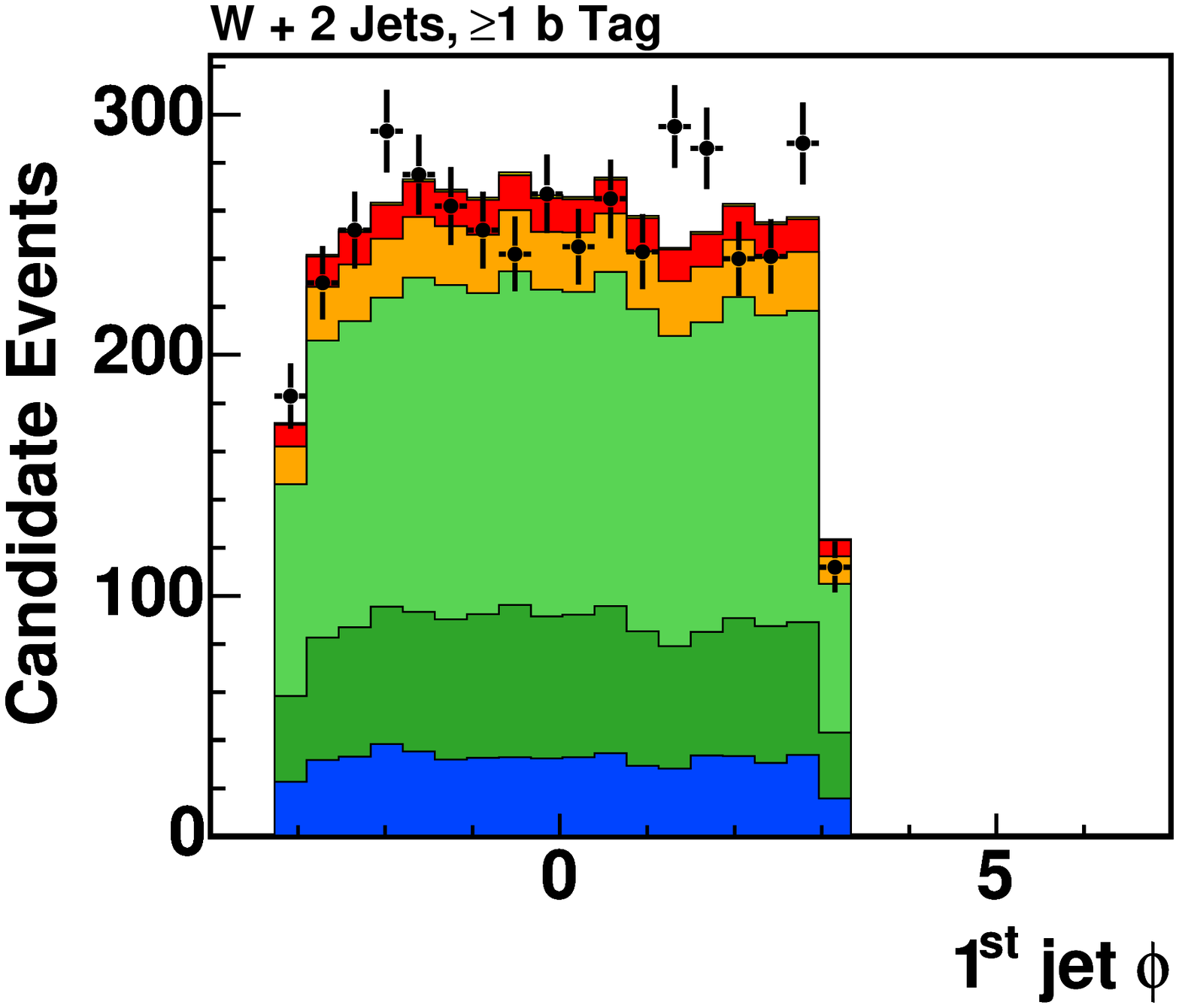}
\includegraphics[width=0.49\columnwidth]{figures_WH_ME_PRD/VAR_jets0.lv.l5.Eta_CUT_jet2bin_gr1tag_All_scaleData.eps}
\includegraphics[width=0.49\columnwidth]{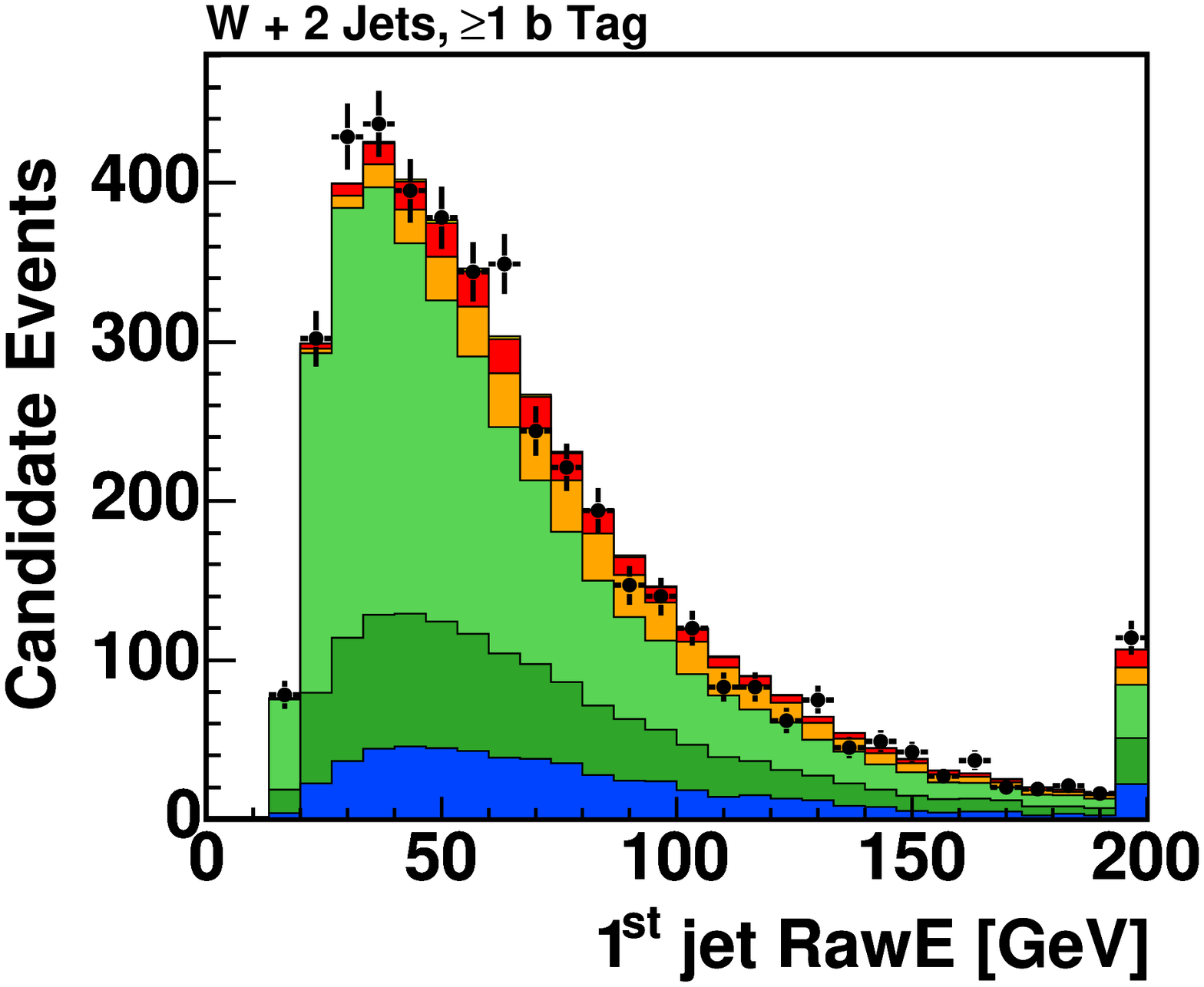}
\includegraphics[width=0.49\columnwidth]{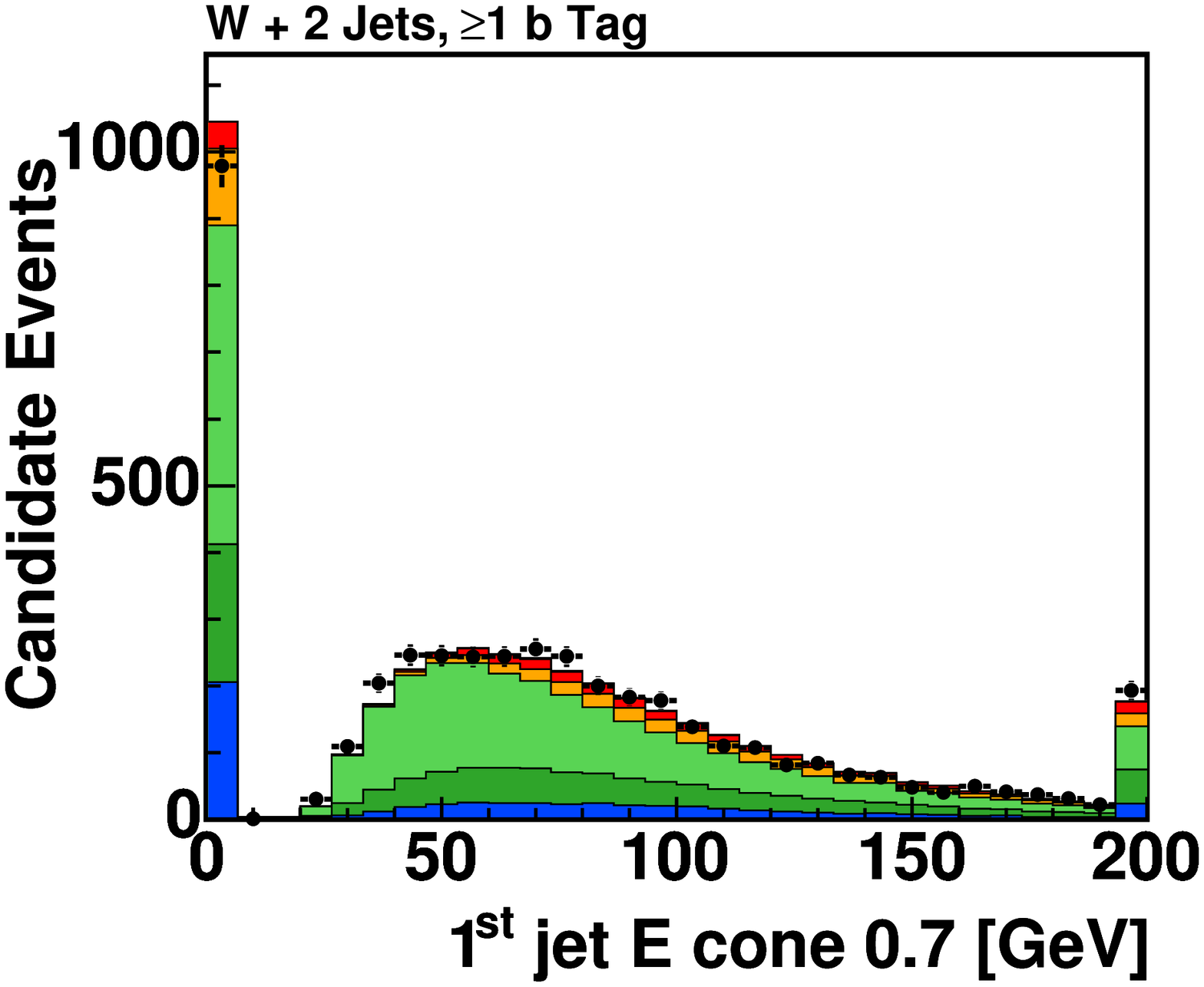}
\includegraphics[width=0.49\columnwidth]{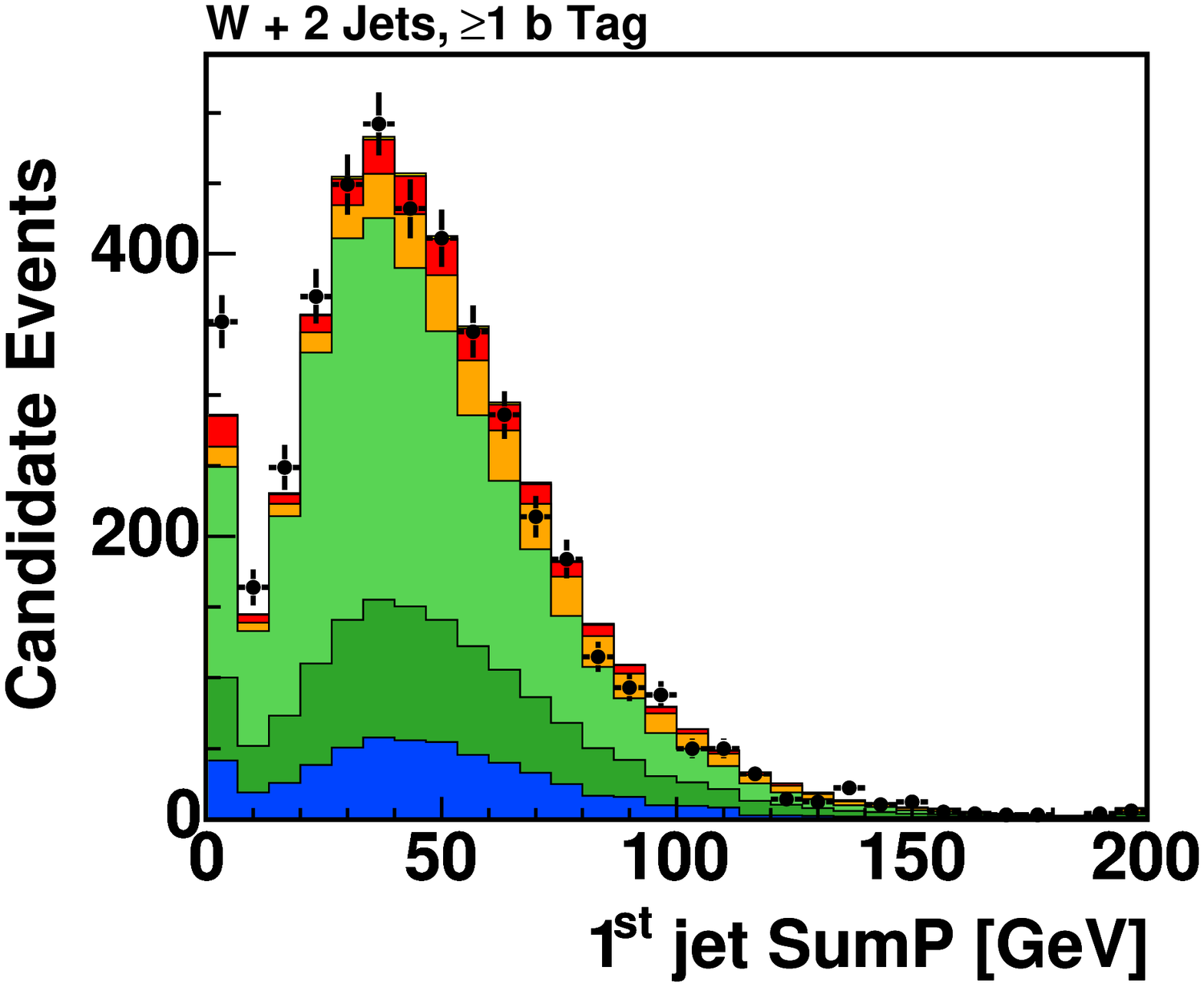}
\caption{\label{fig:NNTF_validation}Validation plots comparing observed
and Monte Carlo simulated events for the seven input variables of the neural network 
transfer function for the first leading jet for events with two jets 
and at least one $b$~tag. The observed events are indicated with points. }
\end{figure}

Figure~\ref{fig:NNTF_validation} shows the data-MC comparison of the
seven input variables for the leading jet in two-jet events where at
least one of the jets has been tagged by {\sc SecVtx} which also
validates the MC expectations in this signal region.

\begin{figure}
  \subfigure[]{\includegraphics[width=0.49\columnwidth]{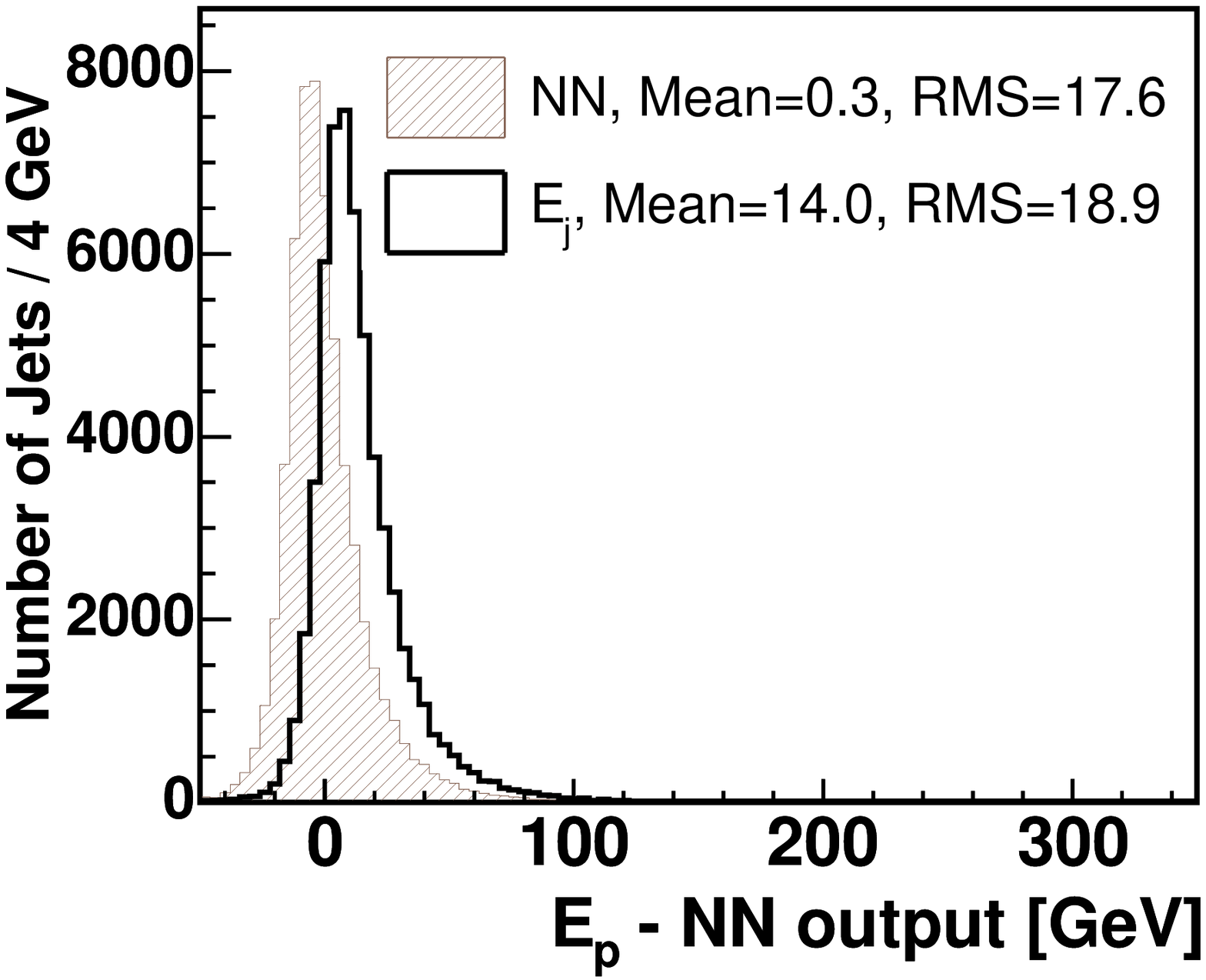}}
  \subfigure[]{\includegraphics[width=0.49\columnwidth]{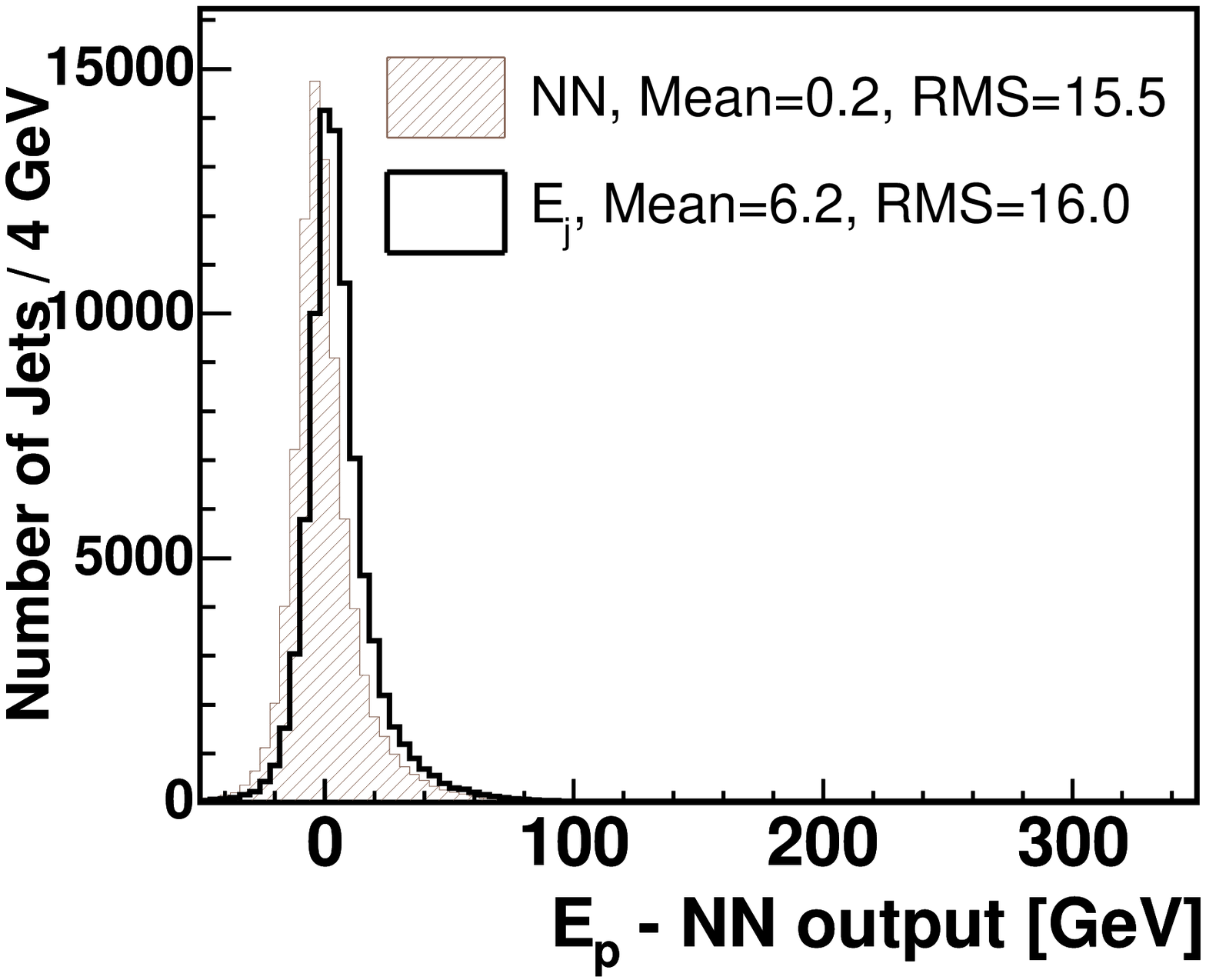}}
  \subfigure[]{\includegraphics[width=0.49\columnwidth]{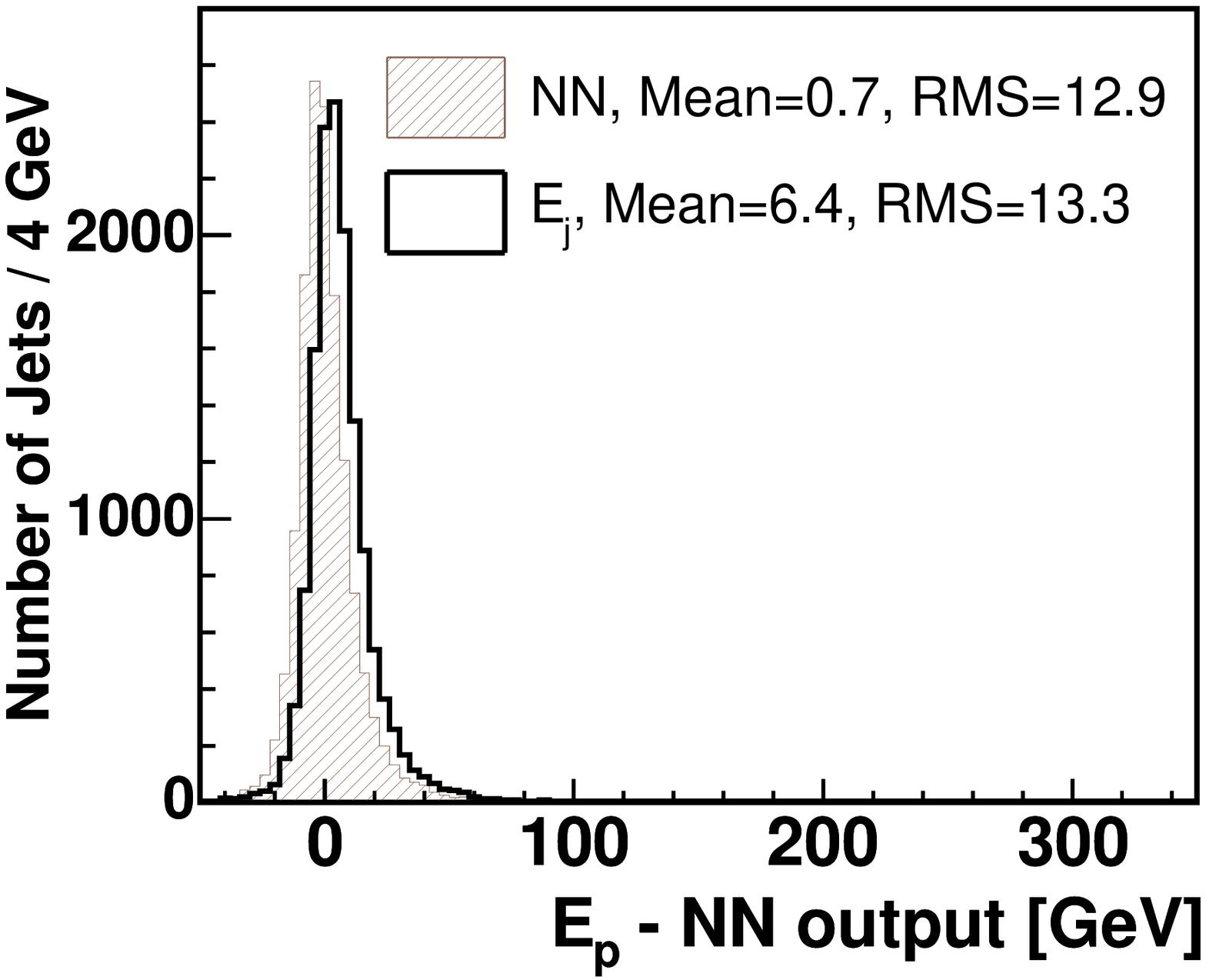}}
  \subfigure[]{\includegraphics[width=0.49\columnwidth]{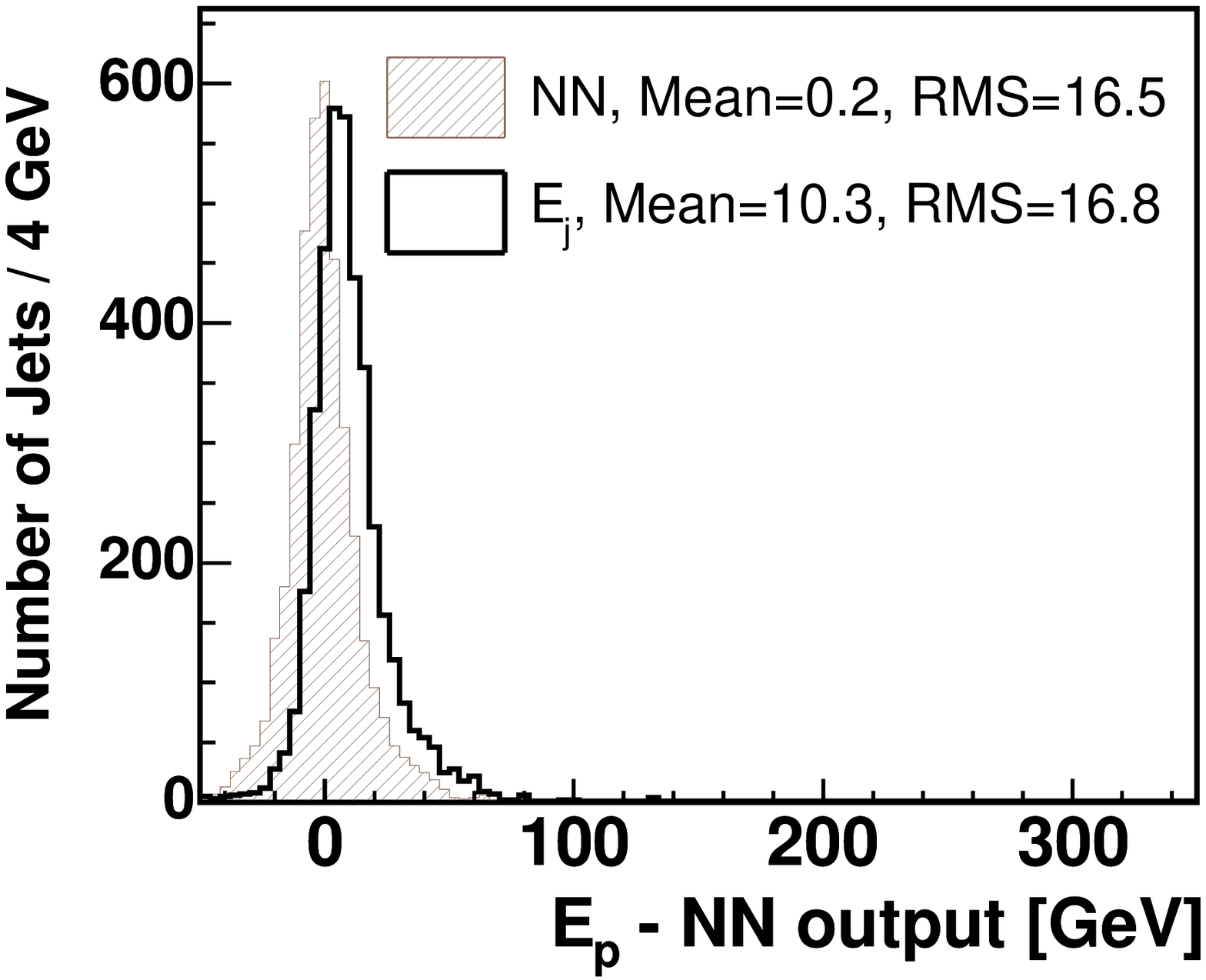}}
  \caption{\label{fig:epNNdiff} Difference between the parton energy
and the measured jet energy (empty histogram) and the
$O_{\mathrm{NN}}$ (dashed histogram) for $b$-jets in $WH$ (m$_H$ = 115
GeV/$c^2$) events (a), $b$-jets in $Wb\bar b$ events (b), light jets
in diboson ($WW$, $WZ$) events (c) and $Wgg$ events for gluon (d).}
\end{figure}

Figure~\ref{fig:epNNdiff} shows the difference between the parton
energy and the corrected jet energy and between the parton energy and
the $O_{\mathrm{NN}}$ for four different physics processes, $WH$,
diboson ($WW$, $WZ$), $Wb\bar b$, and $Wgg$.  In all
cases the average $O_{\mathrm{NN}}$ is closer to the parton energy
than the average corrected jet energy and that the distributions are
more narrow. Therefore, since the $O_{\mathrm{NN}}$ provides a better
jet resolution, using it as an input to the transfer function should
help to improve the performance of the transfer function.

The functional form used to parametrize
$E_{\mathrm{parton}}$-$O_{\mathrm{NN}}$ is the same as the one
described above for $\delta_E$ (Eq.~\ref{eqn:wxyjet}).  More details
on the performance of the NN TF can be found in
Ref.~\cite{bib:barbara_thesis}.

\begin{figure}[h!]
  \includegraphics[width=0.49\columnwidth]{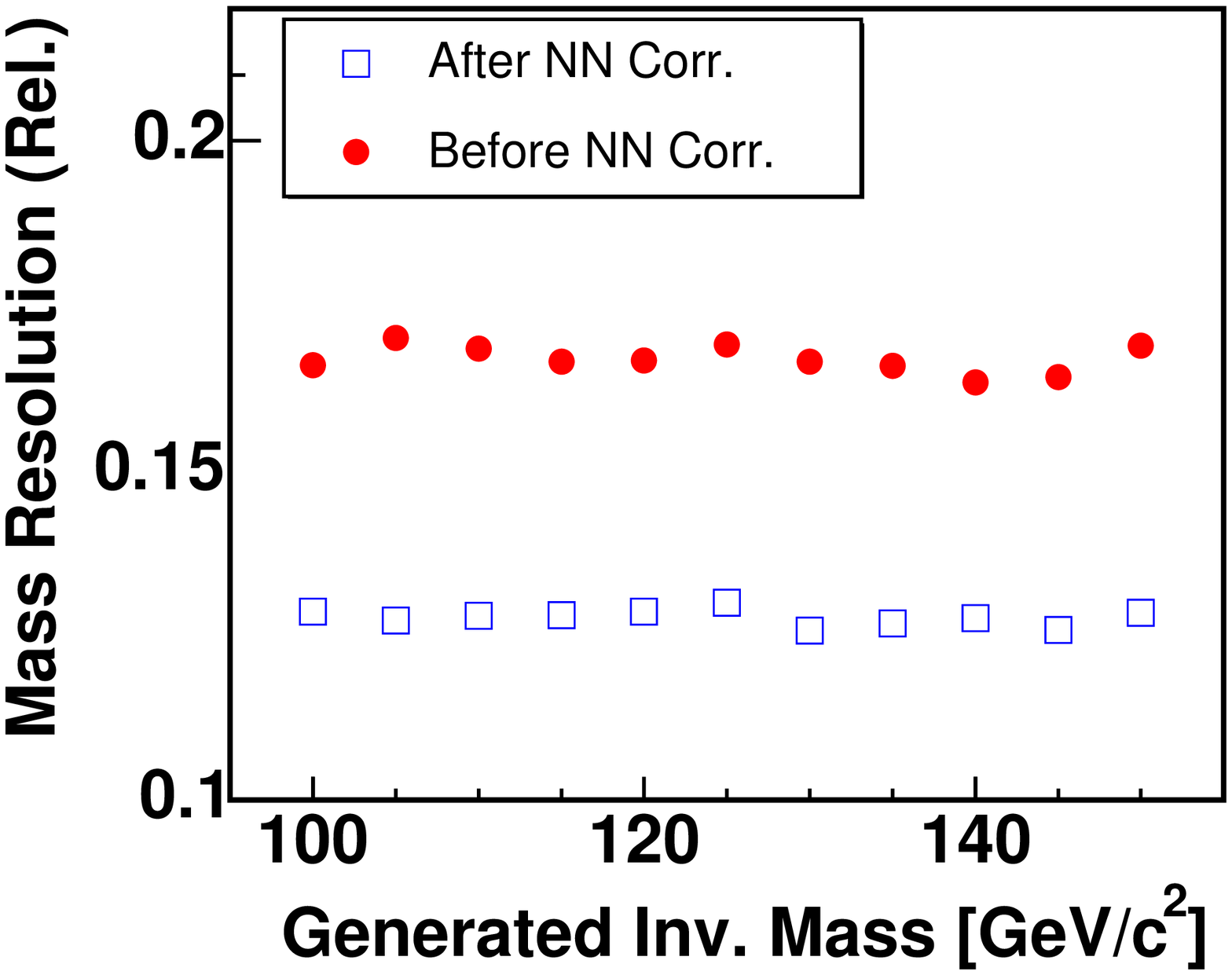}
  \includegraphics[width=0.49\columnwidth]{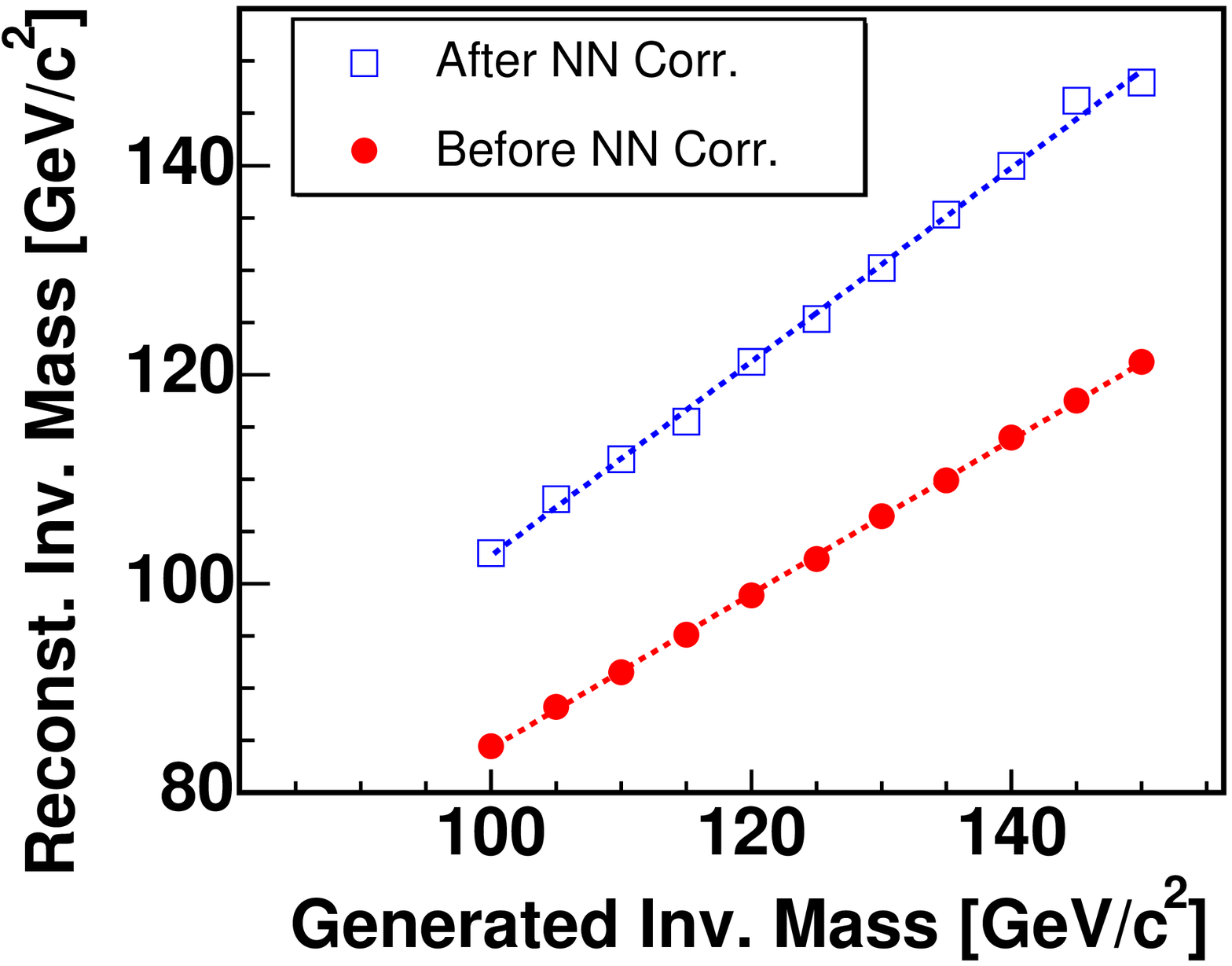}
  \caption{\label{fig:nncorr} Left (right): Relative resolution, sigma
divided by the mean of the Gaussian fit to the invariant mass
distributions, as a function of the invariant mass (reconstructed vs
generated invariant mass) before and after applying the NN correction
to the measured jets.}
\end{figure}

The output of the neural network is used to correct the measured
energy of all the jets from the events that pass the analysis
selection. As a cross-check, a comparison of the invariant mass
resolution of the dijet system in $WH$ signal events before and after
applying this correction is performed. A way to do this is to fit the
invariant mass distribution to a Gaussian function and compare the
resolution, defined as the sigma divided by the mean of the fit, for
all Higgs boson masses. The results are shown in Fig.~\ref{fig:nncorr}
(left). As expected, the invariant mass resolution is better (smaller
sigma) after correcting by the $O_{\mathrm{NN}}$.  The linearity of
the correction is also checked, see Fig.~\ref{fig:nncorr}
(right). Both functions are linear. The only difference is that the
reconstructed invariant mass is closer to the generated one once the
correction is applied.

%----------------------------------------------------------------
\subsection{Event probability discriminant}
\label{sec:epd}
%----------------------------------------------------------------
The event probability densities are used as inputs to build an
event probability discriminant, a variable for which 
the distributions of signal events and background events are
maximally different.

An intuitive discriminant which relates the signal and
background probability densities is the ratio of signal probability
over signal plus background probability,
$EPD=P_{signal}/(P_{signal}+P_{background})$. By construction, this
discriminant is close to zero for background-like events
($P_{background} \gg P_{signal}$) and close to unity for signal-like
events ($P_{signal} \gg P_{background}$).  Expressions
\ref{eqn:epd_SV} and \ref{eqn:epd_SVSV} are the definitions of the
event probability discriminants used in this analysis for single and
double $b$-tagged events, respectively:
\begin{widetext}
\begin{equation}
\label{eqn:epd_SV}
\footnotesize
EPD \equiv\frac{ b~ \hat{P}_{WH}}{ b~( \hat{P}_{WH}+
\hat{P}_{Wb{\bar b}}+ \hat{P}_{t{\bar t}}+ \hat{P}_{s}+
\hat{P}_{t})+(1-b)(\hat{P}_{Wc{\bar c}}+\hat{P}_{Wcj}+
\hat{P}_{W+l}+ \hat{P}_{Wgg}+ \hat{P}_{dib})}
\end{equation}

\begin{equation}
\label{eqn:epd_SVSV}
\footnotesize
  EPD \equiv \frac{ b_1~ b_2~ \hat{P}_{WH}}{ b_1
     b_2(\hat{P}_{WH}+\hat{P}_{Wb{\bar b}}+\hat{P}_{t{\bar t}}+
     \hat{P}_{s})+b_1(1-b_2)
     \hat{P}_{t}+(1-b_1)(1-b_2)(\hat{P}_{Wc{\bar
     c}}+\hat{P}_{Wcj}+\hat{P}_{W+l}+\hat{P}_{Wgg}+\hat{P}_{dib})}
     \end{equation}\\
\end{widetext}
where $\hat{P}_i$ = $C_i\cdot P_i$, $P_i$ is the event probability of
a given physics process ($WH$, $s$-channel, W$b\bar b$, ...), $C_i$
are additional coefficients (to be defined below), and $b$ (defined as
the $b$-jet probability) is a transformation of the output of the neural network jet flavor separator
($b_{\mathrm{NN}}$)~\cite{stprd,Richter:2007zzc}.

Extra non-kinematic information is introduced into the event
probability discriminant by using $b_{\mathrm{NN}}$, and $C_i$.  The
$C_i$ coefficients are included into the EPD and used to optimize the
discrimination power between signal and background.  This set of
coefficients is obtained by an iterative technique that involves the
repeated generation of different sets of parameters and the
computation of the expected limit for each set. However, because the
calculation of limits with the inclusion of systematic uncertainties
is computationally intensive, the optimization is implemented by
performing a faster calculation for a figure of merit based only on
statistical uncertainties. This has been successfully used in previous
versions of this analysis and in the most recent measurement of the
$WW$~+~$WZ$ production cross section~\cite{bib:WWWZxsec}.

For any given set of coefficients $C_i$, the Monte Carlo templates of
the EPD variable are generated normalized to the corresponding number
of expected signal and background events calculated in
Section~\ref{sec:bkgEstim}. The figure of merit is obtained from these
templates using a maximum likelihood fit to extract $\xi$ and its
error $\sigma_{\xi}$, where $\xi$ is a multiplicative factor to the
expected $WH$ cross section. The negative logarithm of the likelihood
used is:

\begin{equation}
\label{eqn:likelihood}
-\log\left(\mathcal{ L(\xi)}\right) = \sqrt{\sum_{k=1}^{n_{bin}}
 {\frac{(\xi S_k)^2}{\xi S_k + B_k + (\xi \Delta S_k)^2 + (\Delta
 B_k)^2}}},
 \end{equation}\\
where $S_k$ and $B_k$ are the expected number of signal and background
events in the $k^{th}$ bin and $\Delta S_k$ and $\Delta B_k$ are the
statistical uncertainty on $S_k$ and $B_k$, respectively.  The
variable $\xi$ represents the most likely value of signal, in units of
the expected signal cross section, that can be fitted on the
background templates and should be always close to zero after the
minimization. The error on the value of $\xi$ is obtained from the
minimization and is related to the strength by which the signal can be
differentiated from the background templates in units of the expected
signal cross section; the larger the error the smaller the strength
and vice versa. For each set of EPD templates the figure of merit is
defined as $1/\sigma_{\xi}$.

The best set of coefficients is then obtained using an iterative
technique, where at the beginning the current best set of coefficients
is initially set to the maximum matrix element probability values
obtained in the respective samples.  For every iteration a trial set
of coefficients is formed by introducing random changes in some of the
coefficients from the current best set, creating new EPD templates and
calculating the corresponding figure of merit of these new EPDs. The
set of coefficients that produces the best figure of merit based on
$\sim$ 2000 iterations is considered optimal and used to for the
analysis.

After the event selection and applying $b$-tagging, several of the
sizable background processes do not have a $b$-quark in the final
state, but are falsely identified as such. This happens either because
a light quark jet is falsely identified to have a displaced secondary
vertex from the primary vertex due to tracking resolution (mistag) or
because charm quark decays happen to have a sufficiently long lifetime
to be tagged.  Therefore, it would be desirable to have better
separation of $b$-quark jets from charm or light quark jets. The
neural network jet flavor separator is used to achieve this
separation.  As mentioned before, the $b$ variable used in the EPD is
a transformation of the $b_{\mathrm{NN}}$ in such a way that it goes
from 0 to 1.  The neural network jet flavor separator is a continuous
variable and the result of a neural network training that uses a broad
range of variables in order to identify $b$-quark jets with high
purity~\cite{Richter:2007zzc}.  A variety of variables is suitable to
exploit the lifetime, mass, and decay multiplicity of
$b$-hadrons. Many of them are related to the reconstructed secondary
vertex; some are reflected by the properties of the tracks in the {\sc
SecVtx} tagged jet.  Including this factor helps to discriminate
signal from background events and improves the final sensitivity.

The event probability discriminants are defined for all the MC events
that pass the analysis selection (see Sect.~\ref{sec:reconstr})
including events with at least one jet tagged by {\sc SecVtx}. This
provides sufficient MC statistics except for $W$~+~LF and non-$W$
events, so in these cases events with no tagged jets are also
included.

The EPDs, for MC events, are defined independently of the tagging
category of the event, but later on, when making the final templates,
the events are weighted by the corresponding tagging probability.
These tagging probabilities are the $b$-tagging correction factor
($\varepsilon_\mathrm{tag}$) used in Eqs.~\ref{eq:evt} and
\ref{wplushf}.  They are functions of the flavor of the quark, the
tagging scale factor and the mistag matrix, a parametrization of the
mistag rate.  If a jet is matched to a heavy-flavor hadron ($\Delta
R$(jet, HF hadron)~$<$~0.4) and tagged by one of the $b$-tagging
algorithms, the weight is the corresponding tagging scale factor
(shown in Table~\ref{tab:tageff1}). If it is matched to a heavy flavor
hadron but the jet is not tagged by any of the $b$-tagging algorithms,
the weight is set to zero. If the jet is not matched to heavy flavor,
it is assigned a weight equal to its mistag probability
(Section~\ref{sec:mistag}), regardless of whether or not it was
tagged, because the Monte Carlo simulation does not properly model
mistagging.  On the other hand, for observed events, tagging is
required and the events are not weighted by any tagging probability.

Since the neural network jet flavor separator $b_{\mathrm{NN}}$ is
defined only for {\sc SecVtx} tagged jets, it requires a special
treatment for the events where any of the jets is not tagged.
$b_{\mathrm{NN}}$ is used for each type of event, in the cases where
the jet is not tagged the value of the $b_{\mathrm{NN}}$ is randomized
using the light or non-$W$ flavor separator template.

In the case of three-jet events (for two-jet events the same idea
applies), for Eq.~\ref{eqn:epd_SV} (EPD for the SVJP and SVnoJP
categories) the criteria for choosing $b$ are:

  \begin{itemize}
     \item[$\bullet$] if the three jets are {\sc SecVtx} tagged, the
     $b$-jet probability of one of them is chosen randomly;
     \item[$\bullet$] if two jets are {\sc SecVtx} tagged, the $b$-jet
     probability of one of them is chosen randomly;
     \item[$\bullet$] if one jet is {\sc SecVtx} tagged, the $b$-jet
     probability of that jet is used;
     \item[$\bullet$] if no jet is {\sc SecVtx} tagged, the $b$-jet
     probability is randomized (a random value is taken from the light
     flavor template for $W$~+~light events and from the non-$W$
     template for non-$W$ events) for each of the 3 jets and one of
     them is chosen randomly.
  \end{itemize}
For Eq.~\ref{eqn:epd_SVSV} (EPD for the SVSV category), the criteria
for choosing $b_1$ and $b_2$ are:

  \begin{itemize}
     \item[$\bullet$] if the three jets are {\sc SecVtx} tagged, the
     $b$-jet probabilities of two of them are chosen randomly;
     \item[$\bullet$] if two jets are {\sc SecVtx} tagged, the $b$-jet
     probability of both of them is used (in random order);
     \item[$\bullet$] if one jet is {\sc SecVtx} tagged, the $b$-jet
     probability of the tagged jet and a random value out of the other
     jets are used (in random order);
     \item[$\bullet$] if no jet is {\sc SecVtx} tagged, the $b$-jet
     probability of the three jets is randomized and two of them are
     randomly chosen.
  \end{itemize}

\begin{figure*}
\begin{center}
\includegraphics[width=0.65\columnwidth]{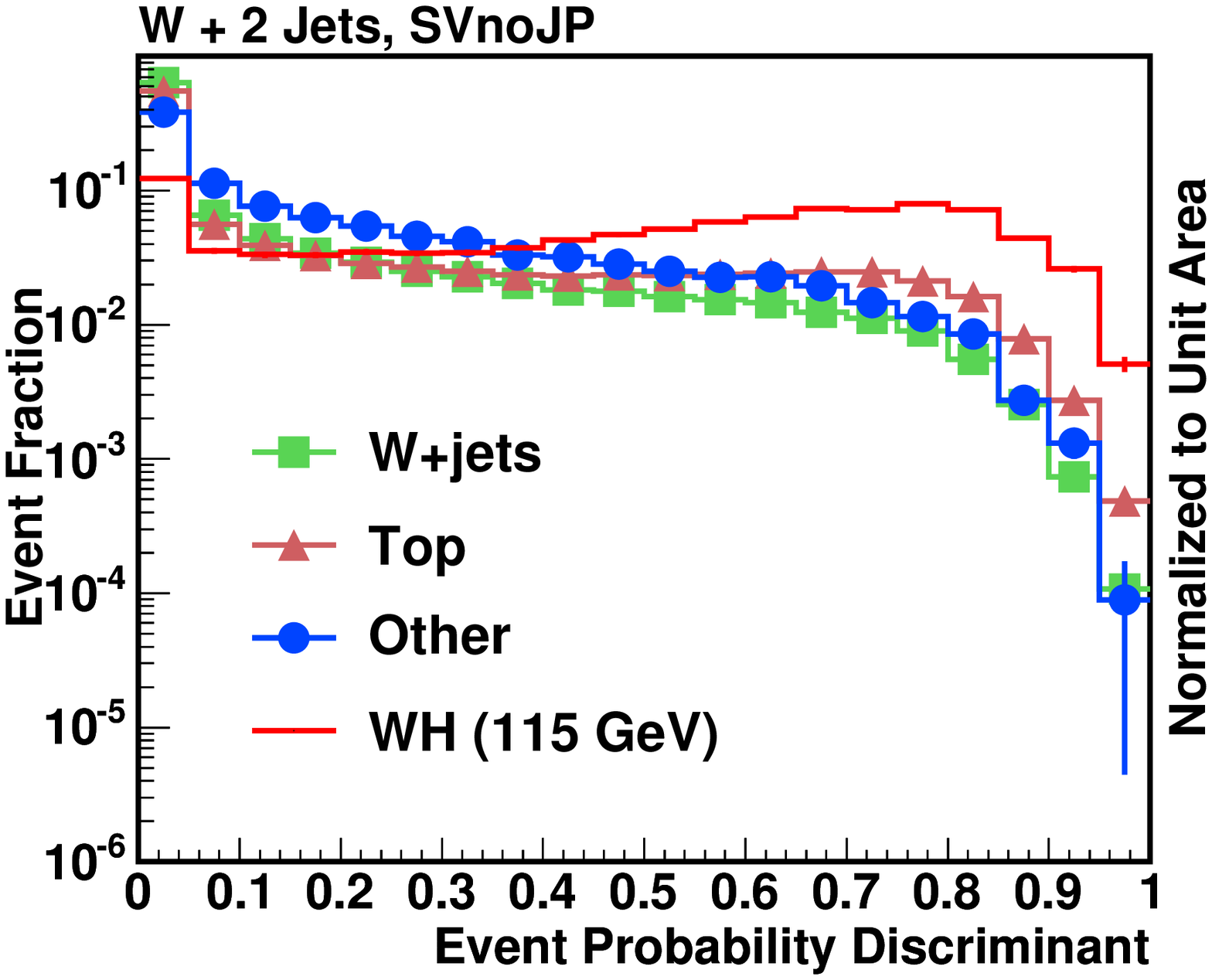}
\includegraphics[width=0.65\columnwidth]{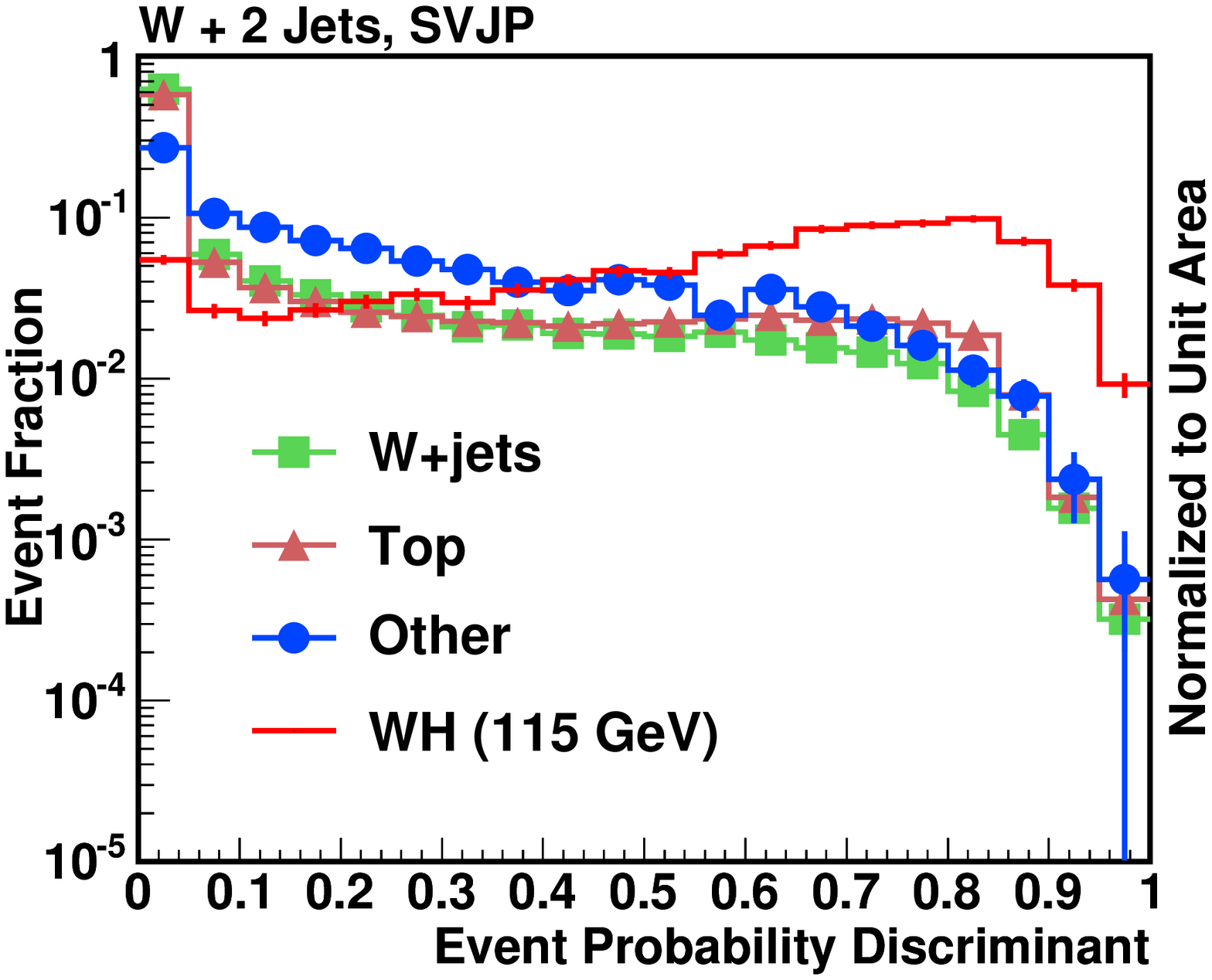}
\includegraphics[width=0.65\columnwidth]{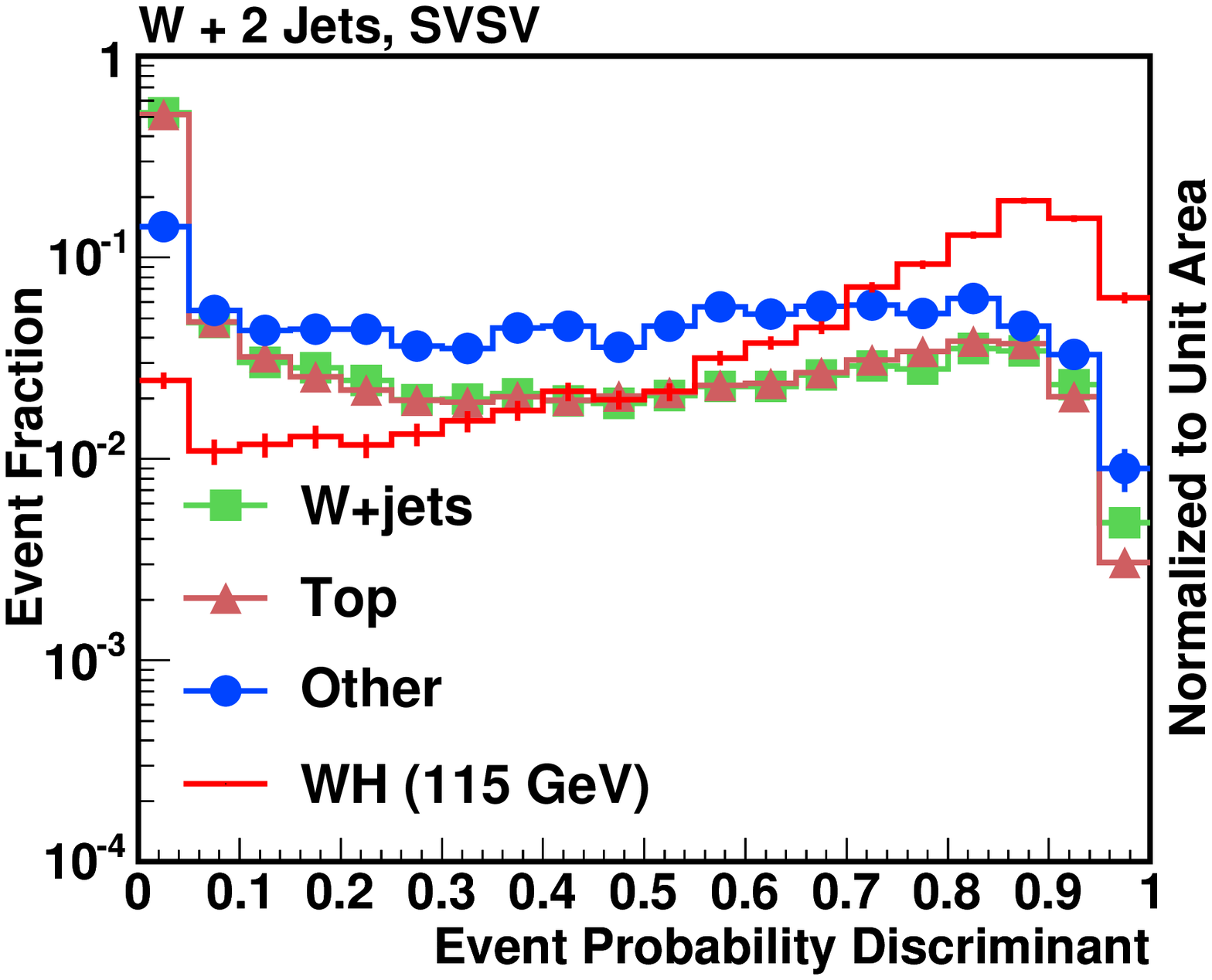}
\end{center}
\caption{\label{fig:EPD_2jets}
Templates of predictions for the signal (m$_H$~=~115~GeV/$c^2$) and
background processes, each scaled to unit area, of the
ME~discriminant, $EPD$, for 2-jet events for each signal region.}
\end{figure*}

\begin{figure*}
\begin{center}
\includegraphics[width=0.65\columnwidth]{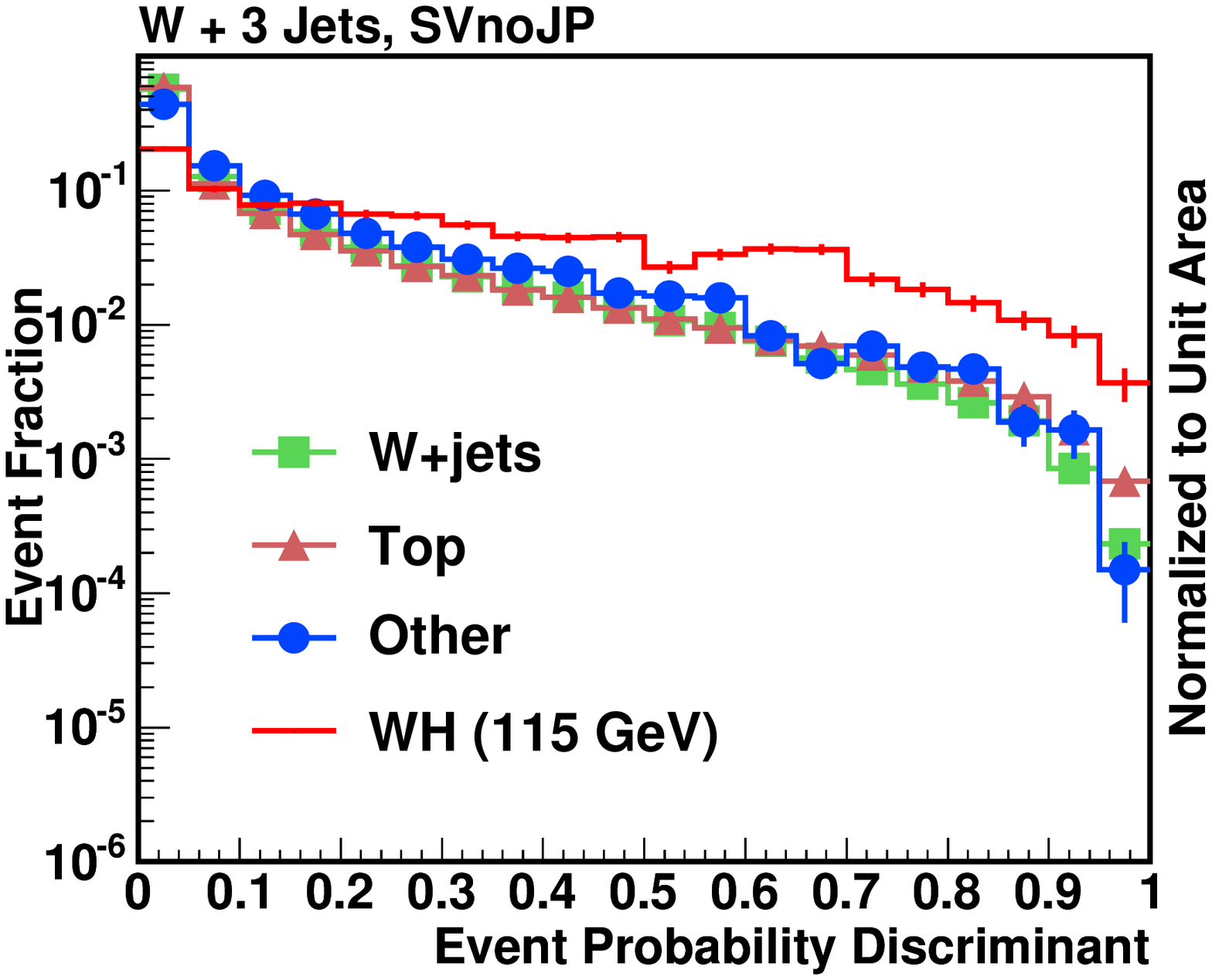}
\includegraphics[width=0.65\columnwidth]{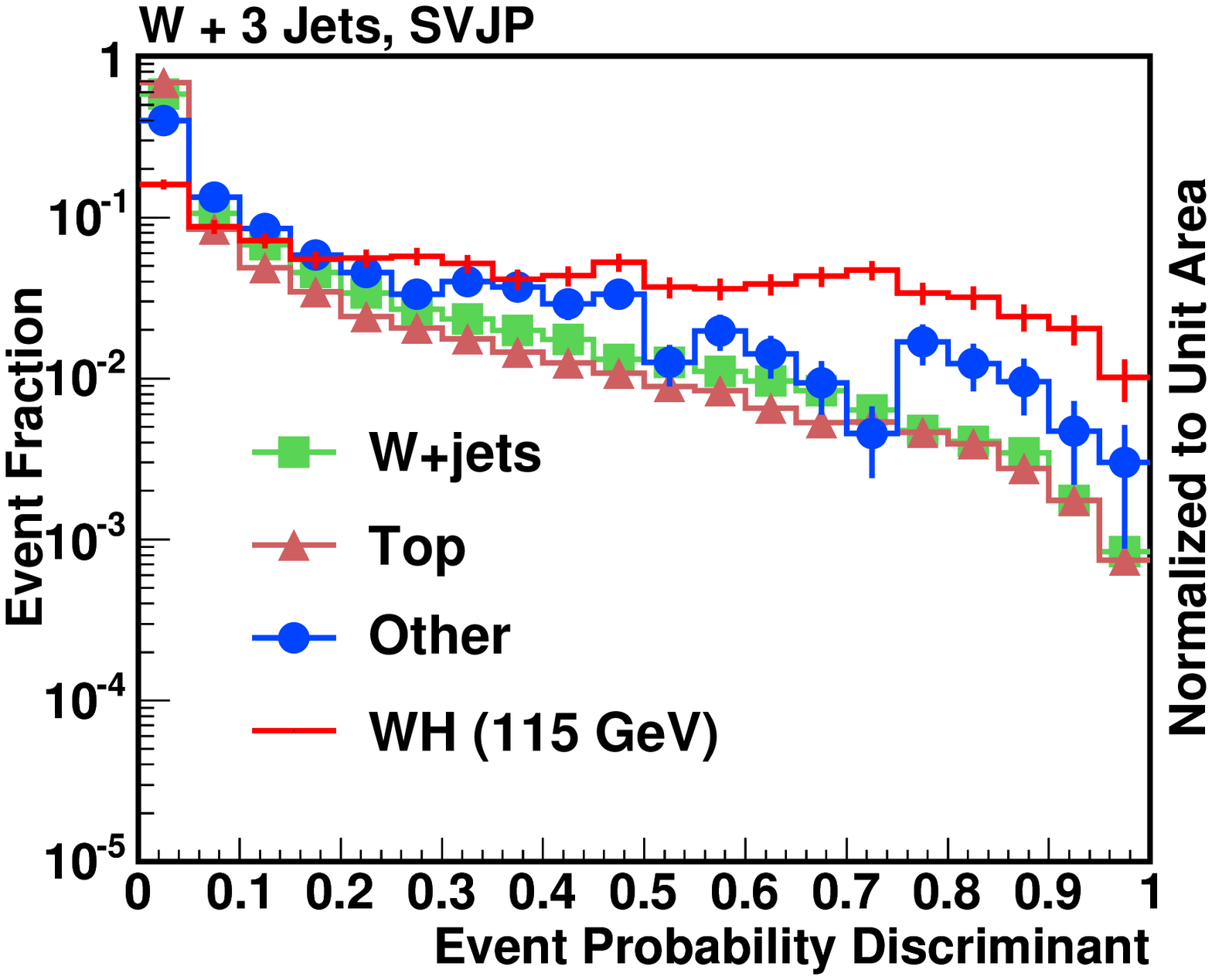}
\includegraphics[width=0.65\columnwidth]{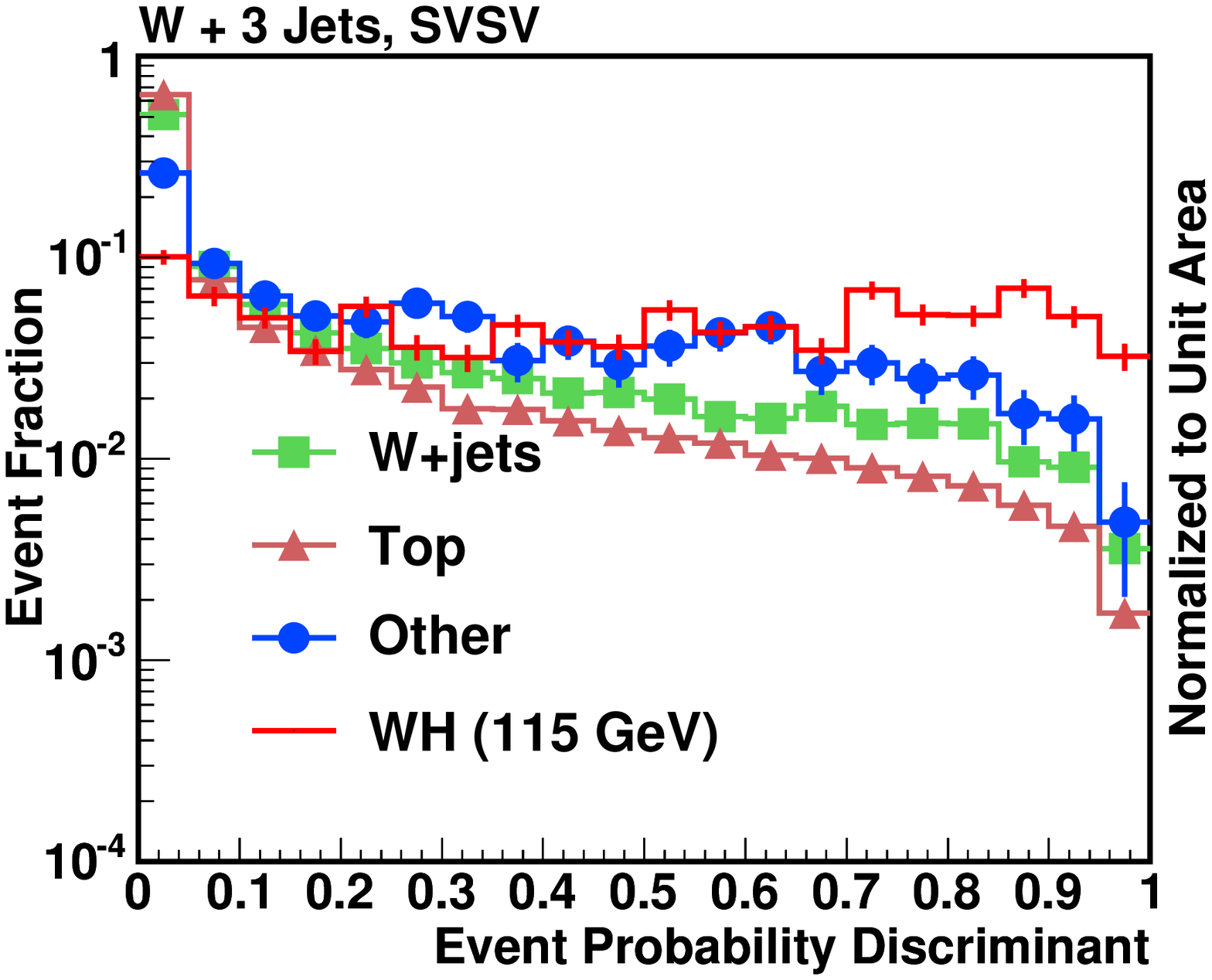}
\end{center}
\caption{\label{fig:EPD_3jets}
Templates of predictions for the signal (m$_H$~=~115~GeV/$c^2$) and
background processes, each scaled to unit area, of the
ME~discriminant, $EPD$, for 3-jet events for each signal region. }
\end{figure*}

In the search for SM Higgs boson production, twelve separate EPD
discriminants are created for each Higgs boson mass point, given by
the different $b$-tagging categories (SVnoJP, SVJP, SVSV), the number
of jets in the final state (2 and 3 jets), and the type of leptons
(tight and EMC leptons). This gives the ability to tune the
discriminants independently.  Figures~\ref{fig:EPD_2jets} and
\ref{fig:EPD_3jets} show the signal and background templates, scaled
to unit area, for two and three-jet events, respectively, for each
signal region.  Note that in these figures all of the lepton categories have been combined.

\subsection{Validation of the discriminant output}
\begin{figure*}
\begin{center}
\includegraphics[width=0.8\columnwidth]{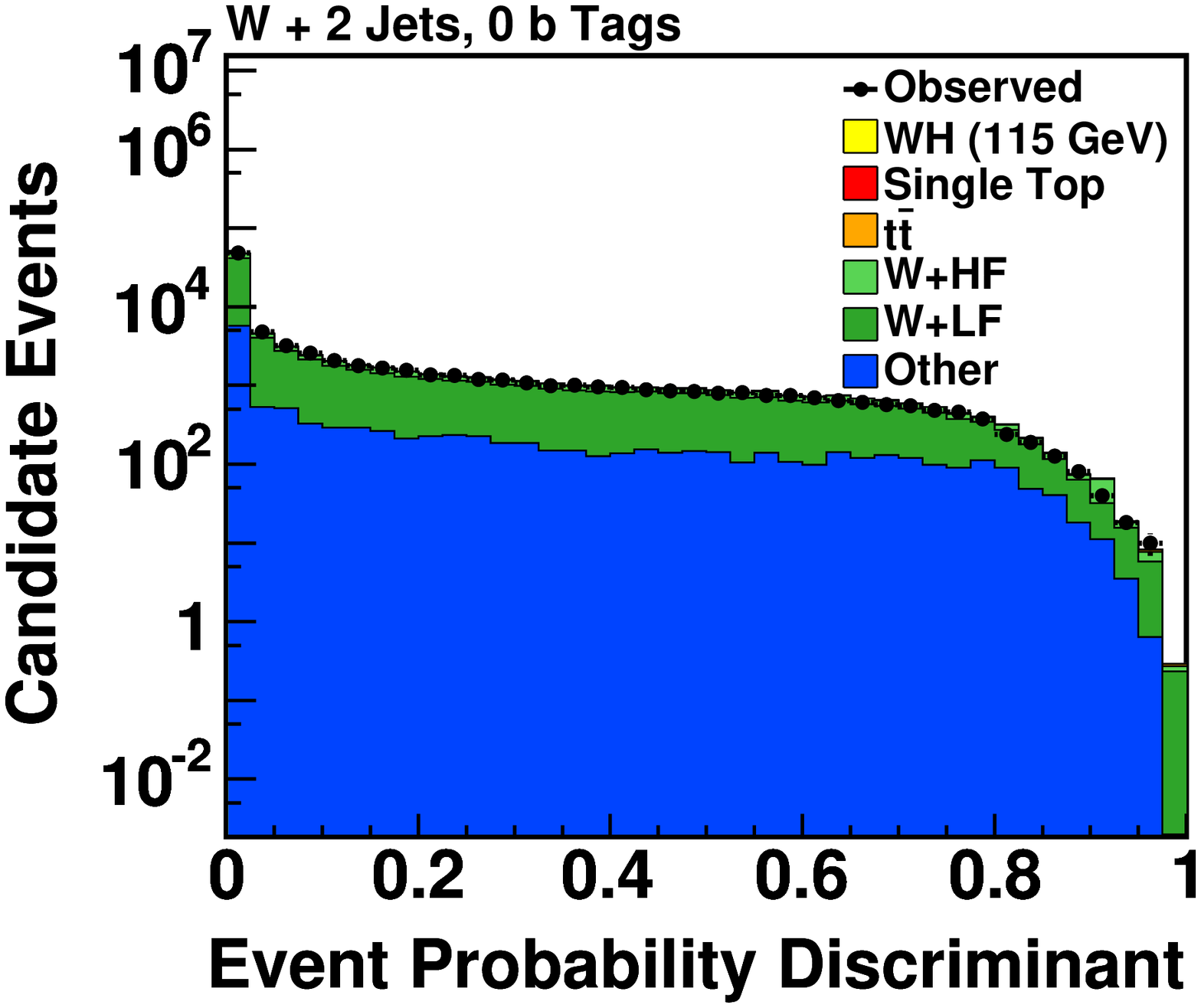}
\includegraphics[width=0.8\columnwidth]{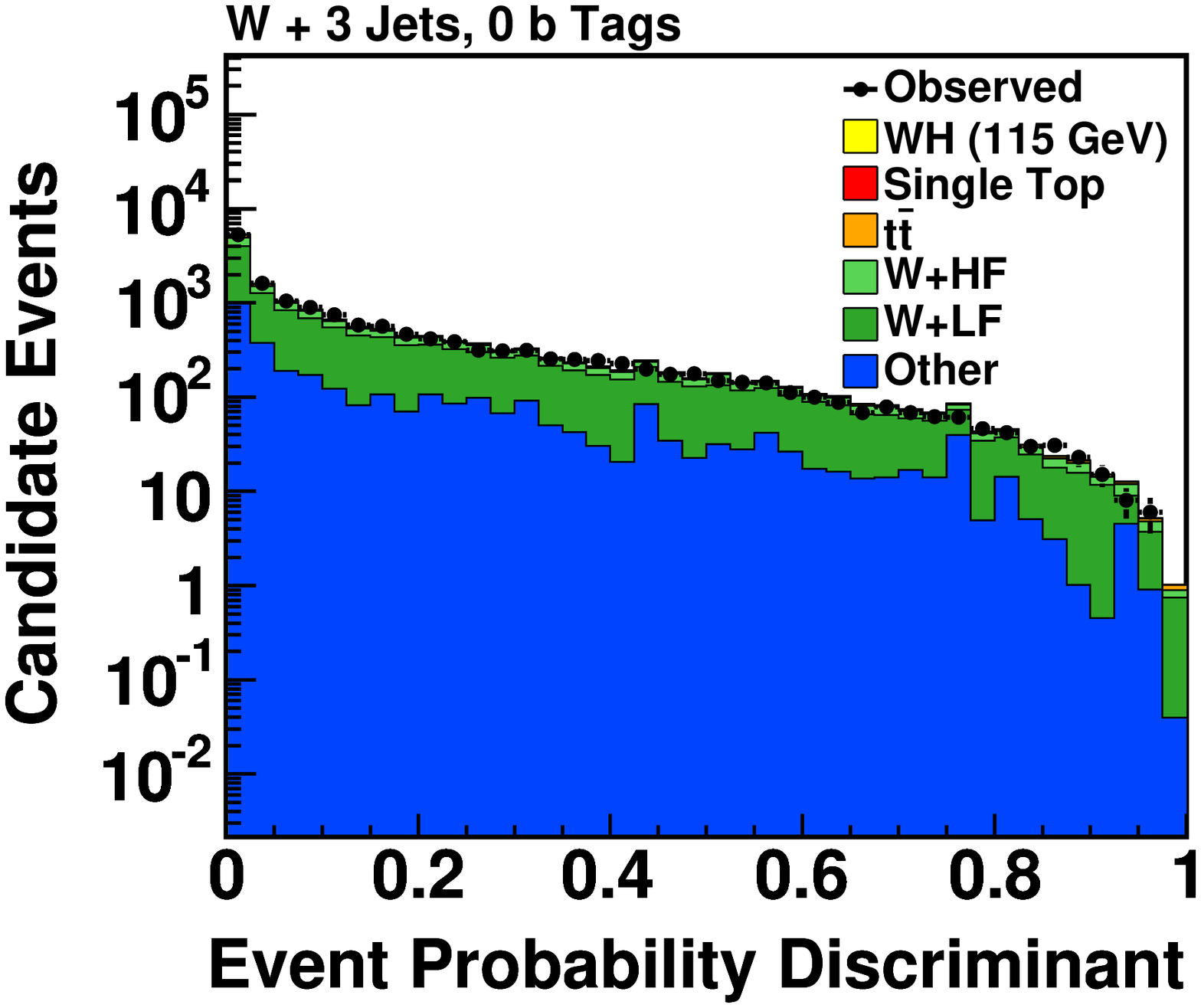}
\end{center}
\caption{\label{fig:allME} Left (right): The discriminant output for
untagged $W$+two (three) jets control sample show that the Monte Carlo
$W$+ two (three) jets samples model the ME distribution of the
observed events well.}
\end{figure*}

The performance of the Monte Carlo to predict the distribution of 
each $EPD$ is validated by checking the untagged
$W+$jets control samples, setting $b_{\mathrm{NN}}=0.5$ so
that it does not affect the $EPD$.  An example is shown in
Fig.~\ref{fig:allME}, for $W$+2-jet and $W$+3-jet events.  
The agreement in this control sample gives confidence that the 
information used in this analysis is well modeled by the Monte 
Carlo simulation.

  The ME method used here is further validated through its successful
  use in previous analyses at the CDF experiment to observe small
  signals with large backgrounds in similar final states to the one
  used here for the Higgs boson search.  The method was used in the
  untagged $W+$jet sample to measure the cross section of diboson
  production~\cite{bib:WWWZxsec}.  In addition, it was used
  successfully in the tagged sample to measure the single top
  production cross section~\cite{stprd}.  In the latter, the modeling
  was also checked for the discriminant output for a second control
  region -- events with four jets.  In this sample, dominated by top
  pair production, the EPD was also found to be well
  modeled~\cite{bib:peter_thesis}.

%-----------------------------
% SYSTEMATICS
%-----------------------------
%\input{./includes/prdSysts}
%-------------------------------------------------------------------------------
\section{Systematic uncertainties}
\label{sec:systs}
%\setlength{\footskip}{2cm}
%-------------------------------------------------------------------------------
Systematic uncertainties can bias the outcome of this analysis and
have to be incorporated into the result. The dominant systematic
uncertainties addressed are from several different sources: 
jet energy scale (JES), initial state radiation (ISR), final 
state radiation (FSR), parton distribution functions, lepton 
identification, luminosity, and $b$-tagging scale factors. 

Systematic uncertainties can influence both the expected event yield
(normalization) and the shape of the discriminant distribution. The
dominant rate uncertainties have been included for each category. Shape uncertainties have only been applied for the JES, which has a
small impact on the final sensitivity.  Other shape
uncertainties are expected to be small. When the sensitivity to
signal events gets closer to the SM prediction the result will be more
affected by sources of systematic uncertainties; currently, this
analysis is statistically limited.

Normalization uncertainties are estimated by recalculating the
acceptance using Monte Carlo samples altered due to a specific
systematic effect. The normalization uncertainty is the
difference between the systematically shifted acceptance and the
default one. The normalization uncertainties for signal and background
processes are shown in Tables~\ref{tab:sys_2jets} (for two-jet events) and 
\ref{tab:sys_3jets} (for three-jet events)~\footnote{Note that empty
entries in the table either mean that the systematic is not relevant
for that process (for example background rates that are derived from
data are not affected by the uncertainty on the luminosity
measurement), or that it was studied and found to be negligible (for
example effect of the JES uncertainty was studied for dibosons and top
production and found to have a negligible impact on the final
sensitivity).}.

\renewcommand{\arraystretch}{1.1}
\begin{table*}[tbh]
\caption{Normalization systematic uncertainties on the signal and
background contributions for the 2 jets channel.  Some uncertainties
are listed as ranges, as the impacts of the uncertain parameters
depend on the tagging category.  Systematic uncertainties for $WH$
shown in this table are obtained for $m_H=115$ GeV/$c^2$. }
\label{tab:sys_2jets}\vspace{.2cm}
\begin{tabular}{lcccccc}
\hline
\hline
\multicolumn{7}{c}{Relative Uncertainties (\%)}\\
\hline
Contribution              & $W$+HF & Mistags & Top& Diboson & Non-$W$ & $WH$  \\ 
\hline
Luminosity ($\sigma_{\mathrm{inel}}(p{\bar{p}})$)
                          &     &      & 3.8      & 3.8      &      & 3.8   \\
Luminosity monitor        &     &      & 4.4      & 4.4      &      & 4.4   \\
Lepton ID                 &     &      & 2        & 2        &      & 2   \\
Jet energy scale          &     &      &          &          &      & 2    \\
ISR+FSR+PDF               &     &      &          &          &      & 3.1-5.6 \\
$b$-tag efficiency        &     &      & 3.5-8.4 & 3.5-8.4 &      & 3.5-8.4    \\
Cross section             &     &      & 10       & 10          &      & 10 \\
HF fraction in $W$+jets   & 30  &      &          &          &      &  \\
Mistag rate               &     & 9-13.3 &       &          &      &  \\
Non-$W$ rate              &     &      &          &          & 40   &  \\
\hline\hline
\end{tabular}
\end{table*}

\renewcommand{\arraystretch}{1.1}
\begin{table*}[tbh]
\caption{Normalization systematic uncertainties on the signal and
background contributions for the 3 jets channel.  Some uncertainties
are listed as ranges, as the impacts of the uncertain parameters
depend on the tagging category.  Systematic uncertainties for $WH$
shown in this table are obtained for $m_H=115$ GeV/$c^2$.}
\label{tab:sys_3jets}\vspace{.2cm}
\begin{tabular}{lcccccc}
\hline
\hline
\multicolumn{7}{c}{Relative Uncertainties (\%)}\\
\hline
Contribution              & $W$+HF & Mistags & Top      & Diboson & Non-$W$ & $WH$  \\ \hline
Luminosity ($\sigma_{\mathrm{inel}}(p{\bar{p}})$)       
                          &     &      & 3.8      & 3.8      &      & 3.8   \\
Luminosity monitor        &     &      & 4.4      & 4.4      &      & 4.4   \\
Lepton ID                 &     &      & 2        & 2        &      & 2   \\
Jet energy scale          &     &      &       &       &      & 13.5-15.8   \\
ISR+FSR+PDF               &     &      &       &       &      & 13.1-21.4  \\
$b$-tag efficiency        &     &      & 3.5-8.4 & 3.5-8.4 &  & 3.5-8.4    \\
Cross section             &     &      & 10    & 10    &      & 10 \\
HF fraction in $W$+jets   & 30  &      &       &       &      &  \\
Mistag rate               &     & 9-13.3 &     &       &      &  \\
Non-$W$ rate              &     &      &       &       & 40   &   \\
\hline\hline
\end{tabular}
\end{table*}

\begin{figure}[h]
\begin{center}
    \includegraphics[width=.85\columnwidth,clip=]{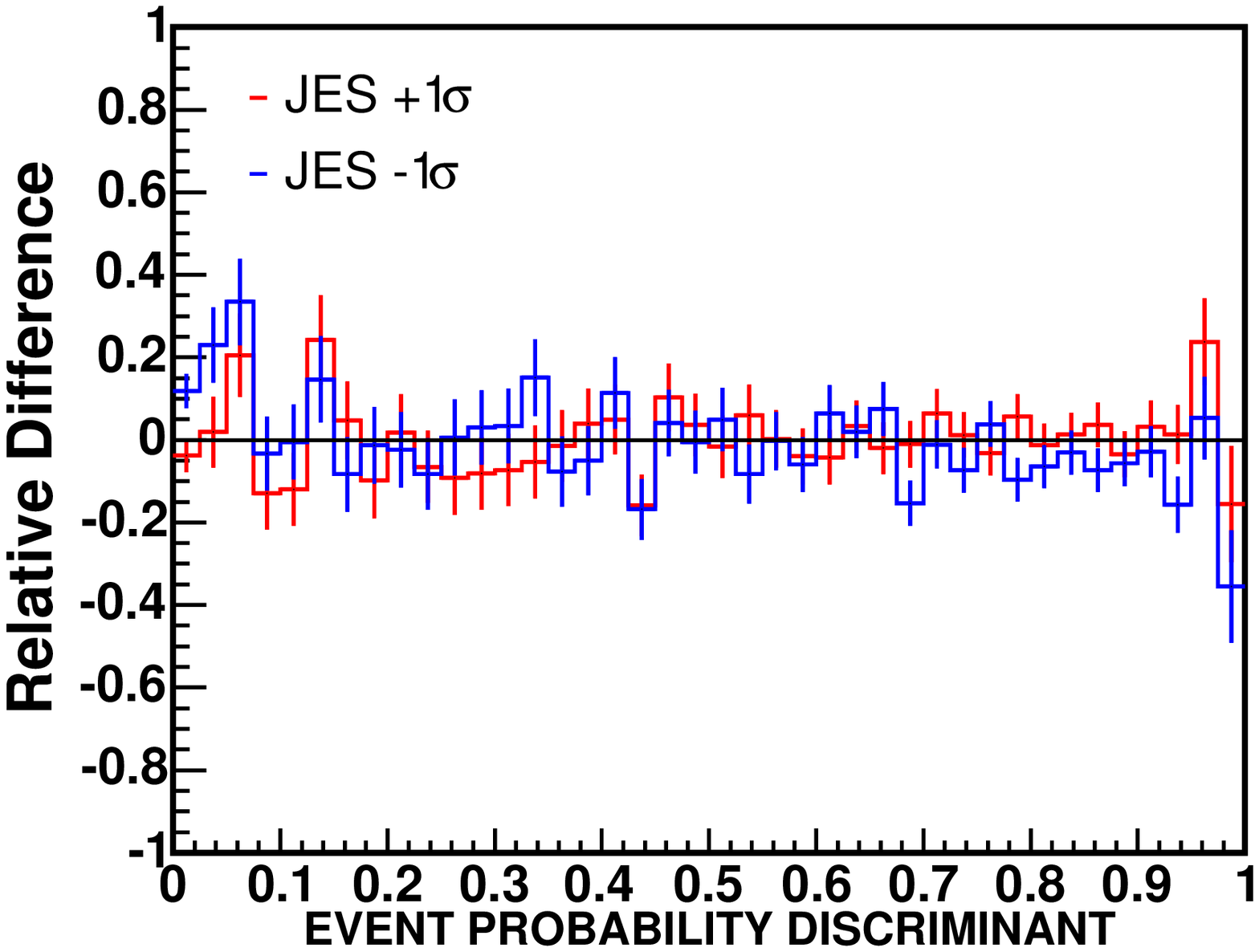}
    \includegraphics[width=.85\columnwidth,clip=]{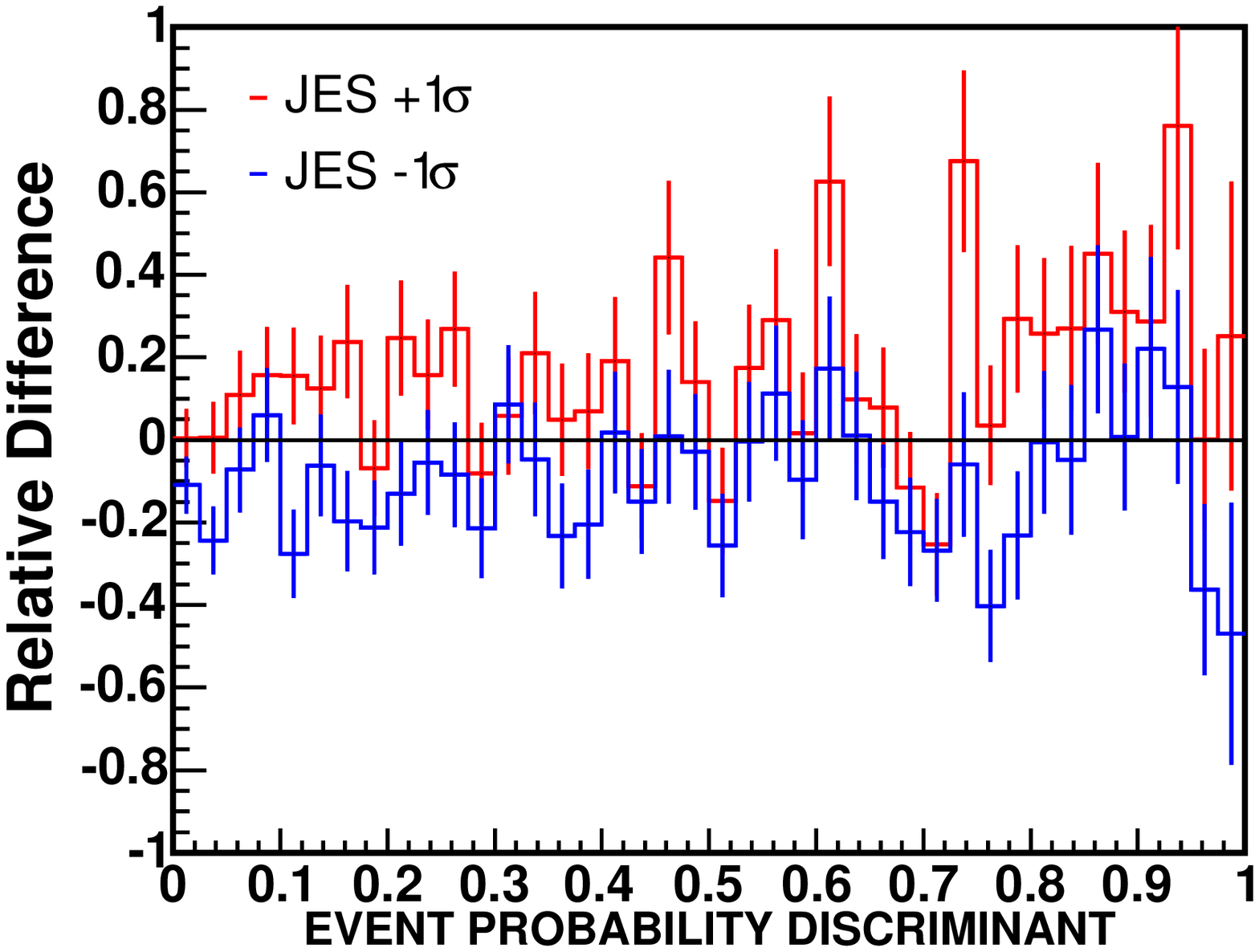}
\caption[$WH$ JES shape systematic]
	{\label{fig:WH_shapesyst} Top (bottom):
	 $WH$ (m$_H$ = 115 GeV/$c^2$) JES shape systematic for two (three) jet events. 
	  The plots show the relative difference of one $\sigma$ up and one $\sigma$ down 
	  jet energy correction with respect to the nominal correction. }
\end{center}
\end{figure}

The effect of the uncertainty in the jet energy scale is evaluated by
applying jet-energy corrections that describe $\pm 1\sigma$ variations
to the default correction factor.  The JES shape uncertainty has been
only applied to the event probability discriminant for the two and
three jet events in the samples with the biggest contribution, for the
$WH$ signal sample, and the $W$~+~jets and $t\bar{t}$ background
samples.  Shape variations due to the jet energy scale for two and
three jet $WH$ signal events are shown in Fig.~\ref{fig:WH_shapesyst}.
The effect of the JES shape uncertainty on the final sensitivity is
small, on the order of only a few percent. This is small compared to
the effect of normalization uncertainties.

Systematic uncertainties due to the modeling of ISR and FSR are
obtained from dedicated Monte Carlo samples for $WH$ signal events
where the strength of ISR/FSR is increased and decreased in the parton
showering to represent $\pm 1\sigma$ variations~\cite{isrfsr}.  The
effects of variations in ISR and FSR are treated as 100\% correlated
with each other.

To evaluate the uncertainty on the signal acceptance associated with
the specific choice of parton distribution functions, events are
reweighted based on different PDF schemes. The twenty independent
eigenvectors of the {\sc cteq}~\cite{Lai:1999wy} PDFs are varied and
compared to the {\sc mrst}~\cite{Martin:1998sq} PDFs. The uncertainty
from the {\sc cteq} and {\sc mrst} PDF uncertainty are summed in
quadrature if the difference between the {\sc cteq} and {\sc mrst}
PDFs is larger than the {\sc cteq} uncertainty.

The estimate of the lepton ID uncertainty is a result of varying the
lepton ID correction factors. The results are then compared to the
nominal prediction for an estimate of the fractional uncertainty. All
lepton ID correction factors are varied either all up or all down
simultaneously.  The yield is then calculated for each sample and
compared to the nominal prediction. The lepton ID uncertainty is
applied to the signal sample and all Monte Carlo based samples.

For the signal sample and all Monte Carlo based samples a systematic
uncertainty is applied for the uncertainty in the CDF luminosity
measurement which is correlated across all samples and channels. This
uncertainty includes the uncertainty in the $p\bar{p}$ inelastic cross
section (3.8\%) as well as the uncertainty in the acceptance of CDF's
luminosity monitor (4.4\%)~\cite{CLC}.

The effect of the $b$-tagging scale factor uncertainty is determined 
from the background estimate. The systematic uncertainty 
on the event tagging efficiency is estimated by varying the tagging scale factor 
and mistag prediction by $\pm 1 \sigma$ and calculating the difference 
between the systematically shifted acceptance and the default one.
 
For all background processes the normalization uncertainties are
represented by the uncertainty on the predicted number of background
events and are incorporated in the analysis as Gaussian constraints
$G(\beta_j|1,\Delta_j)$ in a likelihood function~\cite{stprd}.  The
systematic uncertainties in the normalizations of each source,
$\beta_j$, are incorporated into the likelihood as nuisance
parameters, conforming with a fully Bayesian treatment~\cite{bayes}.
The correlations between normalizations for a given source are taken
into account.  The likelihood function is marginalized by integrating
over all nuisance parameters for many possible values of the $WH$
cross section $\beta_1=\beta_{WH}$. The resulting reduced likelihood
$\mathcal{L}(\beta_{WH})$ is a function of the $WH$ cross section
$\beta_{WH}$ only. More details on the statistical treatment of the
limit calculation are included in Refs.~\cite{bib:pdgstat,stprd}.

%-----------------------------
% RESULTS
%-----------------------------
%\input{./includes/prdResults}
%-------------------------------------------------------------------------------
\section{Results}
\label{sec:results}
\begin{figure*}
\begin{center}
\subfigure[]{\includegraphics[width=.65\columnwidth]{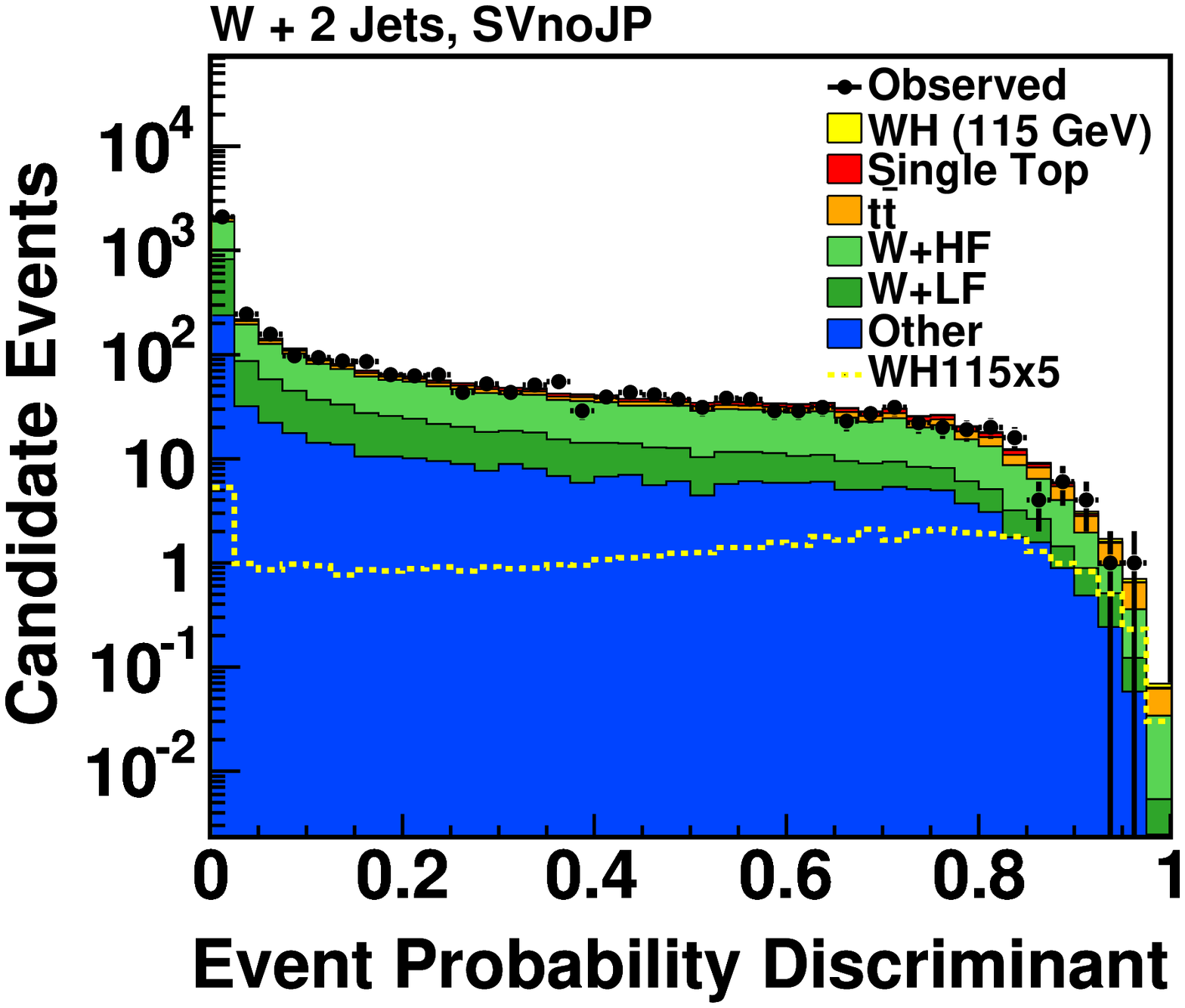}}
\subfigure[]{\includegraphics[width=.65\columnwidth]{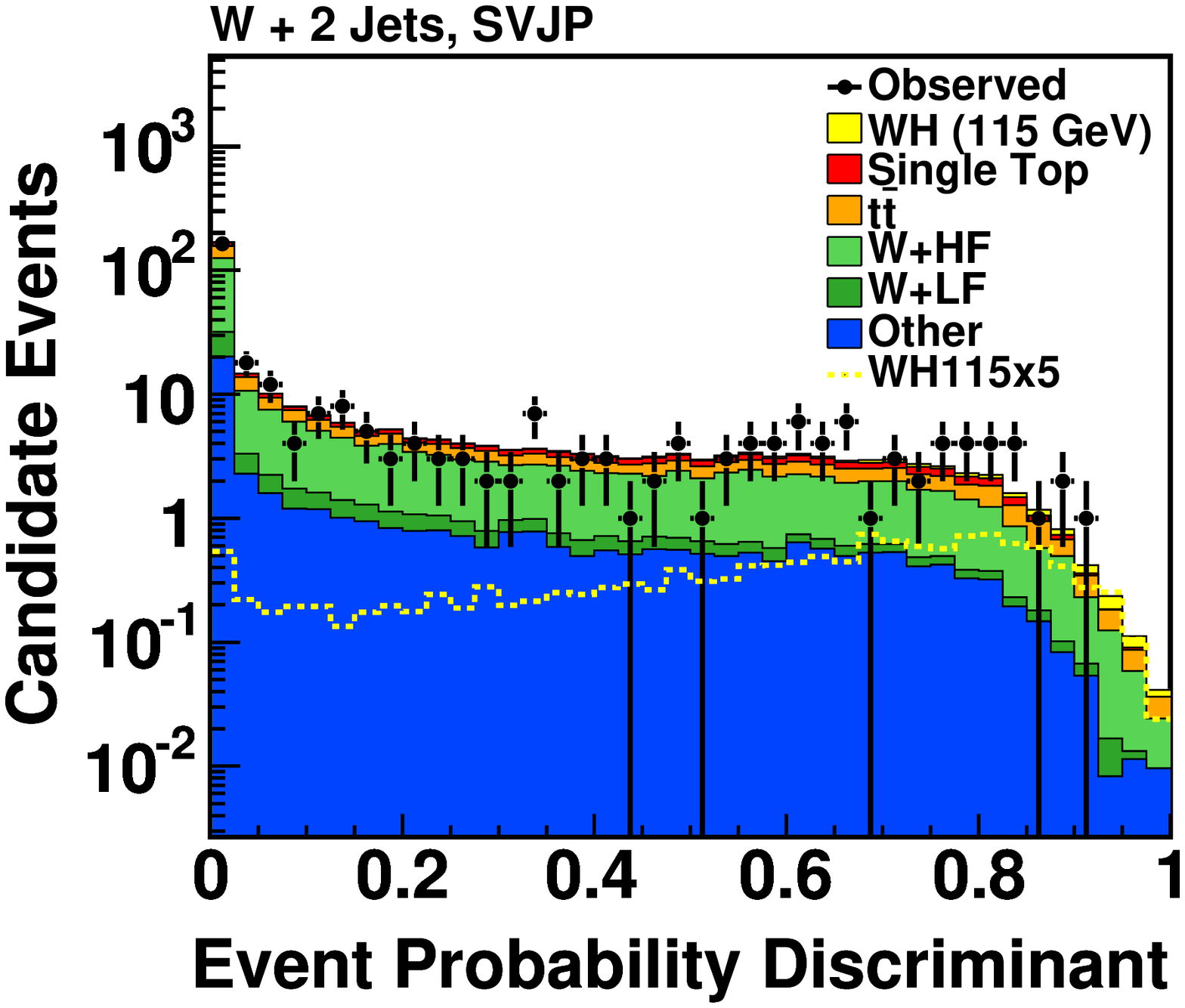}}
\subfigure[]{\includegraphics[width=.65\columnwidth]{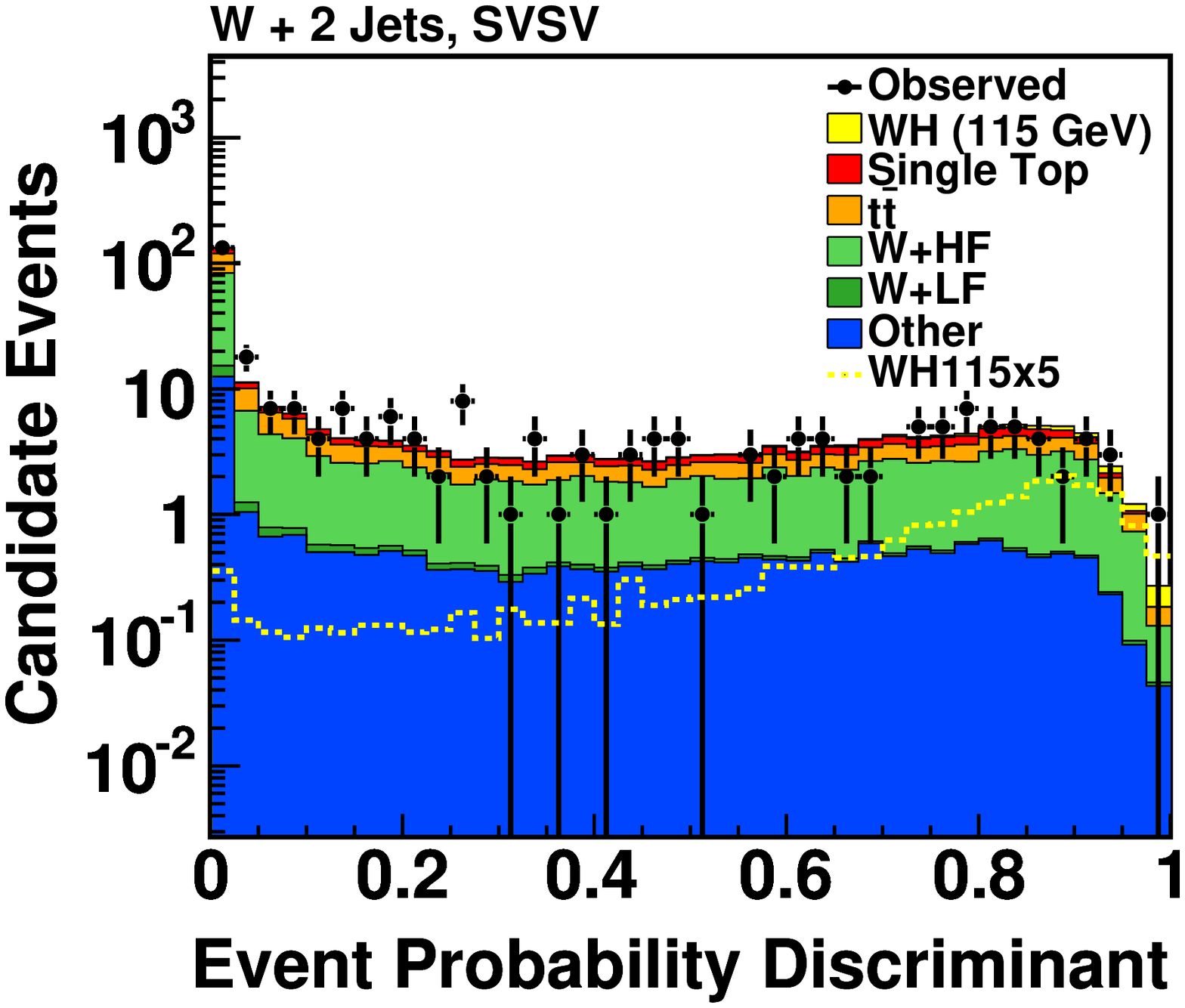}}
\subfigure[]{\includegraphics[width=.65\columnwidth]{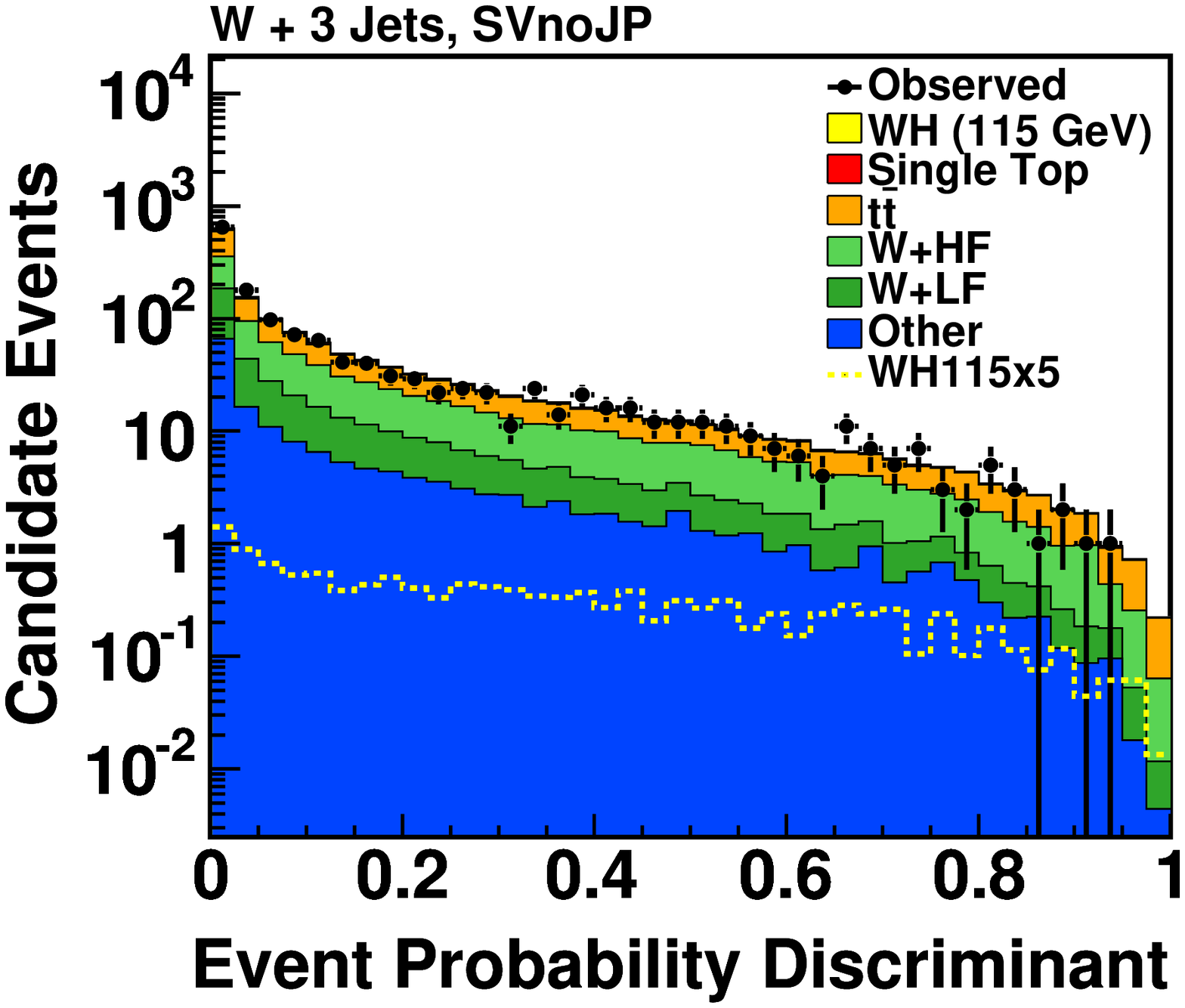}}
\subfigure[]{\includegraphics[width=.65\columnwidth]{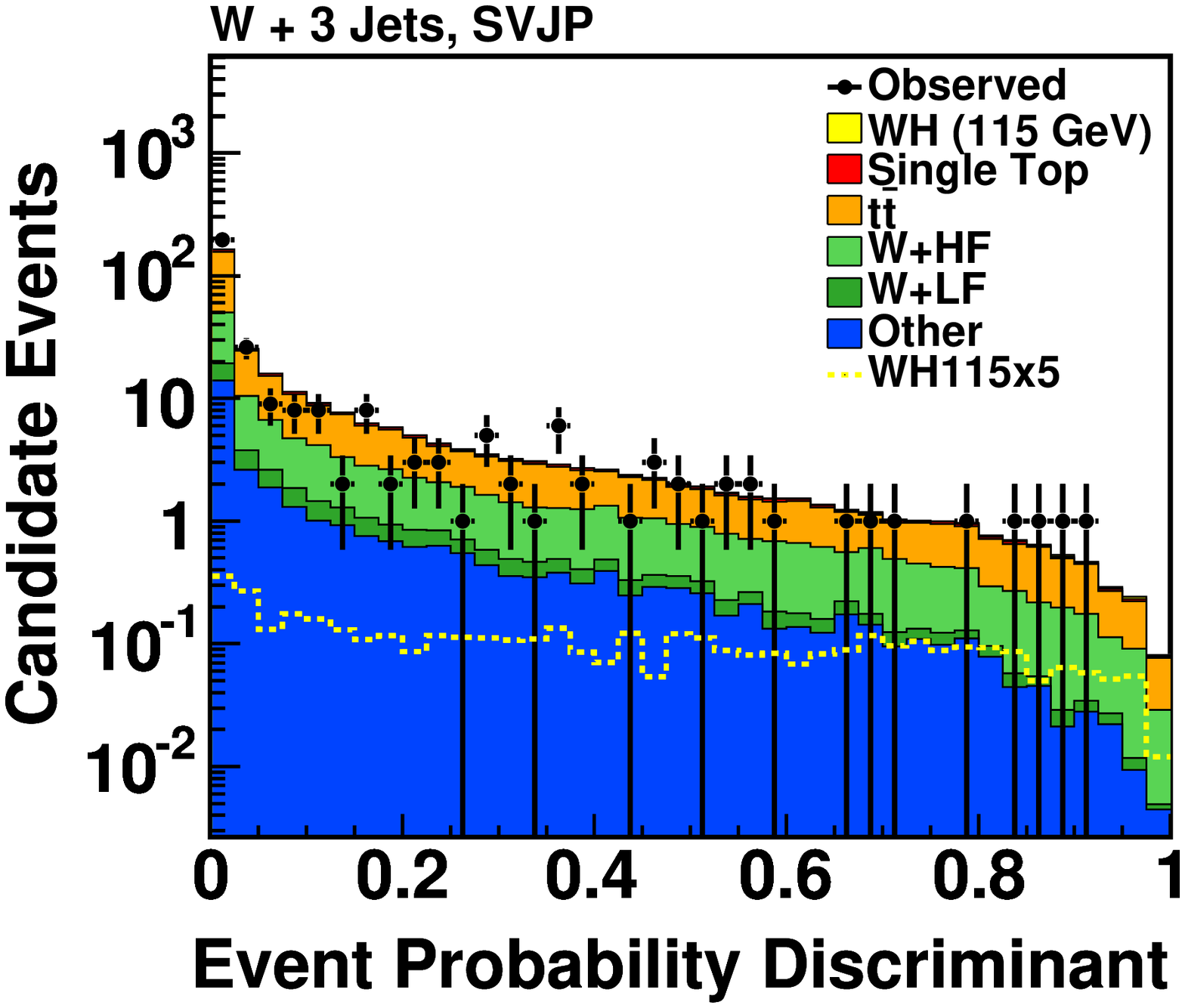}}
\subfigure[]{\includegraphics[width=.65\columnwidth]{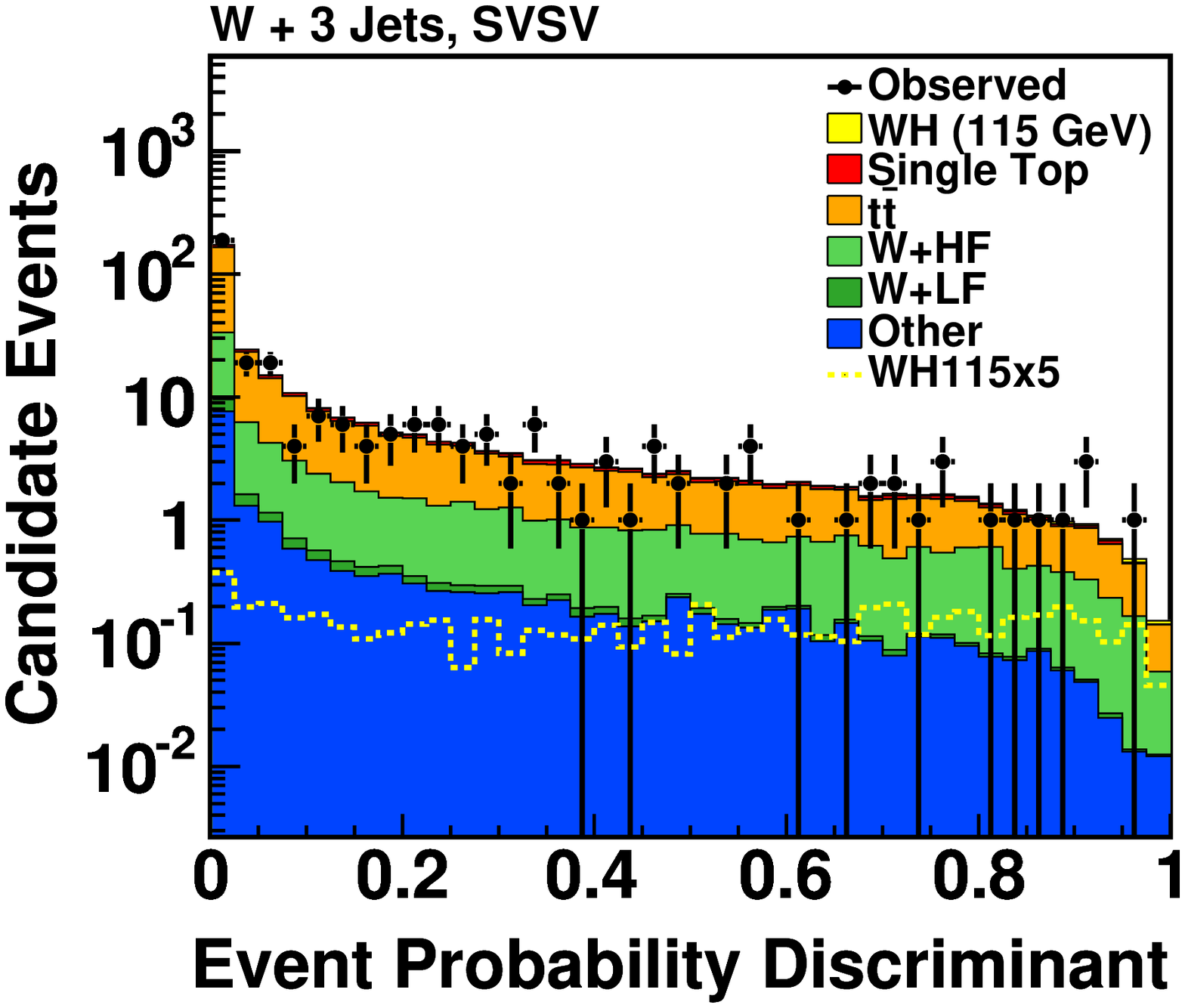}}
\caption{\label{fig:epdData_3j} Top (bottom): Comparison of the EPD
  output for lepton + 2 (3) jets observed events compared to the Monte
  Carlo simulated events for $WH$ ($m_{H}=$~115~GeV/$c^2$) signal and
  background. From left to right: SVnoJP, SVJP, and SVSV tagged
  observed events, respectively.  Note that the signal is twice in
  these plots, as a stacked plot and as a histogram multiplied by 5
  (x5). }
\end{center}
\end{figure*}

%-------------------------------------------------------------------------------
The analysis is applied to observed events in a sample corresponding
to an integrated luminosity of 5.6 fb$^{-1}$.  The EPD output
distribution, for a Higgs boson mass of 115 GeV/$c^2$, of our
candidate events is compared with the sum of predicted $WH$ signal and
background distributions as shown in Fig.~\ref{fig:epdData_3j}.

We search for an excess of Higgs boson signal events in the EPD distributions, 
but no evidence of a signal excess is found in the observed events. 
Thus, we perform a binned likelihood fit to the EPD output distributions 
to set an upper limit on SM Higgs boson production associated with a $W$ 
boson for eleven values of m$_H$, 100~$\leq$~m$_H$~$\leq$~150~GeV/$c^2$ in 
5~GeV/$c^2$ steps. 

In order to extract the most probable $WH$ signal content in the
 observed events the maximum likelihood method described before is
 performed.  A marginalization using the likelihood function is
 performed with all systematic uncertainties included in the
 likelihood function.  The posterior p.d.f is obtained by using Bayes'
 theorem:
\vspace{.5cm}
\[p(\beta_1|EPD)=\frac{\mathcal{L}^{*}(EPD|\beta_{WH})\pi(\beta_{WH})}{\int \mathcal{L}^{*}(EPD|\beta_{WH}')\pi(\beta_{WH}')d\beta_{WH}'}\]
where $\mathcal{L}^{*}(EPD|\beta_{WH})$ is the reduced likelihood and
$\pi(\beta_{WH})$ is the prior p.d.f. for $\beta_{WH}$. 
A flat prior is adopted, $\pi(\beta_{WH})=H(\beta_{WH})$, in this analysis, with 
$H$ being the Heaviside step function.
To set an upper limit on the $WH$ production cross section, 
the posterior probability density is integrated to cover 95\% \cite{bib:pdgstat}. 

\begin{figure}[tbh]
  \centering
  \includegraphics[width=0.95\columnwidth]{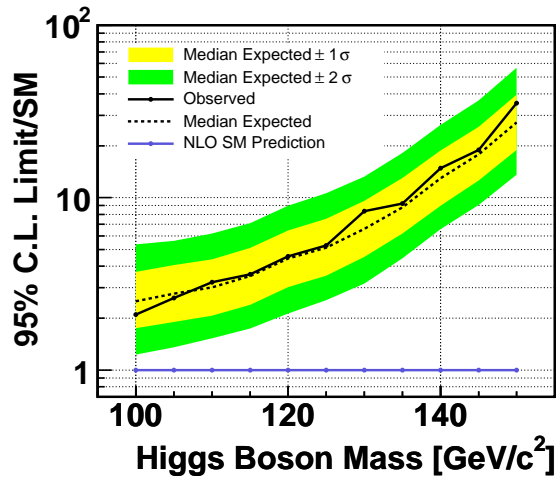}
  \caption{\label{fig:limits} 95\% C.L. upper limits on the $WH$ production 
  cross sections times branching ratio for $H\rightarrow b\bar{b}$ for Higgs
  boson masses between $m_H=$~100~GeV/$c^2$ to $m_H=$~150~GeV/$c^2$.
  The plot shows the limit normalized to the cross section predictions from
  the standard model. }
\end{figure}

The observed and expected limits on
$\sigma(p\bar{p}~\rightarrow~WH)\times~{\cal{B}}(H\rightarrow~b\bar{b})$,
for each Higgs boson mass point from 100 to 150~GeV/$c^2$ in
5~GeV/$c^2$ steps, all $b$-tagging categories, and 2- and 3-jet events
together are shown in Table~\ref{tab:limits} and in
Fig.~\ref{fig:limits}.  The observed and expected limits in SM cross
section units are shown in Table~\ref{tab:limits_xc}.

\begin{table}[tbh]
%\begin{center}
\caption{\label{tab:limits} Expected and observed upper limit cross
sections, relative to the SM prediction, for different Higgs boson
mass points for 2- and 3-jet events.}\vspace{.2cm}
\begin{tabular}{lccccccccccc}
\hline\hline
\multicolumn{12}{c}{2, 3 jets}\\
\hline
$\sigma$ / SM & 100 & 105 & 110 & 115 & 120 & 125 & 130 & 135 &  140 & 145 & 150 \\
\hline
Expected     & 2.5 & 2.7 & 3.0 & 3.5 & 4.4 & 5.1 & 6.6 & 8.7 & 13.0 & 17.8 & 27.5 \\
Observed     & 2.1 & 2.6 & 3.2 & 3.6 & 4.6 & 5.3 & 8.3 & 9.2  & 14.8 & 18.9 & 35.3 \\
\hline \hline
\end{tabular}
%\end{center}
\end{table}

\begin{table}[tbh]
%\begin{center}
\caption{\label{tab:limits_xc} Expected and observed upper limit on
$\sigma(p\bar{p}~\rightarrow~WH)\times~{\cal{B}}(H\rightarrow~b\bar{b})$
in units of pb for different Higgs boson mass points for 2- and 3-jet
events.}\vspace{.2cm}
\begin{tabular}{lccccccccccc}
\hline\hline
\multicolumn{12}{c}{2, 3 jets}\\
\hline
$\sigma$ & 100 & 105 & 110 & 115 & 120 & 125 & 130 & 135 &  140 & 145 & 150 \\
\hline
Exp     & 0.72 & 0.68 & 0.66 & 0.65 & 0.67 & 0.69 & 0.79 & 0.90 & 1.12 & 1.39 & 1.93 \\
Obs     & 0.60 & 0.66 & 0.70 & 0.67 & 0.70 & 0.72 & 1.00 & 0.95 & 1.27 & 1.47 & 2.47 \\
\hline \hline
\end{tabular}
%\end{center}
\end{table}

Tables~\ref{tab:limits_2jets} and \ref{tab:limits_3jets} show the expected and observed 
limits, for each Higgs boson mass point, for events with 2 and 3 jets, respectively.
Including 3 jet events improves the limit by 3 to 10\%, depending on the Higgs boson mass, 
with respect to the result using 2 jet events only.

\begin{table}[h]
%\begin{center}
\caption{\label{tab:limits_2jets} Expected and observed upper limit
cross sections, relative to the SM prediction, for different Higgs
boson mass points in the 2 jets channel.}\vspace{.2cm}
\begin{tabular}{lccccccccccc}
\hline\hline
\multicolumn{12}{c}{2 jets}\\
\hline
$\sigma$ / SM & 100 & 105 & 110 & 115 & 120 & 125 & 130 & 135 &  140 & 145 & 150 \\
\hline 
Expected     & 2.6  & 2.8 & 3.2 & 3.7 & 4.7 & 5.5 & 7.1 & 9.5  & 14.2 & 19.7 & 30.7 \\
Observed     & 2.7 & 3.3 & 3.7 & 4.5 & 5.9 & 6.8 & 9.6 & 12.0 & 19.3 & 24.0 & 43.2 \\
\hline \hline
\end{tabular}
\end{table}

\begin{table}[h]
\caption{\label{tab:limits_3jets} Expected and observed upper limit
cross sections, relative to the SM prediction, for different Higgs
boson mass points in the 3 jets channel.}
\vspace{.2cm}
\begin{tabular}{lccccccccccc}
\hline 
\hline
\multicolumn{12}{c}{3 jets }\\
\hline
$\sigma$ / SM & 100 & 105 & 110 & 115 & 120 & 125 & 130 & 135 &  140 & 145 & 150 \\
\hline 
Exp.     & 12.2 & 12.9 & 13.9 & 15.8 & 19.5 & 23.0 & 28.1 & 39.5 & 56.1 & 77.9 & 120 \\
Obs.     &  5.1 &  5.6 &  8.6 & 8.5 & 10.8 & 12.4 & 17.3 & 22.9 & 33.7 & 42.5 &  81 \\
\hline \hline
\end{tabular}
\end{table}

%-----------------------------
% CONCLUSIONS
%-----------------------------
%\input{./includes/prdConclusions}
%-------------------------------------------------------------------------------
\section{Conclusions}
\label{sec:concl}
%-------------------------------------------------------------------------------
A search for the Higgs boson production in association with a $W$
boson using a matrix element technique has been performed using 5.6
fb$^{-1}$ of CDF data.  A maximum likelihood technique has been
applied to extract the most probable $WH$ content in observed events.
No evidence is observed for a Higgs boson signal using observed events
corresponding to an integrated luminosity of 5.6~\fb~ and 95\%
confidence level upper limits are set.  The limits on the $WH$
production cross section times the branching ratio, relative to the SM
prediction, of the Higgs boson to decay to $b\bar{b}$ pairs are
$\sigma(p\bar{p}~\rightarrow~WH)\times~{\cal{B}}(H\rightarrow~b\bar{b})/SM~<~2.1$
to $35.3$ for Higgs boson masses between $m_H=$~100~GeV/$c^2$ and
$m_H=$~150~GeV/$c^2$.  The expected (median) sensitivity estimated in
pseudoexperiments is
$\sigma(p\bar{p}~\rightarrow~WH)\times~{\cal{B}}(H\rightarrow~b\bar{b})/SM~<~2.5$
to $27.5$ at 95\% C.L.

The search results in this channel at the CDF experiment are the most
sensitive low-mass Higgs boson search at the Tevatron.  While the LHC
experiments will soon have superior sensitivity to the low-mass Higgs
boson, this sensitivity comes primarily from searches in the diphoton
final state.  Therefore, we expect that the searches in the $H
\rightarrow $ \bbbar~ at the Tevatron will provide crucial information
on the existence and nature of the low-mass Higgs boson for years to
come.

%-----------------------------
% ACKNOWLEDGEMENTS
%-----------------------------
\begin{acknowledgments}
We thank the Fermilab staff and the technical staffs of the
participating institutions for their vital contributions. This work
was supported by the U.S. Department of Energy and National Science
Foundation; the Italian Istituto Nazionale di Fisica Nucleare; the
Ministry of Education, Culture, Sports, Science and Technology of
Japan; the Natural Sciences and Engineering Research Council of
Canada; the National Science Council of the Republic of China; the
Swiss National Science Foundation; the A.P. Sloan Foundation; the
Bundesministerium f\"ur Bildung und Forschung, Germany; the Korean
World Class University Program, the National Research Foundation of
Korea; the Science and Technology Facilities Council and the Royal
Society, UK; the Russian Foundation for Basic Research; the Ministerio
de Ciencia e Innovaci\'{o}n, and Programa Consolider-Ingenio 2010,
Spain; the Slovak R\&D Agency; the Academy of Finland; and the
Australian Research Council (ARC).
\end{acknowledgments}

\appendix

\newpage %Just because of unusual number of tables stacked at end

%\bibliographystyle{apsrev}
%\bibliography{prd}

%-----------------------------
% BIBLIOGRAPHY
%-----------------------------
%\input{bib_WH_ME_PRD}

\end{document}

%% file: September2011_Authors.tex
\affiliation{Institute of Physics, Academia Sinica, Taipei, Taiwan 11529, Republic of China}
\affiliation{Argonne National Laboratory, Argonne, Illinois 60439, USA}
\affiliation{University of Athens, 157 71 Athens, Greece}
\affiliation{Institut de Fisica d'Altes Energies, ICREA, Universitat Autonoma de Barcelona, E-08193, Bellaterra (Barcelona), Spain}
\affiliation{Baylor University, Waco, Texas 76798, USA}
\affiliation{Istituto Nazionale di Fisica Nucleare Bologna, $^{ee}$University of Bologna, I-40127 Bologna, Italy}
\affiliation{University of California, Davis, Davis, California 95616, USA}
\affiliation{University of California, Los Angeles, Los Angeles, California 90024, USA}
\affiliation{Instituto de Fisica de Cantabria, CSIC-University of Cantabria, 39005 Santander, Spain}
\affiliation{Carnegie Mellon University, Pittsburgh, Pennsylvania 15213, USA}
\affiliation{Enrico Fermi Institute, University of Chicago, Chicago, Illinois 60637, USA}
\affiliation{Comenius University, 842 48 Bratislava, Slovakia; Institute of Experimental Physics, 040 01 Kosice, Slovakia}
\affiliation{Joint Institute for Nuclear Research, RU-141980 Dubna, Russia}
\affiliation{Duke University, Durham, North Carolina 27708, USA}
\affiliation{Fermi National Accelerator Laboratory, Batavia, Illinois 60510, USA}
\affiliation{University of Florida, Gainesville, Florida 32611, USA}
\affiliation{Laboratori Nazionali di Frascati, Istituto Nazionale di Fisica Nucleare, I-00044 Frascati, Italy}
\affiliation{University of Geneva, CH-1211 Geneva 4, Switzerland}
\affiliation{Glasgow University, Glasgow G12 8QQ, United Kingdom}
\affiliation{Harvard University, Cambridge, Massachusetts 02138, USA}
\affiliation{Division of High Energy Physics, Department of Physics, University of Helsinki and Helsinki Institute of Physics, FIN-00014, Helsinki, Finland}
\affiliation{University of Illinois, Urbana, Illinois 61801, USA}
\affiliation{The Johns Hopkins University, Baltimore, Maryland 21218, USA}
\affiliation{Institut f\"{u}r Experimentelle Kernphysik, Karlsruhe Institute of Technology, D-76131 Karlsruhe, Germany}
\affiliation{Center for High Energy Physics: Kyungpook National University, Daegu 702-701, Korea; Seoul National University, Seoul 151-742, Korea; Sungkyunkwan University, Suwon 440-746, Korea; Korea Institute of Science and Technology Information, Daejeon 305-806, Korea; Chonnam National University, Gwangju 500-757, Korea; Chonbuk National University, Jeonju 561-756, Korea}
\affiliation{Ernest Orlando Lawrence Berkeley National Laboratory, Berkeley, California 94720, USA}
\affiliation{University of Liverpool, Liverpool L69 7ZE, United Kingdom}
\affiliation{University College London, London WC1E 6BT, United Kingdom}
\affiliation{Centro de Investigaciones Energeticas Medioambientales y Tecnologicas, E-28040 Madrid, Spain}
\affiliation{Massachusetts Institute of Technology, Cambridge, Massachusetts 02139, USA}
\affiliation{Institute of Particle Physics: McGill University, Montr\'{e}al, Qu\'{e}bec, Canada H3A~2T8; Simon Fraser University, Burnaby, British Columbia, Canada V5A~1S6; University of Toronto, Toronto, Ontario, Canada M5S~1A7; and TRIUMF, Vancouver, British Columbia, Canada V6T~2A3}
\affiliation{University of Michigan, Ann Arbor, Michigan 48109, USA}
\affiliation{Michigan State University, East Lansing, Michigan 48824, USA}
\affiliation{Institution for Theoretical and Experimental Physics, ITEP, Moscow 117259, Russia}
\affiliation{University of New Mexico, Albuquerque, New Mexico 87131, USA}
\affiliation{The Ohio State University, Columbus, Ohio 43210, USA}
\affiliation{Okayama University, Okayama 700-8530, Japan}
\affiliation{Osaka City University, Osaka 588, Japan}
\affiliation{University of Oxford, Oxford OX1 3RH, United Kingdom}
\affiliation{Istituto Nazionale di Fisica Nucleare, Sezione di Padova-Trento, $^{ff}$University of Padova, I-35131 Padova, Italy}
\affiliation{University of Pennsylvania, Philadelphia, Pennsylvania 19104, USA}
\affiliation{Istituto Nazionale di Fisica Nucleare Pisa, $^{gg}$University of Pisa, $^{hh}$University of Siena and $^{ii}$Scuola Normale Superiore, I-56127 Pisa, Italy}
\affiliation{University of Pittsburgh, Pittsburgh, Pennsylvania 15260, USA}
\affiliation{Purdue University, West Lafayette, Indiana 47907, USA}
\affiliation{University of Rochester, Rochester, New York 14627, USA}
\affiliation{The Rockefeller University, New York, New York 10065, USA}
\affiliation{Istituto Nazionale di Fisica Nucleare, Sezione di Roma 1, $^{jj}$Sapienza Universit\`{a} di Roma, I-00185 Roma, Italy}
\affiliation{Rutgers University, Piscataway, New Jersey 08855, USA}
\affiliation{Texas A\&M University, College Station, Texas 77843, USA}
\affiliation{Istituto Nazionale di Fisica Nucleare Trieste/Udine, I-34100 Trieste, $^{kk}$University of Udine, I-33100 Udine, Italy}
\affiliation{University of Tsukuba, Tsukuba, Ibaraki 305, Japan}
\affiliation{Tufts University, Medford, Massachusetts 02155, USA}
\affiliation{University of Virginia, Charlottesville, Virginia 22906, USA}
\affiliation{Waseda University, Tokyo 169, Japan}
\affiliation{Wayne State University, Detroit, Michigan 48201, USA}
\affiliation{University of Wisconsin, Madison, Wisconsin 53706, USA}
\affiliation{Yale University, New Haven, Connecticut 06520, USA}

\author{T.~Aaltonen}
\affiliation{Division of High Energy Physics, Department of Physics, University of Helsinki and Helsinki Institute of Physics, FIN-00014, Helsinki, Finland}
\author{B.~\'{A}lvarez~Gonz\'{a}lez$^z$}
\affiliation{Instituto de Fisica de Cantabria, CSIC-University of Cantabria, 39005 Santander, Spain}
\author{S.~Amerio}
\affiliation{Istituto Nazionale di Fisica Nucleare, Sezione di Padova-Trento, $^{ff}$University of Padova, I-35131 Padova, Italy}
\author{D.~Amidei}
\affiliation{University of Michigan, Ann Arbor, Michigan 48109, USA}
\author{A.~Anastassov$^x$}
\affiliation{Fermi National Accelerator Laboratory, Batavia, Illinois 60510, USA}
\author{A.~Annovi}
\affiliation{Laboratori Nazionali di Frascati, Istituto Nazionale di Fisica Nucleare, I-00044 Frascati, Italy}
\author{J.~Antos}
\affiliation{Comenius University, 842 48 Bratislava, Slovakia; Institute of Experimental Physics, 040 01 Kosice, Slovakia}
\author{G.~Apollinari}
\affiliation{Fermi National Accelerator Laboratory, Batavia, Illinois 60510, USA}
\author{J.A.~Appel}
\affiliation{Fermi National Accelerator Laboratory, Batavia, Illinois 60510, USA}
\author{T.~Arisawa}
\affiliation{Waseda University, Tokyo 169, Japan}
% ...

\collaboration{CDF Collaboration\footnote{With visitors from
$^a$Istituto Nazionale di Fisica Nucleare, Sezione di Cagliari, 09042 Monserrato (Cagliari), Italy,
$^b$University of CA Irvine, Irvine, CA 92697, USA,
$^c$University of CA Santa Barbara, Santa Barbara, CA 93106, USA,
$^d$University of CA Santa Cruz, Santa Cruz, CA 95064, USA,
$^e$Institute of Physics, Academy of Sciences of the Czech Republic, Czech Republic,
$^f$CERN, CH-1211 Geneva, Switzerland,
$^g$Cornell University, Ithaca, NY 14853, USA,
$^h$University of Cyprus, Nicosia CY-1678, Cyprus,
$^i$Office of Science, U.S. Department of Energy, Washington, DC 20585, USA,
$^j$University College Dublin, Dublin 4, Ireland,
$^k$ETH, 8092 Zurich, Switzerland,
$^l$University of Fukui, Fukui City, Fukui Prefecture, Japan 910-0017,
$^m$Universidad Iberoamericana, Mexico D.F., Mexico,
$^n$University of Iowa, Iowa City, IA 52242, USA,
$^o$Kinki University, Higashi-Osaka City, Japan 577-8502,
$^p$Kansas State University, Manhattan, KS 66506, USA,
$^q$Korea University, Seoul, 136-713, Korea,
$^r$University of Manchester, Manchester M13 9PL, United Kingdom,
$^s$Queen Mary, University of London, London, E1 4NS, United Kingdom,
$^t$University of Melbourne, Victoria 3010, Australia,
$^u$Muons, Inc., Batavia, IL 60510, USA,
$^v$Nagasaki Institute of Applied Science, Nagasaki, Japan,
$^w$National Research Nuclear University, Moscow, Russia,
$^x$Northwestern University, Evanston, IL 60208, USA,
$^y$University of Notre Dame, Notre Dame, IN 46556, USA,
$^z$Universidad de Oviedo, E-33007 Oviedo, Spain,
$^{aa}$CNRS-IN2P3, Paris, F-75205 France,
$^{bb}$Texas Tech University, Lubbock, TX 79609, USA,
$^{cc}$Universidad Tecnica Federico Santa Maria, 110v Valparaiso, Chile,
$^{dd}$Yarmouk University, Irbid 211-63, Jordan,
}}
\noaffiliation